\documentclass[a4paper,11pt]{article}
\pdfoutput=1
\usepackage{style}
\usepackage{subfigure}
\usepackage{relsize}
\usepackage{graphicx}
\usepackage{caption}

\def  \bcen   {\begin{center}}
\def  \ecen   {\end{center}}
\def  \beq    {\begin{equation}}
\def  \eeq    {\end{equation}}
\def  \beqa   {\begin{eqnarray}}
\def  \eeqa   {\end{eqnarray}}
\def  \nn     {\nonumber }
\def\bea{\begin{eqnarray}}
\def\eea{\end{eqnarray}}

\title{Consistency of perturbed Tribimaximal, Bimaximal and Democratic mixing
 with Neutrino mixing data}

\author{Sumit K. Garg\textsuperscript{1}}
\affiliation{\textsuperscript{1}\textit{Department of Physics, CMR University Bengaluru 562149, India}}

\emailAdd{sumit.k@cmr.edu.in}

\abstract{We scrutinize corrections to tribimaximal (TBM), bimaximal (BM) and democratic (DC) mixing matrices for explaining recent
global fit neutrino mixing data. These corrections are parameterized in terms of small orthogonal 
rotations (R) with corresponding modified PMNS matrices of the forms \big($R_{ij}^l\cdot U,~U\cdot R_{ij}^r,~U \cdot R_{ij}^r \cdot R_{kl}^r,~R_{ij}^l \cdot R_{kl}^l \cdot U$\big )  
where $R_{ij}^{l, r}$ is rotation in ij sector and  U is any one of these special matrices. We showed that for perturbative   
schemes dictated by single rotation, only \big($ R_{12}^l\cdot U_{BM},~R_{13}^l\cdot U_{BM},~U_{TBM}\cdot R_{13}^r$ \big ) can fit the mixing data at 
$3\sigma$ level. However for $R_{ij}^l\cdot R_{kl}^l\cdot U$ type rotations, only \big ($R_{23}^l\cdot R_{13}^l \cdot U_{DC} $\big ) is successful to fit 
all neutrino mixing angles within $1\sigma$ range. For $U\cdot R_{ij}^r\cdot R_{kl}^r$ perturbative scheme, only \big($U_{BM} \cdot 
R_{12}^r\cdot R_{13}^r$,~$U_{DC} \cdot R_{12}^r\cdot R_{23}^r$,~$U_{TBM} \cdot R_{12}^r\cdot R_{13}^r$\big ) are consistent at $1\sigma$ level. The remaining
double rotation cases are either excluded at 3$\sigma$ level or successful in producing 
mixing angles only at $2\sigma-3\sigma$ level. We also updated our previous analysis on PMNS matrices of the form \big($R_{ij}\cdot U \cdot R_{kl}$\big ) with recent mixing data. We showed that the results modifies 
substantially with fitting accuracy level decreases for all of  the permitted cases except \big($R_{12}\cdot U_{BM}\cdot R_{13}$ 
, $R_{23}\cdot U_{TBM}\cdot R_{13}$ and $R_{13}\cdot U_{TBM} \cdot R_{13}$\big ) in this rotation scheme.} 


\begin{document}
\maketitle
\section{Introduction}
The Standard Model(SM) which governs the dynamics
of interactions between fundamental particles contain massless neutrinos. But now its well established fact from the  solar, atmospheric and reactor neutrino 
experiments~\cite{Dayabay, T2K, Doublechooz, Minos, RENO} 
that neutrino switches flavor while traveling because of their tiny mass and flavor mixing. Thus its a clear evidence of Physics Beyond Standard Model. The understanding
of extremely small neutrino mass($\sim meV$) and magnitude of mixing among different neutrino flavors are much deliberated issues in Particle Physics. 
The smallness of neutrino mass scale is well explained by seesaw mechanism~\cite{seesaw} which links it to a new physical scale in Nature. However neutrino mixing indicate interesting 
pattern in which two mixing angles of a  three flavor scenario seems to be maximal while third one remains small. Different mixing schemes like 
tribimaximal(TBM)~\cite{scott}, bimaximal(BM)~\cite{BM} and democratic mixing(DC)~\cite{DM} were explored to explain experimental neutrino mixing data. All these mixing scenarios have 
same predictions for the reactor mixing angle viz $\theta_{13}=0$. The atmospheric mixing angle, $\theta_{23}=45^{\circ}$ for BM and TBM while for DC it takes the value 54.7$^{\circ}$.
The solar mixing angle is maximal (i.e. $45^{\circ}$) for BM and DC while takes the value of 35.3$^{\circ}$ for TBM mixing. As far origin of these special structures is 
concerned they can easily come from various discrete symmetries like $A_4$~\cite{A4}, $S_4$~\cite{S4} etc. These  issues are extensively discussed in literature.  

However, reactor based Daya Bay~\cite{Dayabay} experiment in China presented first conclusive result of non zero $\theta_{13}$ using $\bar{\nu}_e$ beam with
corresponding significance of $5.2\sigma$. The  value of 1-3 mixing angle consistent with data at 90\% CL is reported to be  in the range 
$\sin^2 2\theta_{13}= 0.092\pm 0.016(stat)\pm 0.05(syst)$. Previously long baseline neutrino oscillation T2K experiment~\cite{T2K} 
observed the events corresponding to $\nu_{\mu}\rightarrow \nu_e$ transition 
which is also consistent with non zero $\theta_{13}$ in a three flavor scenario. The  value of 1-3 mixing angle consistent with data 
at 90\% CL is reported to be  in the range $ 5^\circ(5.8^\circ) < \theta_{13} < 16^\circ(17.8^\circ)$ for Normal (Inverted) neutrino mass hierarchy. These results goes
in well agreement with other oscillation experiments like Double Chooz~\cite{Doublechooz}, Minos~\cite{Minos} and RENO~\cite{RENO}. Moreover it is
evident from recent global fit~\cite{Gonzalez-Garcia:2014bfa, Capozzi:2017ipn} for 
neutrino masses and mixing angles (given in Table~\ref{singRot1}) that these mixing scenarios can't be taken at their face value and thus should be
investigated for possible perturbations. This issue has been taken up many times~\cite{largeth13, models} in literature. In particular, ref.~\cite{dcpertbs} and 
ref.~\cite{tbmpertbs} discussed the perturbations that are parametrized in terms of mixing angles to DC and TBM mixing respectively for 
obtaining large $\theta_{13}$. Here we looked into possible perturbations which are parameterized by one and two rotation matrices and are of the forms
 \big($R_{ij}\cdot U, U\cdot R_{ij}, R_{ij}\cdot R_{kl}\cdot U,  U\cdot R_{ij}\cdot R_{kl}$\big). These corrections show strong correlations among neutrino mixing angles which are 
weakened with full perturbation matrix. Since the form of PMNS matrix is given by $U_{PMNS} = U_l^{\dagger} U_\nu$ so these modifications may originate from 
charged lepton and neutrino sector respectively. Unlike previous studies~\cite{dcpertbs, tbmpertbs, Chaoetal}, we used $\chi^2$ approach~\cite{skgetal} to investigate the situation of mixing 
angle fitting in parameter space and thus capture essential information about corresponding correction parameters. Also to study correlations among neutrino
mixing angles, we allowed to vary all mixing angles in their permissible limits instead of fixing one of them at a particular value to study correlation between remaining two mixing angles. 
This in our view show a full picture and thus we present our results in terms of  2 dimensional scatter plots instead of line plots~\cite{Chaoetal}.  Moreover we confined 
ourselves to small rotation limit which in turn justify them to pronounce as perturbative corrections. We also 
looked into new cases apart from the ones that were presented in~\cite{Chaoetal} and thus presenting a complete analysis on these rotation schemes. Here
we also take opportunity to update our previous analysis~\cite{skgetal} on \big($R_{ij}\cdot U\cdot R_{kl}$\big) PMNS 
matrix for new mixing data. We find ($\chi^2_{min}$, Best fit level) for different mixing cases~\cite{skgetal} using
previous mixing data~\cite{Gonzalez-Garcia:2014bfa} and compare  it with the results that are obtained with new mixing data~\cite{Capozzi:2017ipn}. In this study, we works in
CP conserving limit and thus all phases are taken to be zero. However we would like to emphasize that CP violation can have a deep impact on these studies. But including CP violation for
the characterized perturbations is a elaborative task which in our opinion needs separate attention. So we leave the discussion
of corrections coming from non zero CP phase for future discussions. These results can help in understanding the structure of corrections that these well known mixing scenarios require in order to be consistent with
neutrino mixing data. Thus this study may turn out to be useful in restricting vast number of possible models which offers different corrections to these mixing schemes in neutrino model building physics. It would 
also be interesting to inspect the origin of these perturbations in a model dependent framework. However the investigation of all 
such issues is left for future consideration.

The main outline of the paper is as follows. In section 2, we will give general discussion about our work. In sections 3-7, we will present our 
numerical results for various possible perturbation cases. Finally in section 8, we will give the summary and conclusions of our study.


\section{General Setup}

A $3\times 3$ Unitary matrix can be parametrized by 3 mixing angles and 6 phases. However 5 phases can be 
moved away leaving behind only 1 physical phase. Thus light neutrino mixing is given in standard form as~\cite{upmns}
\begin{eqnarray}
U &=& \left( \begin{array}{ccc} 1 & 0 & 0 \\ 0 & c^{}_{23}  & s^{}_{23} \\
0 & -s^{}_{23} & c^{}_{23} \end{array} \right)
\left( \begin{array}{ccc} c^{}_{13} & 0 & s^{}_{13} e^{-i\delta} \\ 0 & 1 & 0 \\
- s^{}_{13} e^{i\delta} & 0 & c^{}_{13} \end{array} \right)
\left( \begin{array}{ccc} c^{}_{12} & s^{}_{12}
 & 0 \\ -s^{}_{12}  & c^{}_{12}  & 0 \\
0 & 0 & 1 \end{array} \right) \left( \begin{array}{ccc} 1 & 0
 & 0 \\ 0  & e^{i\rho}  & 0 \\
0 & 0 & e^{i\sigma} \end{array} \right)
,\label{standpara}
\end{eqnarray}
where $c_{ij}\equiv \cos\theta_{ij}$, $s_{ij}\equiv \sin\theta_{ij}$ and $\delta$ is the Dirac CP violating phase.
Here two additional phases $\rho$ and $\sigma$ are only pertinent if neutrinos turn out to be Majorana particles. The Majorana phases however 
do not affect the neutrino oscillations and thus are not relevant here. In this study, we will consider the case where all the CP violating phases
i.e. $\delta, \rho, \sigma$ are zero. The effects  of CP violation will be discussed somewhere else.

The form of mixing matrix for the three mixing scenarios under consideration is given as follows:
\begin{eqnarray} \nn
U_{\rm TBM} = \left ( \begin{array}{rrr}
\sqrt{2\over 3}&\sqrt{1\over 3}&0\\
-\sqrt{{1\over 6}}&\sqrt{{1\over 3}}&\sqrt{1\over 2}\\
-\sqrt{{1\over 6}}&\sqrt{{1\over 3}}&-\sqrt{{1\over 2}}
\end{array}
\right )\; , \hspace{0.2cm}
U_{\rm BM}=\left(
\begin{array}{rrr}
\sqrt{1\over 2 } & \sqrt{1\over 2 } & 0 \\
-{1 \over 2 }& {1 \over 2 } & \sqrt{1\over 2 } \\
{1 \over 2 } &  -{ 1 \over 2 } &\sqrt{1\over 2 }
\end{array}\right) \; , \hspace{0.2cm}
U_{\rm DC} = \left ( \begin{array}{rrr}
\sqrt{\frac{1}{2}}&\sqrt{\frac{1}{2}}&0\\
\sqrt{\frac{1}{6}}&-\sqrt{\frac{1}{6}}&-\sqrt{\frac{2}{3}}\\
-\sqrt{\frac{1}{3}}&\sqrt{\frac{1}{3}}&-\sqrt{\frac{1}{3}}
\end{array}
\right )\;. \label{vtri}
\end{eqnarray}

\begin{table} 
\begin{center}
\begin{tabular}{lccc}

\hline

\hline

Mixing Angle & Bimaximal(BM) & Democratic(DC) & Tribimaximal(TBM)  \\

\hline

$\theta_{23}^\circ$ & 45  & 54.7 & 45\\

\hline

$\theta_{12}^\circ$ & 45  & 45 & 35.3\\

\hline

$\theta_{13}^\circ$  & 0  & 0 & 0   \\

\hline

\label{mixingangles} 

\end{tabular}
\end{center}
\vspace{-1cm}
\caption{\it{Mixing angle values from special matrices.  All angles are in degrees($\theta^\circ$).}}

\label{mixingangles} 
\end{table}

\begin{center}
\begin{tabular}{|l|l||l|l|l|}
\hline
  \multicolumn{1}{|l|}{Neutrino Mixing} & \multicolumn{4}{|l|}{Normal Hierarchy(NH)}\\
\cline{2-5} 
Angle($\theta^\circ$) & Best fit  & $1\sigma$ & $2\sigma$  & $3\sigma$   \\ 
 \hline\hline
$\theta_{12}^\circ$  & $33.02$& 32.01-34.08& $30.98-35.30$  & $30.0-36.51$   \\
\hline
$\theta_{23}^\circ$  &$40.68$ & 39.81-41.9 &$38.93-43.28$ &$38.11-51.64$  \\
\hline
$\theta_{13}^\circ$  &$8.43$ & 8.29-8.56 &$8.10-8.74$&$7.92-8.91$  \\
 \hline
\end{tabular}
\captionof{table}{\it{Three-flavor oscillation neutrino mixing angles from fit to global data~\cite{Capozzi:2017ipn}. All mixing angles are in degrees($\theta^\circ$).}}
\label{singRot1}
\end{center}


The resulting value of mixing angles from above matrices is given in Table~\ref{mixingangles}.
All of them predict vanishing value of $\theta_{13}=0^{\circ}$. 
The atmospheric mixing angle($\theta_{23}$) is maximal in TBM and BM scenarios
while it takes the larger value of 54.7 $^{\circ}$ for DC mixing. The value of solar mixing angle ($\theta_{12}$) is maximal in BM and DC scenarios while its
 value is $35.3^{\circ}$ for TBM case. However these mixing angles are in conflict with recent experimental observations which provide best fit values at 
 $\theta_{13} \sim 8^{\circ}$, $\theta_{12} \sim 33^{\circ}$ and $\theta_{23} \sim 41^{\circ}$.
Thus these well studied structures need corrections~\cite{chrgdleptcrrs, neutrinocrrs, bthsctrcrrs} in order to be consistent with current neutrino mixing data.\\
From theoretical point of view, neutrino mixing matrix U comes from the mismatch between diagonalization of charged lepton and neutrino mass matrix and  is given as \\
\beq \nn
U = U_l^{\dagger} U_\nu
\eeq
where U$_l$ and U$_\nu$ are the unitary matrices that diagonalizes the charged lepton (M$_l$) and neutrino mass matrix (M$_\nu$).
Thus corrections that can modify the original predictions of above special structures can originate from following
sources.
\begin{flushleft}
(i) Charged lepton sector i.e. $U_{PMNS}^{'} = U_{crr}^l \cdot U_{PMNS}$\\
(ii) Neutrino sector i.e. $U_{PMNS}^{'} = U_{PMNS}\cdot U_{crr}^{\nu}$\\
(iii) Both sectors i.e. $U_{PMNS}^{'} = U_{crr}^l \cdot U_{PMNS}\cdot U_{crr}^{\nu}$
\end{flushleft}

Here $U_{crr}^{l}$ and $U_{crr}^{\nu}$ are the correction matrices corresponding to charged lepton and neutrino sector respectively. For
CP conserving case, $U_{crr}^{l}$ and $U_{crr}^{\nu}$ are the real orthogonal matrices which can be parametrized in terms of 3 
mixing angles as discussed previously. 

In this study, we are considering first two cases i.e. possible modifications that comes from charged lepton and neutrino sector. The 3rd case is already investigated~\cite{skgetal} 
in detail before. We tested different possibilities for fitting
neutrino mixing data that are governed by either one or two mixing angles of perturbation matrix. These cases exhibit strong correlations among mixing angles which are weakened or hidden in situation where all angles
are non zero. Here  corresponding PMNS matrix is of the forms  $R_X \cdot U$, $U\cdot R_X$, $ R_X \cdot R_Y \cdot U$ and $U \cdot R_X \cdot R_Y$ where $R_X$ 
and $R_Y$ denote generic perturbation matrices and U is any of these special matrix. 
The perturbation matrices $R_X$ and $R_Y$ can be expressed in terms of mixing matrices
as $R_X(R_Y)= \{ R^{}_{23}, R^{}_{13}, R^{}_{12} \}$ in general, where
$R^{}_{23}$, $R^{}_{13}$ and $R^{}_{12}$ represent the rotations in 23, 13 and 12 sector respectively and are given by
\begin{eqnarray} \nn
&&R^{}_{12} = \left (\begin{array}{ccc}
\cos \alpha & \sin \alpha &0\\
-\sin \alpha &\cos \alpha &0\\
0&0&1
\end{array}
\right )\;,  R^{}_{23} = \left (\begin{array}{ccc}
1&0&0\\
0&\cos \beta  &\sin \beta \\
0&-\sin \beta & \cos \beta
\end{array}\right )\;, R^{}_{13} = \left ( \begin{array}{ccc}
\cos \gamma &0&\sin \gamma  \\
0&1&0\\
-\sin \gamma  &0& \cos \gamma
\end{array}
\right )\; \label{vb}
\end{eqnarray}
where $\alpha$, $\beta$, $\gamma$ denote rotation angles. The corresponding PMNS
matrix for single rotation case is given by:
\begin{eqnarray}
&& U^{TBML}_{\rm \alpha\beta}= R_{\alpha\beta}^{l}  \cdot U_{TBM}^{}  \; , \label{p1}\\
&& U^{BML}_{\rm \alpha\beta}= R_{\alpha\beta}^{l}  \cdot U_{BM}^{}  \; ,\label{p2} \\
&& U^{DCL}_{\rm \alpha\beta}= R_{\alpha\beta}^{l}  \cdot U_{DC}^{} \; ,\label{p3} 
\end{eqnarray}
and 
\begin{eqnarray}
&& U^{TBMR}_{\rm \alpha\beta}=  U_{TBM}^{} \cdot R_{\alpha\beta}^{r}  \; , \label{p1}\\
&& U^{BMR}_{\rm \alpha\beta}= U_{BM}^{} \cdot R_{\alpha\beta}^{r}  \; ,\label{p2} \\
&& U^{DCR}_{\rm \alpha\beta}= U_{DC}^{} \cdot R_{\alpha\beta}^{r}  \; ,\label{p3} 
\end{eqnarray}
where  $(\alpha\beta) =(12), (13),
(23)$ respectively. The corresponding PMNS matrix for two rotation matrix thus becomes:
\begin{eqnarray}
&& U^{TBML}_{\rm ijkl}= R_{ij}^{l} \cdot R_{kl}^{l} \cdot U_{TBM}^{}  \; , \label{p1}\\
&& U^{BML}_{\rm ijkl}= R_{ij}^{l} \cdot R_{kl}^{l} \cdot U_{BM}^{}  \; ,\label{p2} \\
&& U^{DCL}_{\rm ijkl}= R_{ij}^{l} \cdot R_{kl}^{l}  \cdot U_{DC}^{} \; ,\label{p3} 
\end{eqnarray}
and 
\begin{eqnarray}
&& U^{TBMR}_{\rm ijkl}=  U_{TBM}^{} \cdot R_{ij}^{r} \cdot R_{kl}^{r} \; , \label{p1}\\
&& U^{BMR}_{\rm ijkl}= U_{BM}^{} \cdot R_{ij}^{r} \cdot  R_{kl}^{r} \; ,\label{p2} \\
&& U^{DCR}_{\rm ijkl}= U_{DC}^{} \cdot R_{ij}^{r} \cdot R_{kl}^{r} \; ,\label{p3} 
\end{eqnarray}
where  $(ij), (kl) =(12), (13),
(23)$ respectively. Now to get neutrino mixing angles the above matrices
are compared with the standard PMNS matrix.

To investigate numerically the effect of these perturbations we define
a $\chi^2$ function  which is a measure of deviation from the central value of mixing angles:
\begin{equation}
   \chi^2 = \mathlarger{\mathlarger{‎‎\sum}}_{i=1}^{3‎} \{ \frac{\theta_i(P)-\theta_i}{\delta \theta_i} \}^2
\end{equation}
with $\theta_i(P)$ are the theoretical value of mixing angles which are functions  of perturbation parameters ($\alpha,\beta, \gamma$).
$\theta_i$ are the experimental value of neutrino mixing angles with corresponding $1\sigma$ uncertainty $\delta \theta_i$. The corresponding
value of $\chi^2$ for different mixing schemes is given in Table~\ref{chisq}.


\begin{center}
\begin{tabular}{|l||l|l|}
\hline
 \multicolumn{3}{|l|}{$\chi^{2}_{NH}$} \\
\cline{1-3} 
 Bimaximal(BM) & Democratic(DC) & Tribimaximal(TBM)  \\ 
 \hline\hline
 $1112.0$ & $1274.9$ & $965.5$ \\
 \hline
\end{tabular} 
\captionof{table}{\it{$\chi^2$ value for Normal Hierarchy(NH) with different unperturbed mixing schemes.}}
\label{chisq}
\end{center}

We investigated the role of various possible perturbations for lowering the value of $\chi^2$ from its original prediction
in different mixing schemes. A good numerical fit should produce low $\chi^2$ value in parameter space.

\section{Numerical Results}
In this section,  we will discuss numerical findings  of our study for Normal Hierarchy. We
studied the role of various perturbative schemes in producing large $\theta_{13}$~\cite{largeth13} and fitting other mixing angles. 
In Figs.~\ref{fig12L.1}-\ref{fig2313R.3}, we present the results in terms of $\chi^2$ over perturbation parameters  and
$\theta_{13}$ over $\theta_{23}-\theta_{12}$ plane  for various studied cases.
The numerical value of correction parameters $\alpha$, $\beta$ and $\gamma$ are confined in the
range [-0.5, 0.5] so as to keep them in perturbative limits. We enforced the condition $\chi^2 < \chi^2_{i}$ (i = TBM, DC and BM) for 
plotting data points. In double rotation plots of $\chi^2$ vs perturbation parameters ($\theta_1, \theta_2$) red, blue and light green color 
regions corresponds to $\chi^2$ value in the interval $[0, 3]$,  $[3, 10]$  and  $ > 10$ respectively. 
In figures of neutrino mixing angles, light green band corresponds to $1\sigma$ and full color band to $3\sigma$ values
of $\theta_{13}$. Also `$\times$' refers to the case which is unable to fit mixing angles even at $3\sigma$ level while `-' denotes 
the situation where $\theta_{13}=0$.

In order to show the correlations between left and right figures we marked the $\chi^2 < 3, [3, 10]$ regions 
in mixing angle plots with different color codings. The white region corresponds to $ 3 < \chi^2 < 10 $ while yellow region belongs to 
$\chi^2 < 3$. Horizontal and vertical dashed black, dashed pink and thick black lines corresponds to $1\sigma$, $2\sigma$ and $3\sigma$ ranges 
of the other two mixing angles. Now we will take up the case of various possible forms of rotations one by one:

\section{Rotations-$R_{ij}^l.U$}

Here we first consider the perturbations for which the form of modified PMNS matrix is given by $U_{PMNS} = R_{ij}^l.U$. We will 
investigate their role in fitting the neutrino mixing angles. 

\subsection{12 Rotation}

This case corresponds to rotation in 12 sector of  these special matrices. 
Since for small rotation $\sin\theta \approx \theta$ and $\cos\theta \approx 1-\theta^2$, so the neutrino mixing angles
truncated at order O ($\theta^2$) for this rotation are given by

\beqa
 \sin\theta_{13} &\approx&  |\alpha U_{23}  |,\\
 \sin\theta_{23} &\approx& |\frac{ (\alpha^2-1) U_{23}}{\cos\theta_{13}}|,\\
 \sin\theta_{12} &\approx& |\frac{U_{12} + \alpha U_{22} -\alpha^2 U_{12} }{\cos\theta_{13}}|.
\eeqa

\begin{figure}[!t]\centering
\begin{tabular}{c c} 
\hspace{-5mm}
\includegraphics[angle=0,width=80mm]{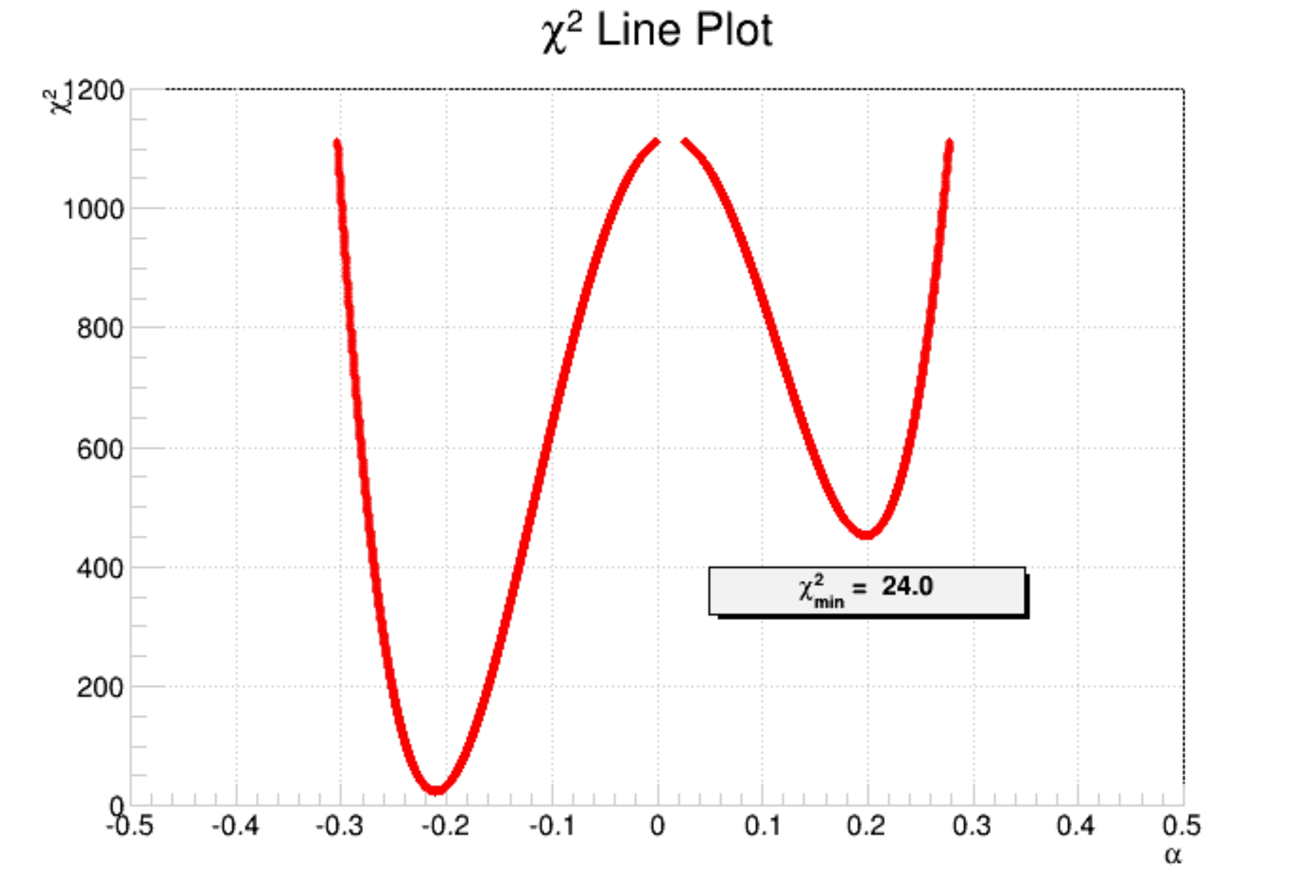} &
\includegraphics[angle=0,width=80mm]{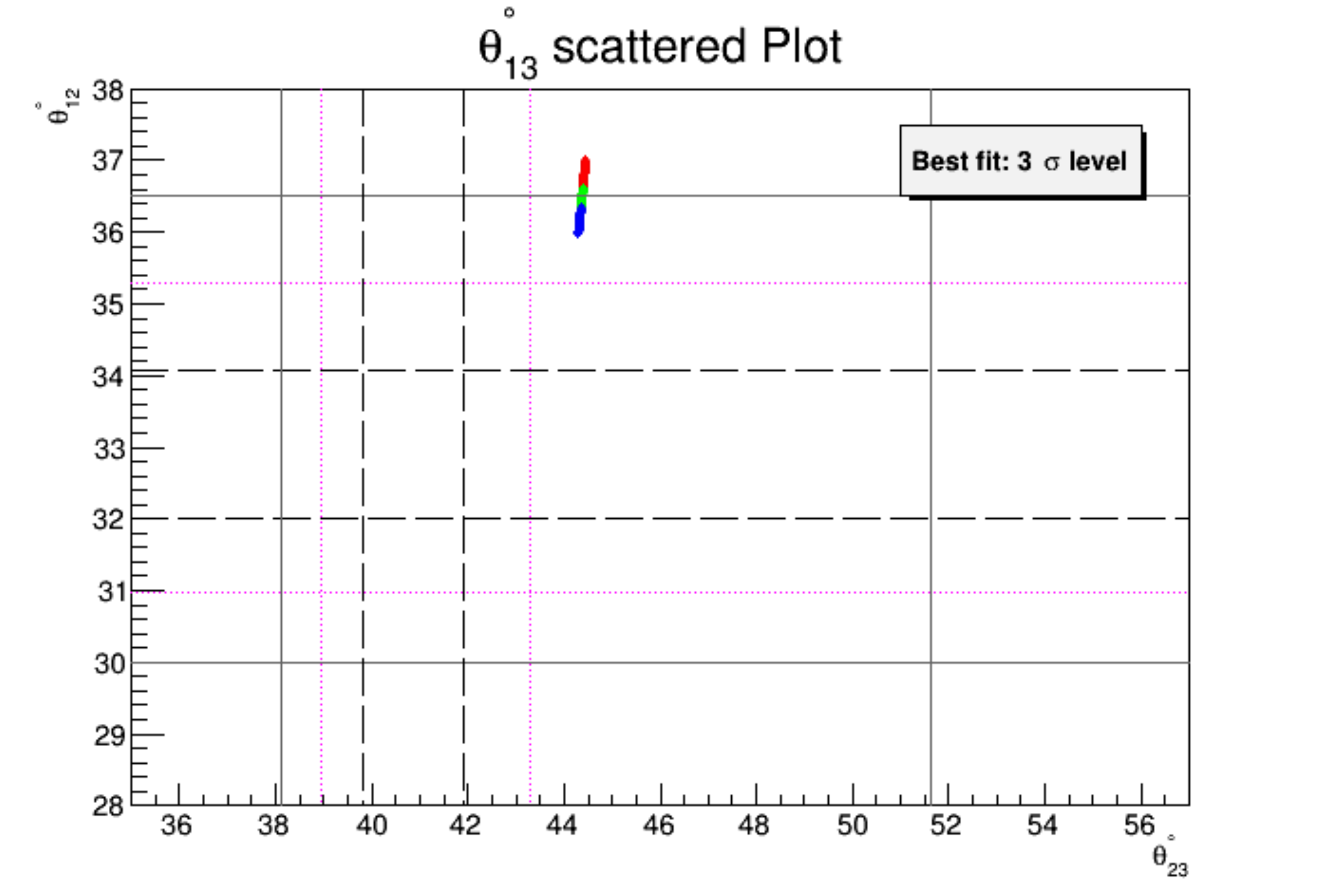}\\
\end{tabular}
\caption{\it{Line plot of $\chi^2$ (left fig.) vs $\alpha$ (in radians) and scattered plot of $\theta_{13}$ (right fig.) 
over $\theta_{23}-\theta_{12}$ (in degrees) plane for $U^{BML}_{12}$ rotation scheme. The discontinuity in left curve corresponds 
to region where $\chi^2_{perturbed} > \chi^2_{original}$. The information about other color coding and various horizontal, vertical 
lines in right fig. is given in text.}}
\label{fig12L.1}
\end{figure}

\begin{figure}[!t]\centering
\begin{tabular}{c c} 
\hspace{-5mm}
\includegraphics[angle=0,width=80mm]{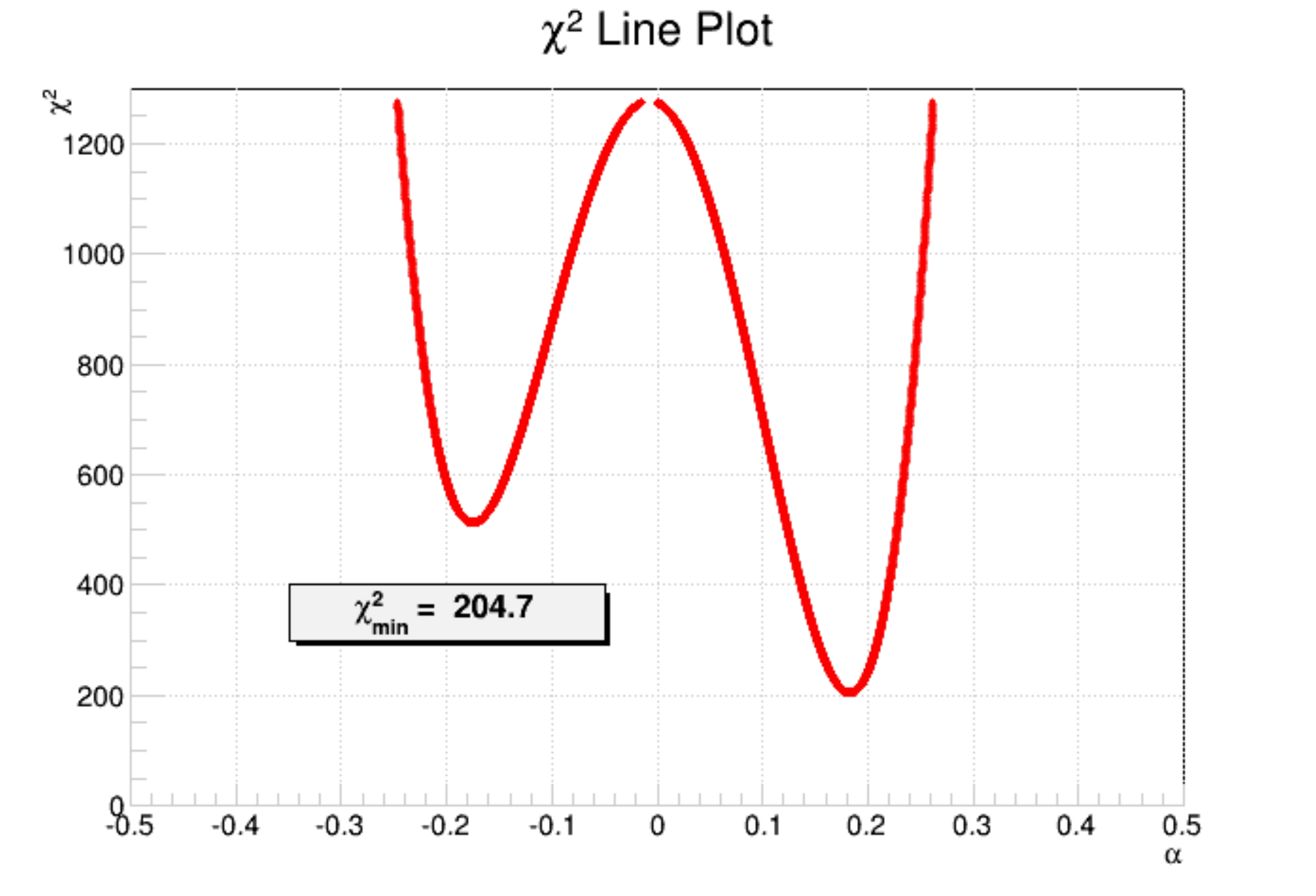} &
\includegraphics[angle=0,width=80mm]{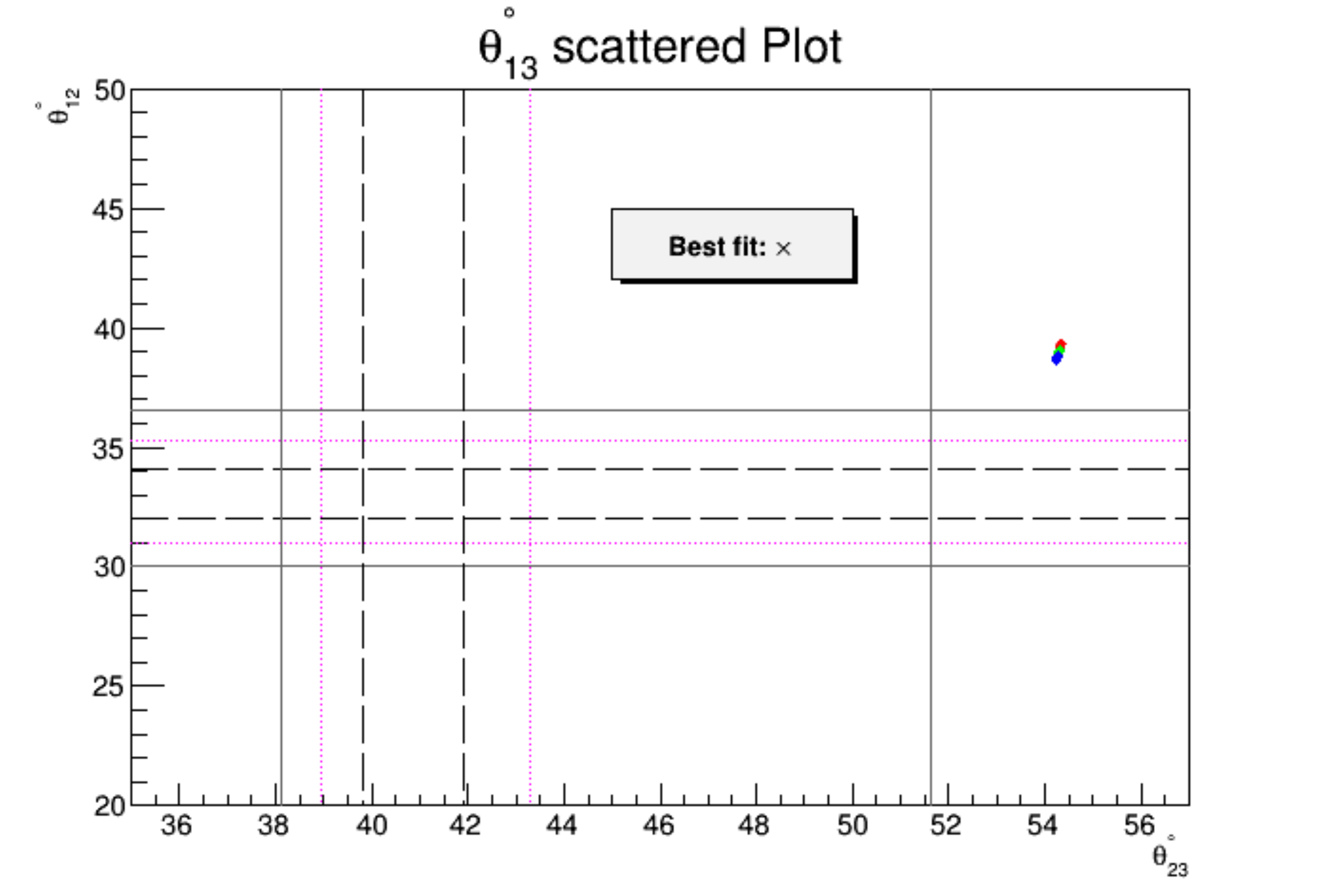} \\
\end{tabular}
\caption{\it{Line plot of $\chi^2$ (left fig.) vs $\alpha$ (in radians) and scattered plot of $\theta_{13}$ (right fig.) 
over $\theta_{23}-\theta_{12}$ (in degrees) plane for $U^{DCL}_{12}$ rotation scheme.}}
\label{fig12L.2}
\end{figure}

\begin{figure}[!t]\centering
\begin{tabular}{c c} 
\hspace{-5mm}
\includegraphics[angle=0,width=80mm]{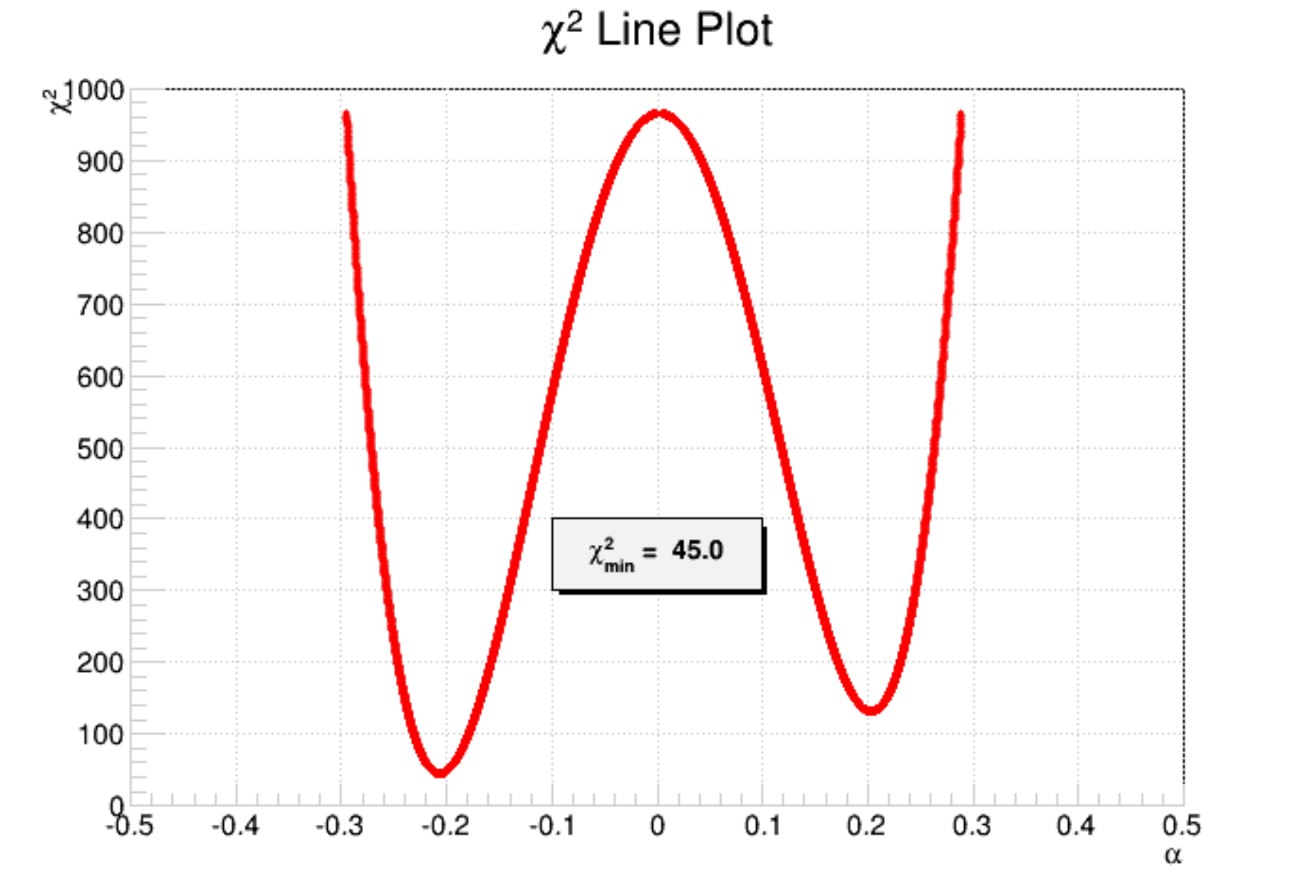} &
\includegraphics[angle=0,width=80mm]{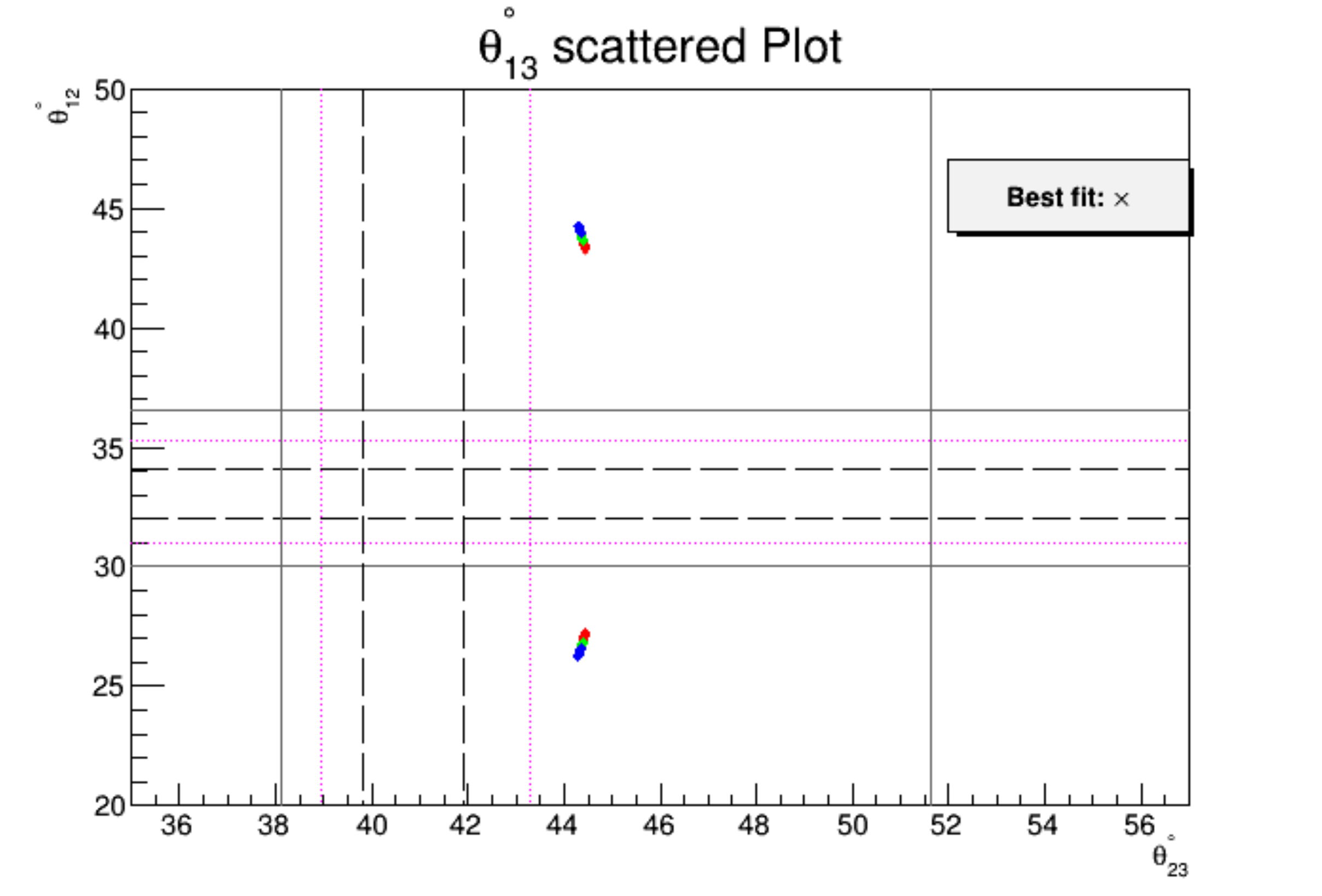} \\
\end{tabular}
\caption{\it{Line plot of $\chi^2$ (left fig.) vs $\alpha$ (in radians) and scattered plot of $\theta_{13}$ (right fig.) 
over $\theta_{23}-\theta_{12}$ (in degrees) plane for $U^{TBML}_{12}$ rotation scheme.}}
\label{fig12L.3}
\end{figure}

Figs.~\ref{fig12L.1}-\ref{fig12L.3} show the numerical results corresponding to TBM, BM 
and DC case. The salient features of this mixing is given by:\\
{\bf{(i)}} The atmospheric mixing angle($\theta_{23}$) receive corrections only of $O(\theta^2)$ so its perturbed value
remain close to its original prediction. \\
{\bf{(ii)}} In BM, for fitting $\theta_{13}$ in its $3\sigma$ range constraints
$|\alpha| \in [0.196, 0.220]$ which in turn fixes $\theta_{12} \in [35.97^\circ, 36.99^\circ]$ for negative and 
$\theta_{12} \in [53.0^\circ, 54.02^\circ]$ for positive $\alpha$ region. The atmospheric angle($\theta_{23}$) remain confined to very narrow
range $\theta_{23} \in [44.29^\circ, 44.44^\circ]$ for mentioned value of $\alpha$. Thus negative $\alpha$ is preferred and 
it can fit all mixing angles at $3\sigma$ level.\\
{\bf{(iii)}} However same window 
of $\alpha$ for TBM (since $U_{23}$ is same in both cases) gives low(high) value of $\theta_{12}$ for its negative(positive) range.
These achieved values are well away from their allowed $3\sigma$ boundary. In DC case, $|\alpha| \in [0.169, 0.190]$ for setting $\theta_{13}$ in 
its $3\sigma$ range which in turn fixes  $\theta_{12} $ away from its allowed $3\sigma$ range. \\
{\bf{(iv)}} The minimum value of $\chi^2 \sim 24.0$, $204.7$ and $45.0$ for BM, DC and TBM case respectively.\\
{\bf{(v)}} Thus perturbed BM can be  consistent at $3\sigma$ level while other two cases are not favored.

\subsection{13 Rotation}

This case pertains to rotation in 13 sector of  these special matrices. 
The neutrino mixing angles for small perturbation parameter $\gamma$ are given by

\beqa
 \sin\theta_{13} &\approx&  |\gamma U_{33}  |,\\
 \sin\theta_{23} &\approx& |\frac{ U_{23}}{\cos\theta_{13}}|,\\
 \sin\theta_{12} &\approx& |\frac{U_{12} + \gamma U_{32} - \gamma^2 U_{12} }{\cos\theta_{13}}|.
\eeqa

\begin{figure}[!t]\centering
\begin{tabular}{c c} 
\hspace{-5mm}
\includegraphics[angle=0,width=80mm]{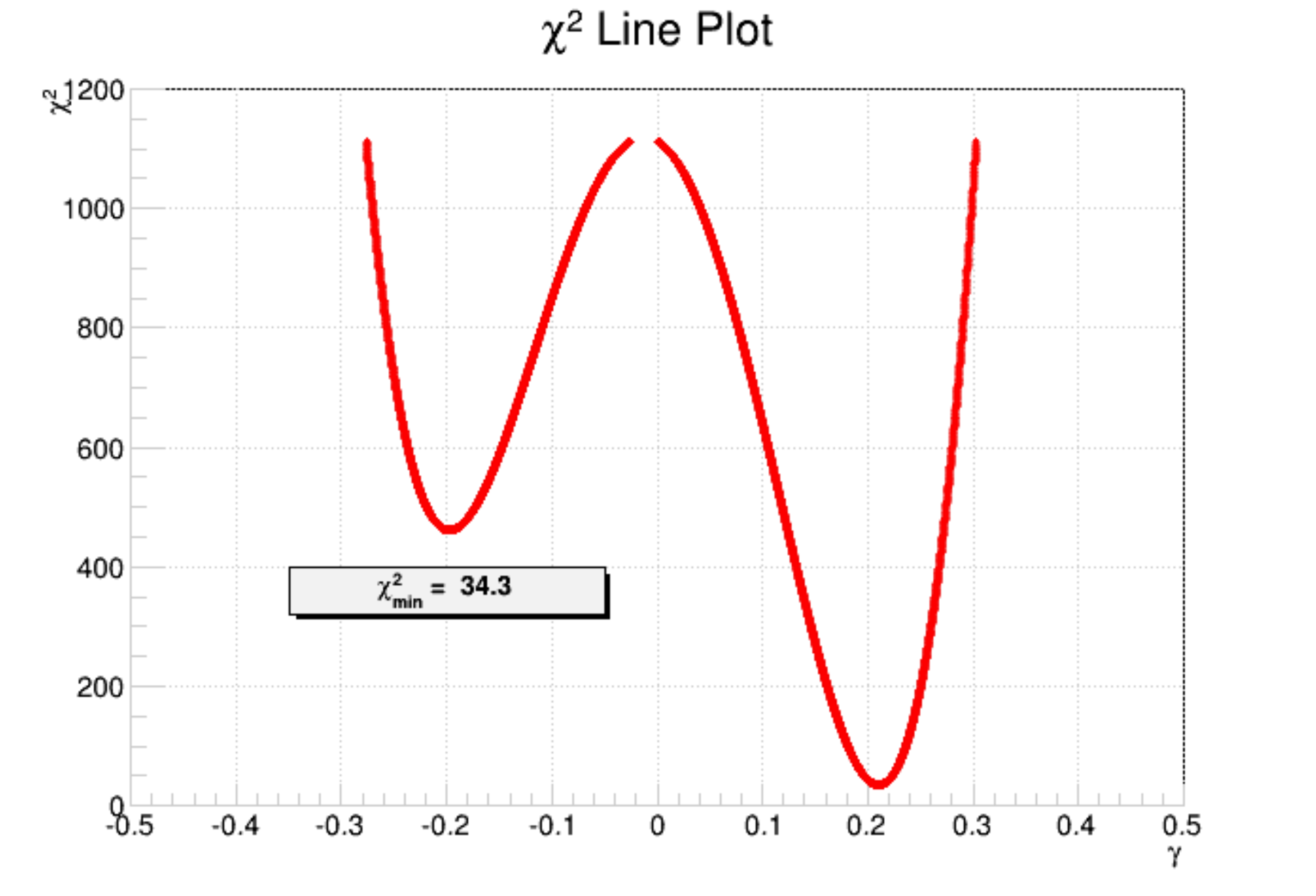} &
\includegraphics[angle=0,width=80mm]{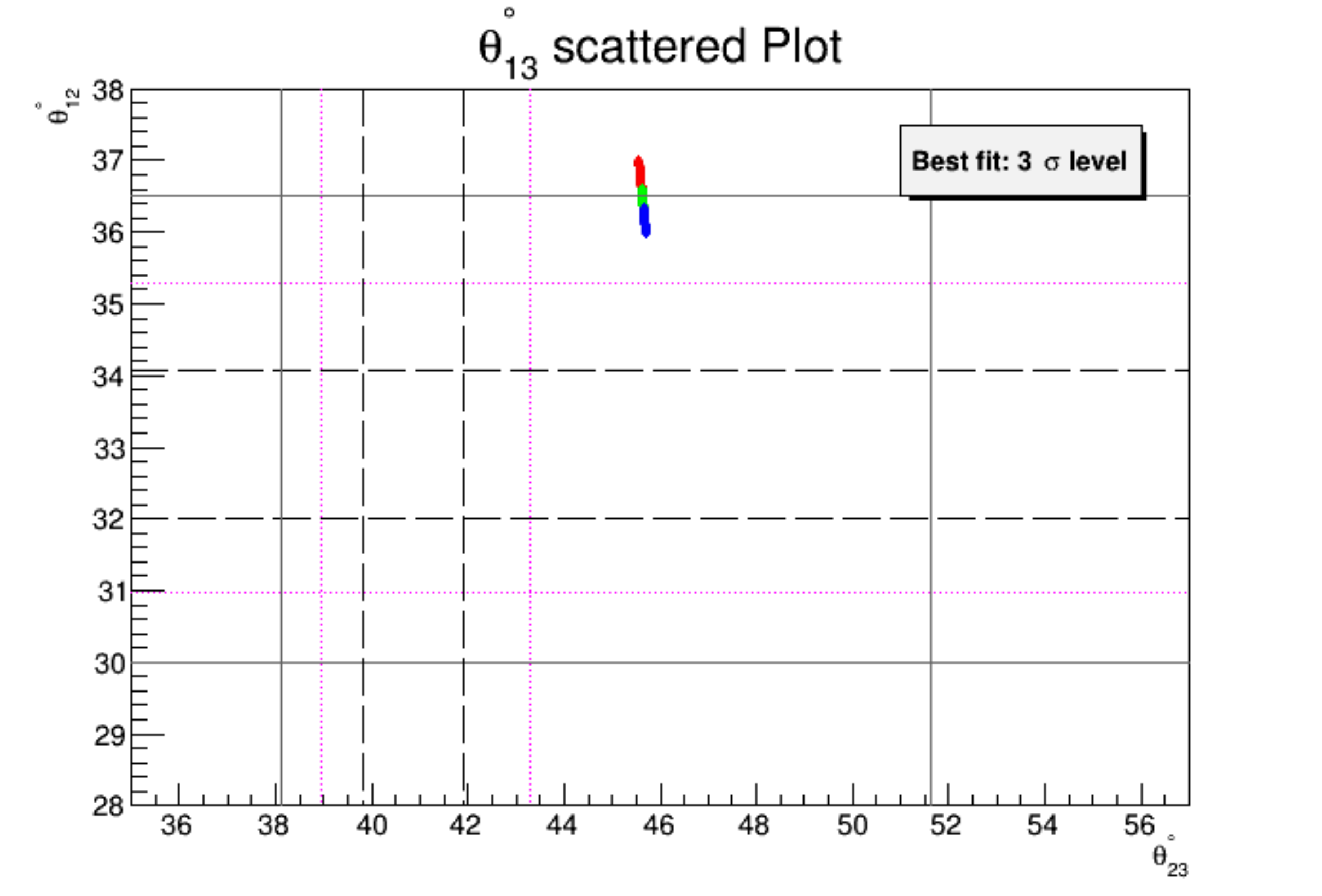}\\
\end{tabular}
\caption{\it{Line plot of $\chi^2$ (left fig.) vs $\gamma$ (in radians) and scattered plot of $\theta_{13}$ (right fig.) 
over $\theta_{23}-\theta_{12}$ (in degrees) plane for $U^{BML}_{13}$ rotation scheme.}}
\label{fig13L.1}
\end{figure}

\begin{figure}[!t]\centering
\begin{tabular}{c c} 
\hspace{-5mm}
\includegraphics[angle=0,width=80mm]{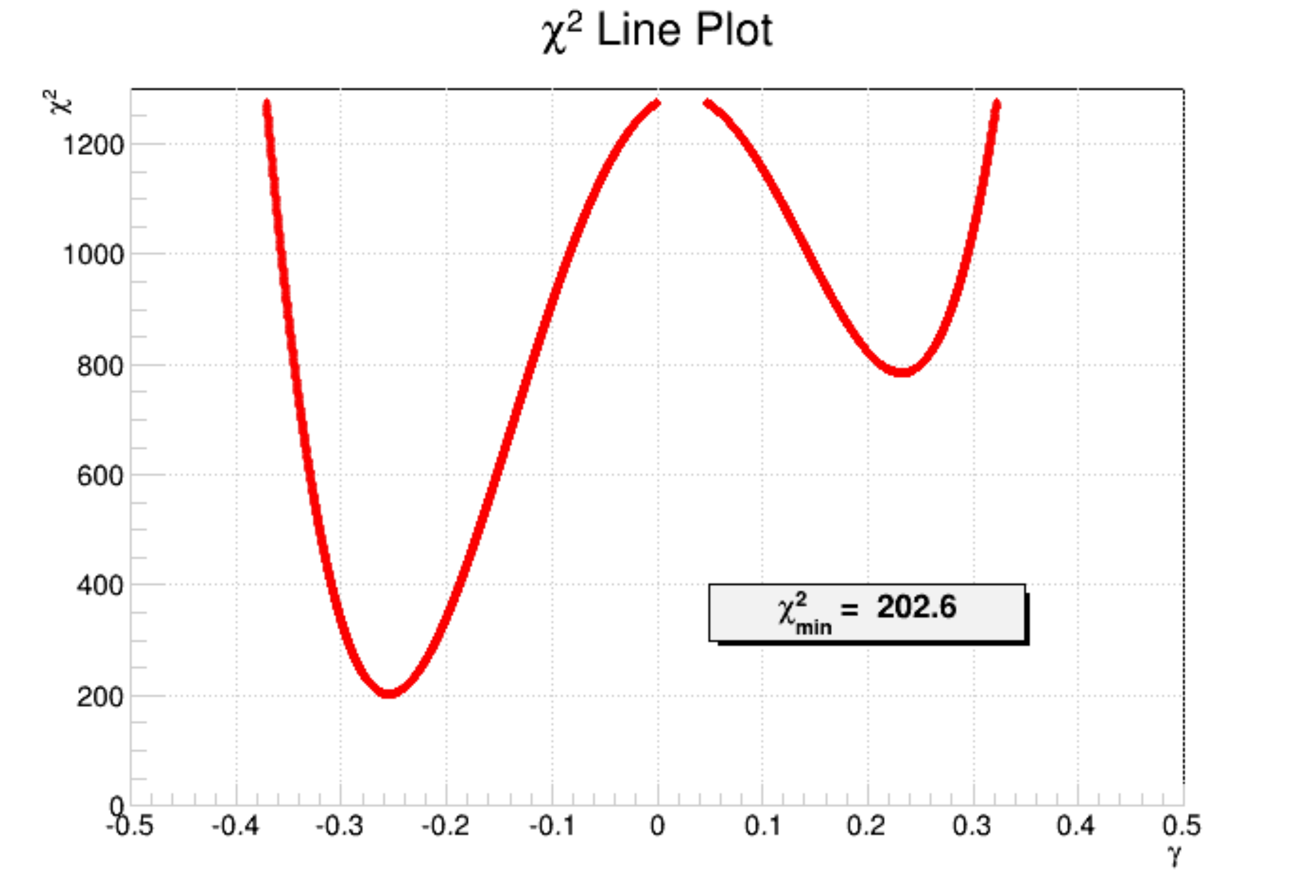} &
\includegraphics[angle=0,width=80mm]{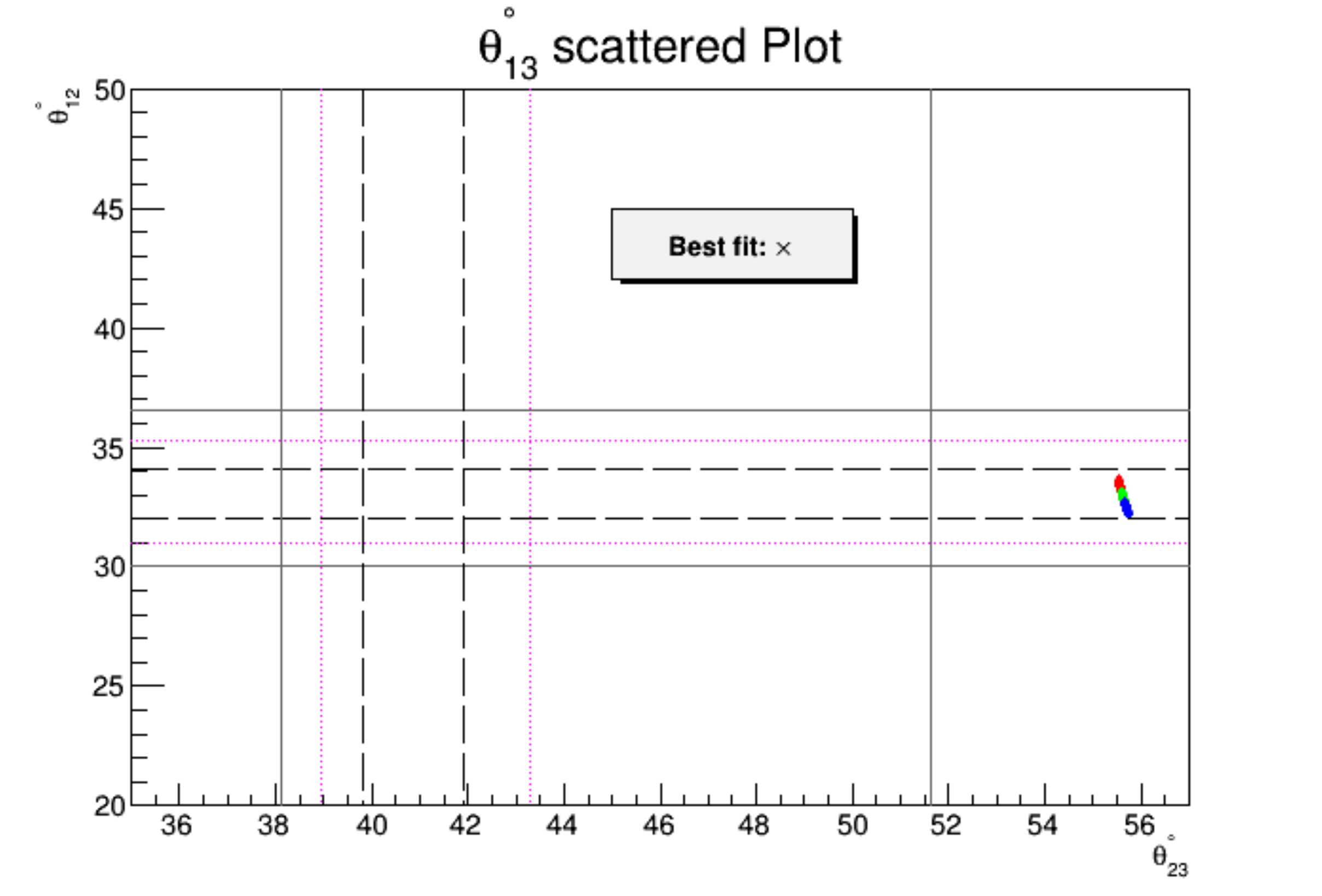}\\
\end{tabular}
\caption{\it{Line plot of $\chi^2$ (left fig.) vs $\gamma$ (in radians) and scattered plot of $\theta_{13}$ (right fig.) 
over $\theta_{23}-\theta_{12}$ (in degrees) plane for $U^{DCL}_{13}$ rotation scheme.}}
\label{fig13L.2}
\end{figure}

\begin{figure}[!t]\centering
\begin{tabular}{c c} 
\hspace{-5mm}
\includegraphics[angle=0,width=80mm]{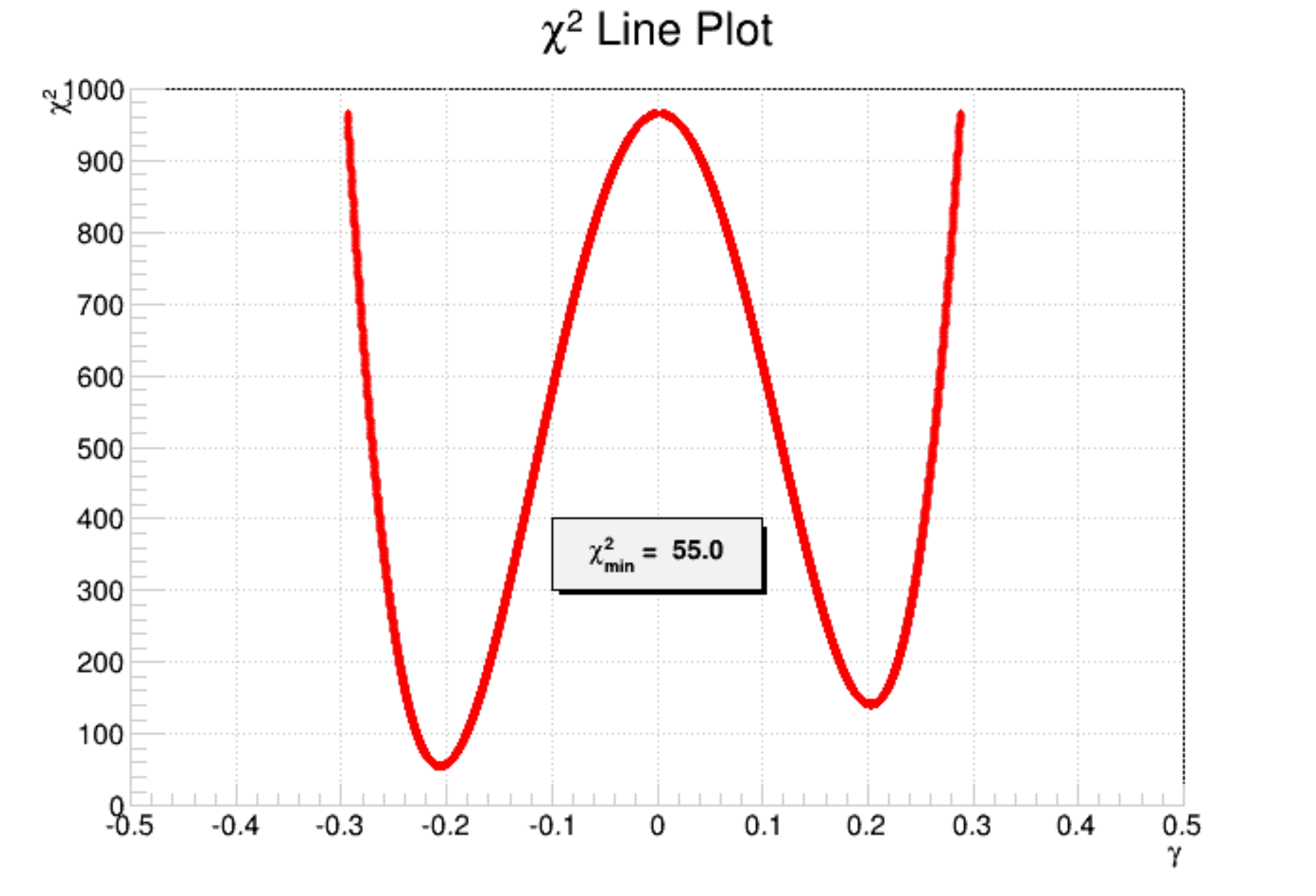} &
\includegraphics[angle=0,width=80mm]{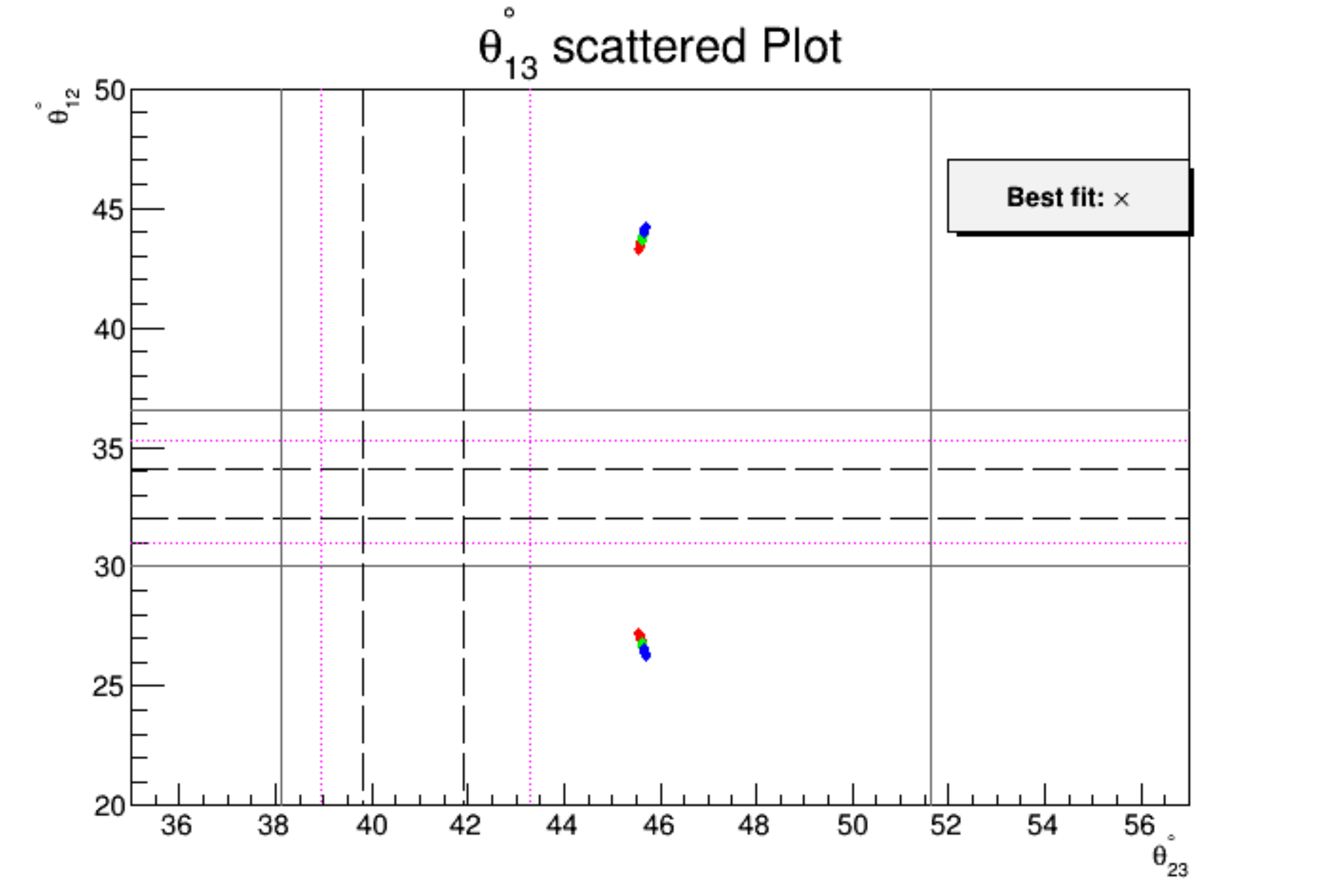}\\
\end{tabular}
\caption{\it{Line plot of $\chi^2$ (left fig.) vs $\gamma$ (in radians) and scattered plot of $\theta_{13}$ (right fig.) 
over $\theta_{23}-\theta_{12}$ (in degrees) plane for $U^{TBML}_{13}$ rotation scheme.}}
\label{fig13L.3}
\end{figure}

\begin{figure}[!t]\centering
\begin{tabular}{c c c} 
\hspace{-5mm}
\includegraphics[angle=0,width=55mm]{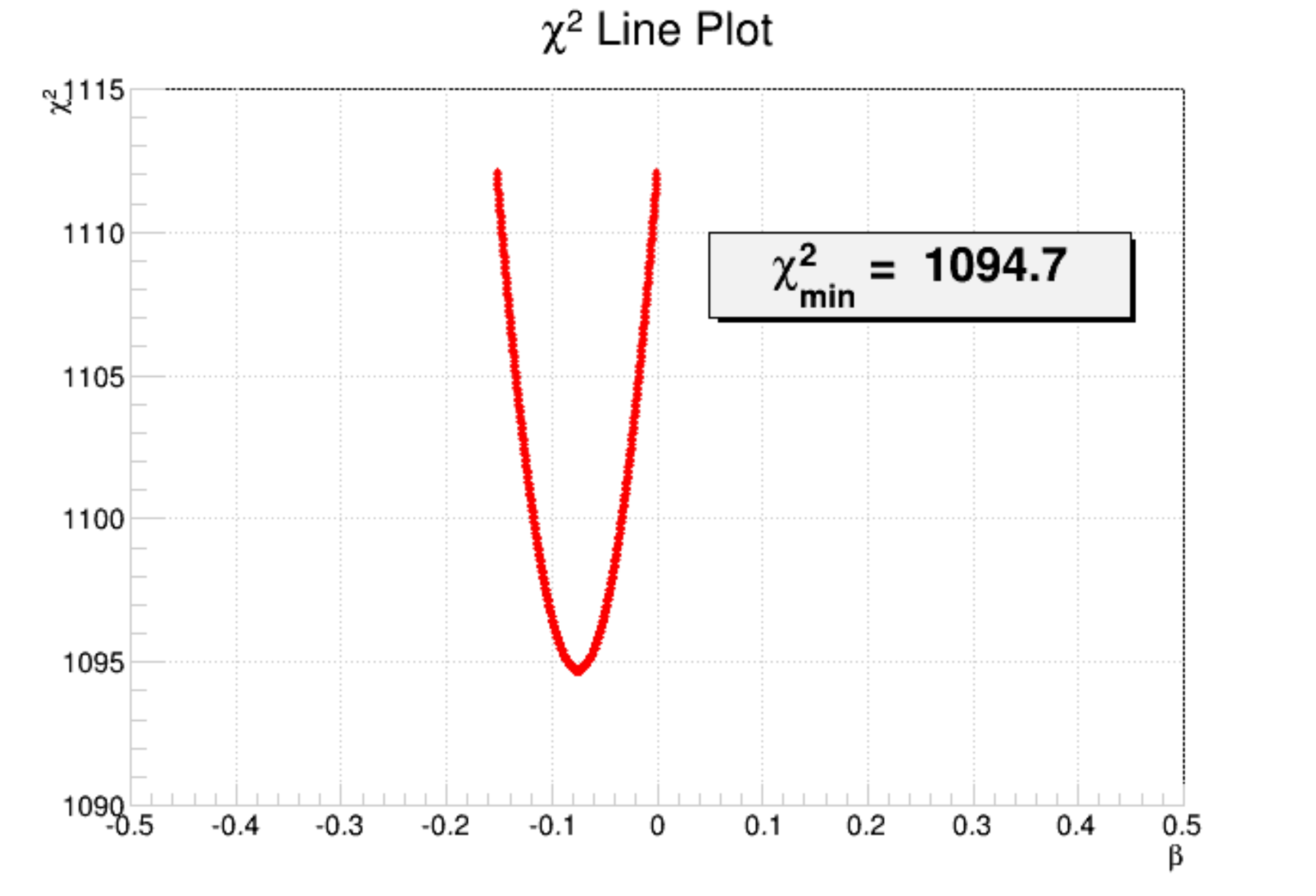} &
\includegraphics[angle=0,width=55mm]{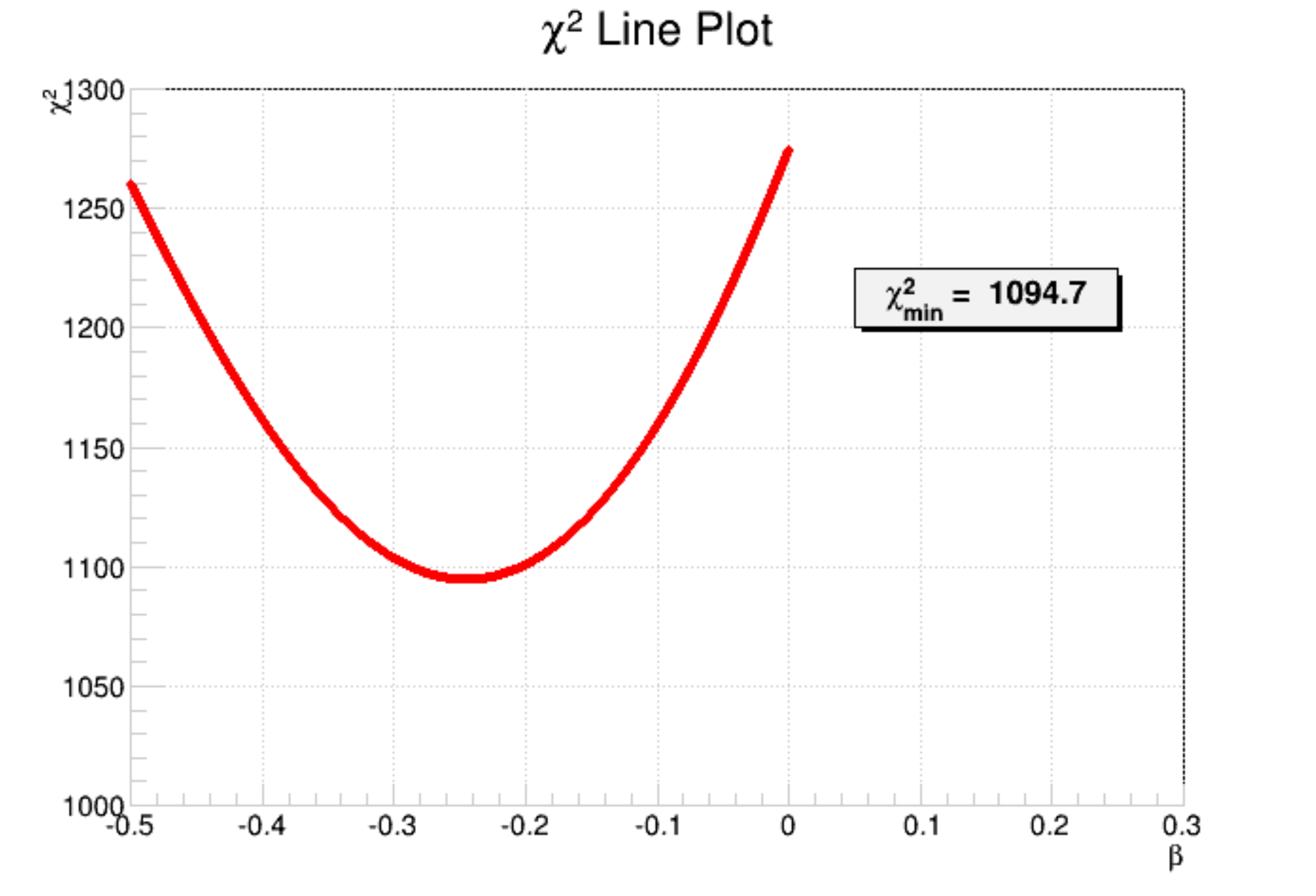}&
\includegraphics[angle=0,width=55mm]{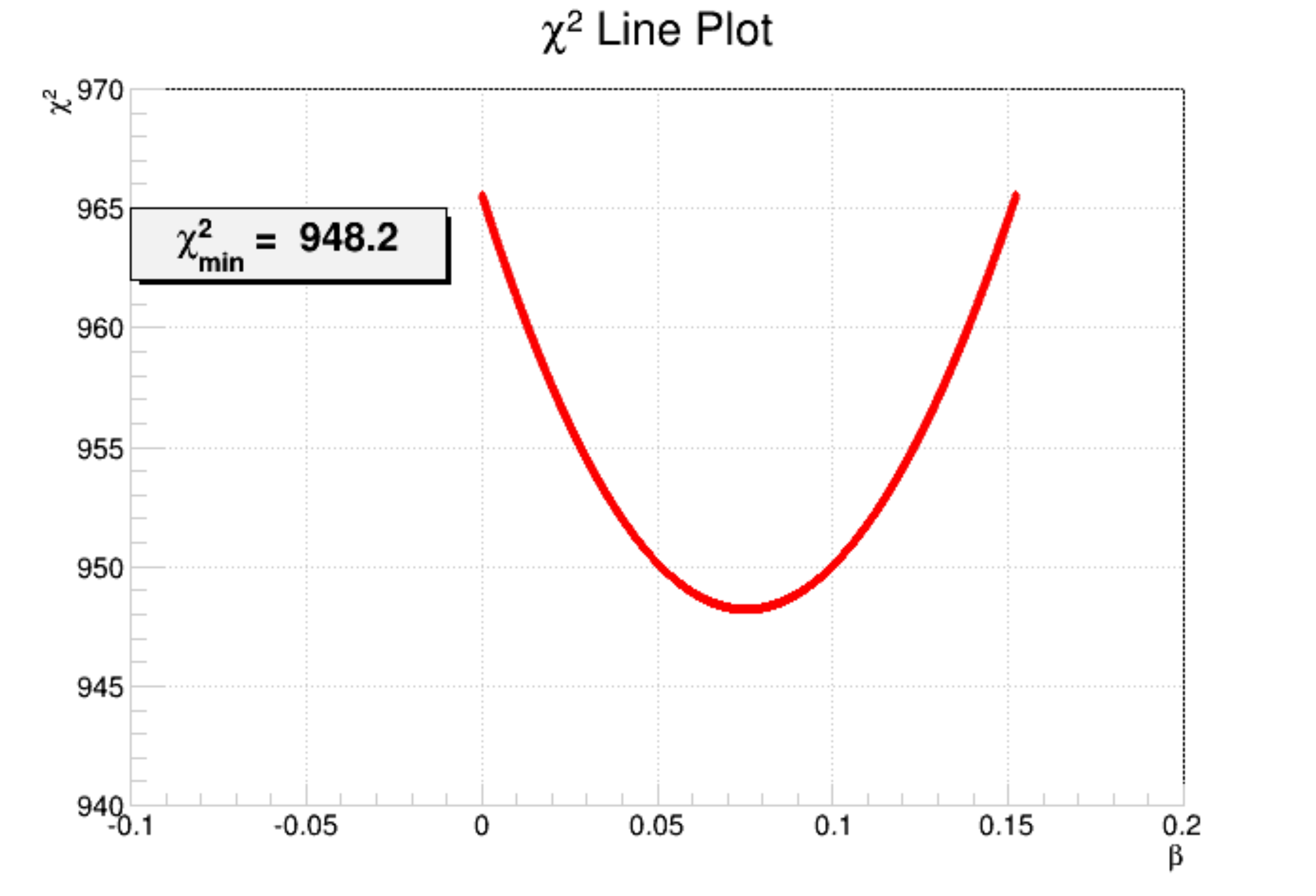}\\
\end{tabular}
\caption{\it{Line plot of $\chi^2$ (left fig.) vs $\gamma$ (in radians) for $U^{BML}_{23}$,  $U^{DCL}_{23}$ and  $U^{TBML}_{23}$ rotation scheme.}}
\label{fig23L.3}
\end{figure}

\begin{figure}[!t]\centering
\begin{tabular}{c c c} 
\hspace{-5mm}
\includegraphics[angle=0,width=55mm]{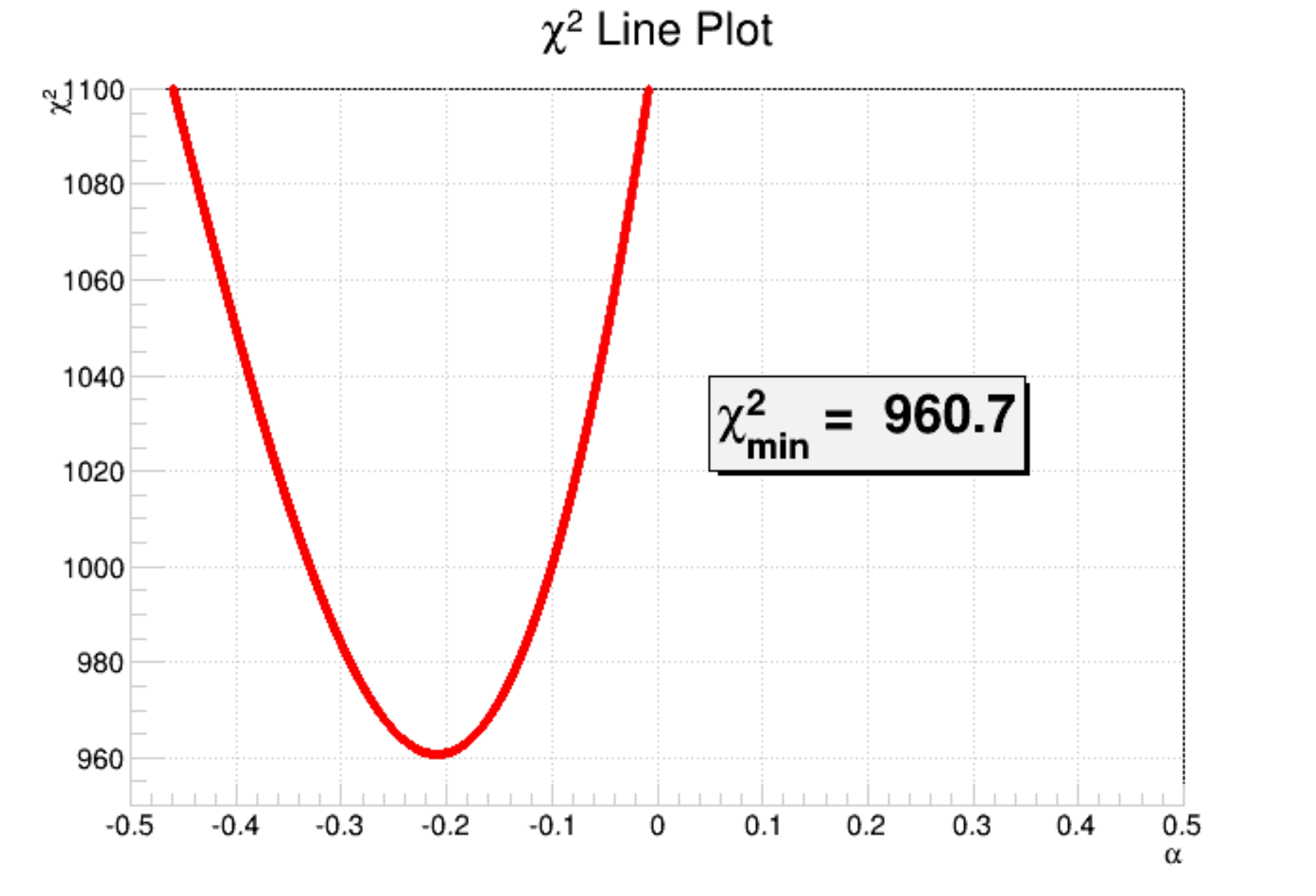} &
\includegraphics[angle=0,width=55mm]{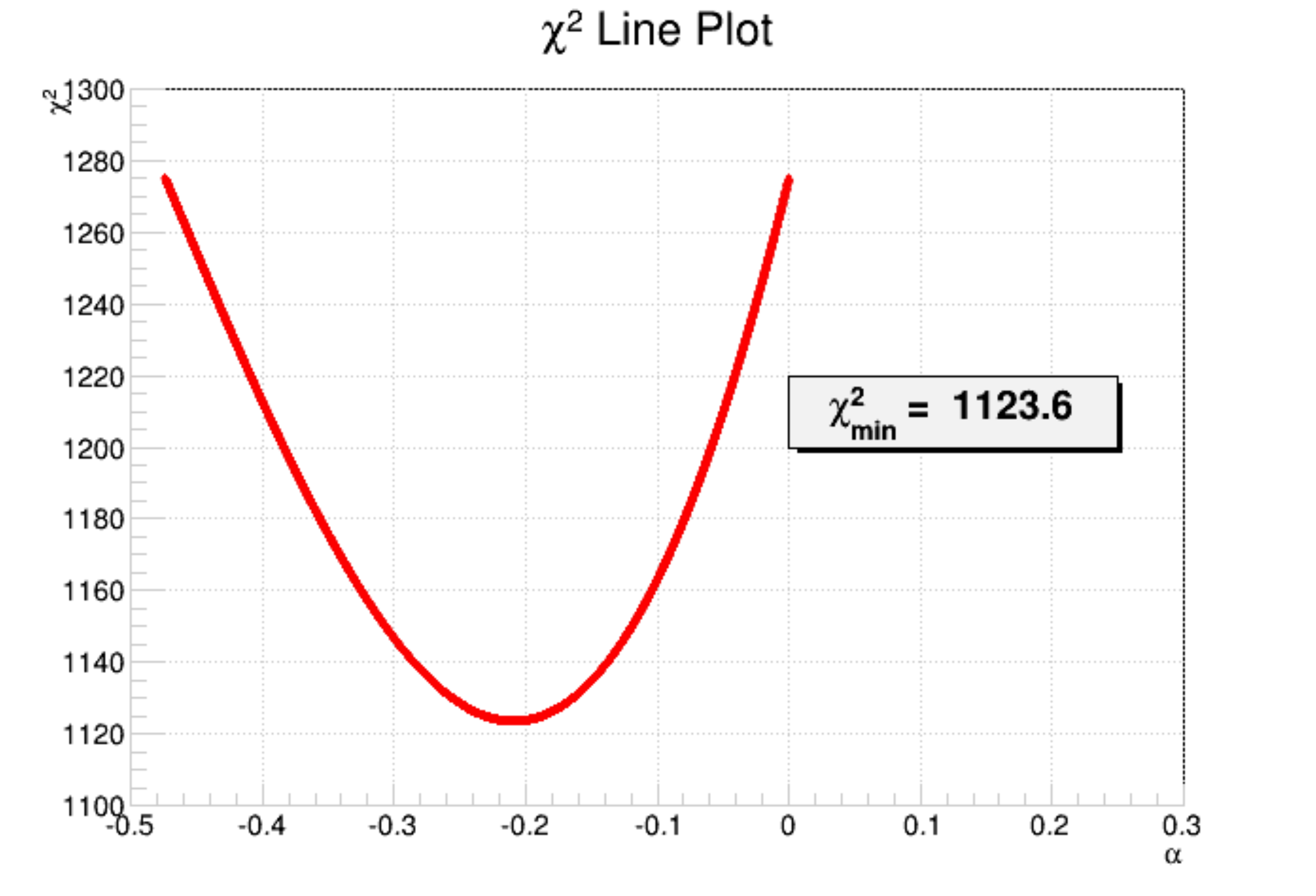}&
\includegraphics[angle=0,width=55mm]{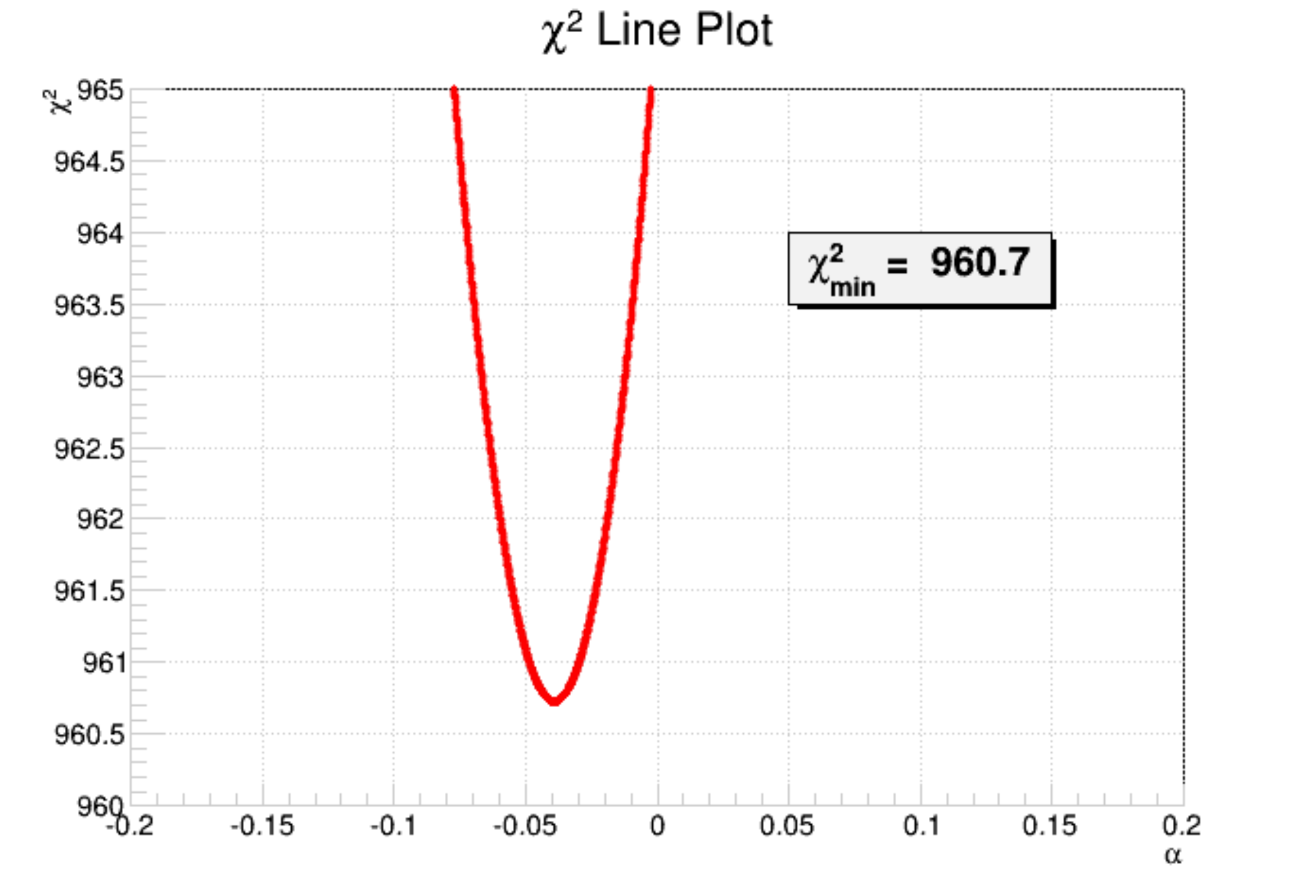}\\
\end{tabular}
\caption{\it{Line plot of $\chi^2$ (left fig.) vs $\gamma$ (in radians) for $U^{BMR}_{12}$,  $U^{DCR}_{12}$ and  $U^{TBMR}_{12}$ rotation scheme.}}
\label{fig12R.3}
\end{figure}

Figs.~\ref{fig13L.1}-\ref{fig13L.3} show the numerical results corresponding to TBM, BM 
and DC case. The main features of this perturbative scheme are given by\\
{\bf{(i)}} Here atmospheric mixing angle($\theta_{23}$) receives only minor corrections through $\sin\theta_{13}$  and thus its value 
doesn't change much from its original prediction. \\ 
{\bf{(ii)}} In BM case, for fitting $\theta_{13}$ in its $3\sigma$ range constraints $|\gamma| \in [0.196, 0.220]$ which in turn fixes 
$\theta_{12} \in [35.97^\circ, 36.99^\circ]$ for positive and $\theta_{12} \in [53.0^\circ, 54.02^\circ]$ for its negative range. The corresponding
$\theta_{23}$ remains close to its original prediction in the range $\theta_{23}\in [45.55^\circ, 45.70^\circ]$.\\
{\bf{(iii)}}However 
same range of $|\gamma|$ for TBM since $|U_{33}|$ 
is same in both cases gives $\theta_{12}$ well away from its allowed $3\sigma$ range. In DC case, $|\gamma| \in [0.241, 0.271]$ for 
$\theta_{13}$ to be in its $3\sigma$ range which fixes $\theta_{12}\in [56.34^\circ, 57.78^\circ]$ for +ve values of $\gamma$ and
$\theta_{12}\in[32.21^\circ, 33.65^\circ]$ for its -ve range. Thus negative $\gamma$ is preferred but this in turn puts $\theta_{23}\in [55.52^\circ, 55.73^\circ]$ which is
well away from its allowed $3\sigma$ range.\\ 
{\bf{(iv)}} The minimum value of $\chi^2 \sim 34.3$, $202.6$ and $55.0$ for BM, DC and TBM case respectively.\\
{\bf{(v)}} Thus BM case can be  consistent at $3\sigma$ level while other two cases are not viable.

\subsection{23 Rotation}

In this case, $\theta_{13}$ doesn't receive any corrections from perturbation matrix (i.e. $\theta_{13}=0$) and
the minimum value of $\chi^2 \sim 1094.7$, $1094.7$ and $948.2$ for BM, DC and TBM case respectively. Thus we left
out this case for any further discussion.

\section{Rotations-$U.R_{ij}^r$}

Here we first consider the perturbations for which modified PMNS matrix is given by $U_{PMNS} = R_{ij}^r.U$. We will 
investigate the role of these perturbations in fitting the neutrino mixing data. 

\subsection{12 Rotation}

This case pertains to the situation where $\theta_{13}=0$ and the minimum value of $\chi^2 \sim 960.7$, $1123.6$ and $960.7$ for BM, 
DC and TBM case respectively. Thus we haven't investigated it any further.

\subsection{13 Rotation}

This case corresponds to rotations in 13 sector of  these special matrices. 
The mixing angles for small perturbation parameter $\gamma$ are given by

\beqa
 \sin\theta_{13} &\approx&  |\gamma U_{11}  |,\\
 \sin\theta_{23} &\approx& |\frac{ U_{23} + \gamma U_{21}- \gamma^2 U_{23}}{\cos\theta_{13}}|,\\
 \sin\theta_{12} &\approx& |\frac{U_{12} }{\cos\theta_{13}}|.
\eeqa

\begin{figure}[!t]\centering
\begin{tabular}{c c} 
\hspace{-5mm}
\includegraphics[angle=0,width=80mm]{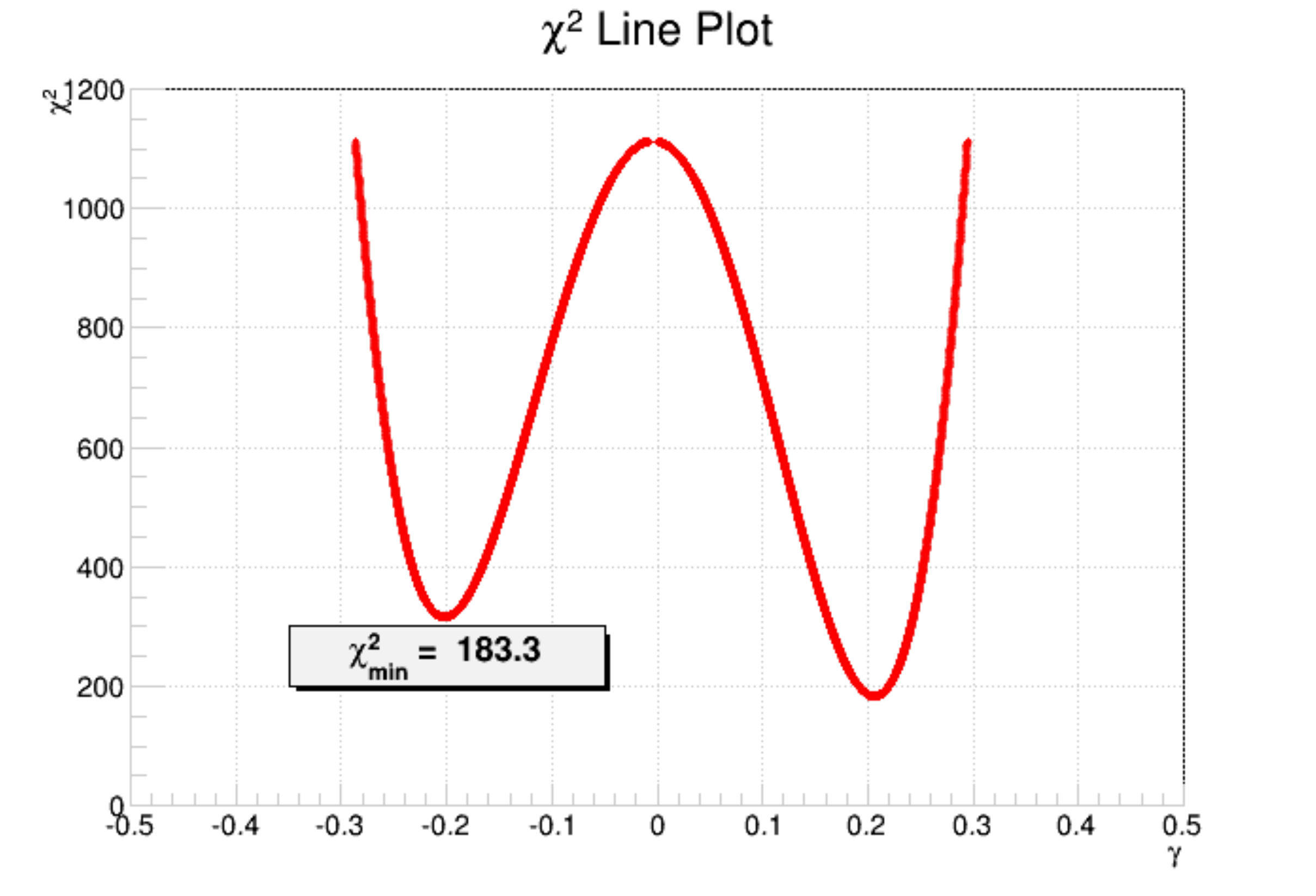} &
\includegraphics[angle=0,width=80mm]{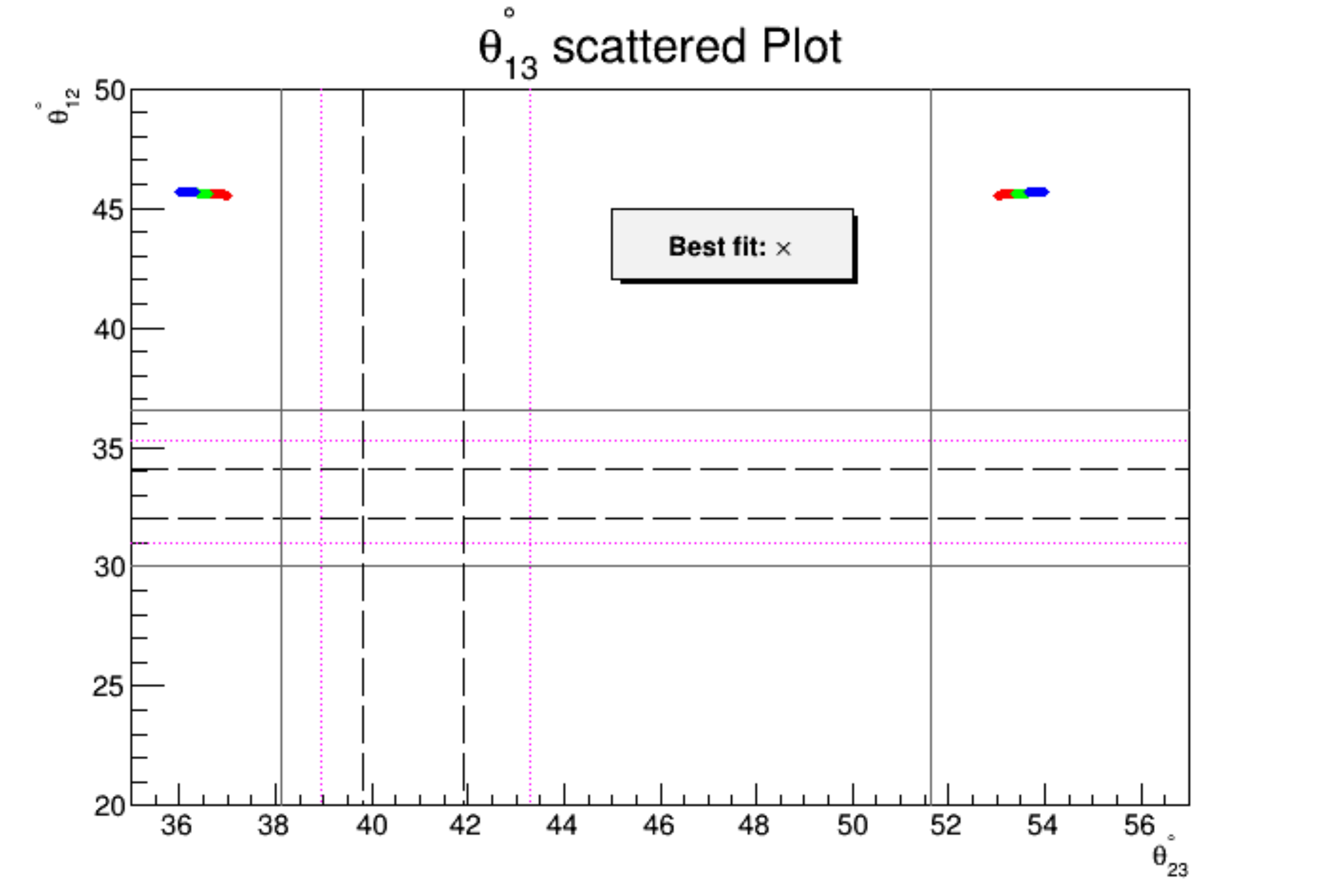}\\
\end{tabular}
\caption{\it{Line plot of $\chi^2$ (left fig.) vs $\gamma$ (in radians) and scattered plot of $\theta_{13}$ (right fig.) 
over $\theta_{23}-\theta_{12}$ (in degrees) plane for $U^{BMR}_{13}$ rotation scheme. }}
\label{fig13R.1}
\end{figure}

\begin{figure}[!t]\centering
\begin{tabular}{c c} 
\hspace{-5mm}
\includegraphics[angle=0,width=80mm]{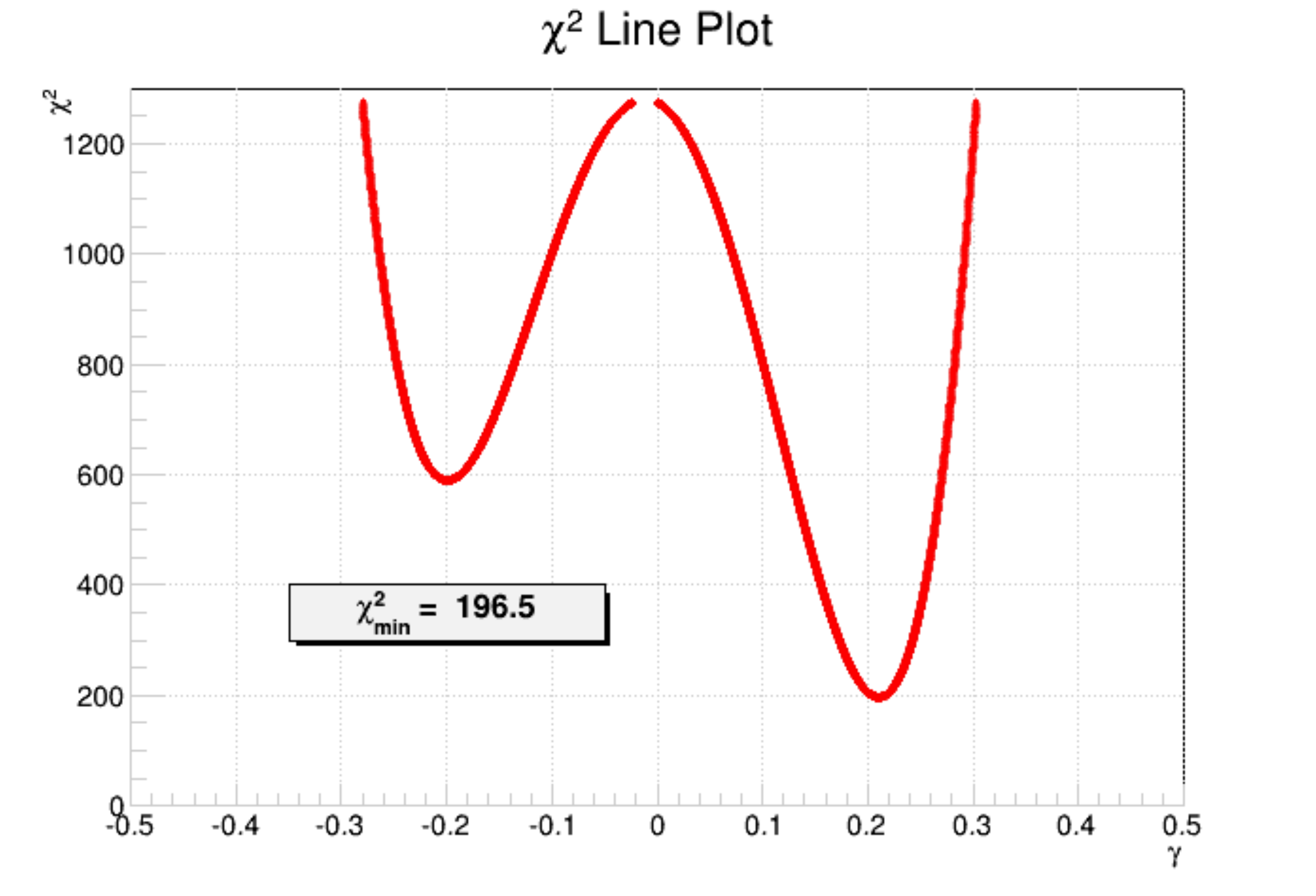} &
\includegraphics[angle=0,width=80mm]{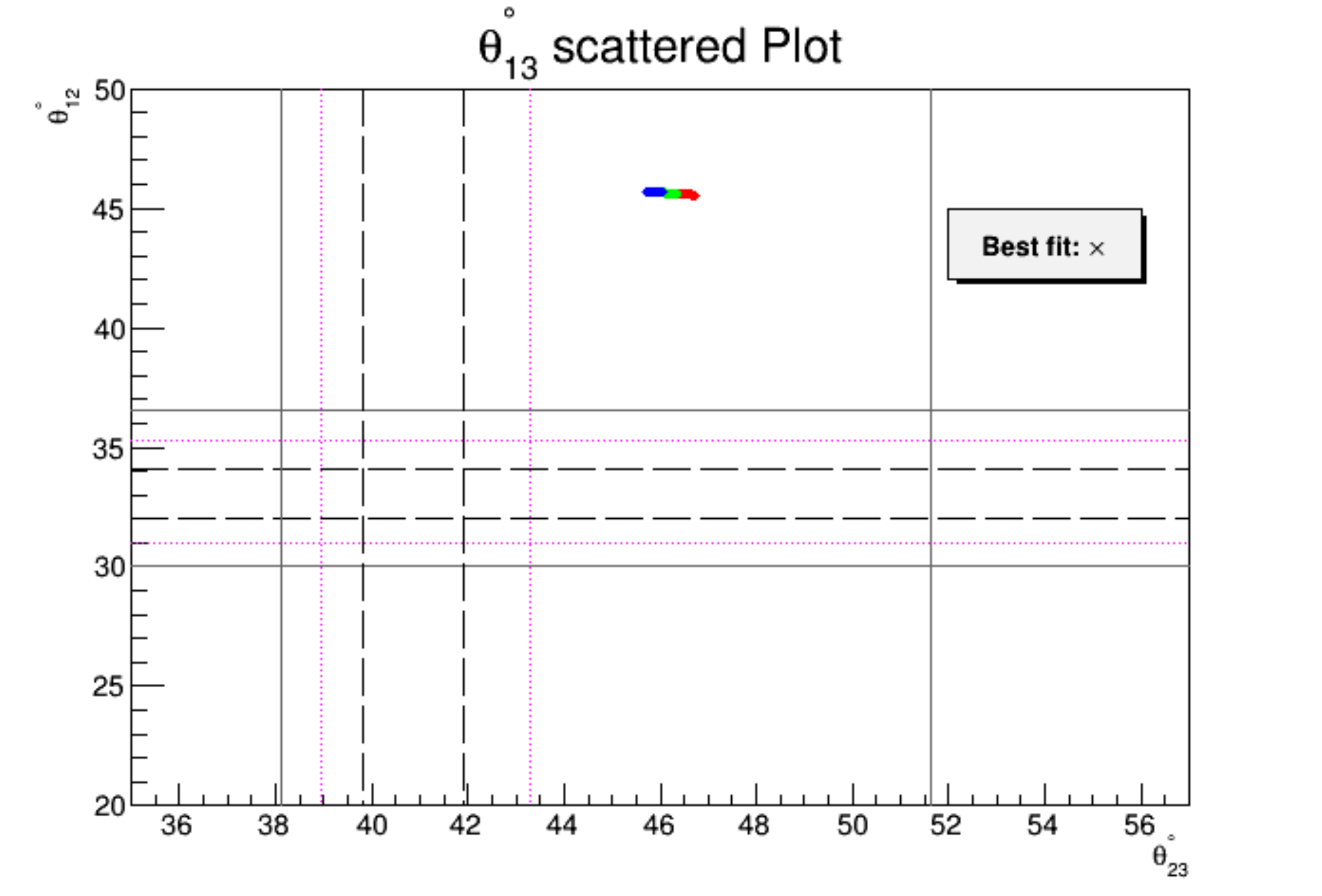}\\
\end{tabular}
\caption{\it{Line plot of $\chi^2$ (left fig.) vs $\gamma$ (in radians) and scattered plot of $\theta_{13}$ (right fig.) 
over $\theta_{23}-\theta_{12}$ (in degrees) plane for $U^{DCR}_{13}$ rotation scheme.
}}
\label{fig13R.2}
\end{figure}

\begin{figure}[!t]\centering
\begin{tabular}{c c} 
\hspace{-5mm}
\includegraphics[angle=0,width=80mm]{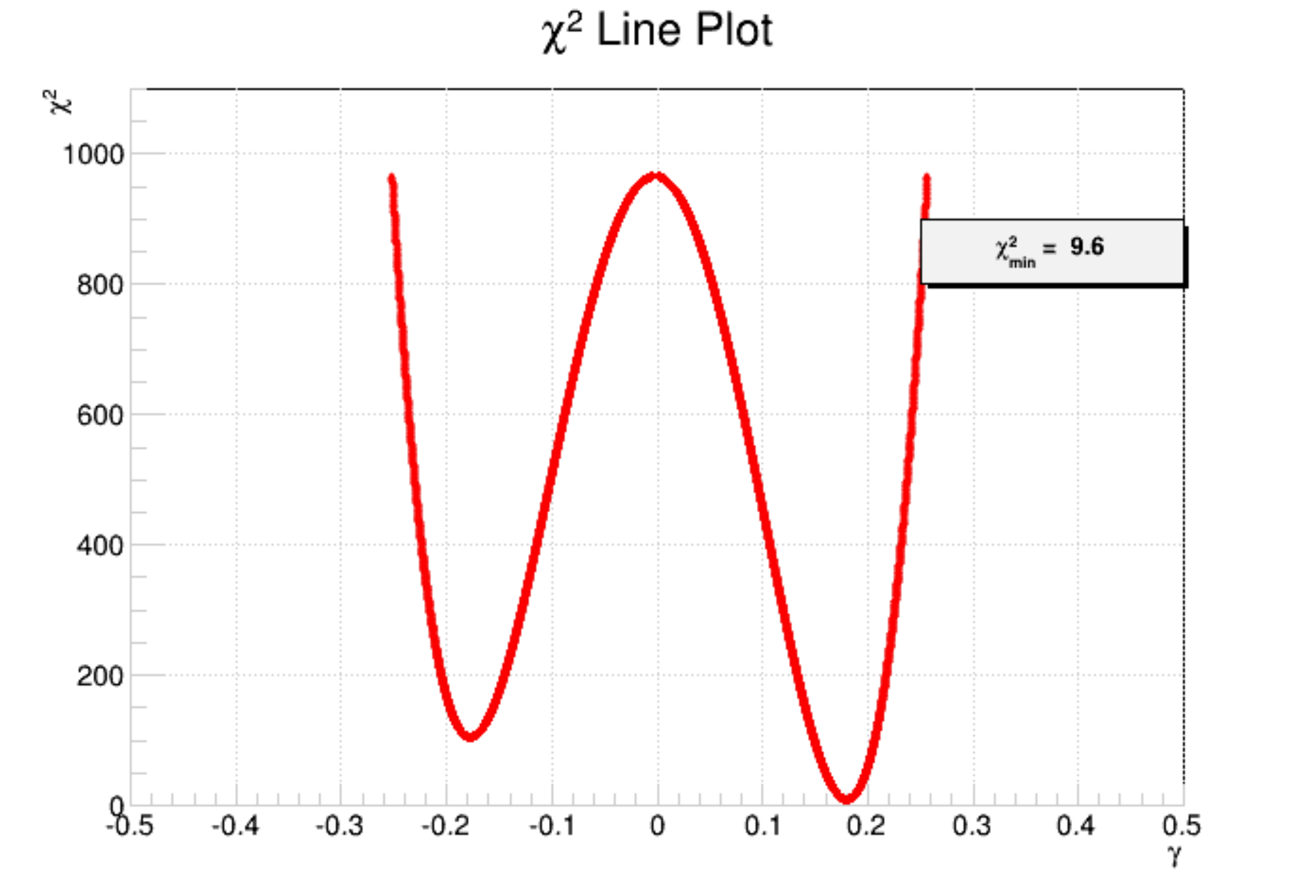} &
\includegraphics[angle=0,width=80mm]{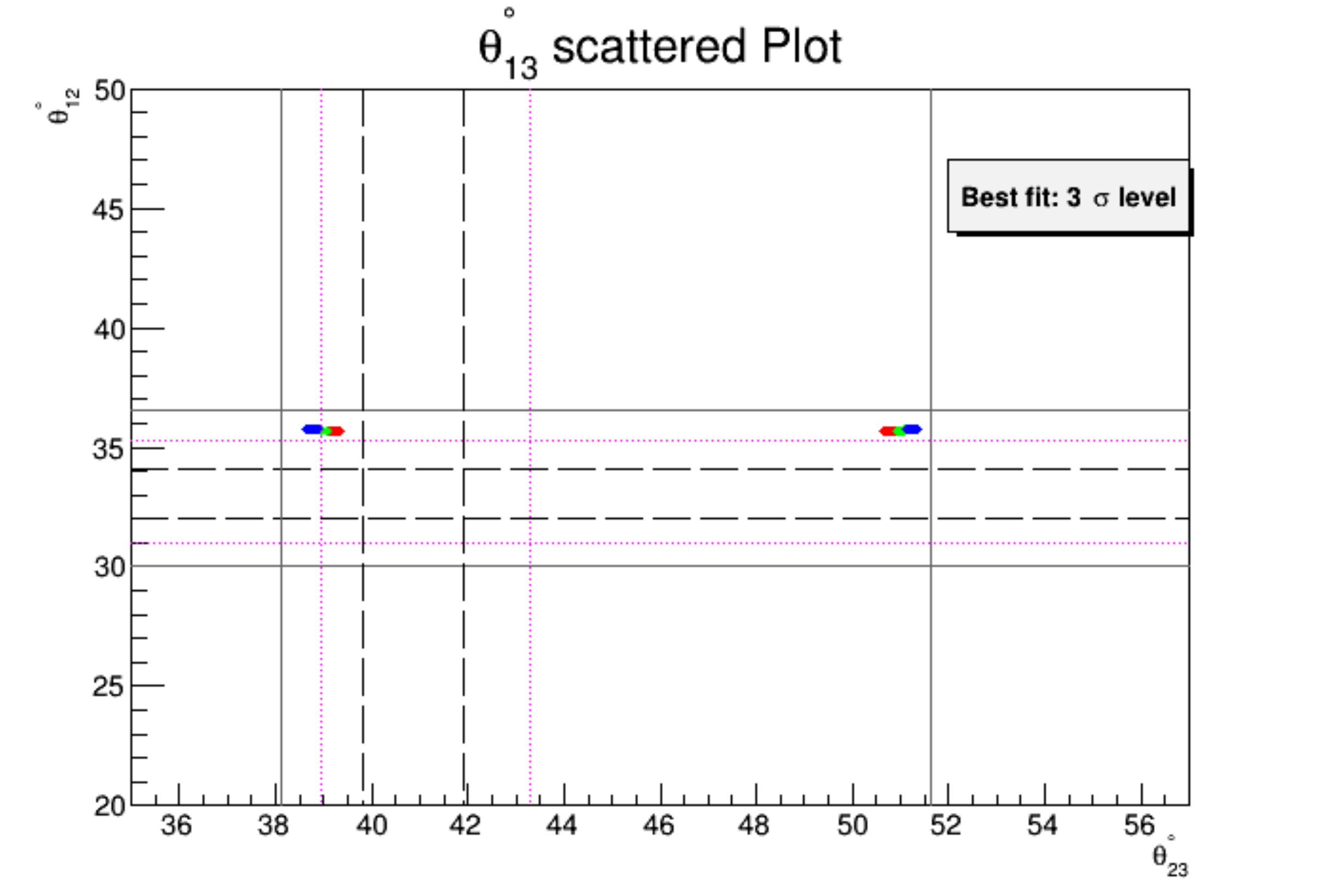}\\
\end{tabular}
\caption{\it{Line plot of $\chi^2$ (left fig.) vs $\gamma$ (in radians) and scattered plot of $\theta_{13}$ (right fig.) 
over $\theta_{23}-\theta_{12}$ (in degrees) plane for $U^{TBMR}_{13}$ rotation scheme. }}
\label{fig13R.3}
\end{figure}

Figs.~\ref{fig13R.1}-\ref{fig13R.3} show the numerical results corresponding to TBM, BM 
and DC case. The main features of these corrections are given as:\\
{\bf{(i)}} Here solar mixing angle($\theta_{12}$) receives only small corrections through $\sin\theta_{13}$ and thus its value 
remain close to its original prediction.\\
{\bf{(ii)}} In TBM case, for fitting $\theta_{13}$ in its $3\sigma$ range constraints $|\gamma| \in [0.169, 0.190]$ which in turn fixes 
$\theta_{23} \in [38.63^\circ, 39.35^\circ]$ for positive and $\theta_{23} \in [50.64^\circ, 51.36^\circ]$ for its negative range. The solar
mixing angle($\theta_{12}$) remains confined in the narrow range $\theta_{12}\in [35.65^\circ, 35.76^\circ]$. 
Thus  positive as well as negative $\gamma$ regions are allowed.\\
{\bf{(iii)}}However for BM and DC, we require $|\gamma| \in [0.196, 0.220]$ to fit $\theta_{13}$ in its $3\sigma$
range. But said range drives $\theta_{12}$ well away from its allowed $3\sigma$ range.\\
{\bf{(iii)}} The minimum value of $\chi^2 \sim 183.3$, $196.5$ and $9.6$ for BM, DC and TBM case respectively.\\
{\bf{(iv)}} Thus TBM case is consistent at $3\sigma$ level while other two cases are not viable.

\subsection{23 Rotation}

This case corresponds to rotation in 23 sector of  these special matrices. The neutrino
mixing angles for small rotation parameter $\beta$ are given by

\beqa
 \sin\theta_{13} &\approx&  |\beta U_{12}  |,\\
 \sin\theta_{23} &\approx& |\frac{ U_{23} + \beta U_{22}- \beta^2 U_{23}}{\cos\theta_{13}}|,\\
 \sin\theta_{12} &\approx& |\frac{(\beta^2-1) U_{12} }{\cos\theta_{13}}|.
\eeqa

\begin{figure}[!t]\centering
\begin{tabular}{c c} 
\hspace{-5mm}
\includegraphics[angle=0,width=80mm]{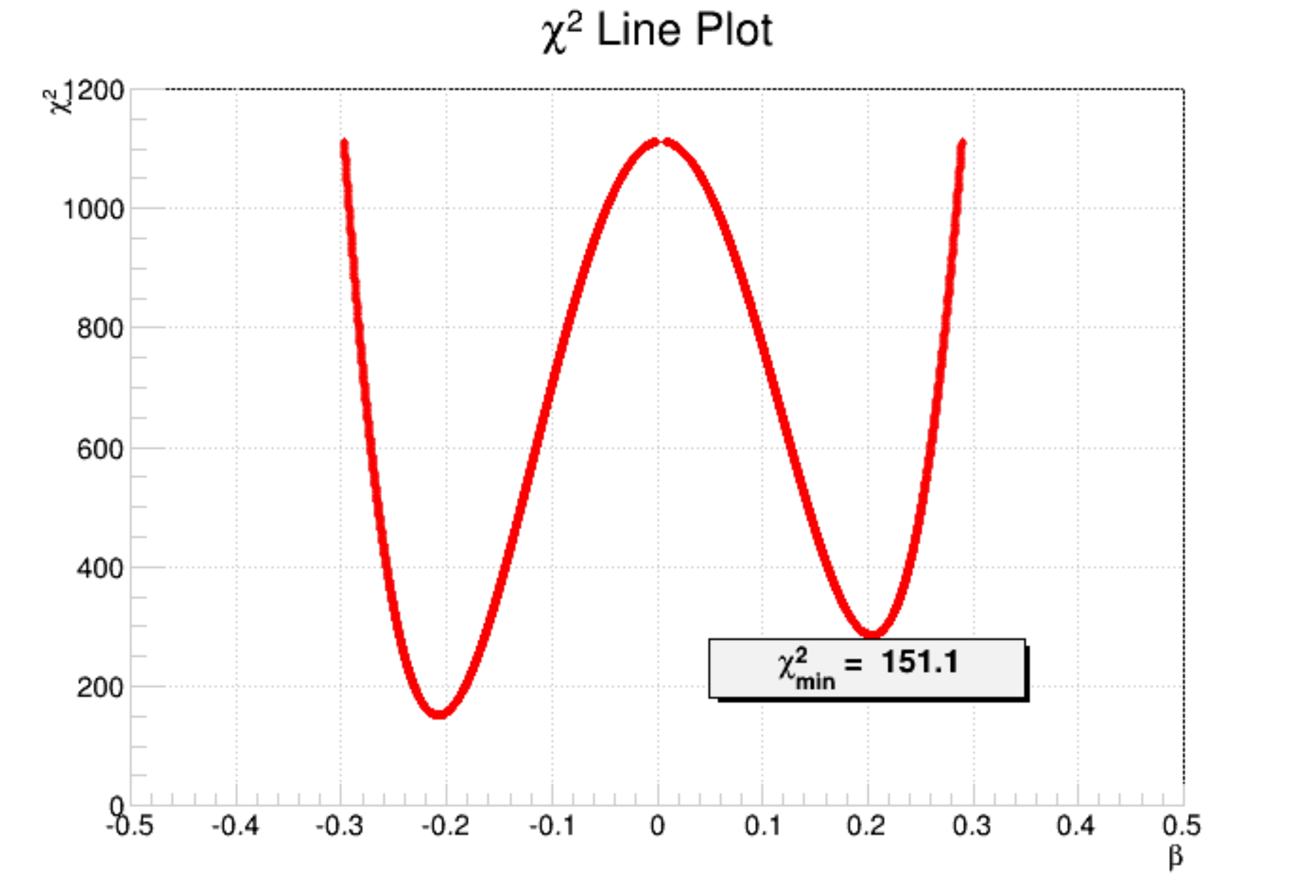} &
\includegraphics[angle=0,width=80mm]{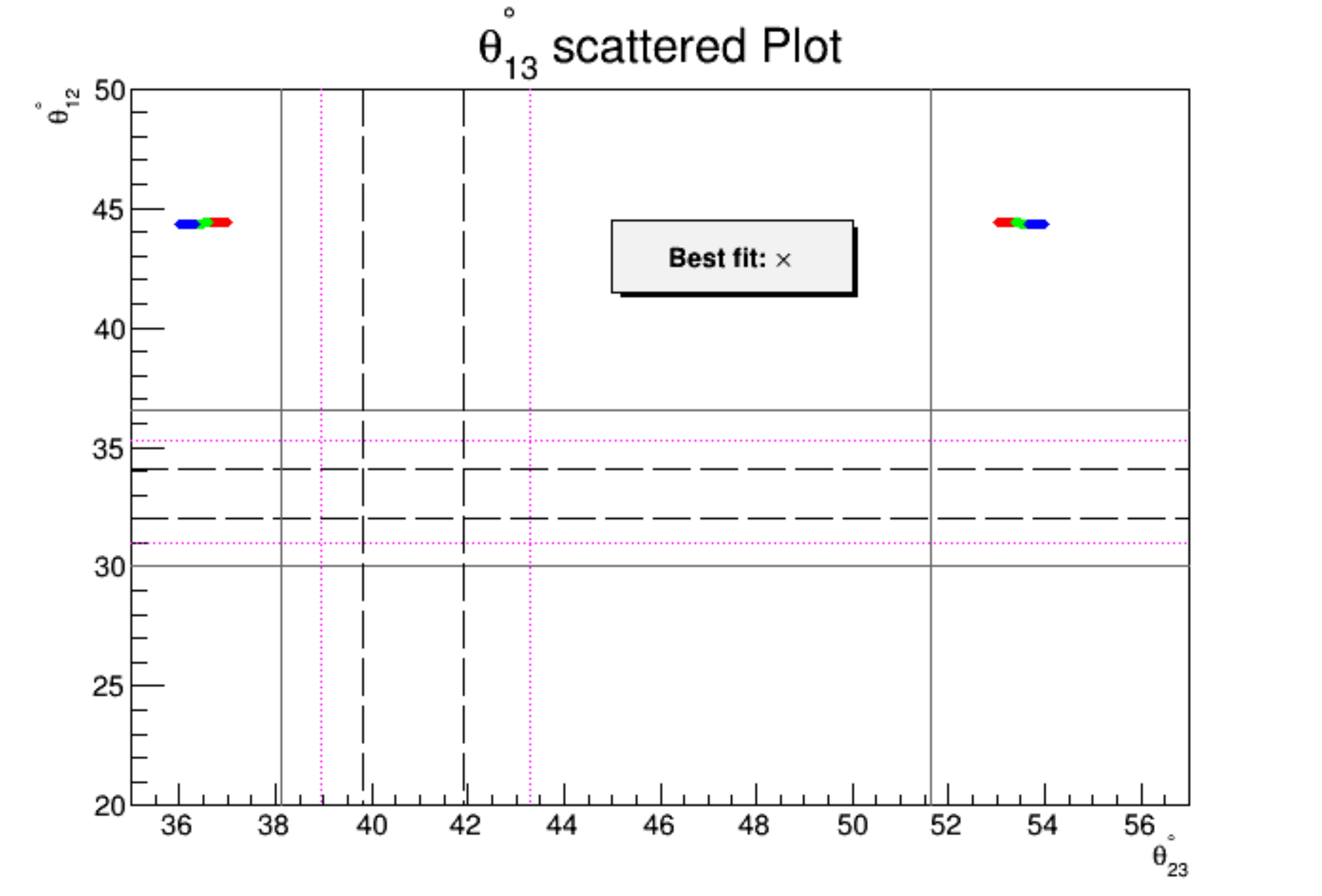}\\
\end{tabular}
\caption{\it{Line plot of $\chi^2$ (left fig.) vs $\beta$ (in radians) and scattered plot of $\theta_{13}$ (right fig.) 
over $\theta_{23}-\theta_{12}$ (in degrees) plane for $U^{BMR}_{23}$ rotation scheme. }}
\label{fig23R.1}
\end{figure}

\begin{figure}[!t]\centering
\begin{tabular}{c c} 
\hspace{-5mm}
\includegraphics[angle=0,width=80mm]{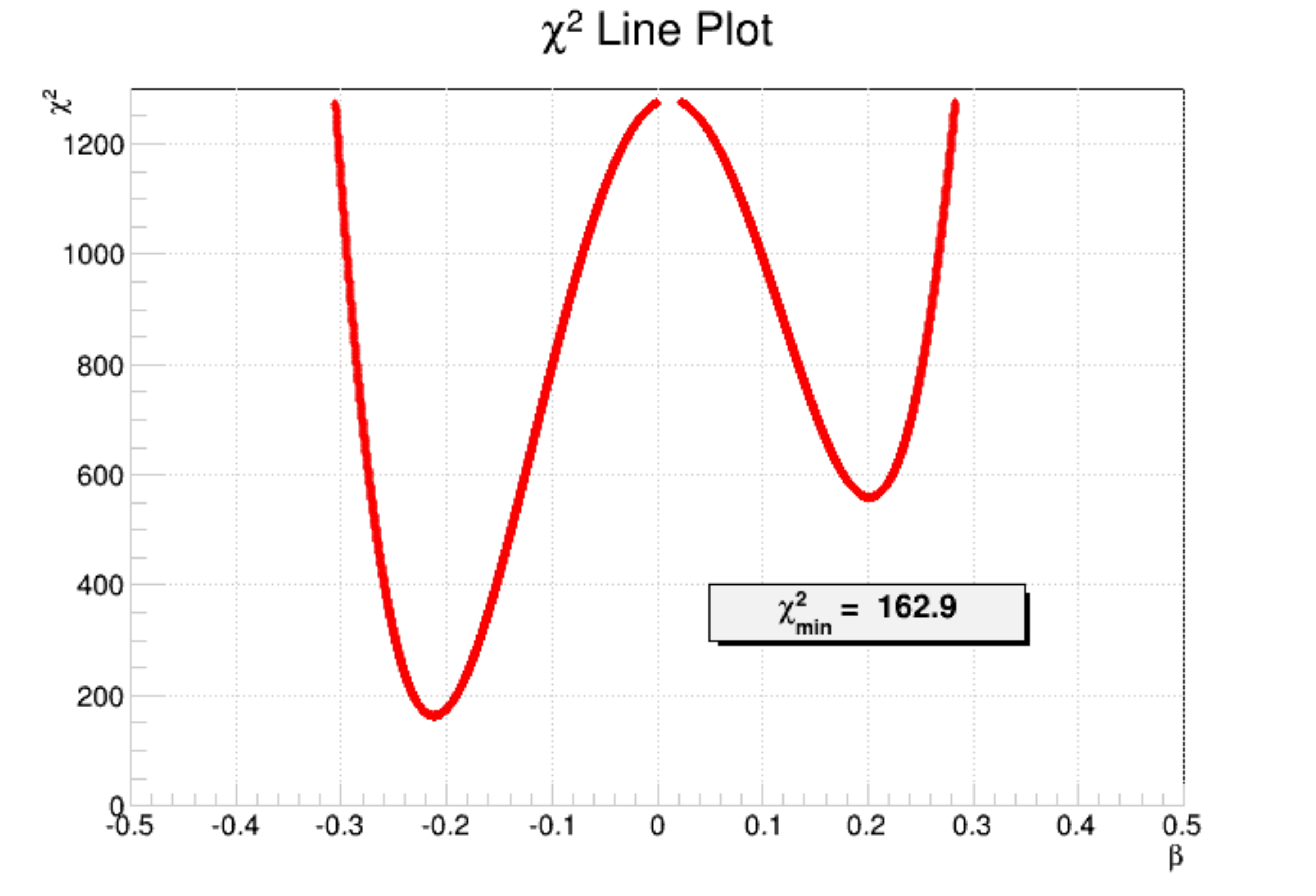} &
\includegraphics[angle=0,width=80mm]{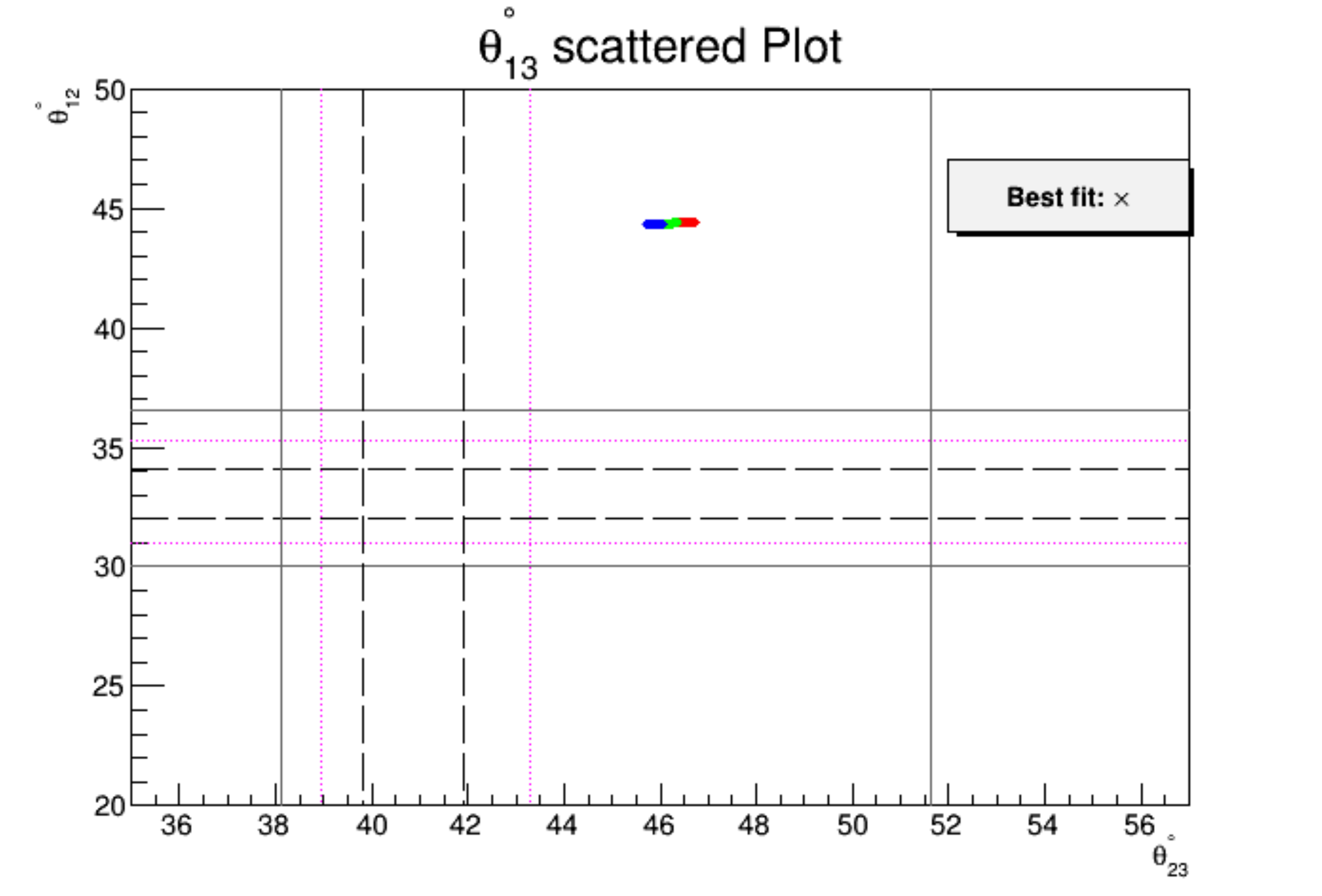}\\
\end{tabular}
\caption{\it{Line plot of $\chi^2$ (left fig.) vs $\beta$ (in radians) and scattered plot of $\theta_{13}$ (right fig.) 
over $\theta_{23}-\theta_{12}$ (in degrees) plane for $U^{DCR}_{23}$ rotation scheme.
}}
\label{fig23R.2}
\end{figure}

\begin{figure}[!t]\centering
\begin{tabular}{c c c} 
\hspace{-5mm}
\includegraphics[angle=0,width=80mm]{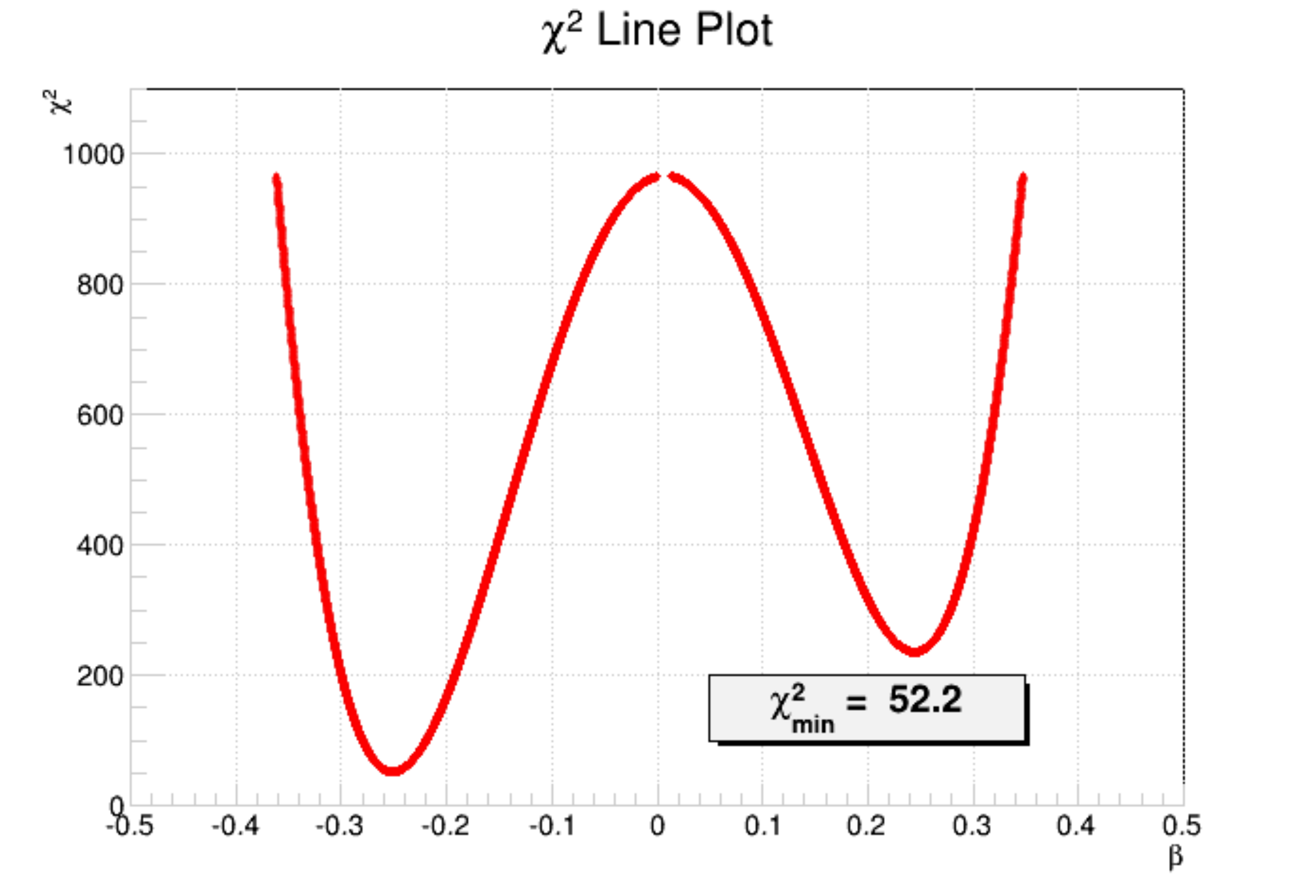}&
\includegraphics[angle=0,width=80mm]{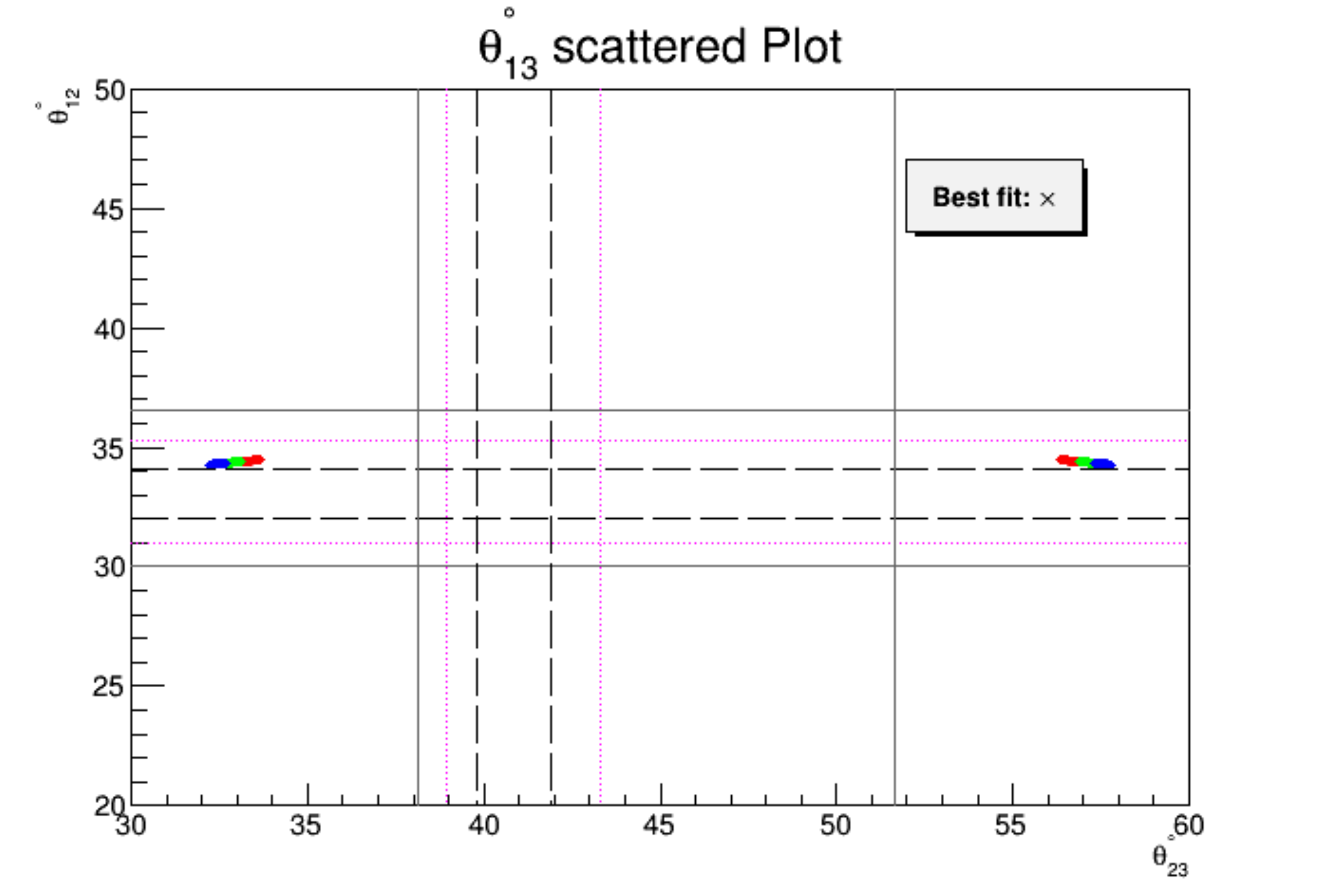}\\
\end{tabular}
\caption{\it{Line plot of $\chi^2$ (left fig.) vs $\beta$ (in radians) and scattered plot of $\theta_{13}$ (right fig.) 
over $\theta_{23}-\theta_{12}$ (in degrees) plane for $U^{TBMR}_{23}$ rotation scheme. }}
\label{fig23R.3}
\end{figure}

Figs.~\ref{fig23R.1}-\ref{fig23R.3} show the numerical results corresponding to TBM, BM 
and DC case. The salient features in this perturbative scheme are given by:\\
{\bf{(i)}} The atmospheric mixing angle($\theta_{12}$) receives corrections only of the $O(\theta^2)$ so its value
remains close to its original prediction. \\
{\bf{(ii)}} In TBM case, for fitting $\theta_{13}$ in its $3\sigma$ range constraints $|\beta| \in [0.241, 0.271]$ which 
in turn fixes $\theta_{23} \in [56.34^\circ, 57.78^\circ]$ for positive and $\theta_{23} \in [32.21^\circ, 33.65^\circ]$ for its negative range. 
However solar mixing angle($\theta_{12}$) lies close to its original prediction $\theta_{12}\in [34.26^\circ, 34.47^\circ]$. Thus this case is ruled out at $3\sigma$ level.\\
{\bf{(iii)}}However for BM and DC, we require $|\beta| \in [0.196, 0.220]$ to fit $\theta_{13}$ in its $3\sigma$
range. But said range drives $\theta_{12}$ well away from its allowed $3\sigma$ range. \\
{\bf{(iii)}} The minimum value of $\chi^2 \sim 151.1$, $162.9$ and $52.2$ for BM, DC and TBM 
case respectively. \\
{\bf{(iv)}} Thus  all cases are ruled out at  $3\sigma$ level.

\section{Rotations-$R_{ij}^l.R_{kl}^l.U$}

Here we first consider the perturbations for which modified PMNS matrix is given by $U_{PMNS} = R_{ij}^l.R_{kl}^l.U$. We will 
investigate the role of these perturbations in fitting the neutrino mixing data. 

\subsection{12-13 Rotation}

This case corresponds to rotations in 12 and 13 sector of  these special matrices. 
Since for small rotation $\sin\theta \approx \theta$ and $\cos\theta \approx 1-\theta^2$, so the neutrino mixing angles
truncated at order O ($\theta^2$) for these rotations are given by

\beqa
 \sin\theta_{13} &\approx&  |\alpha U_{23} + \gamma U_{33} |,\\
 \sin\theta_{23} &\approx& |\frac{ (1-\alpha^2) U_{23}-\alpha\gamma U_{33} }{\cos\theta_{13}}|,\\
 \sin\theta_{12} &\approx& |\frac{U_{12} + \alpha U_{22} + \gamma U_{32}-(\alpha^2 + \gamma^2)U_{12} }{\cos\theta_{13}}|.
\eeqa

\begin{figure}[!t]\centering
\begin{tabular}{c c} 
\includegraphics[angle=0,width=80mm]{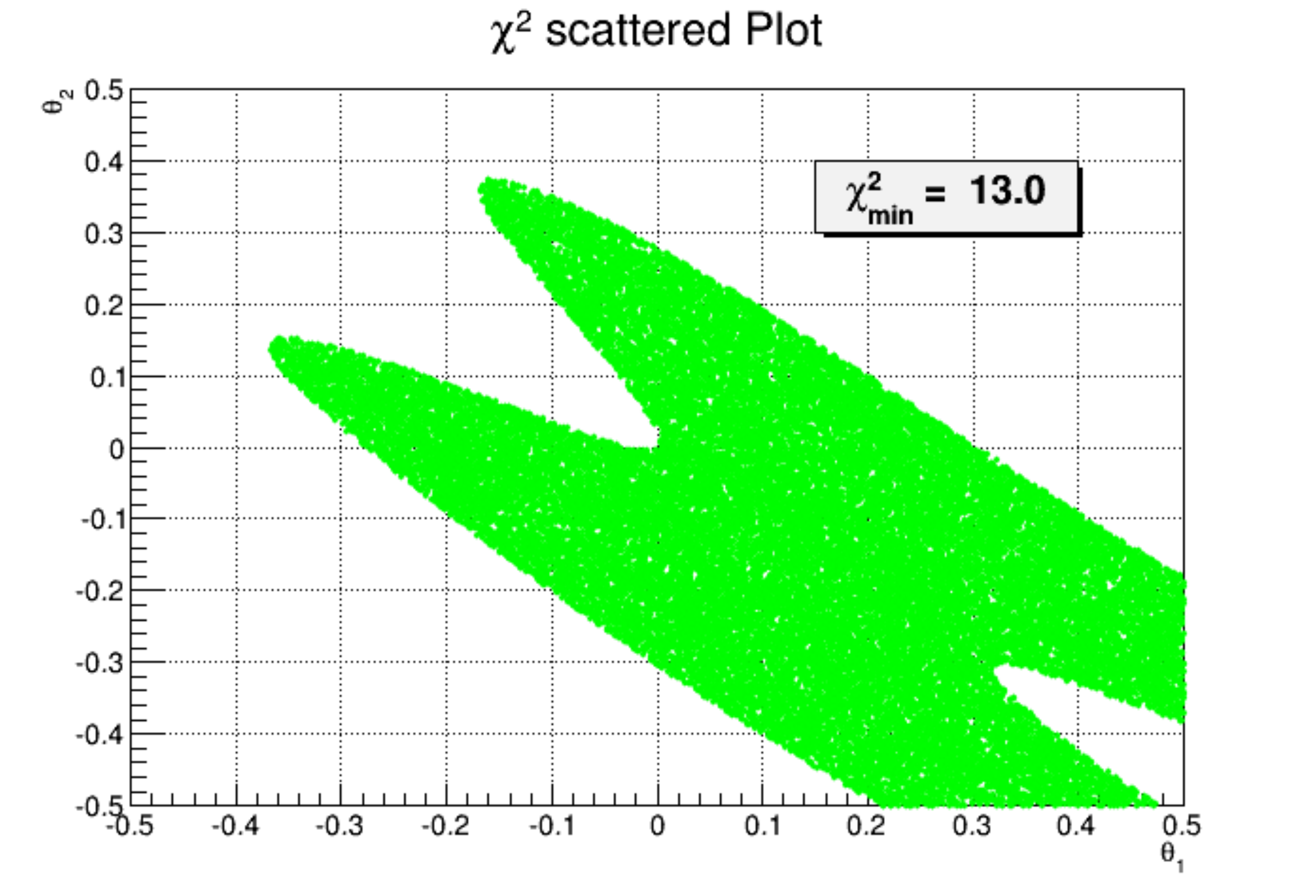} &
\includegraphics[angle=0,width=80mm]{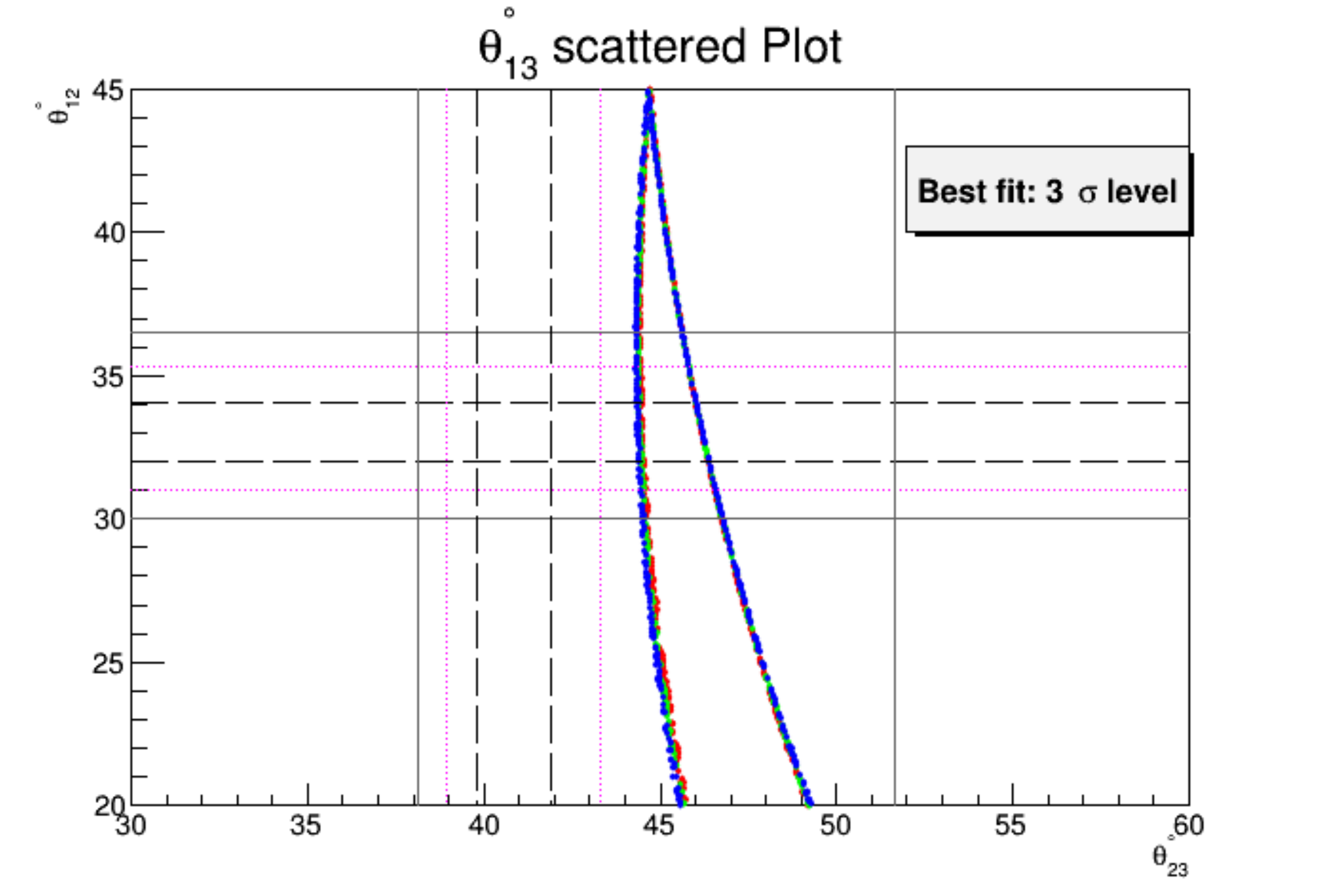}\\
\end{tabular}
\caption{\it{$U^{BML}_{1213}$ scatter plot of $\chi^2$ (left fig.) over $\alpha-\gamma$ (in radians) plane and $\theta_{13}$ (right fig.) 
over  $\theta_{23}-\theta_{12}$ (in degrees) plane. The information about color coding and various horizontal, vertical lines in right fig. is
given in text.}}
\label{fig1213L.1}
\end{figure}

\begin{figure}[!t]\centering
\begin{tabular}{c c} 
\includegraphics[angle=0,width=80mm]{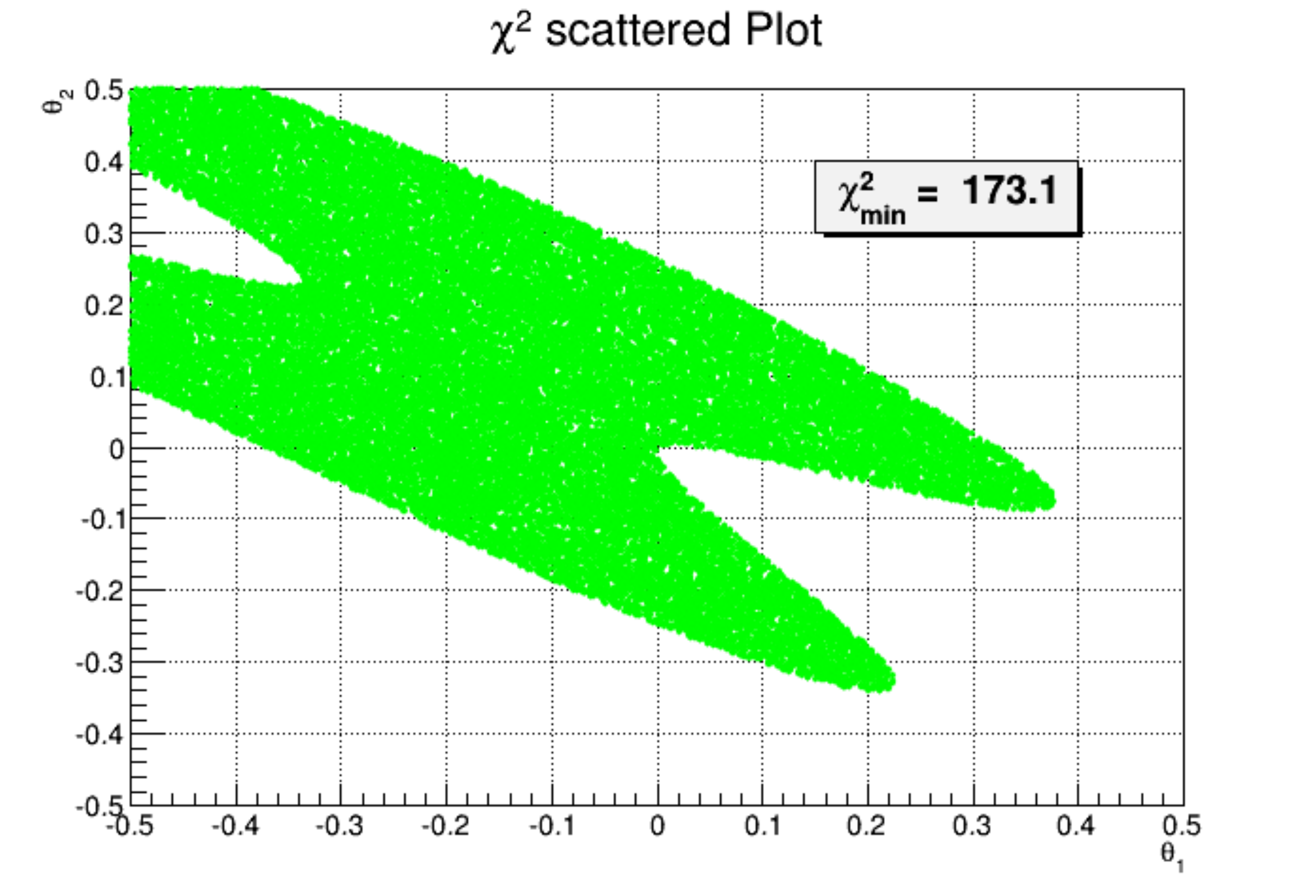} &
\includegraphics[angle=0,width=80mm]{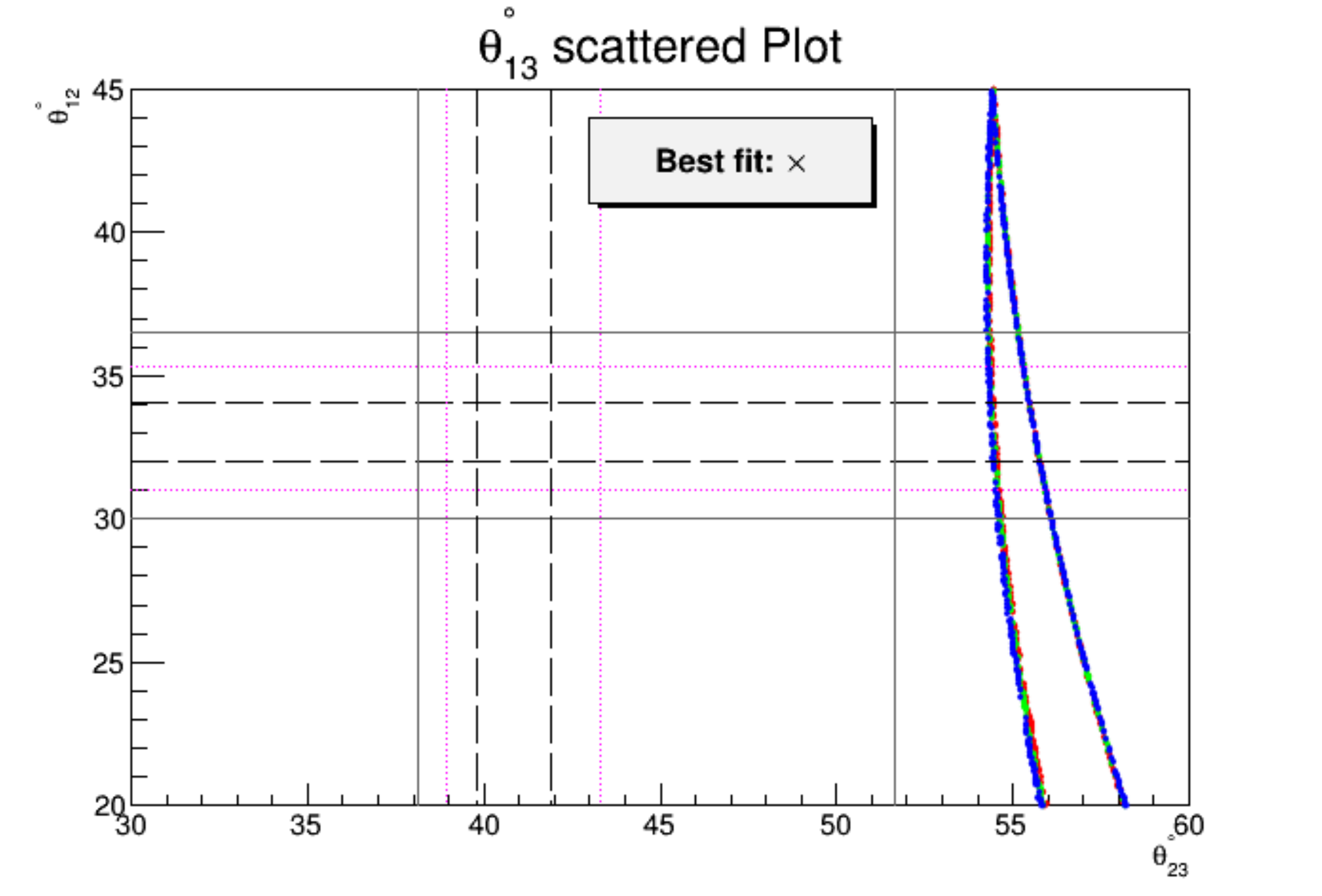}\\
\end{tabular}
\caption{\it{ $U^{DCL}_{1213}$ scatter plot of $\chi^2$ (left fig.) over $\alpha-\gamma$ (in radians) plane and $\theta_{13}$ (right fig.) 
over  $\theta_{23}-\theta_{12}$ (in degrees) plane. }}
\label{fig1213L.2}
\end{figure}

\begin{figure}[!t]\centering
\begin{tabular}{c c} 
\includegraphics[angle=0,width=80mm]{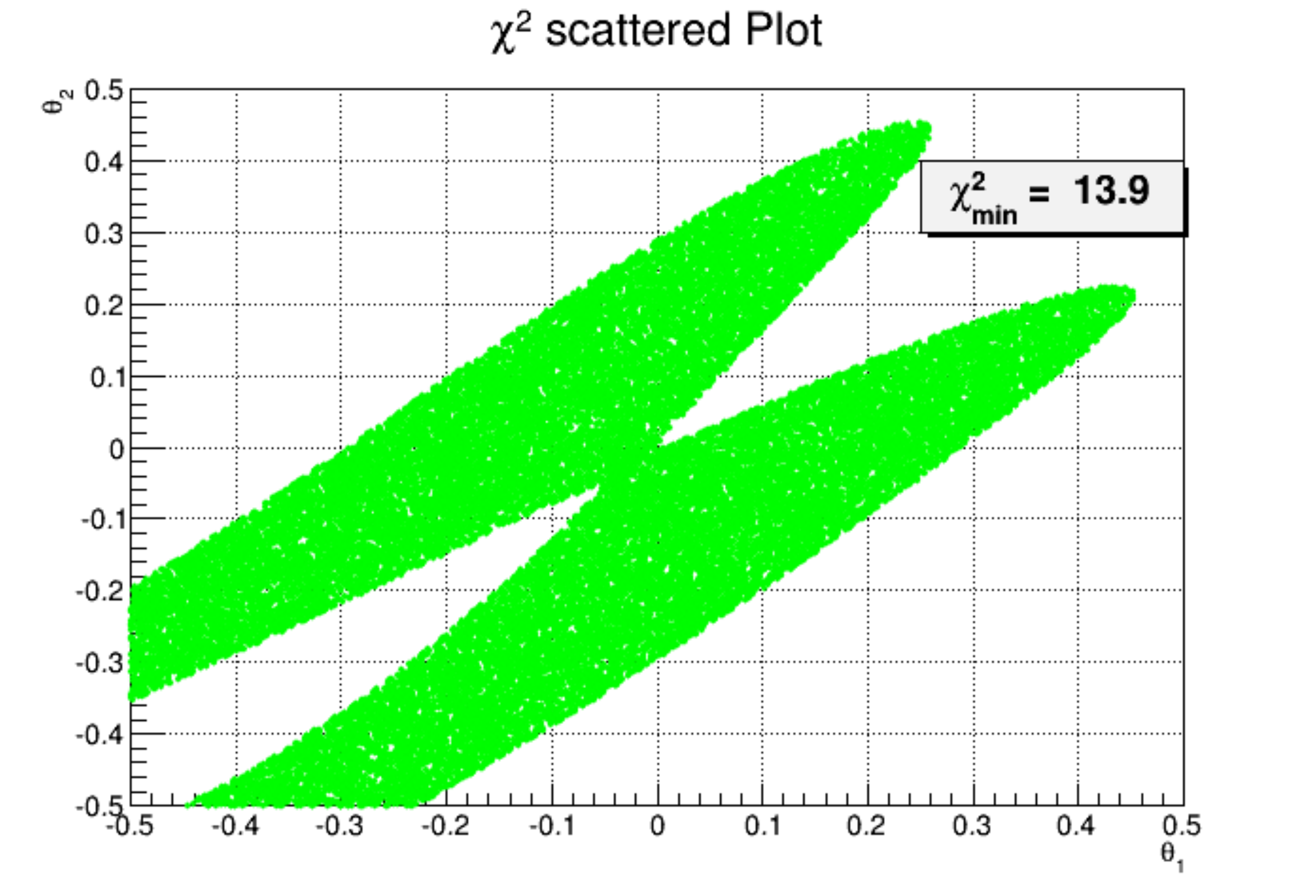} &
\includegraphics[angle=0,width=80mm]{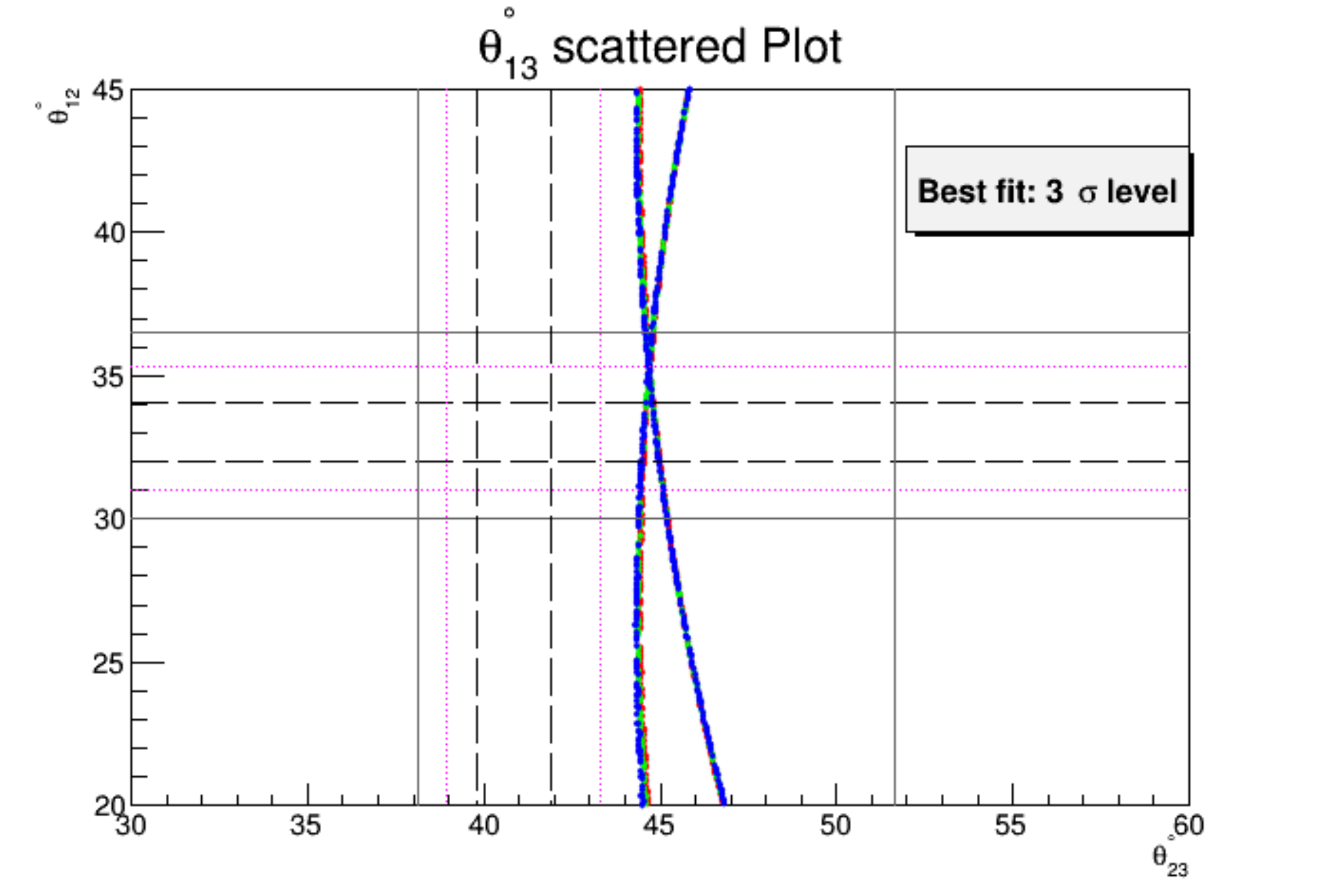}\\
\end{tabular}
\caption{\it{$U^{TBML}_{1213}$ scatter plot of $\chi^2$ (left fig.) over $\alpha-\gamma$ (in radians) plane and $\theta_{13}$ (right fig.) 
over $\theta_{23}-\theta_{12}$ (in degrees) plane. }}
\label{fig1213L.3}
\end{figure}

Figs.~\ref{fig1213L.1}-\ref{fig1213L.3} show the numerical results corresponding to BM, DC and TBM case  with $\theta_1 = \gamma$ and $\theta_2 = \alpha$.
The main features of this perturbative matrix are:\\
{\bf{(i)}} From the expression of $\theta_{23}$ mixing angle, it is clear that the perturbation parameters enter only at O($\theta^2$) 
and thus its value remain close to the original prediction. Thus TBM and BM can become consistent while DC looks off track completely. \\
{\bf{(ii)}} The minimum value of $\chi^2 \sim 13.0$, $173.1$ and $13.9$ for BM, DC and TBM case respectively.\\
{\bf{(iii)}} In perturbative TBM and BM case its possible to fit all mixing angles at $3\sigma$ level while DC is not consistent.

\subsection{12-23 Rotation}

This case corresponds to rotations in 12 and 23 sector of  these special matrices.  
The neutrino mixing angles for small perturbation 
parameters $\alpha$ and $\beta$ are given by

\beqa
 \sin\theta_{13} &\approx&  |\alpha U_{23} + \alpha\beta U_{33} |,\\
 \sin\theta_{23} &\approx& |\frac{U_{23} + \beta U_{33} -(\alpha^2 + \beta^2)U_{23} }{\cos\theta_{13}}|,\\
 \sin\theta_{12} &\approx& |\frac{ U_{12} + \alpha U_{22}-\alpha^2 U_{12}+ \alpha\beta U_{32} }{\cos\theta_{13}}|.
\eeqa

\begin{figure}[!t]\centering
\begin{tabular}{c c} 
\includegraphics[angle=0,width=80mm]{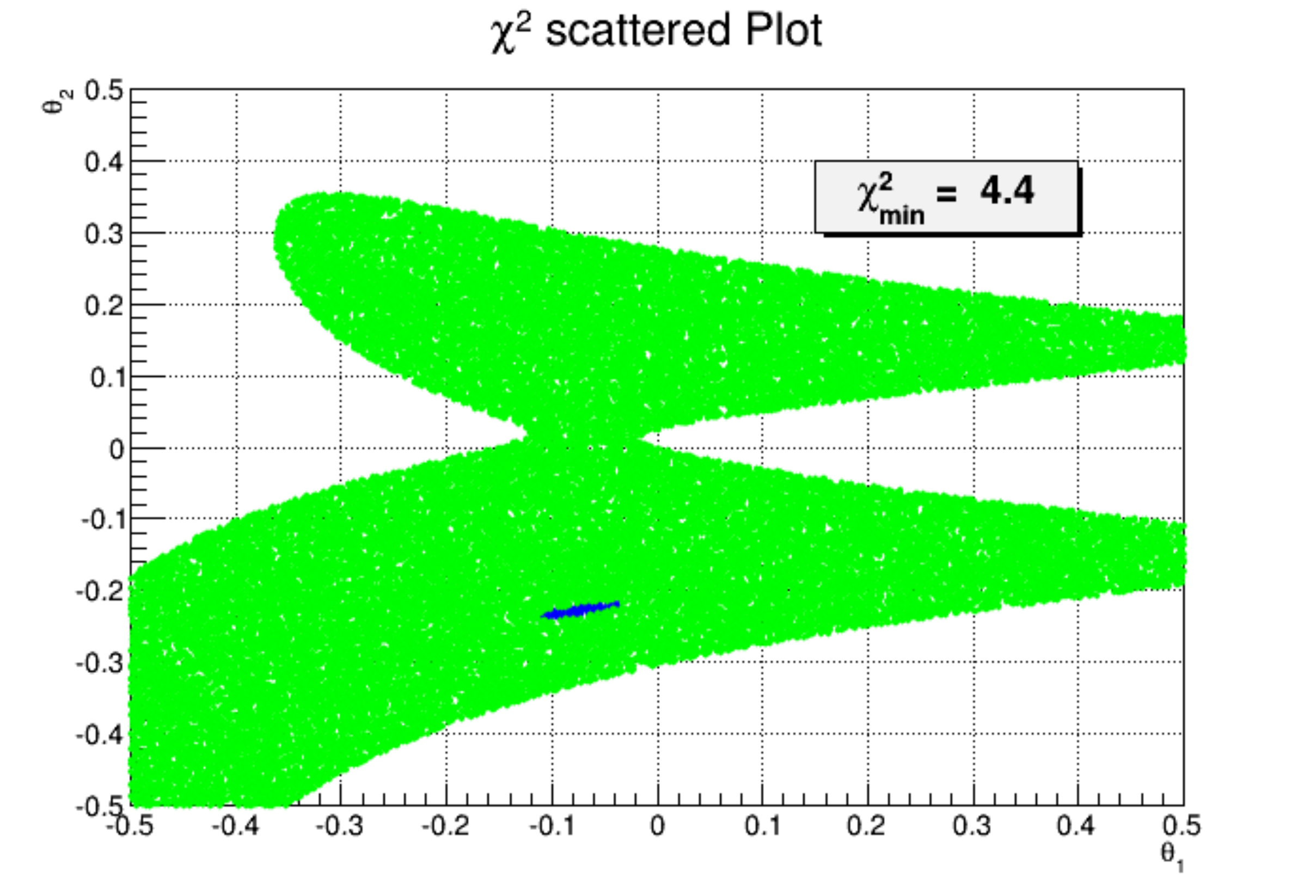} &
\includegraphics[angle=0,width=80mm]{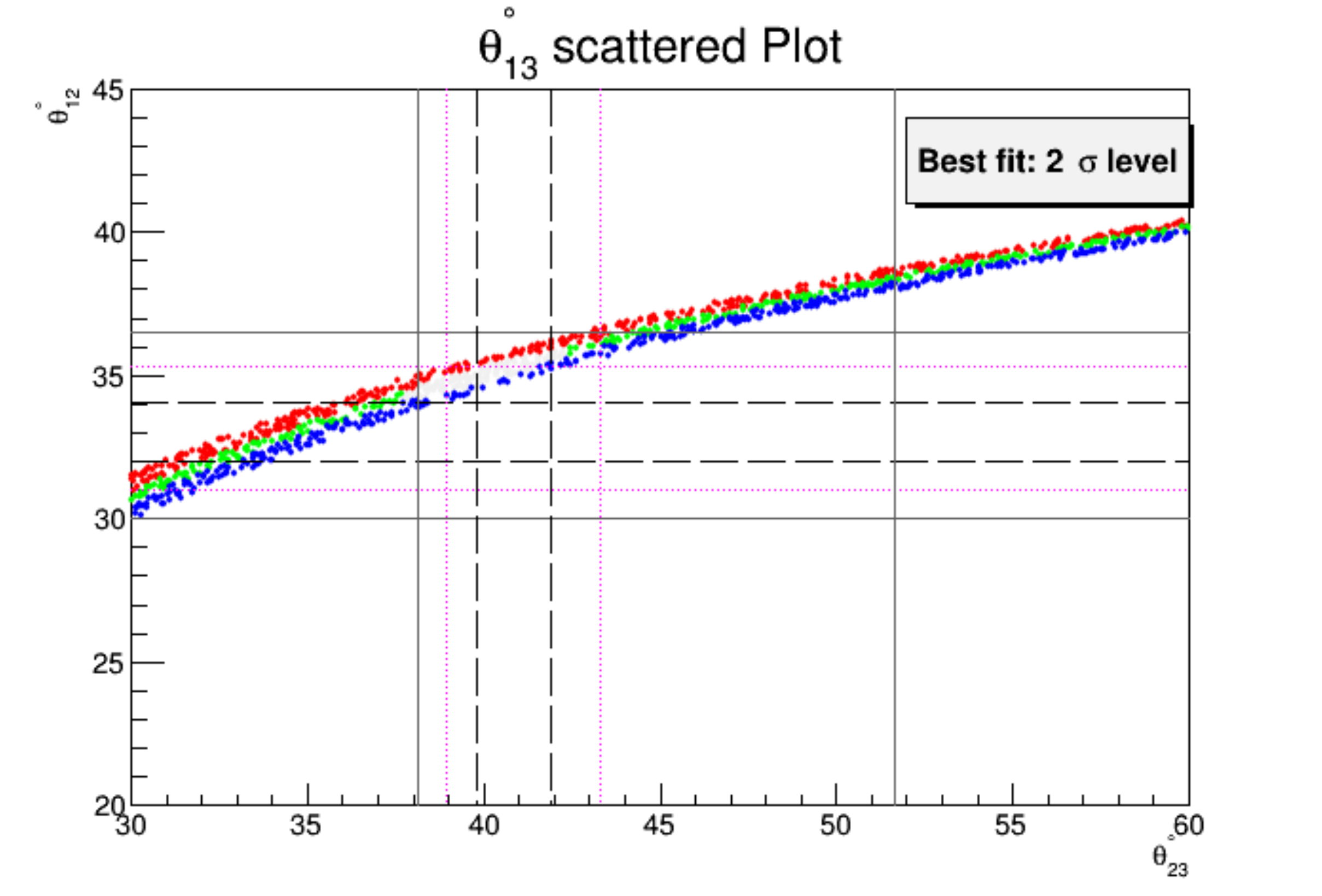}\\
\end{tabular}
\caption{\it{$U^{BML}_{1223}$ scatter plot of $\chi^2$ (left fig.) over $\alpha-\beta$ (in radians) plane and $\theta_{13}$ (right fig.) 
over  $\theta_{23}-\theta_{12}$ (in degrees) plane.}}
\label{fig1223L.1}
\end{figure}

\begin{figure}[!t]\centering
\begin{tabular}{c c} 
\includegraphics[angle=0,width=80mm]{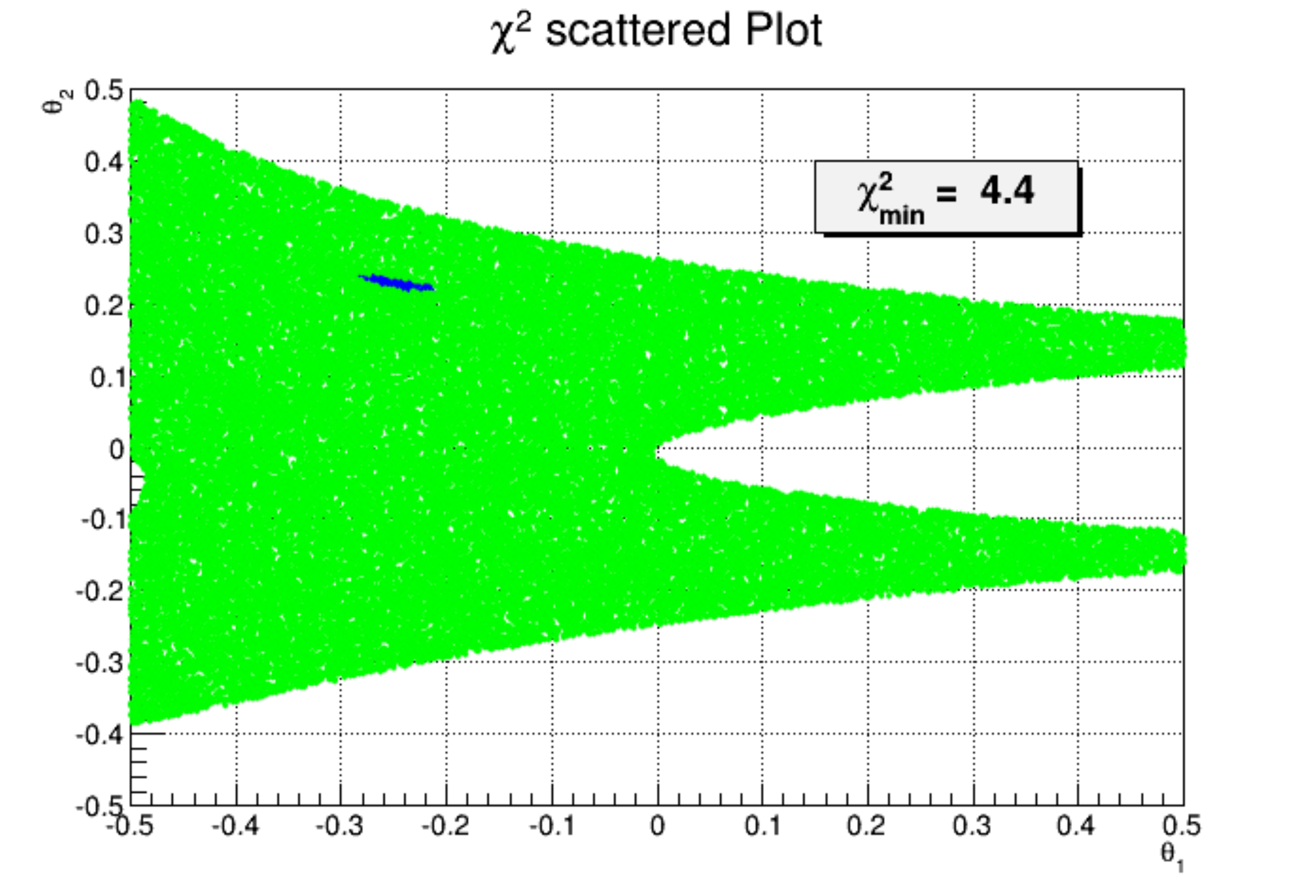} &
\includegraphics[angle=0,width=80mm]{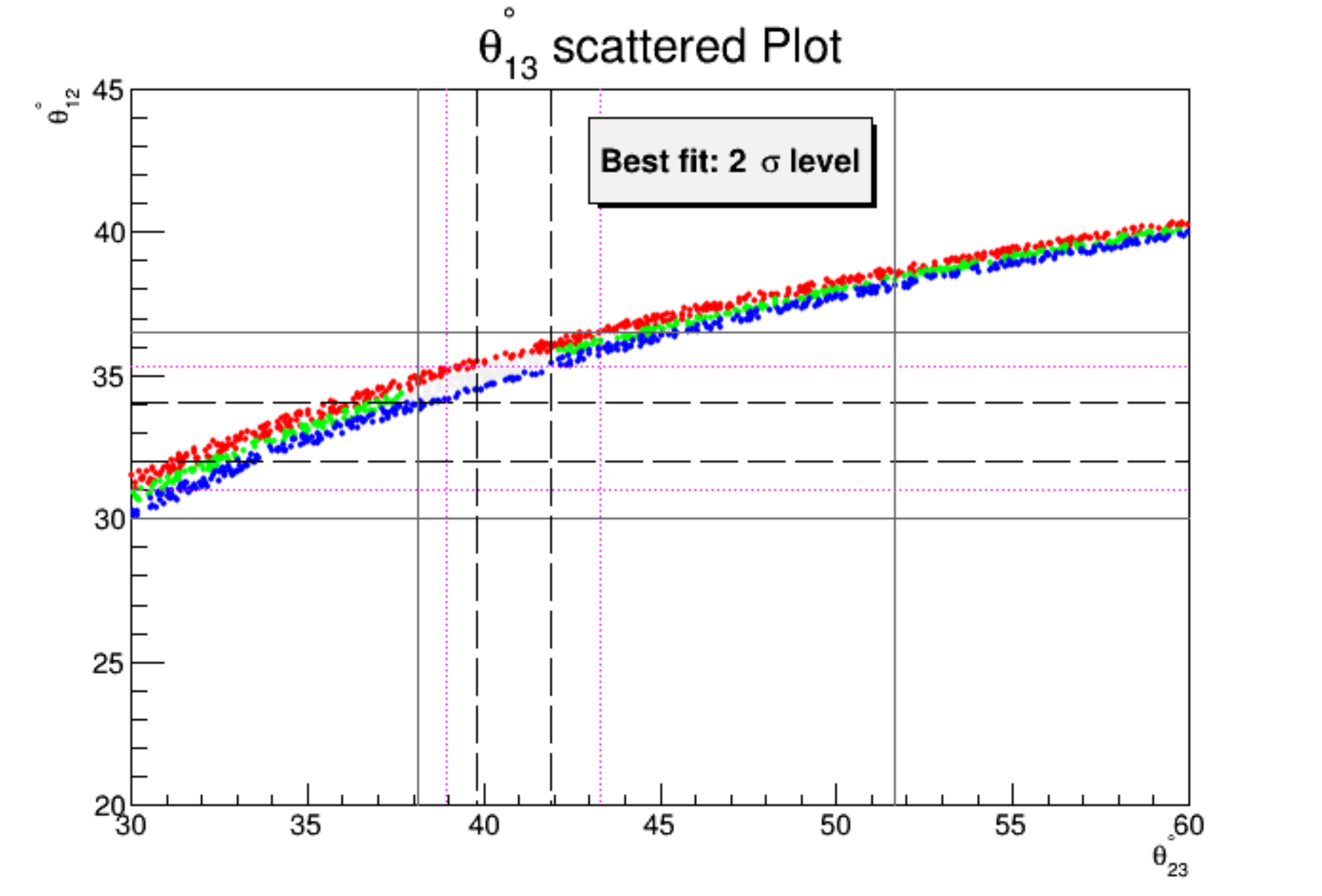}\\
\end{tabular}
\caption{\it{$U^{DCL}_{1223}$ scatter plot of $\chi^2$ (left fig.) over $\alpha-\beta$ (in radians) plane and $\theta_{13}$ (right fig.) 
over  $\theta_{23}-\theta_{12}$ (in degrees) plane. }}
\label{fig1223L.2}
\end{figure}

\begin{figure}[!t]\centering
\begin{tabular}{c c} 
\includegraphics[angle=0,width=80mm]{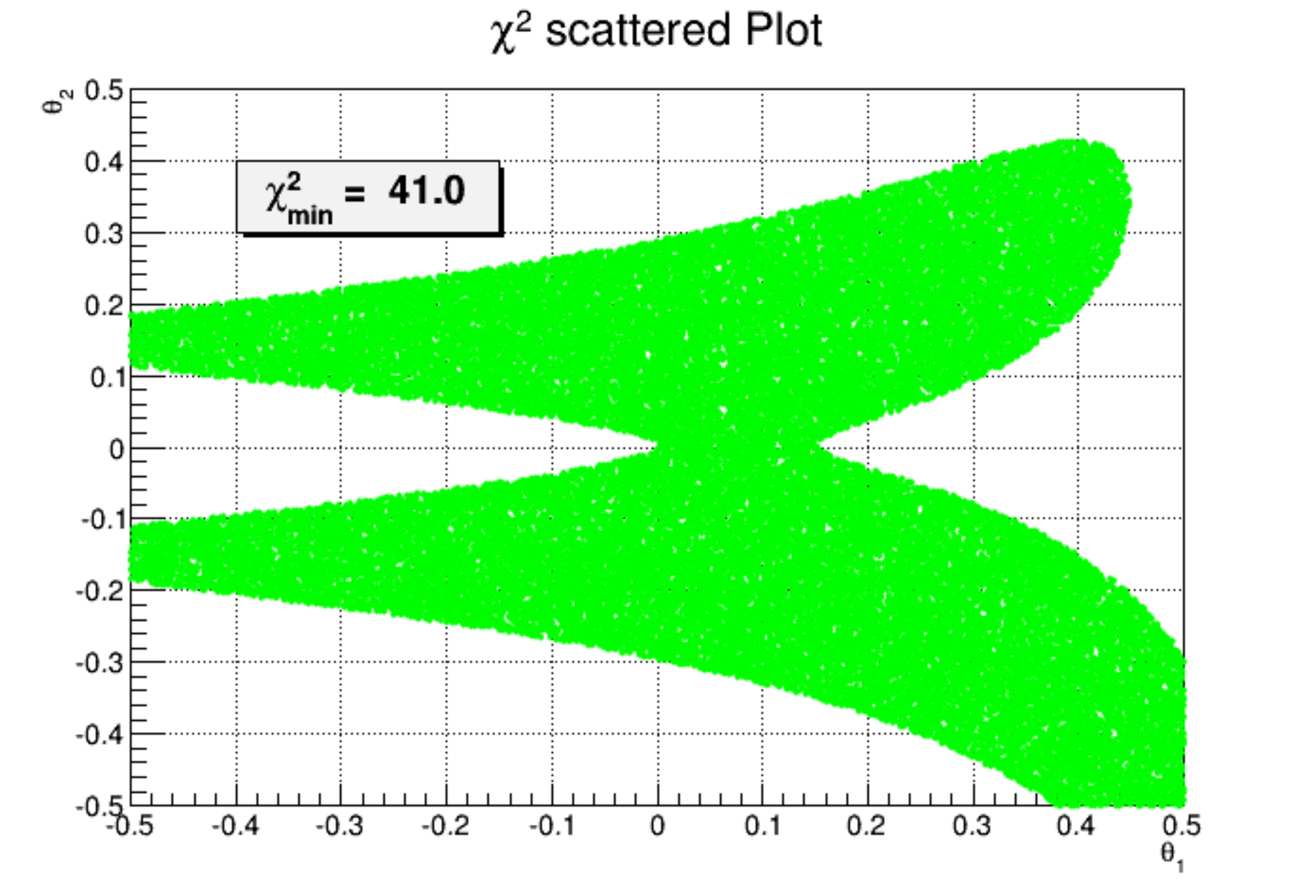} &
\includegraphics[angle=0,width=80mm]{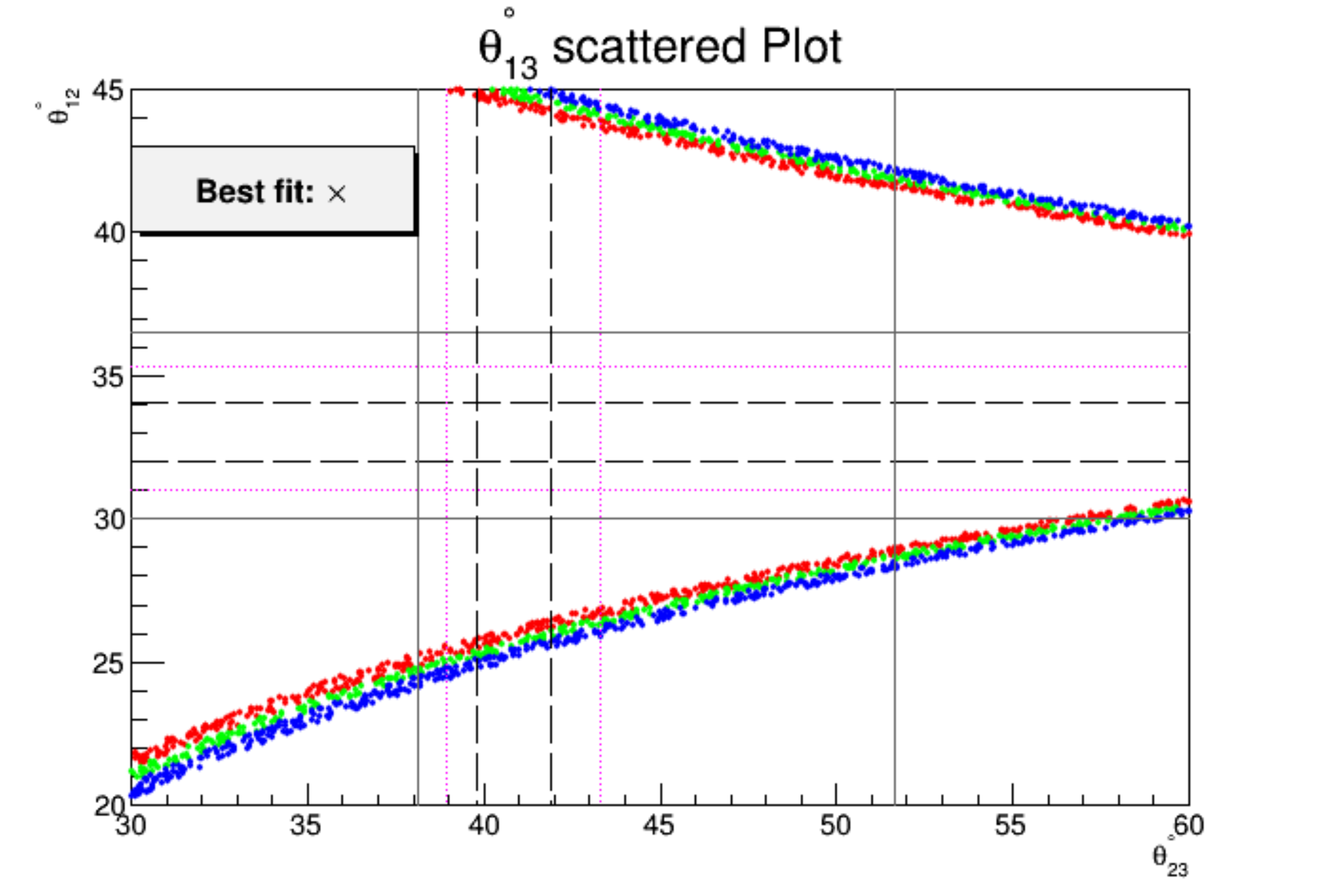}\\
\end{tabular}
\caption{\it{$U^{TBML}_{1223}$ scatter plot of $\chi^2$ (left fig.) over $\alpha-\beta$ (in radians) plane and $\theta_{13}$ (right fig.) 
over $\theta_{23}-\theta_{12}$ (in degrees) plane. }}
\label{fig1223L.3}
\end{figure}

Figs.~\ref{fig1223L.1}-\ref{fig1223L.3} corresponds to BM, DC and TBM case respectively with $\theta_1 = \beta$ and $\theta_2 = \alpha$. The
main features in this mixing scheme are:\\
{\bf{(i)}} In this scheme, perturbation parameters enters at leading order in all mixing angles and hence show interesting 
correlations among themselves. \\
{\bf{(ii)}} It is possible to get $\chi^2 < 5$ for BM and DC case while perturbative TBM is not that promising and corresponding
value $\chi^2 > 41$ in parameter space. \\
{\bf{(iii)}} The minimum value of $\chi^2 \sim 4.4$, $4.4$ and $41.0$ for BM, DC and TBM case respectively.\\
{\bf{(iv)}} In BM and DC case its possible to fit all mixing angles at 2$\sigma$ level while TBM case is completely excluded at 3$\sigma$ level.

\subsection{13-12 Rotation}

This case corresponds to rotations in 13 and 12 sector of  these special matrices. 
The neutrino mixing angles for small perturbation parameters $\alpha$ and $\gamma$ are given by

\beqa
 \sin\theta_{13} &\approx&  |\alpha U_{23} + \gamma U_{33} |,\\
 \sin\theta_{23} &\approx& |\frac{ (1-\alpha^2) U_{23} }{\cos\theta_{13}}|,\\
 \sin\theta_{12} &\approx& |\frac{U_{12} + \alpha U_{22} + \gamma U_{32}-(\alpha^2 + \gamma^2)U_{12} }{\cos\theta_{13}}|.
\eeqa

\begin{figure}[!t]\centering
\begin{tabular}{c c} 
\includegraphics[angle=0,width=80mm]{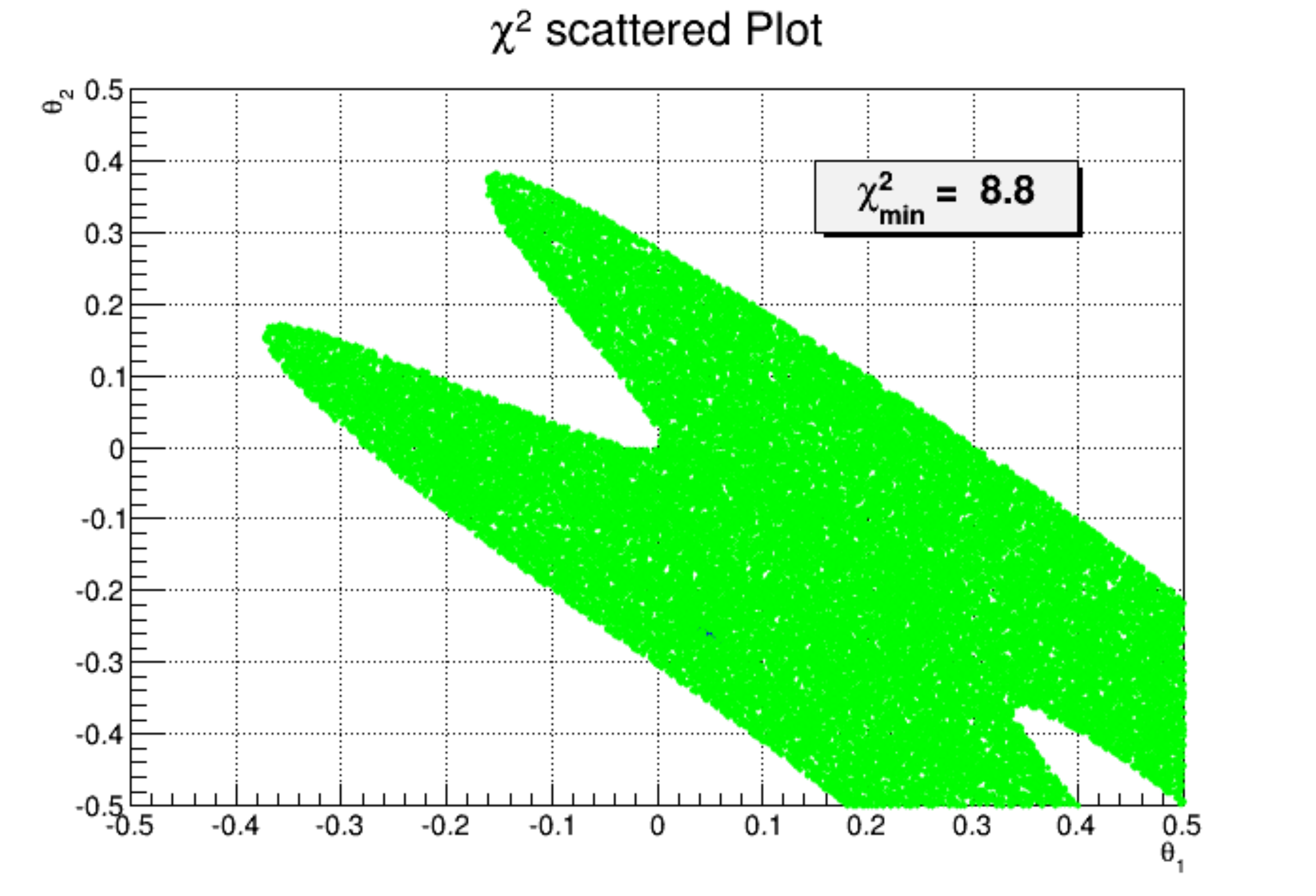} &
\includegraphics[angle=0,width=80mm]{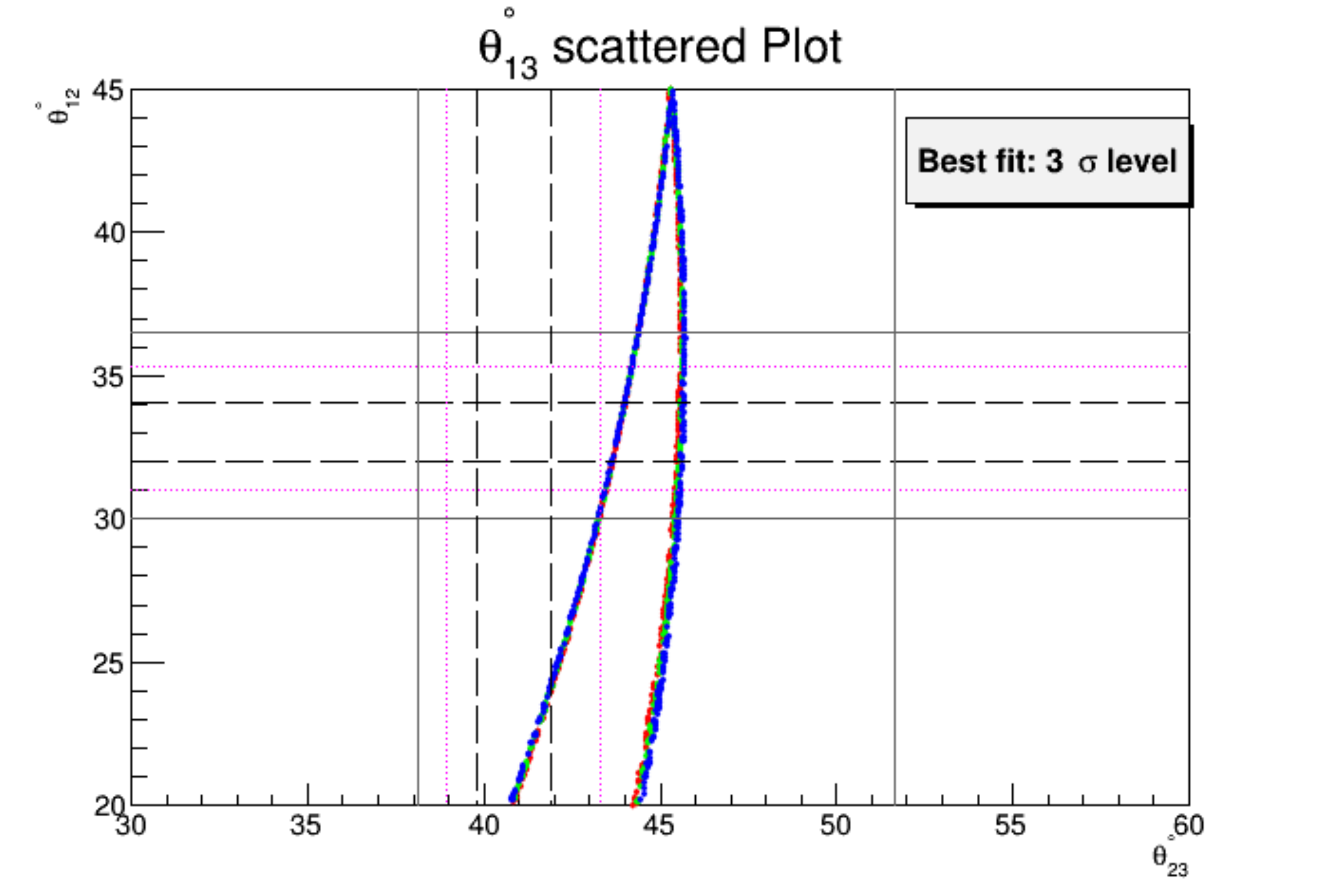}\\
\end{tabular}
\caption{\it{$U^{BML}_{1312}$ scatter plot of $\chi^2$ (left fig.) over $\gamma-\alpha$ (in radians) plane and $\theta_{13}$ (right fig.) 
over  $\theta_{23}-\theta_{12}$ (in degrees) plane. }}
\label{fig1312L.1}
\end{figure}

\begin{figure}[!t]\centering
\begin{tabular}{c c} 
\includegraphics[angle=0,width=80mm]{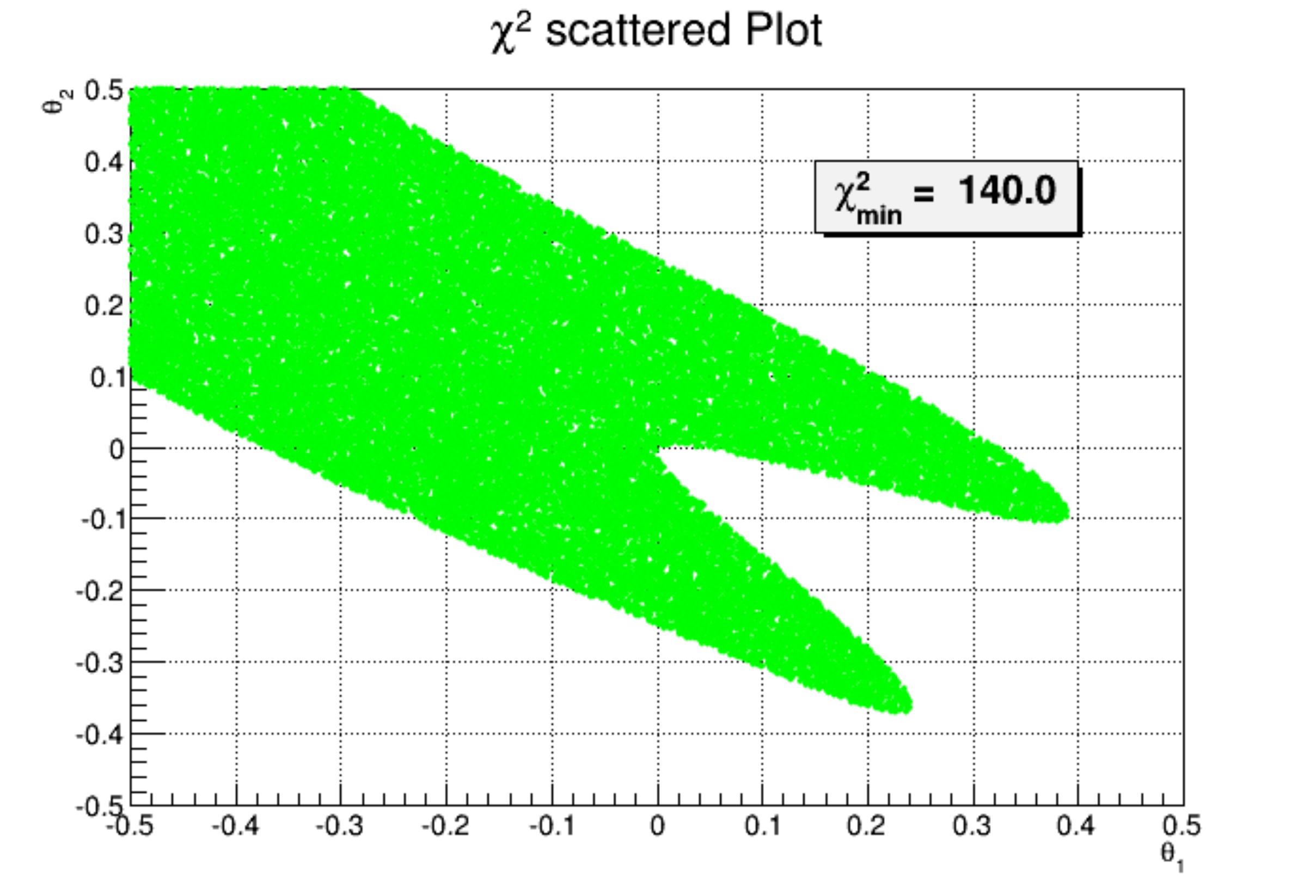} &
\includegraphics[angle=0,width=80mm]{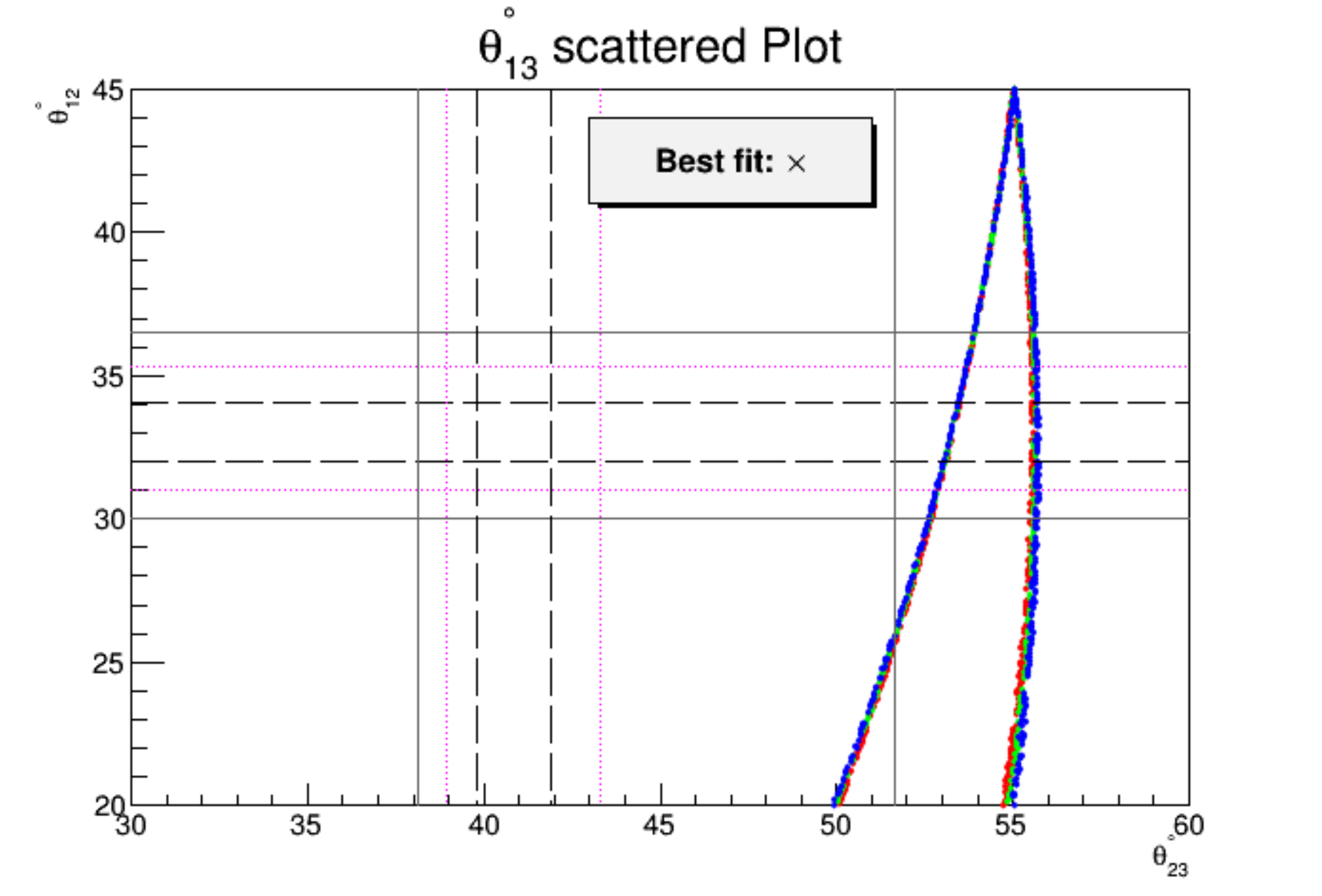}\\
\end{tabular}
\caption{\it{$U^{DCL}_{1312}$ scatter plot of $\chi^2$ (left fig.) over $\gamma-\alpha$ (in radians) plane and $\theta_{13}$ (right fig.) 
over  $\theta_{23}-\theta_{12}$ (in degrees) plane.}}
\label{fig1312L.2}
\end{figure}

\begin{figure}[!t]\centering
\begin{tabular}{c c} 
\includegraphics[angle=0,width=80mm]{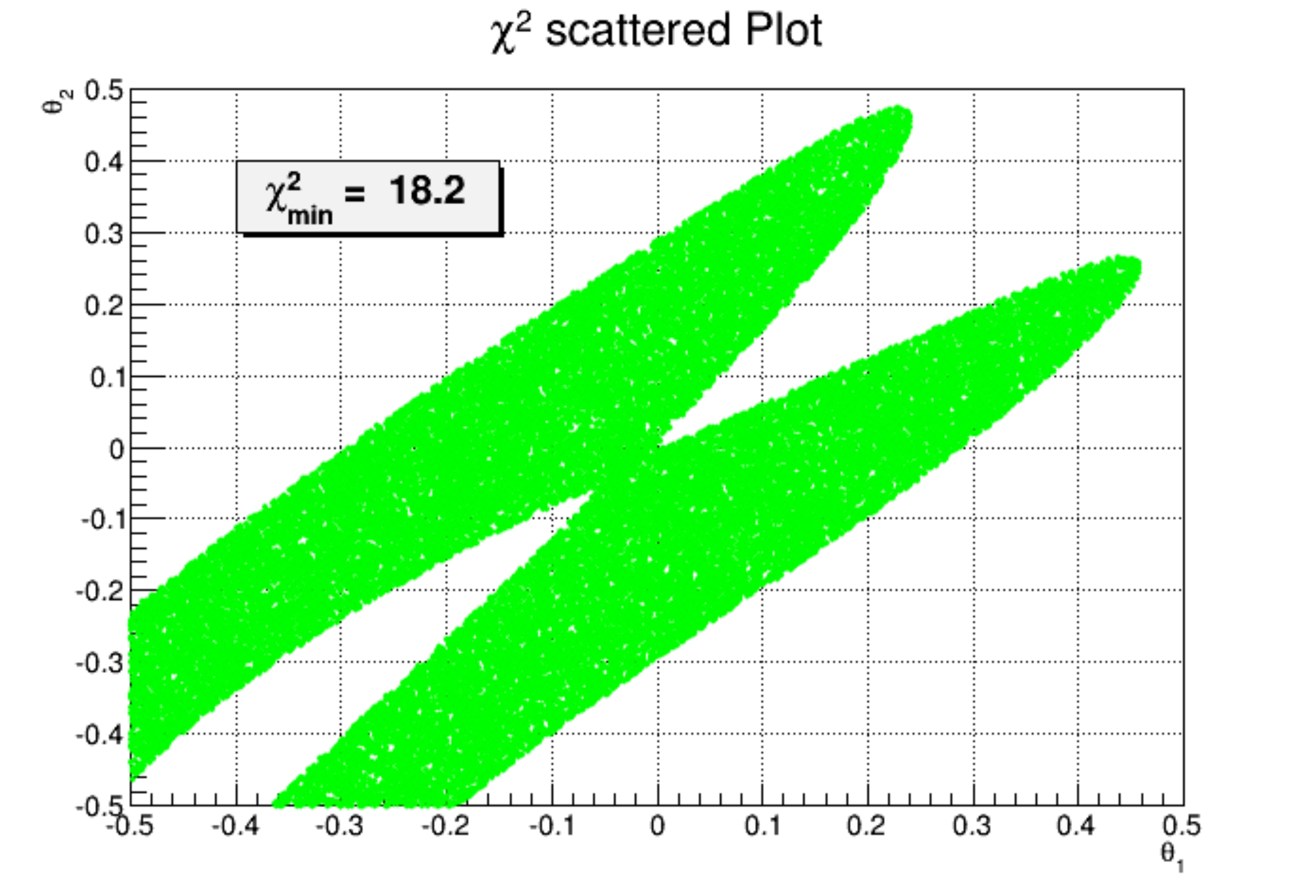} &
\includegraphics[angle=0,width=80mm]{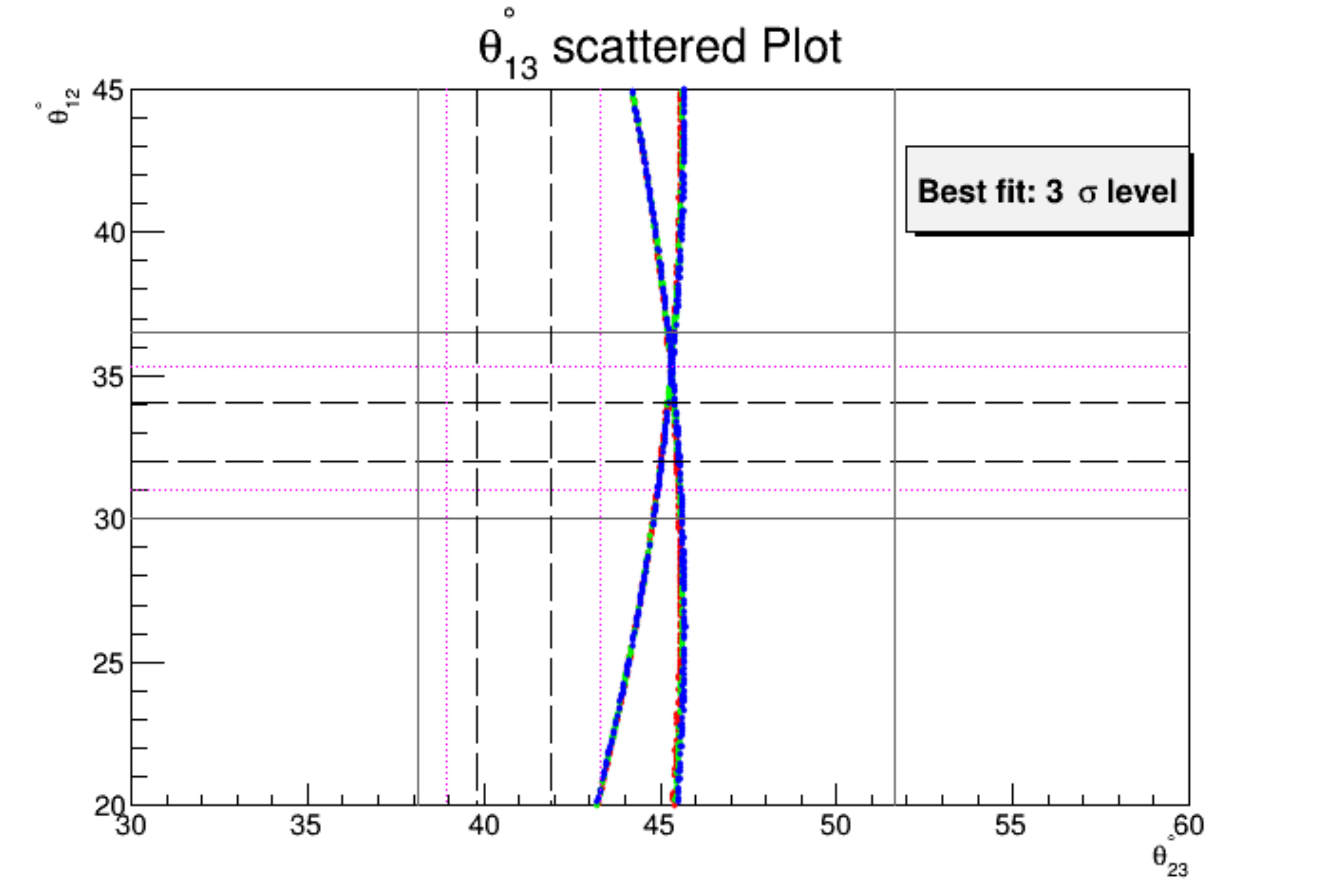}\\
\end{tabular}
\caption{\it{$U^{TBML}_{1312}$ scatter plot of $\chi^2$ (left fig.) over $\gamma-\alpha$ (in radians) plane and $\theta_{13}$ (right fig.) 
over $\theta_{23}-\theta_{12}$ (in degrees) plane. }}
\label{fig1312L.3}
\end{figure}

Figs.~\ref{fig1312L.1}-\ref{fig1312L.3} corresponds to BM, DC and TBM case respectively with $\theta_1 = \gamma$ and $\theta_2 =\alpha$.
The main characteristics in this perturbative scheme are given by:\\
{\bf{(i)}} The case is much similar to 12-13 rotation except for $\theta_{23}$ where in previous case it got additional O($\theta^2$) 
correction term. Thus for this case also $\theta_{23}$ remains close to its unperturbed value.\\
{\bf{(ii)}} The minimum value of $\chi^2 \sim 8.8$, $140.0$ and $18.2$ for perturbative BM, DC and TBM case respectively.\\
{\bf{(iii)}} For TBM and BM case its possible to fit all mixing angles at $3\sigma$ level while 
DC case is not consistent even at 3$\sigma$ level.

\subsection{13-23 Rotation}

\begin{figure}[!t]\centering
\begin{tabular}{c c} 
\includegraphics[angle=0,width=80mm]{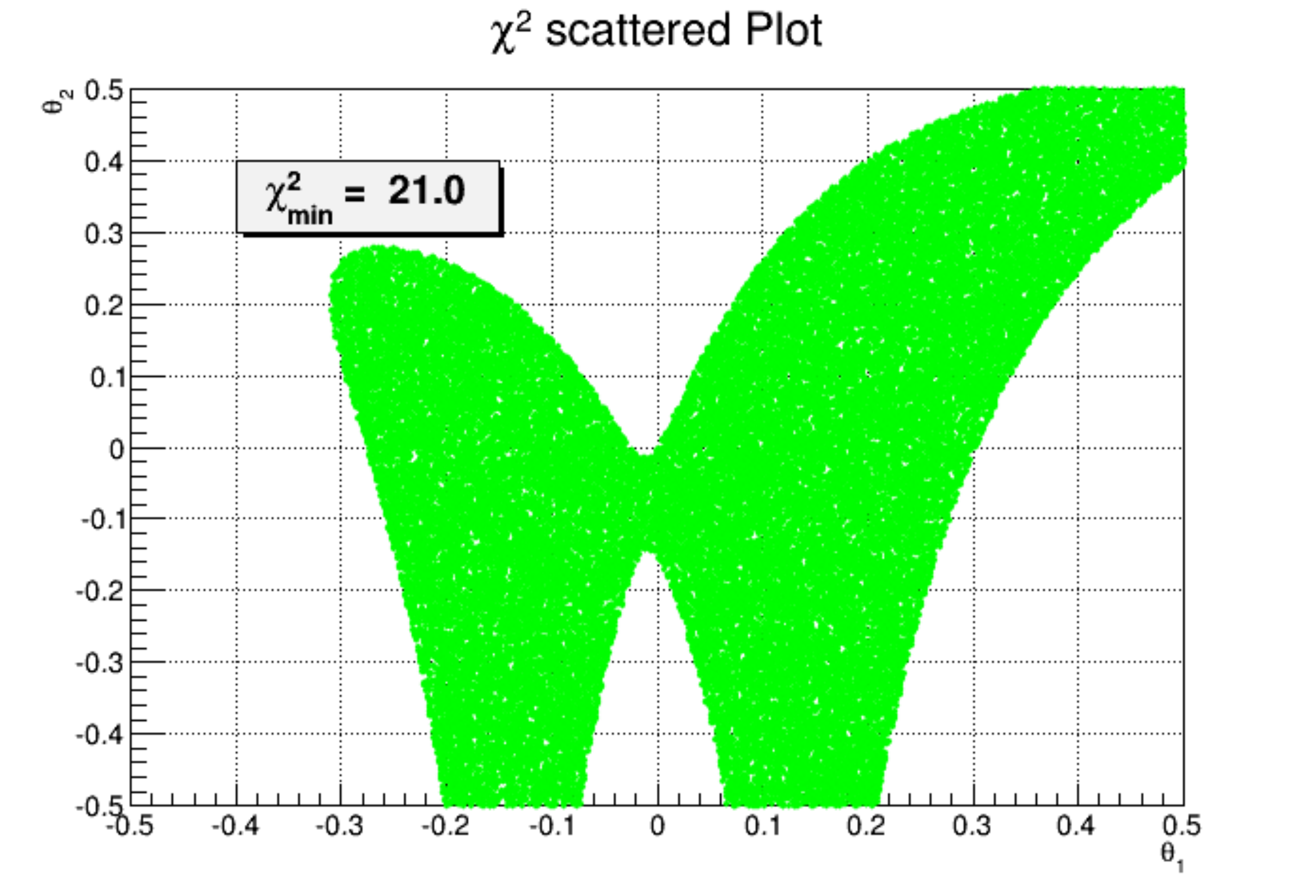} &
\includegraphics[angle=0,width=80mm]{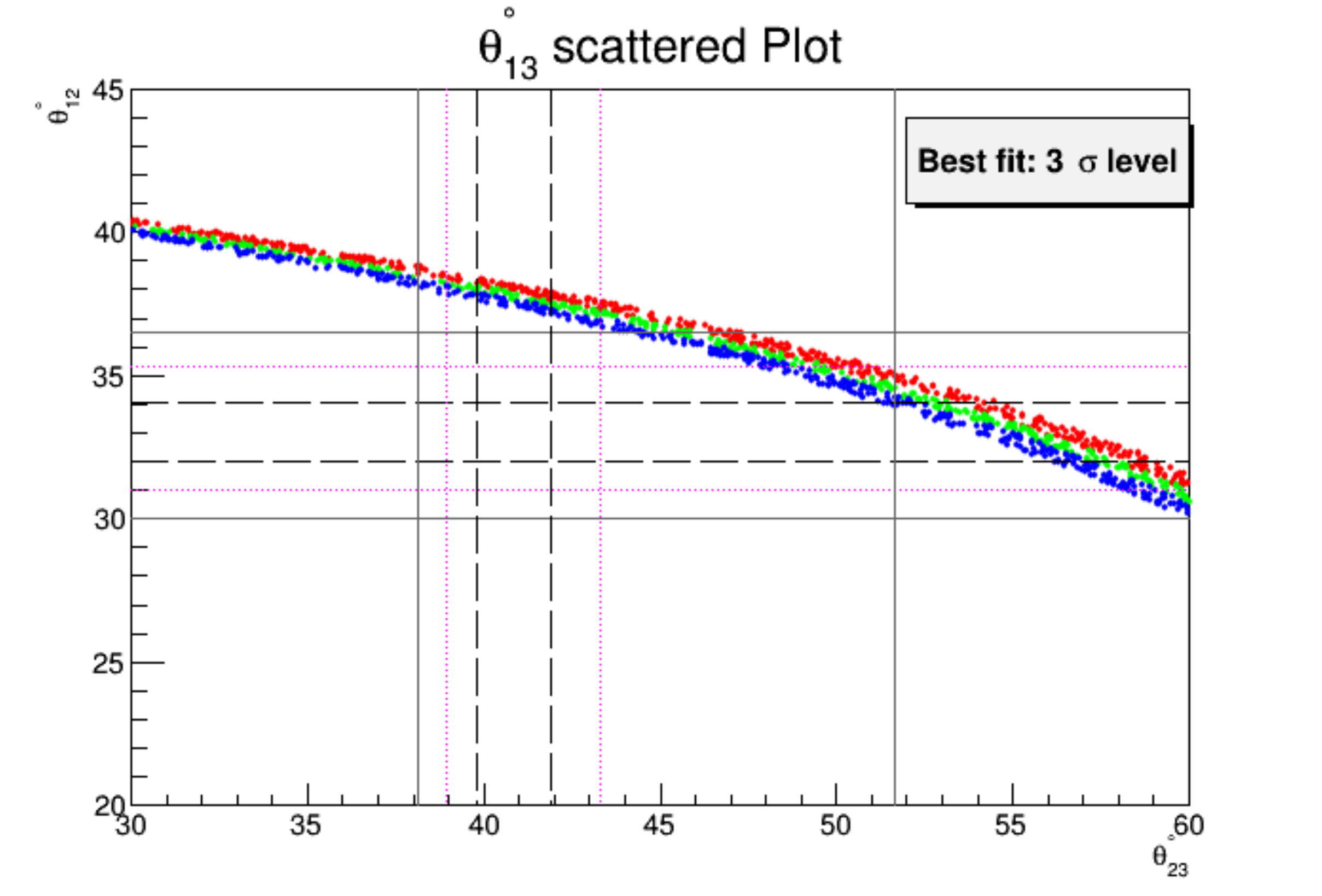}\\
\end{tabular}
\caption{\it{$U^{BML}_{1323}$ scatter plot of $\chi^2$ (left fig.) over $\gamma-\beta$ (in radians) plane and $\theta_{13}$ (right fig.) 
over  $\theta_{23}-\theta_{12}$ (in degrees) plane. }}
\label{fig1323L.1}
\end{figure}

\begin{figure}[!t]\centering
\begin{tabular}{c c} 
\includegraphics[angle=0,width=80mm]{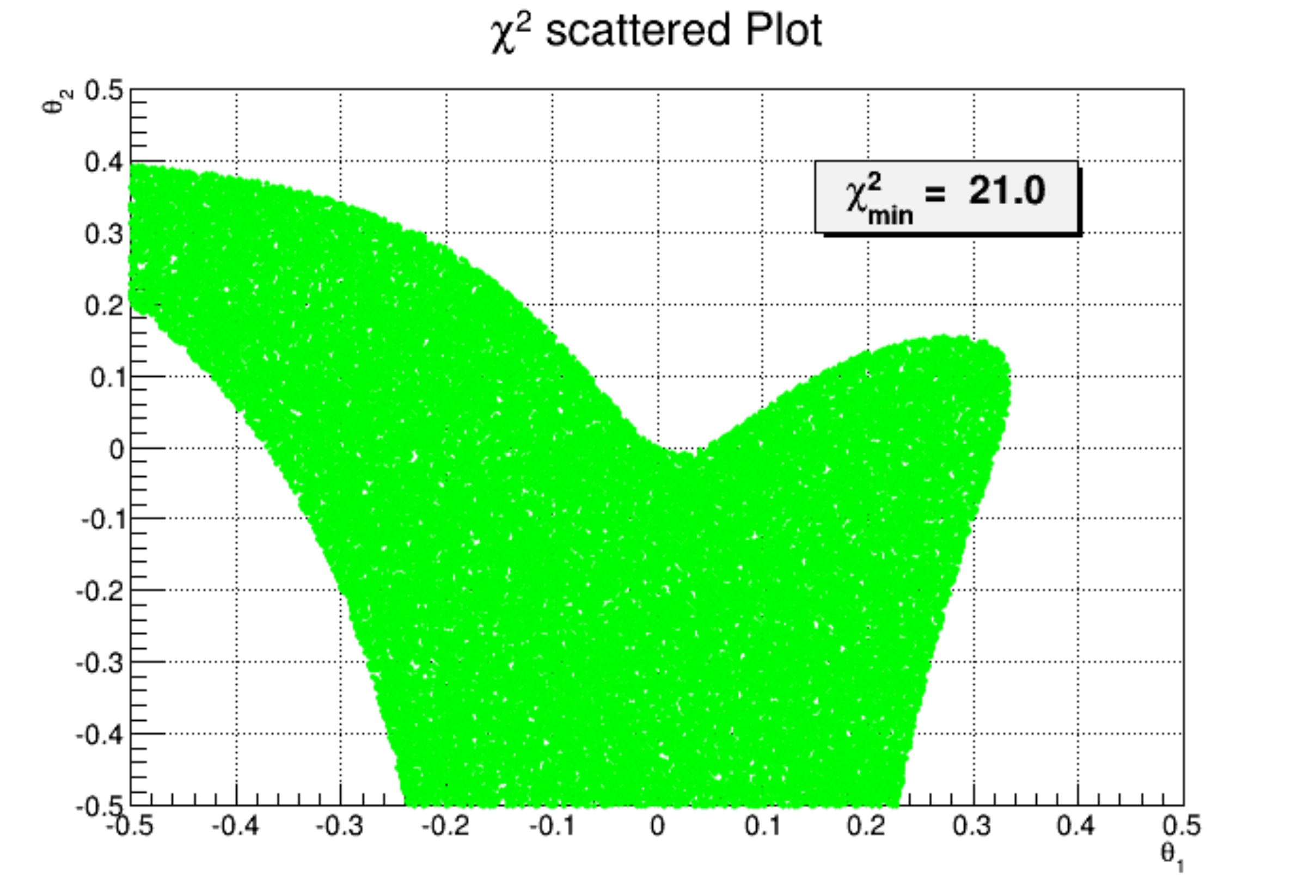} &
\includegraphics[angle=0,width=80mm]{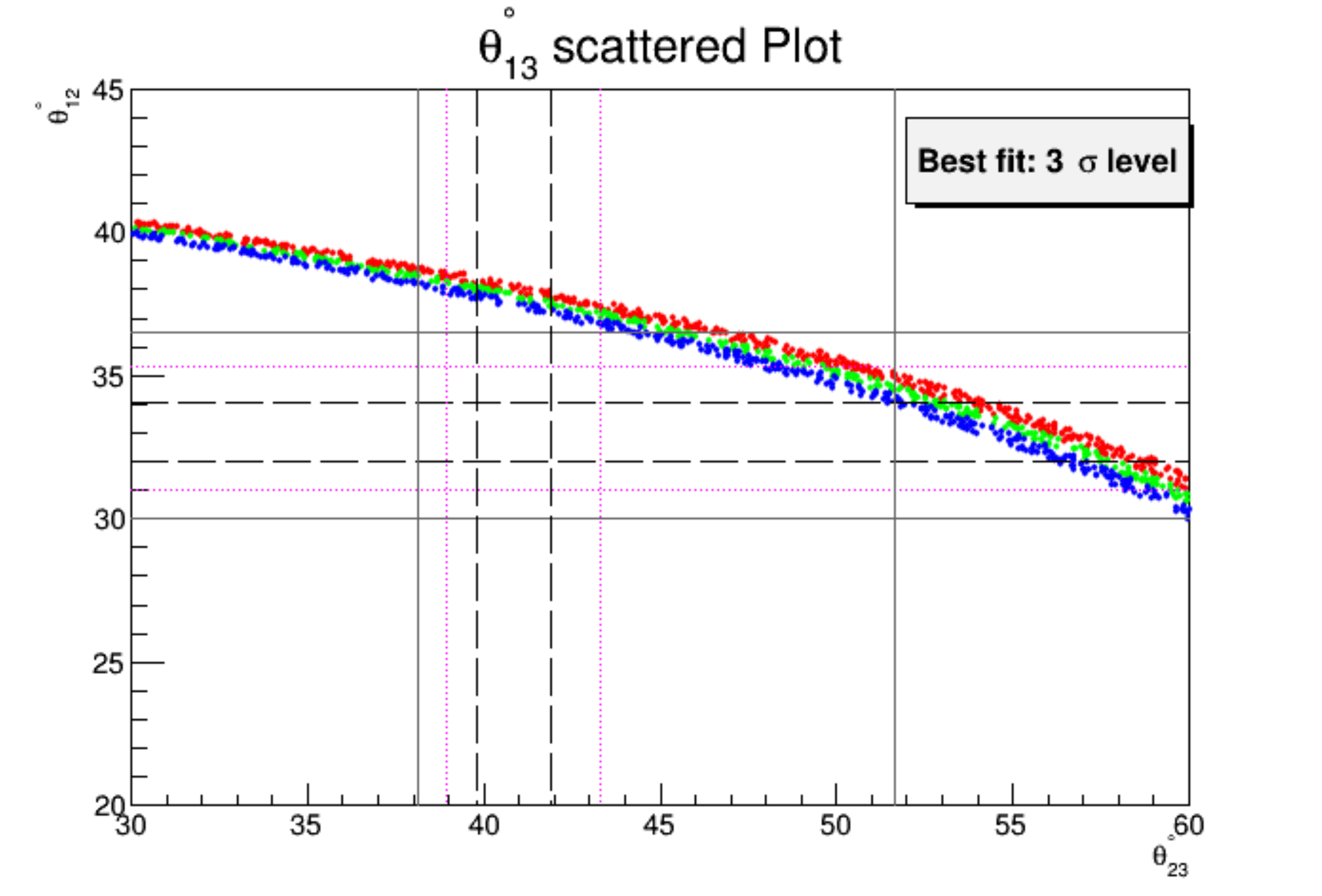}\\
\end{tabular}
\caption{\it{$U^{DCL}_{1323}$ scatter plot of $\chi^2$ (left fig.) over $\gamma-\beta$ (in radians) plane and $\theta_{13}$ (right fig.) 
over  $\theta_{23}-\theta_{12}$ (in degrees) plane. }}
\label{fig1323L.2}
\end{figure}

\begin{figure}[!t]\centering
\begin{tabular}{c c} 
\includegraphics[angle=0,width=80mm]{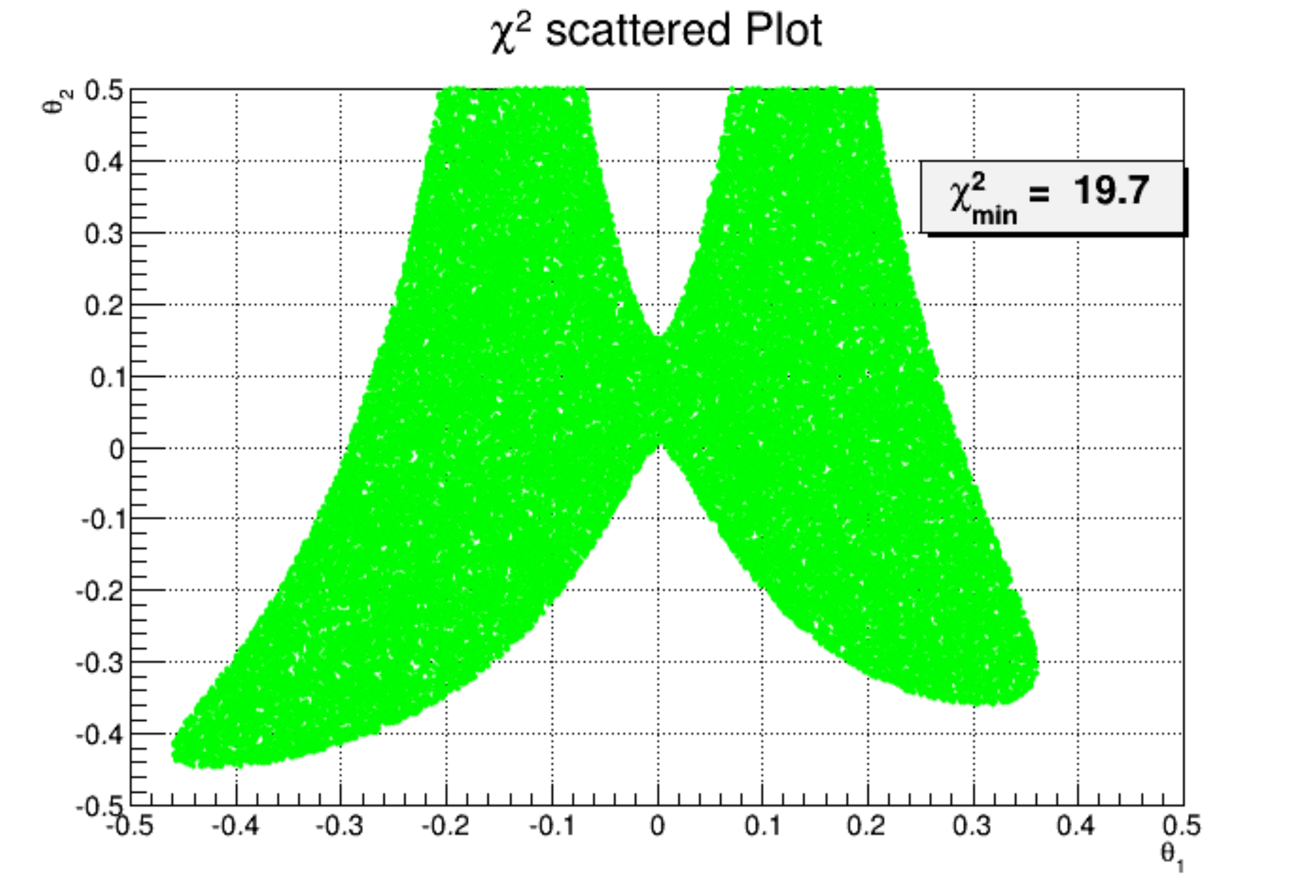} &
\includegraphics[angle=0,width=80mm]{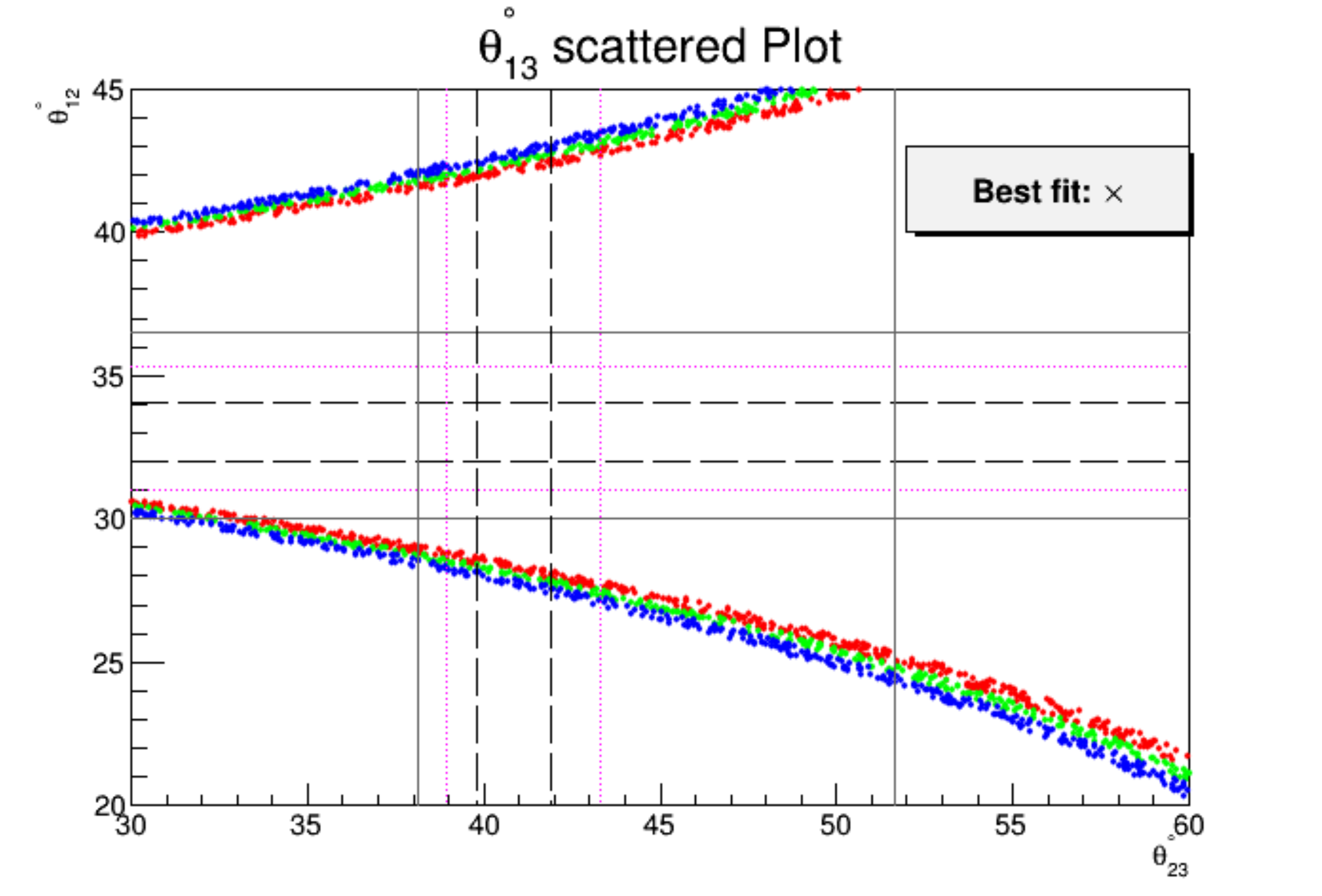}\\
\end{tabular}
\caption{\it{$U^{TBML}_{1323}$ scatter plot of $\chi^2$ (left fig.) over $\gamma-\beta$ (in radians) plane and $\theta_{13}$ (right fig.) 
over $\theta_{23}-\theta_{12}$ (in degrees) plane. }}
\label{fig1323L.3}
\end{figure}

This case corresponds to rotations in 13 and 23 sector of  these special matrices. 
The neutrino mixing angles for small perturbation parameters $\gamma$ and $\beta$ are given by

\beqa
 \sin\theta_{13} &\approx&  |\gamma U_{33} -\beta\gamma U_{23}|,\\
 \sin\theta_{23} &\approx& |\frac{ U_{23} + \beta U_{33} -\beta^2 U_{23}  }{\cos\theta_{13}}|,\\
 \sin\theta_{12} &\approx& |\frac{U_{12} + \gamma U_{32}-\gamma^2 U_{12}-\beta\gamma U_{22}}{\cos\theta_{13}}|.
\eeqa
Figs.~\ref{fig1323L.1}-\ref{fig1323L.3} corresponds to BM, DC and TBM case respectively with $\theta_1 = \gamma$ and $\theta_2 = \beta$.
The following features define this perturbation scheme:\\
{\bf{(i)}} The parameters $\beta$ and $\gamma$ enters into all mixing angles at leading order and thus show good correlations among
themselves. \\ 
{\bf{(ii)}} The minimum value of $\chi^2 \sim 21.0$, $21.0$ and $19.7$ for BM, DC and TBM case respectively.\\
{\bf{(iii)}} For this perturbative scheme, TBM is excluded while BM and DC is only consistent at 3$\sigma$ level.

\subsection{23-12 Rotation}

This case corresponds to rotations in 23 and 12 sector of  these special matrices.

\beqa
 \sin\theta_{13} &\approx&  |\alpha U_{23}  |,\\
 \sin\theta_{23} &\approx& |\frac{ U_{23} + \beta U_{33}-(\alpha^2 + \beta^2 )U_{23} }{\cos\theta_{13}}|,\\
 \sin\theta_{12} &\approx& |\frac{U_{12} + \alpha U_{22} -\alpha^2 U_{12} }{\cos\theta_{13}}|.
\eeqa

\begin{figure}[!t]\centering
\begin{tabular}{c c} 
\includegraphics[angle=0,width=80mm]{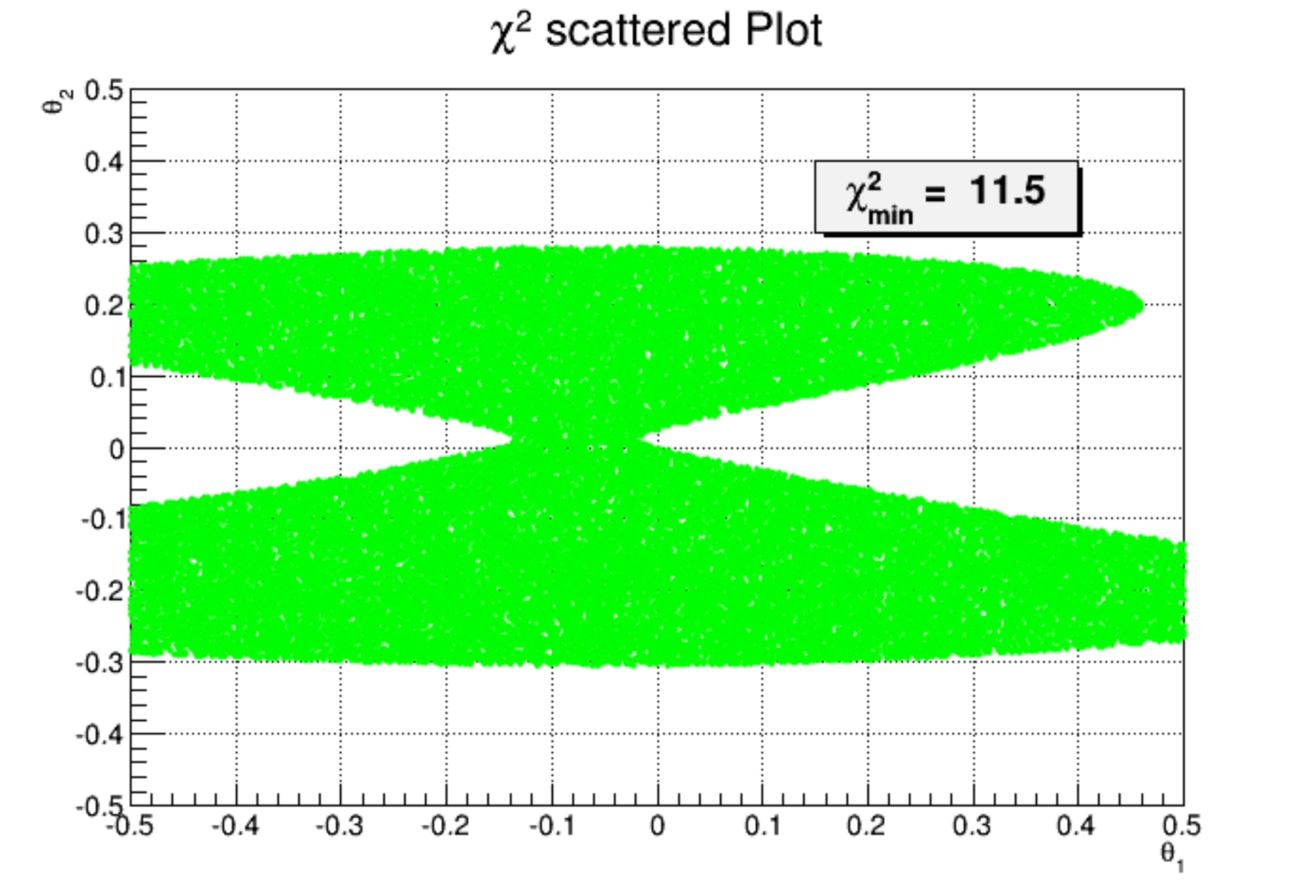} &
\includegraphics[angle=0,width=80mm]{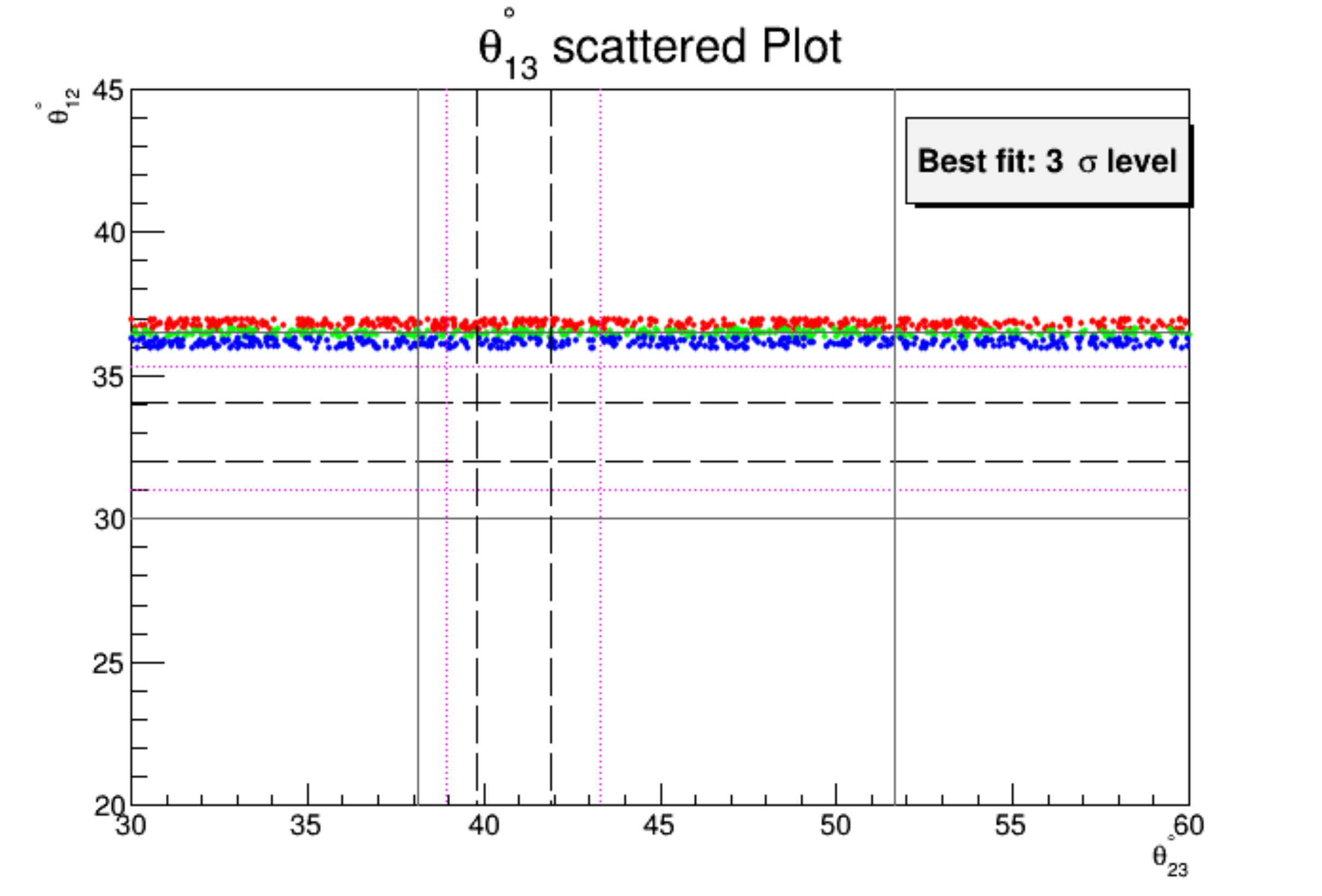}\\
\end{tabular}
\caption{\it{$U^{BML}_{2312}$ scatter plot of $\chi^2$ (left fig.) over $\beta-\alpha$ (in radians) plane and $\theta_{13}$ (right fig.) 
over  $\theta_{23}-\theta_{12}$ (in degrees) plane. }}
\label{fig2312L.1}
\end{figure}

\begin{figure}[!t]\centering
\begin{tabular}{c c} 
\includegraphics[angle=0,width=80mm]{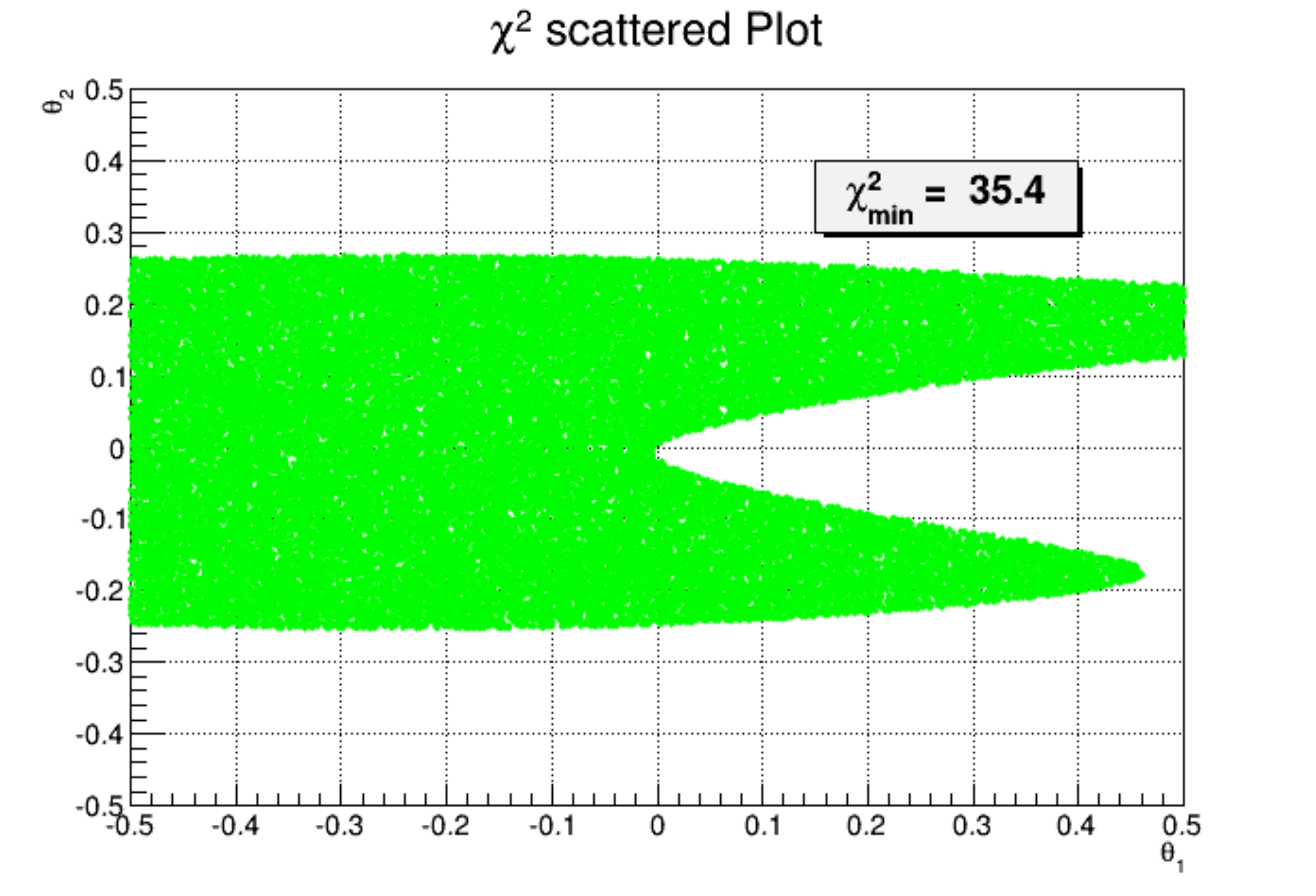} &
\includegraphics[angle=0,width=80mm]{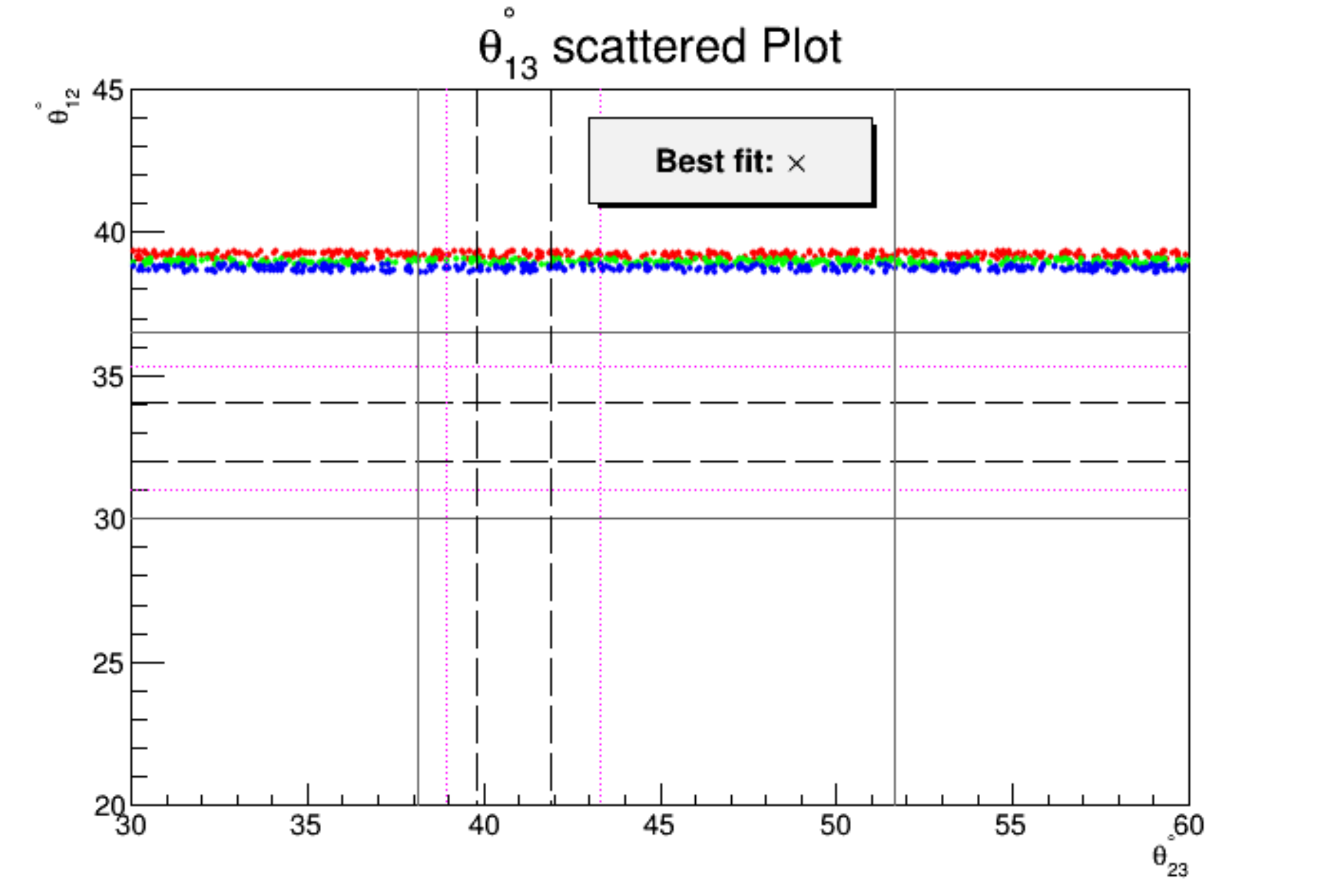}\\
\end{tabular}
\caption{\it{$U^{DCL}_{2312}$ scatter plot of $\chi^2$ (left fig.) over $\beta-\alpha$ (in radians) plane and $\theta_{13}$ (right fig.) 
over  $\theta_{23}-\theta_{12}$ (in degrees) plane. }}
\label{fig2312L.2}
\end{figure}

\begin{figure}[!t]\centering
\begin{tabular}{c c} 
\includegraphics[angle=0,width=80mm]{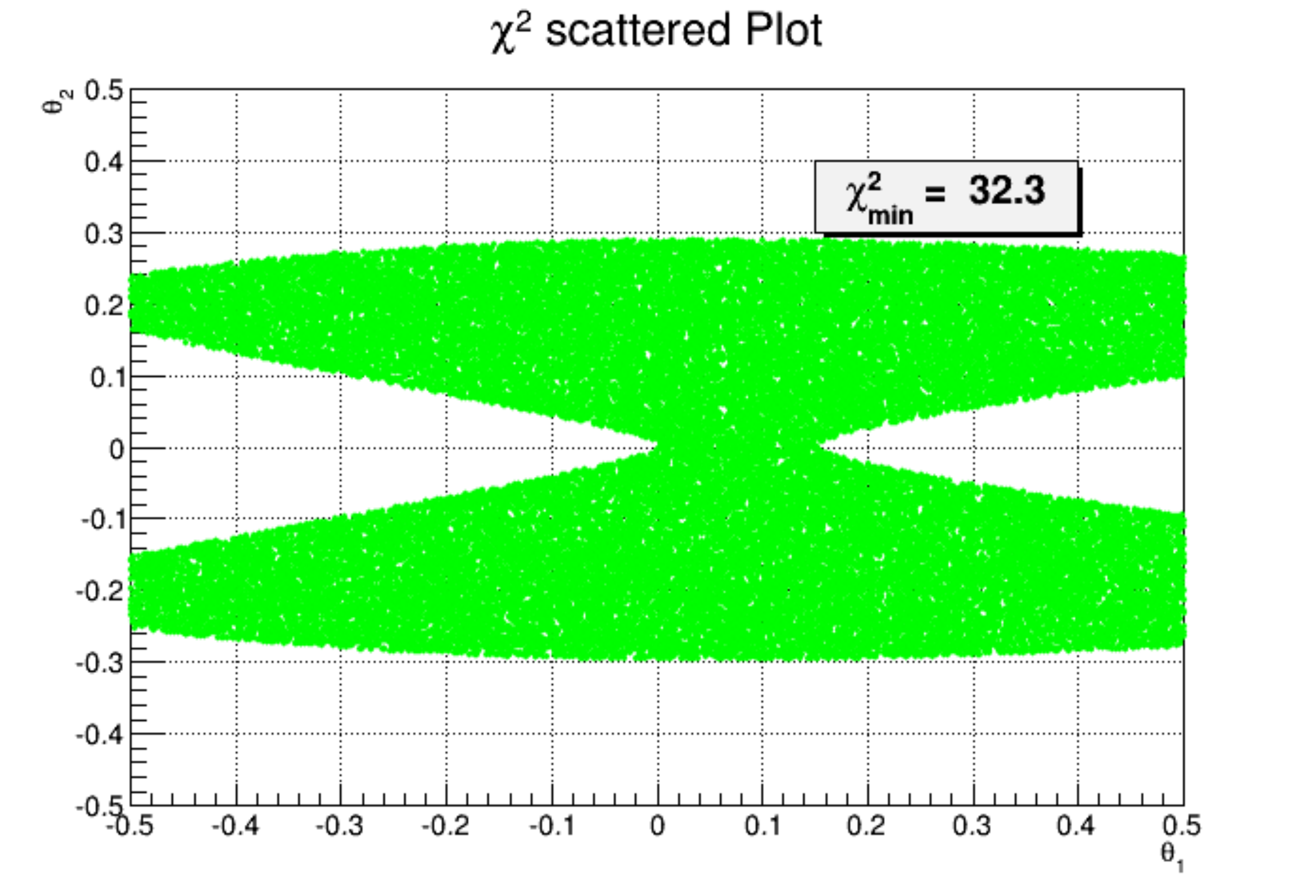} &
\includegraphics[angle=0,width=80mm]{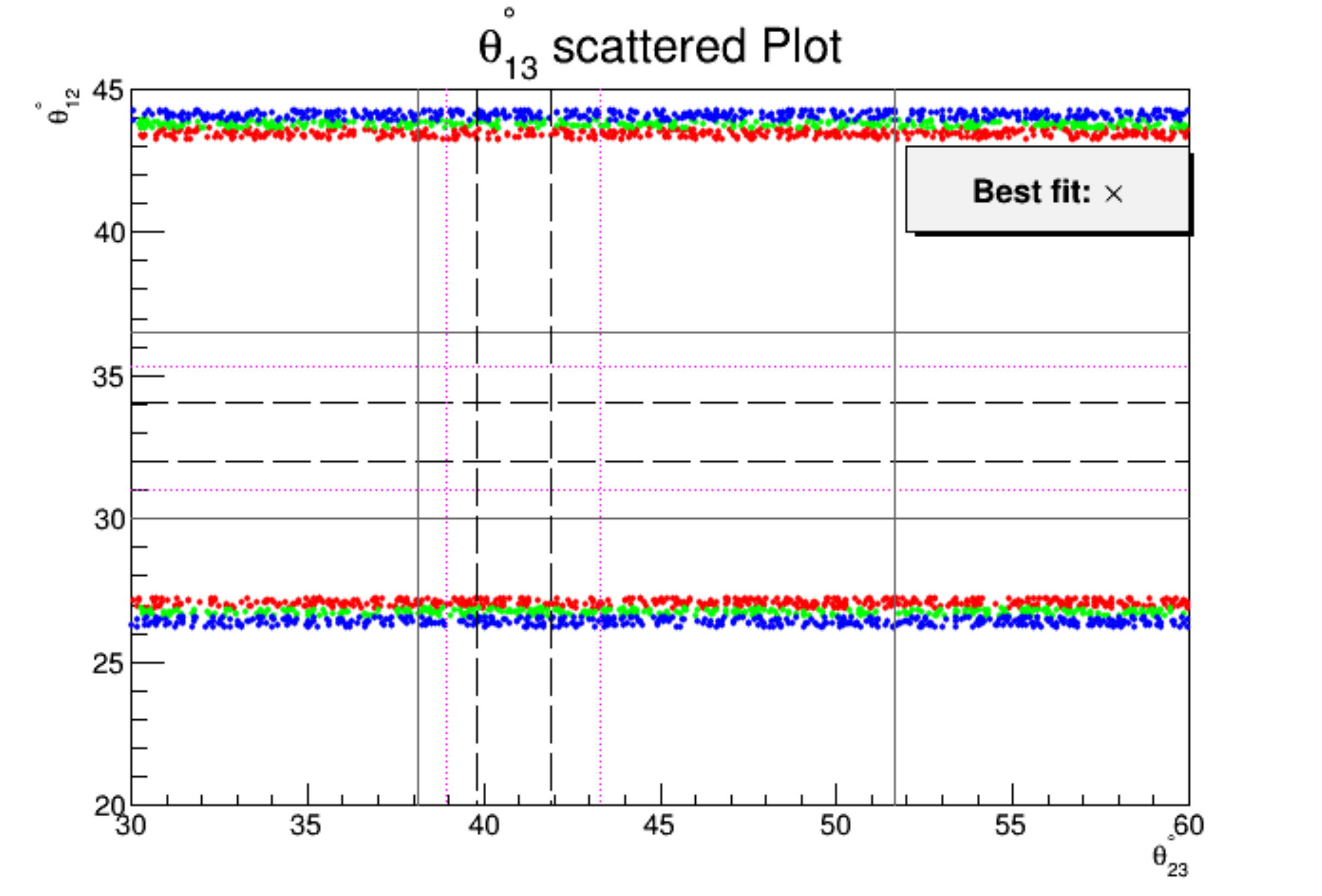}\\
\end{tabular}
\caption{\it{$U^{TBML}_{2312}$ scatter plot of $\chi^2$ (left fig.) over $\beta-\alpha$ (in radians) plane and $\theta_{13}$ (right fig.) 
over $\theta_{23}-\theta_{12}$ (in degrees) plane. }}
\label{fig2312L.3}
\end{figure}

Figs.~\ref{fig2312L.1}-\ref{fig2312L.3} corresponds to BM, DC and TBM case respectively with $\theta_1 = \beta$ and $\theta_2 = \alpha$.
The following are the main characteristics of this pertubative scheme:\\
{\bf{(i)}} The corrections to mixing angle $\theta_{13}$ and $\theta_{12}$ is only governed by perturbation parameter
$\alpha$. Thus magnitude of parameter $\alpha$ is tightly constrained from fitting of $\theta_{13}$. This in turn permits very narrow ranges
for $\theta_{12}$ corresponding to negative and positive values of $\alpha$ in parameter space. However $\theta_{23}$ mainly depends on $\beta$
and thus can have wide range of values in parameter space.\\
{\bf{(ii)}} The minimum value of $\chi^2 \sim 11.5$, $35.4$ and $32.3$ for 
BM, DC and TBM case respectively.\\
{\bf{(iii)}} The case of TBM and DC is completely excluded while BM is still viable at $3\sigma$ level.

\subsection{23-13 Rotation}

For this perturbative scheme, the neutrino mixing angles for small correction parameters $\beta$ and $\gamma$ are given by

\beqa
 \sin\theta_{13} &\approx&  |\gamma U_{33}|,\\
 \sin\theta_{23} &\approx& |\frac{U_{23} + \beta U_{33} -\beta^2 U_{23}}{\cos\theta_{13}}|,\\
 \sin\theta_{12} &\approx& |\frac{U_{12} + \gamma U_{32}-\gamma^2 U_{12} }{\cos\theta_{13}}|.
 \eeqa

\begin{figure}[!t]\centering
\begin{tabular}{c c} 
\includegraphics[angle=0,width=80mm]{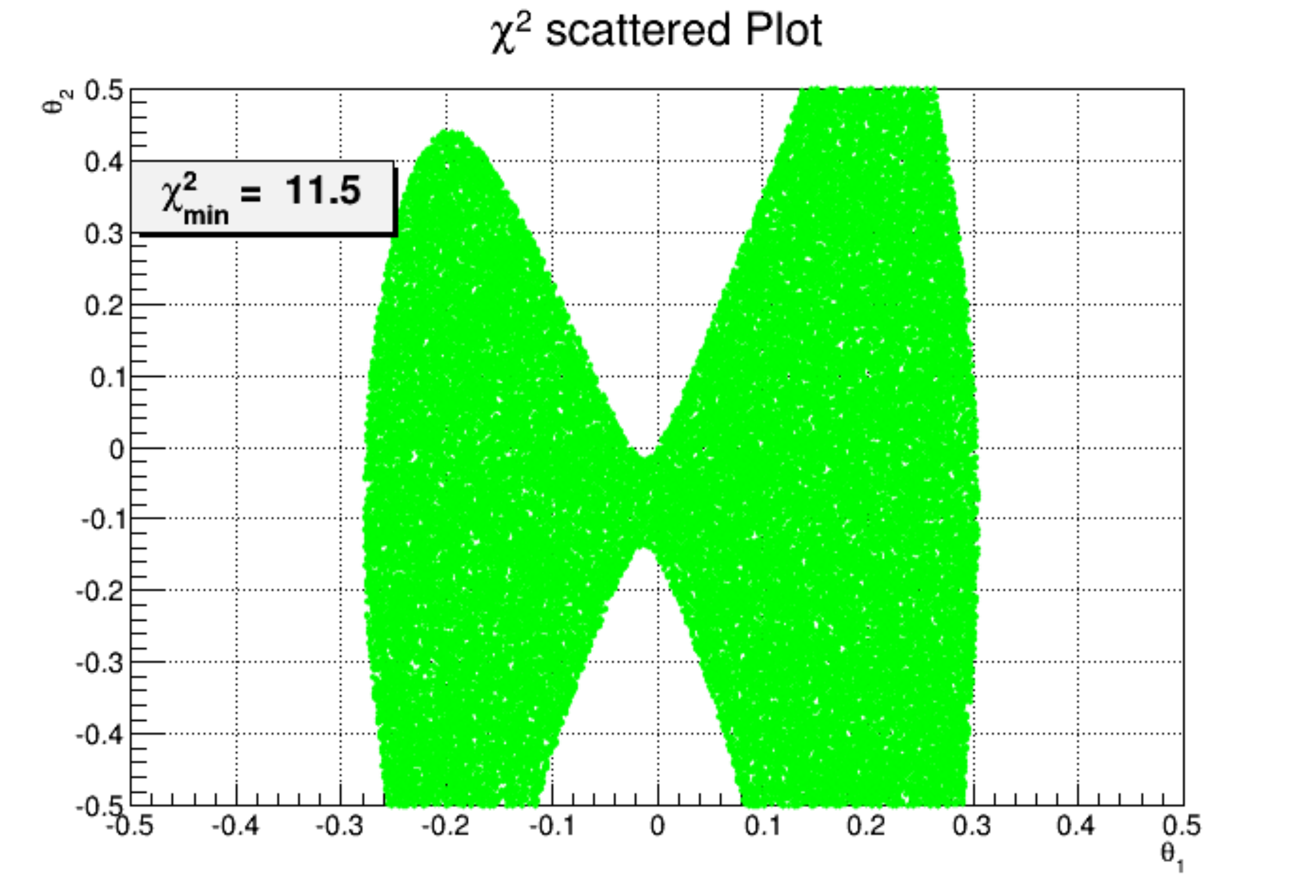} &
\includegraphics[angle=0,width=80mm]{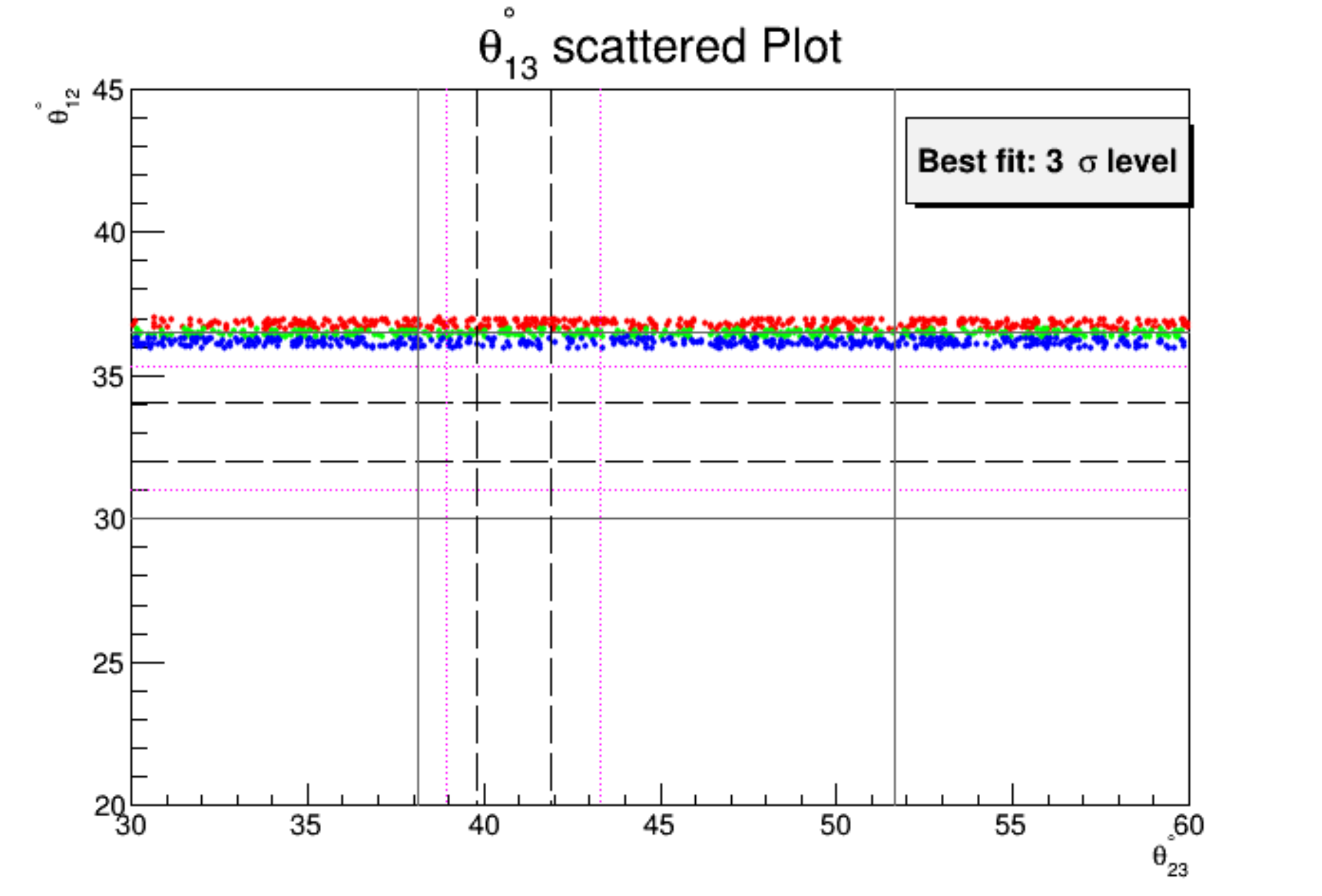}\\
\end{tabular}
\caption{\it{$U^{BML}_{2313}$ scatter plot of $\chi^2$ (left fig.) over $\beta-\gamma$ (in radians) plane and $\theta_{13}$ (right fig.) 
over  $\theta_{23}-\theta_{12}$ (in degrees) plane.}}
\label{fig2313L.1}
\end{figure}

\begin{figure}[!t]\centering
\begin{tabular}{c c} 
\includegraphics[angle=0,width=80mm]{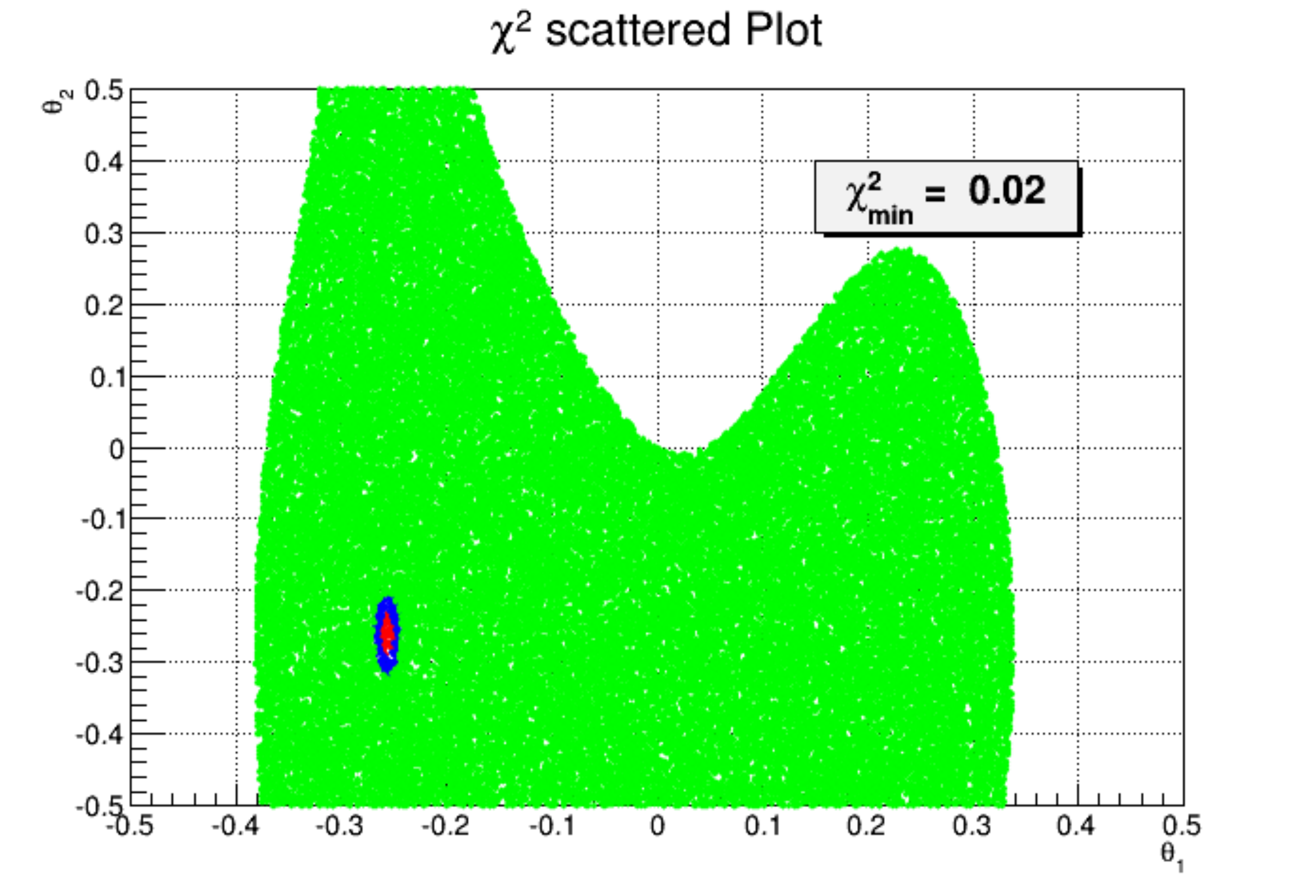} &
\includegraphics[angle=0,width=80mm]{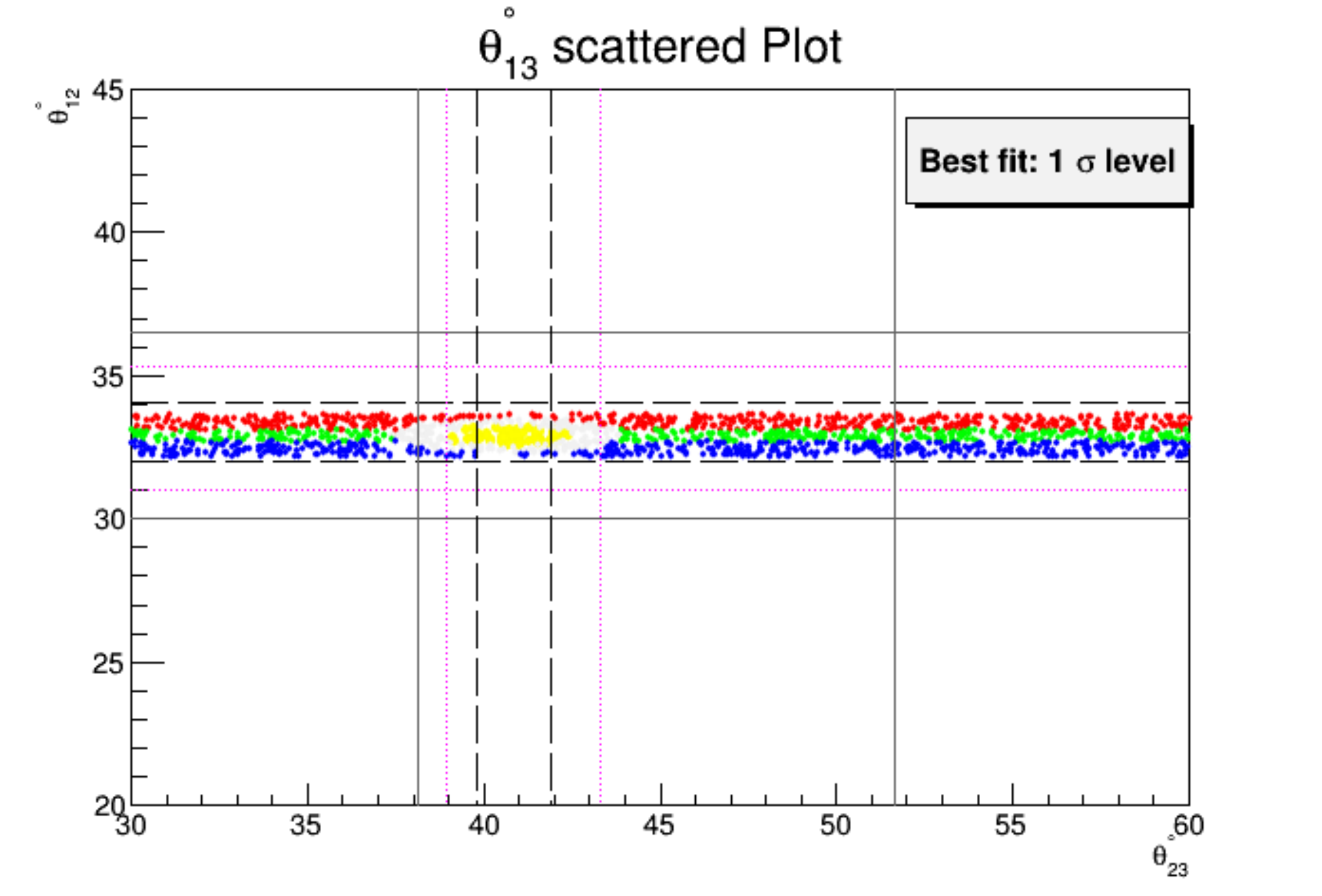}\\
\end{tabular}
\caption{\it{$U^{DCL}_{2313}$ scatter plot of $\chi^2$ (left fig.) over $\beta-\gamma$ (in radians) plane and $\theta_{13}$ (right fig.) 
over  $\theta_{23}-\theta_{12}$ (in degrees) plane. }}
\label{fig2313L.2}
\end{figure}

\begin{figure}[!t]\centering
\begin{tabular}{c c} 
\includegraphics[angle=0,width=80mm]{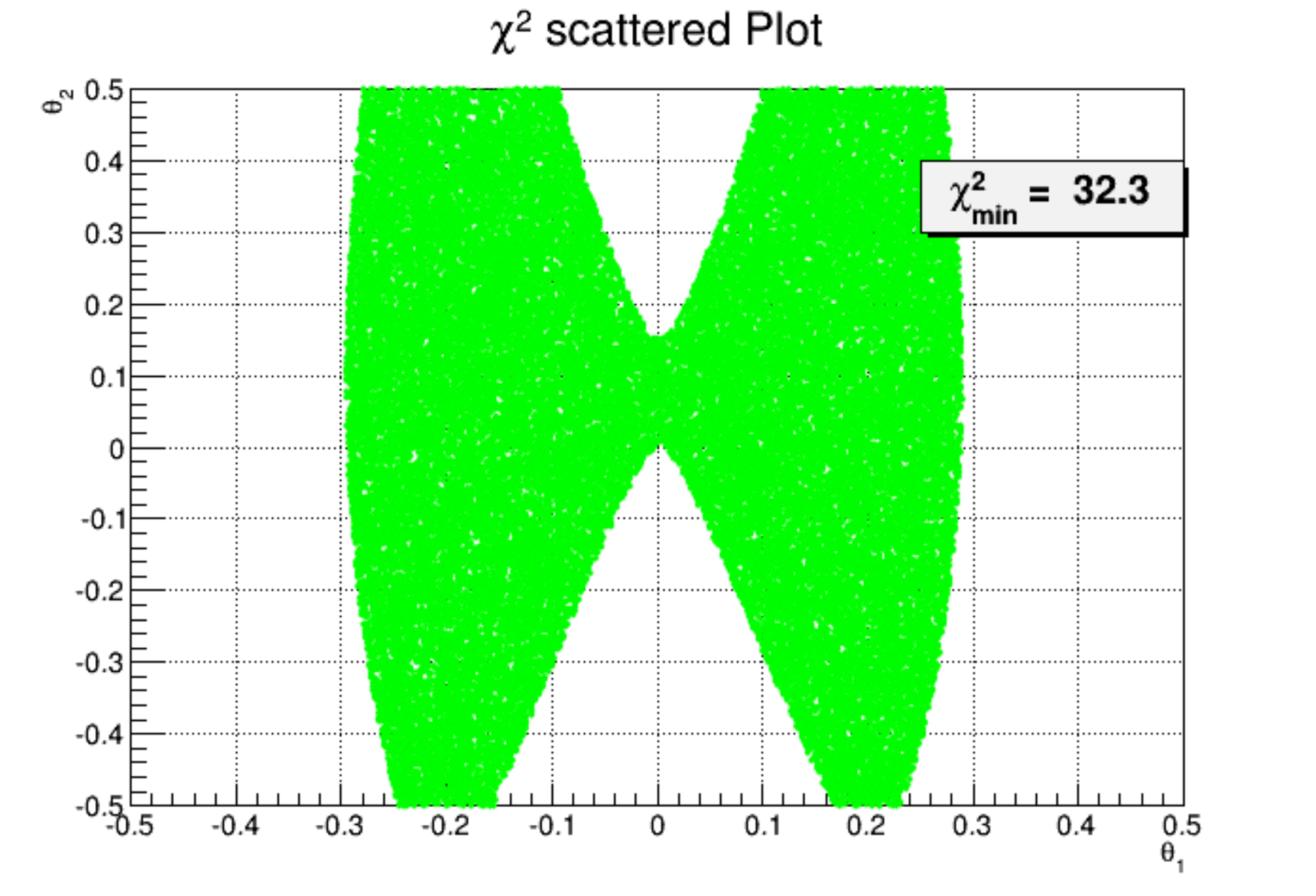} &
\includegraphics[angle=0,width=80mm]{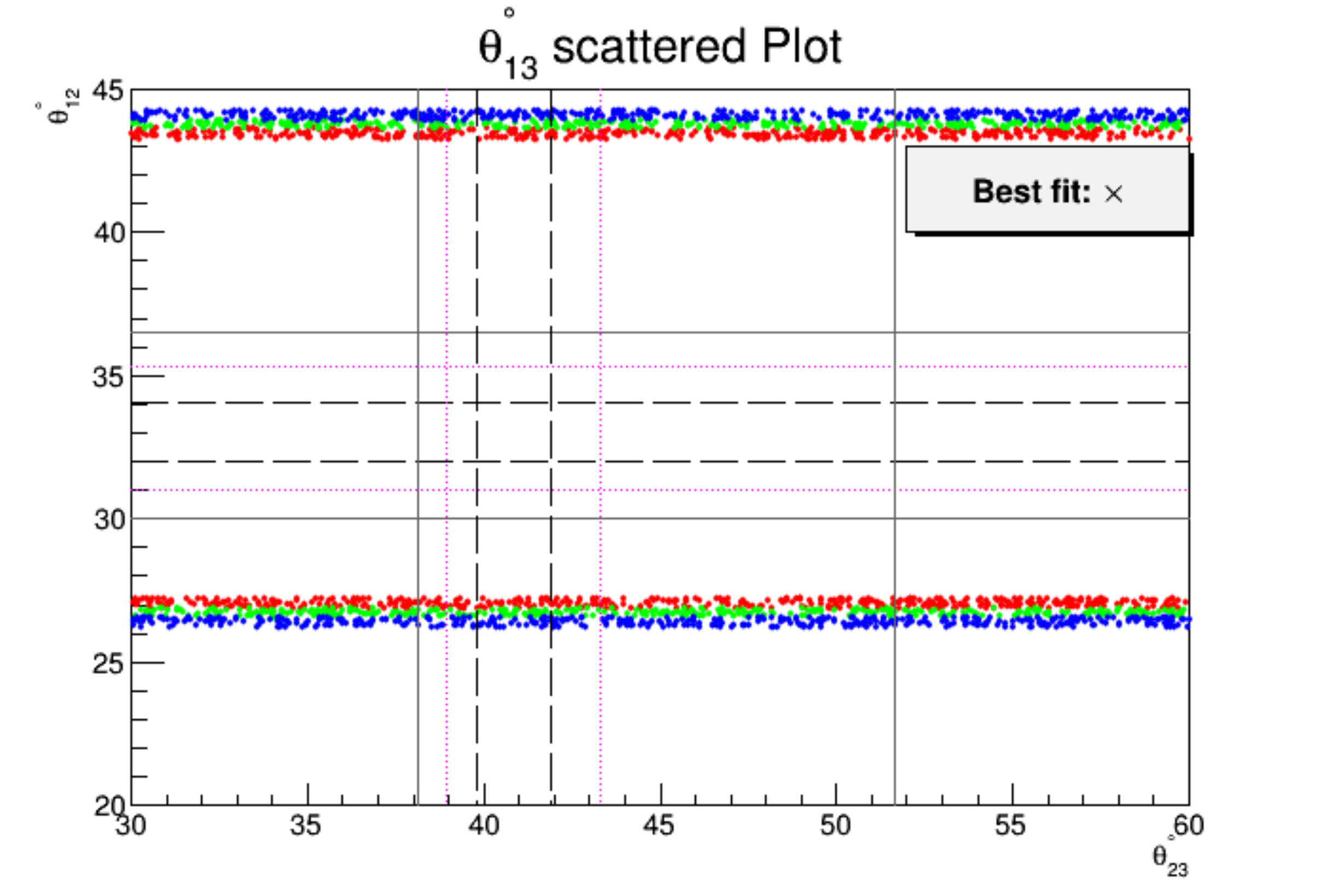}\\
\end{tabular}
\caption{\it{$U^{TBML}_{2313}$ scatter plot of $\chi^2$ (left fig.) over $\beta-\gamma$ (in radians) plane and $\theta_{13}$ (right fig.) 
over  $\theta_{23}-\theta_{12}$ (in degrees) plane. }}
\label{fig2313L.3}
\end{figure}

Figs.~\ref{fig2313L.1}-\ref{fig2313L.3} corresponds to BM, DC and TBM case respectively with $\theta_1 = \gamma$ and $\theta_2 = \beta$.\\
The main characteristics of this scheme are:\\
{\bf{(i)}} The modifications to mixing angle $\theta_{13}$ and $\theta_{12}$ is only governed by perturbation parameter
$\gamma$. Thus magnitude of parameter $\gamma$ is tightly constrainted from fitting of $\theta_{13}$. This in turn allows very narrow ranges
for $\theta_{12}$ corresponding to negative and positive values of $\gamma$ in parameter space.  However $\theta_{23}$ solely depends on $\beta$ and thus
can have wide range of possible values in parameter space.\\
{\bf{(ii)}} For this rotation scheme its possible to get $\chi^2 < 1$ for DC in tiny viable parameter space while for BM and TBM cases $\chi^2 > 11$
and $\chi^2 > 32$ respectively. \\
{\bf{(iii)}} The minimum value of $\chi^2 \sim 11.5$, $0.02$ and $32.3$ for BM, DC and TBM case respectively.\\
{\bf{(iv)}} The TBM case is completely excluded as it prefers much lower value of $\theta_{12}$ while BM can only be consistent
at 3$\sigma$ level. Here DC case is most preferable as it can fit all mixing angles within 1$\sigma$ range.

\section{Rotations-$U.R_{ij}^r.R_{kl}^r$}

Now we comes to the perturbations for which modified PMNS matrix is given by $U_{PMNS} = U.R_{ij}^r.R_{kl}^r$. We will 
investigate the role of these corrections in fitting the neutrino mixing data.  

\subsection{12-13 Rotation}

This case corresponds to rotations in 12 and 13 sector of  these special matrices. The neutrino mixing angles
truncated at order O ($\theta^2$) for this mixing scheme is given by

\beqa
 \sin\theta_{13} &\approx&  |-\gamma U_{11} + \alpha \gamma U_{12}|,\\
 \sin\theta_{23} &\approx& |\frac{ U_{23} + \gamma U_{21}-\gamma^2 U_{23}-\alpha\gamma U_{22}  }{\cos\theta_{13}}|,\\
 \sin\theta_{12} &\approx& |\frac{U_{12} + \alpha U_{11}-\alpha^2 U_{12}}{\cos\theta_{13}}|.
\eeqa

\begin{figure}[!t]\centering
\begin{tabular}{c c} 
\includegraphics[angle=0,width=80mm]{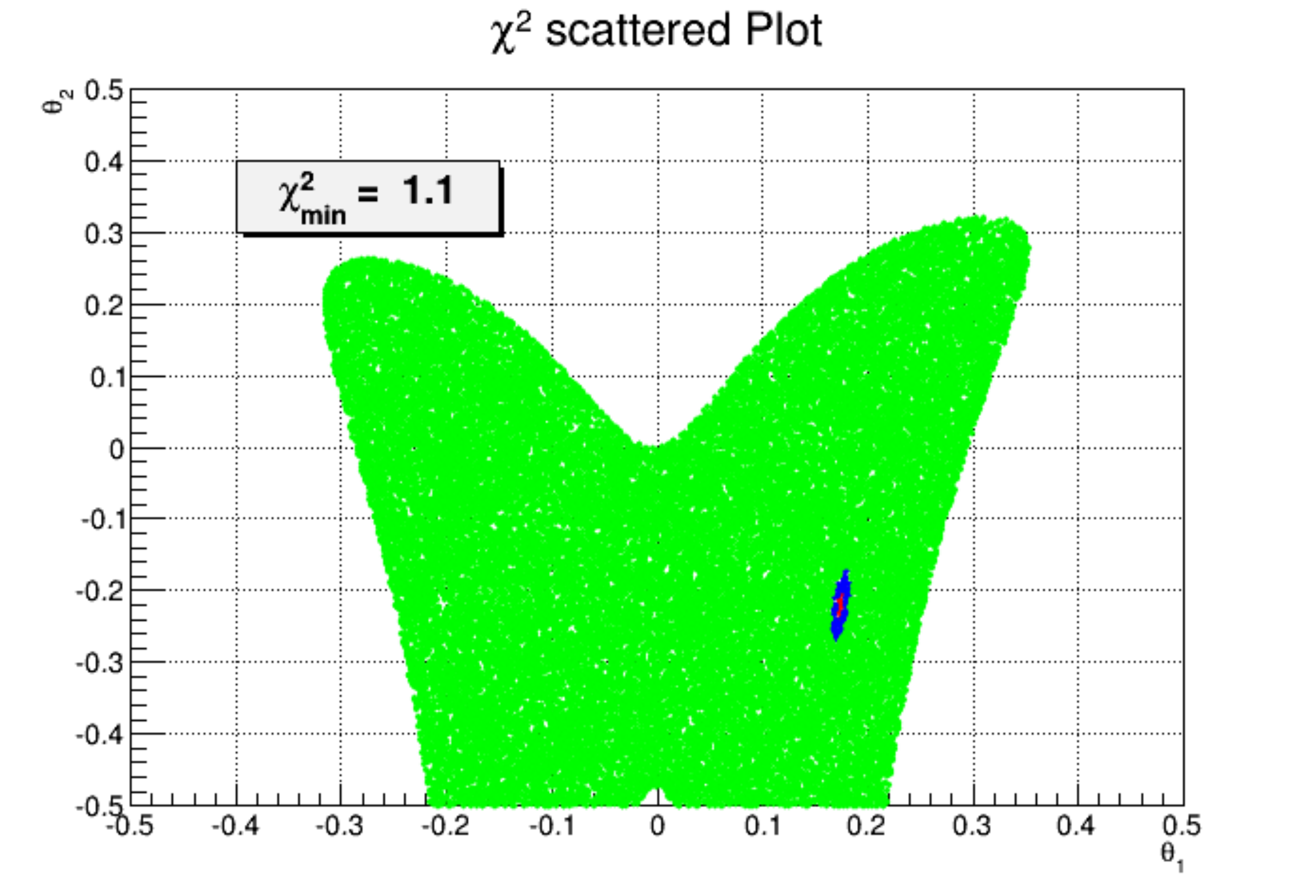} &
\includegraphics[angle=0,width=80mm]{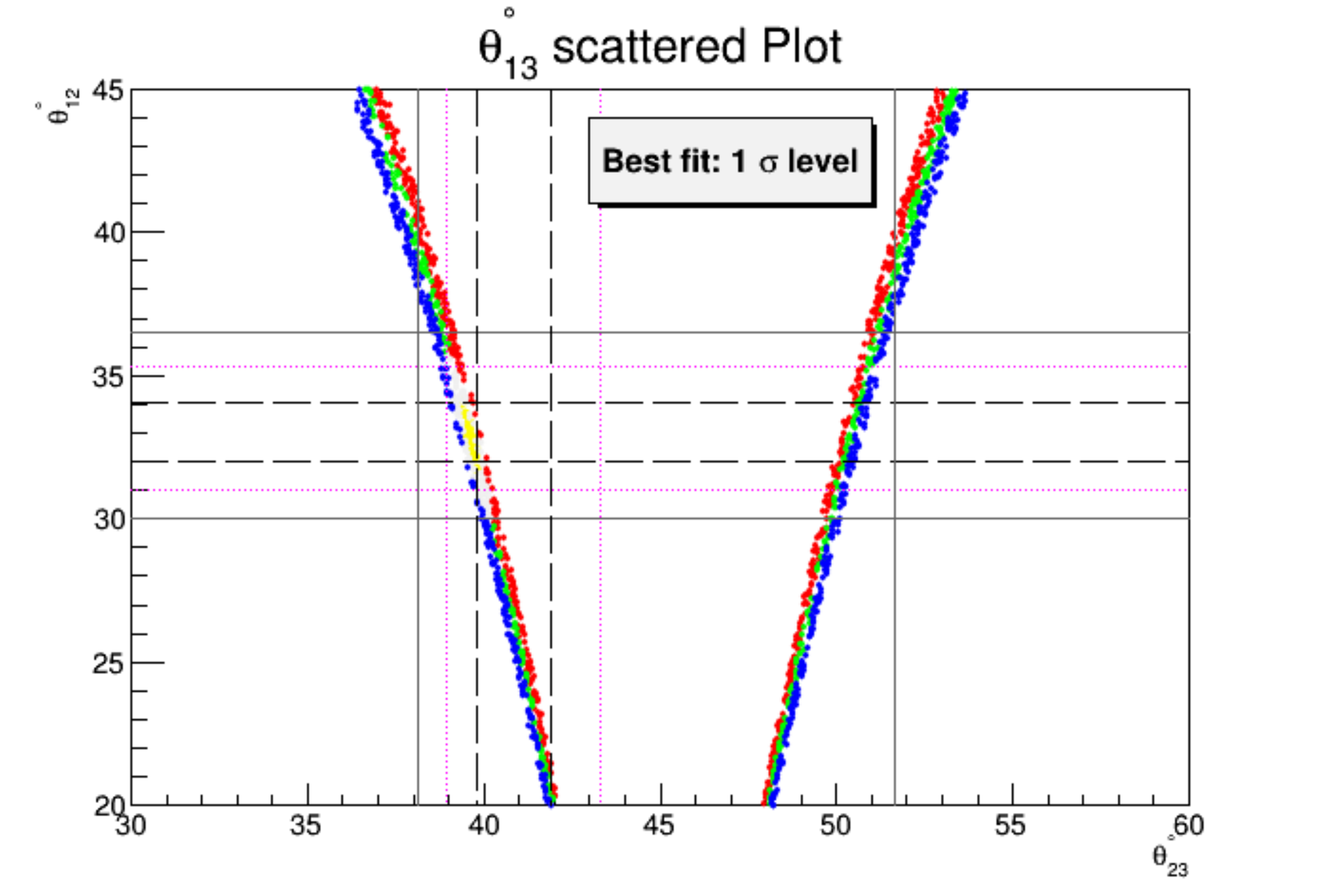}\\
\end{tabular}
\caption{\it{$U^{BMR}_{1213}$ scatter plot of $\chi^2$ (left fig.) over $\alpha-\gamma$ (in radians) plane and $\theta_{13}$ (right fig.) 
over  $\theta_{23}-\theta_{12}$ (in degrees) plane. }}
\label{fig1213R.1}
\end{figure}

\begin{figure}[!t]\centering
\begin{tabular}{c c} 
\includegraphics[angle=0,width=80mm]{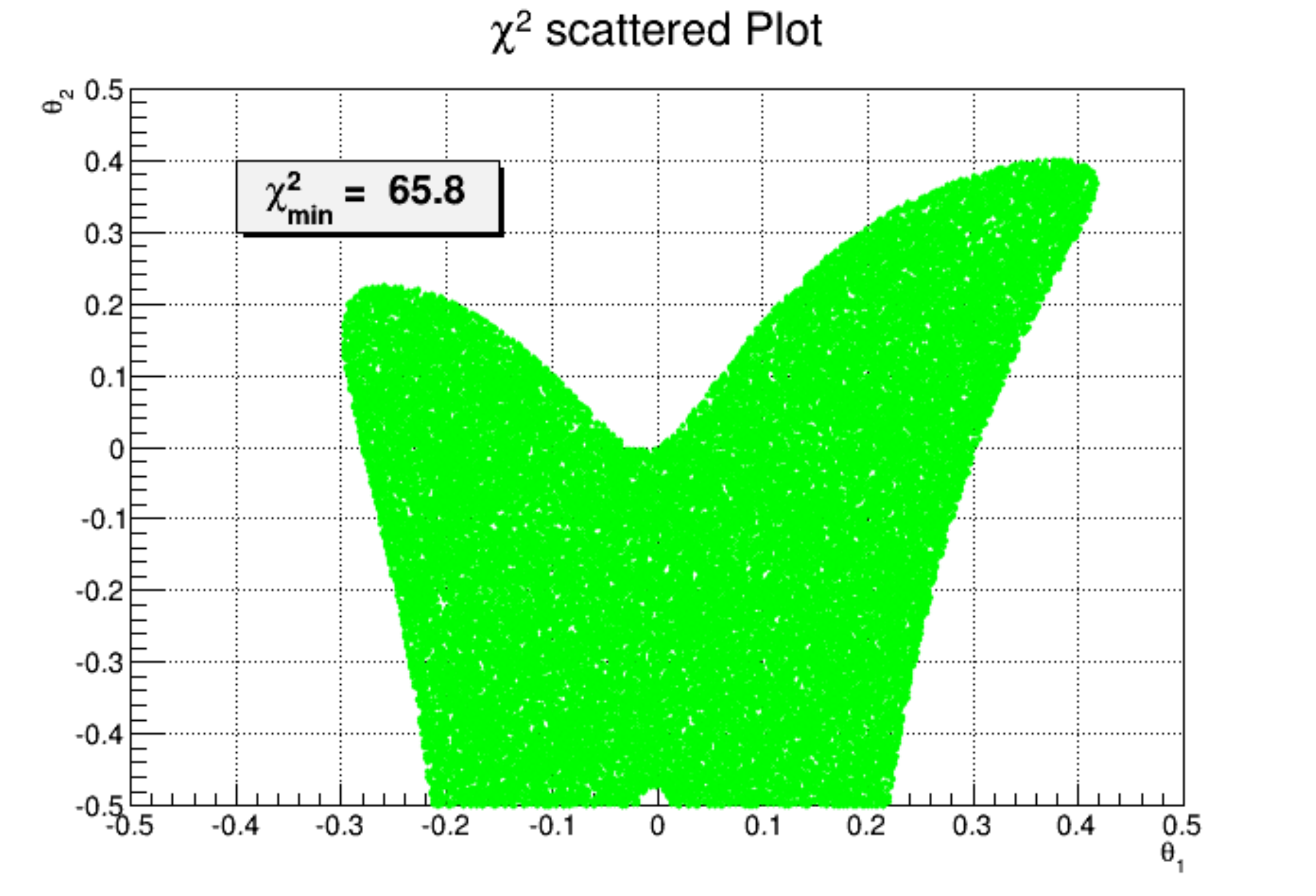} &
\includegraphics[angle=0,width=80mm]{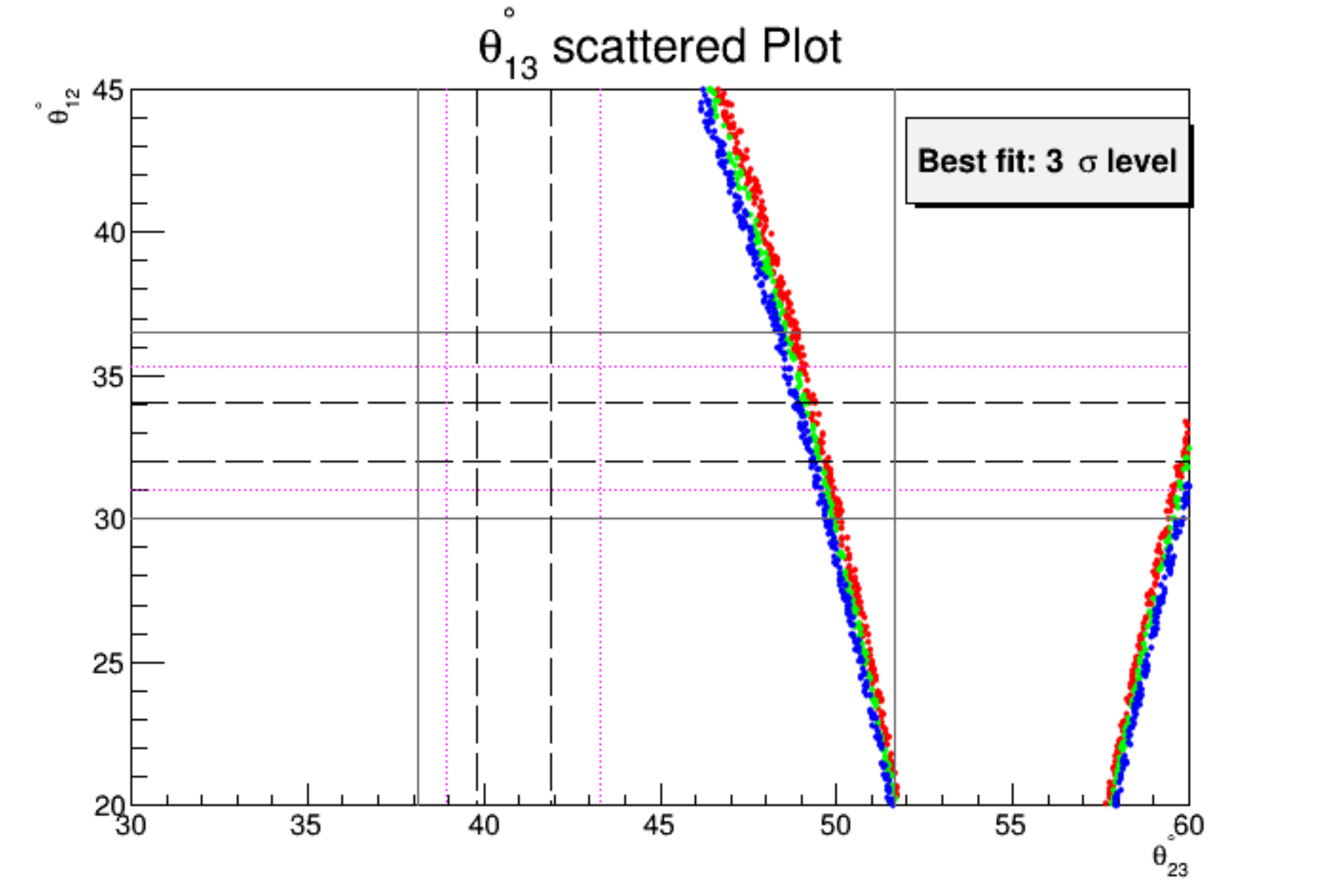}\\
\end{tabular}
\caption{\it{$U^{DCR}_{1213}$ scatter plot of $\chi^2$ (left fig.) over $\alpha-\gamma$ (in radians) plane and $\theta_{13}$ (right fig.) 
over  $\theta_{23}-\theta_{12}$ (in degrees) plane.}}
\label{fig1213R.2}
\end{figure}

\begin{figure}[!t]\centering
\begin{tabular}{c c} 
\includegraphics[angle=0,width=80mm]{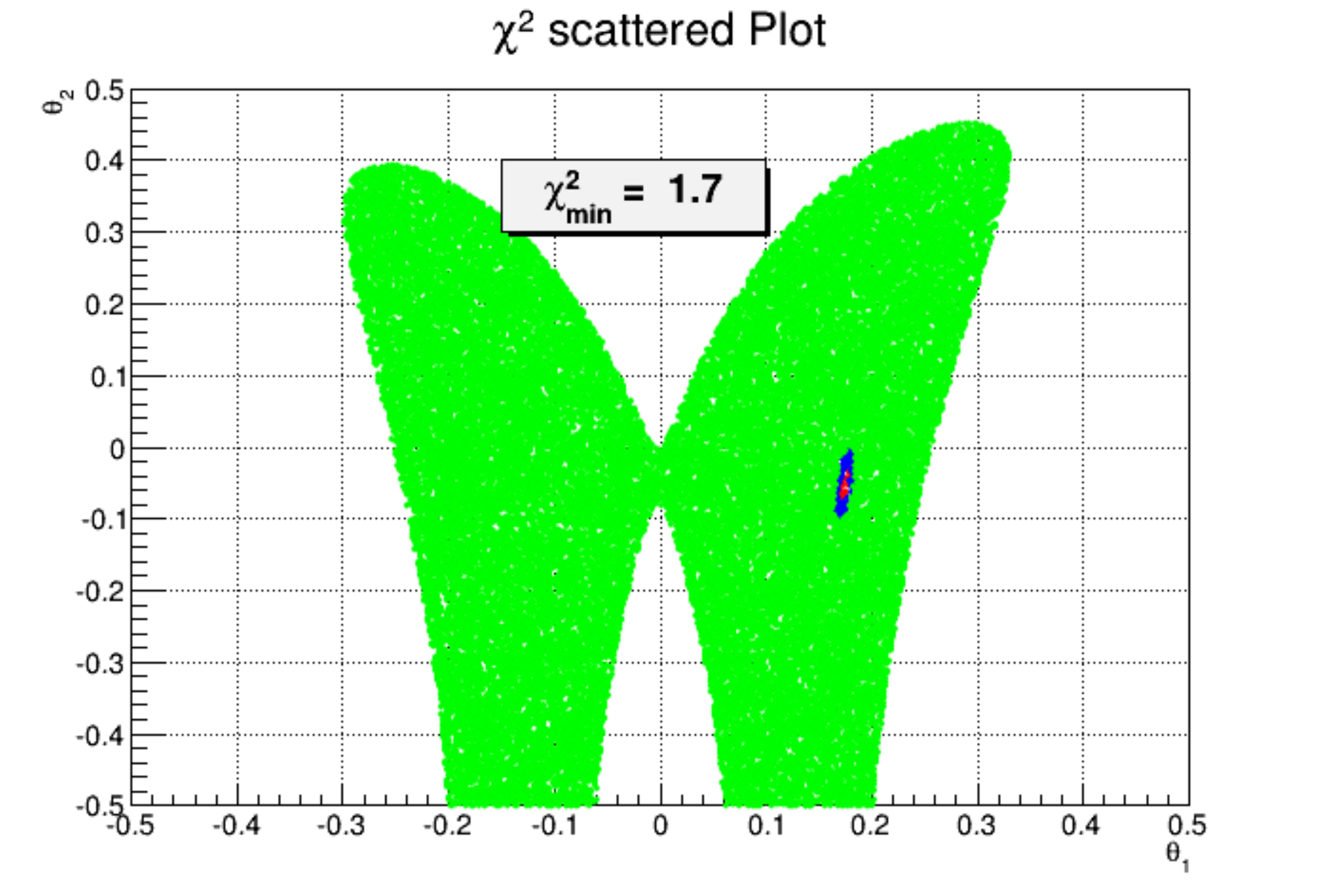} &
\includegraphics[angle=0,width=80mm]{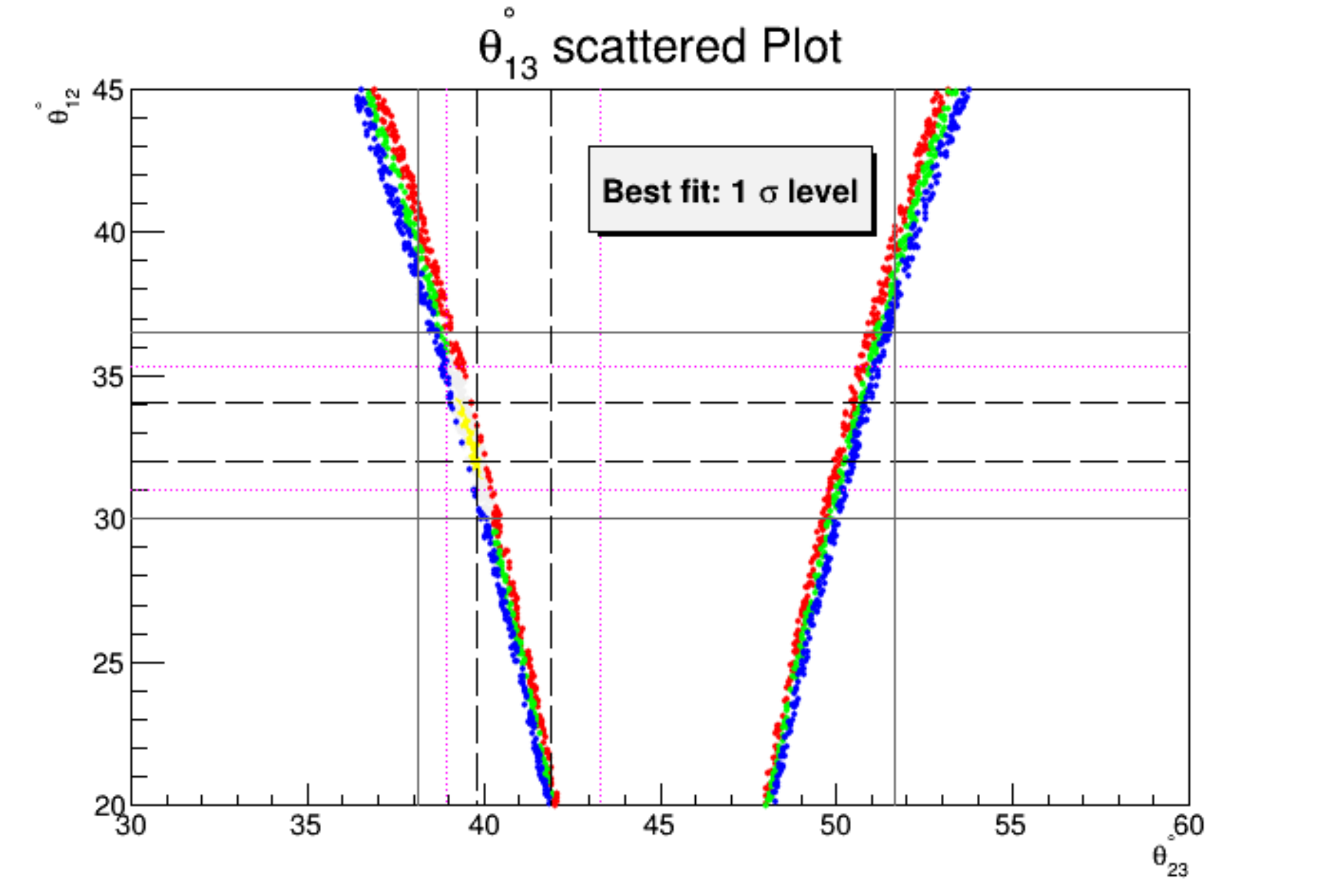}\\
\end{tabular}
\caption{\it{$U^{TBMR}_{1213}$ scatter plot of $\chi^2$ (left fig.) over $\alpha-\gamma$ (in radians) plane and $\theta_{13}$ (right fig.) 
over $\theta_{23}-\theta_{12}$ (in degrees) plane.}}
\label{fig1213R.3}
\end{figure}

In Figs.~\ref{fig1213R.1}-\ref{fig1213R.3}, 
we present the numerical results corresponding to BM, DC and TBM case. 
The main features of this perturbative scheme with $\theta_1 = \gamma$ and $\theta_2 = \alpha$ are given by:\\
{\bf{(i)}} The perturbation parameters ($\alpha, \gamma$) enters into these mixing angles at leading order
and thus exhibit good correlations among themselves.\\
{\bf{(ii)}} Here parameter space prefers two regions for mixing angles. In TBM and BM case for first allowed region 
$\theta_{23} \sim 40^\circ$ while  for 2nd it lies around $50^\circ$. In DC case higher $\theta_{23}$ region remains outside
$3\sigma$ band while lower $\theta_{23}$ remains around $50^\circ$.  \\
{\bf{(iii)}} The minimum value of $\chi^2 \sim 1.1$, $65.8$ and $1.7$ for BM, DC and TBM case respectively.\\
{\bf{(iv)}} It is possible to fit all mixing angles at 1$\sigma$ level for TBM and BM while DC case can 
only be consistent at 3$\sigma$ level.

\subsection{12-23 Rotation}

\begin{figure}[!t]\centering
\begin{tabular}{c c} 
\includegraphics[angle=0,width=80mm]{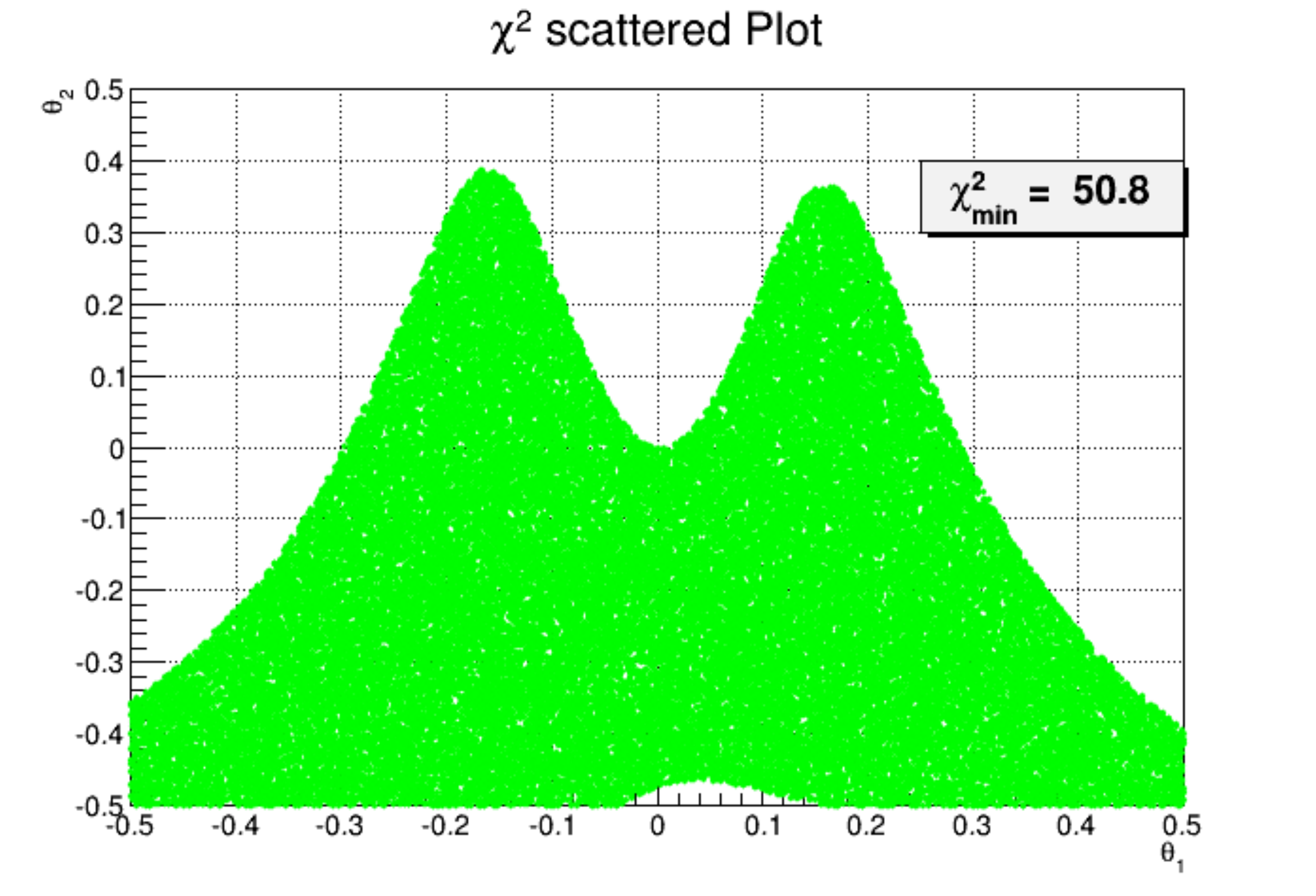} &
\includegraphics[angle=0,width=80mm]{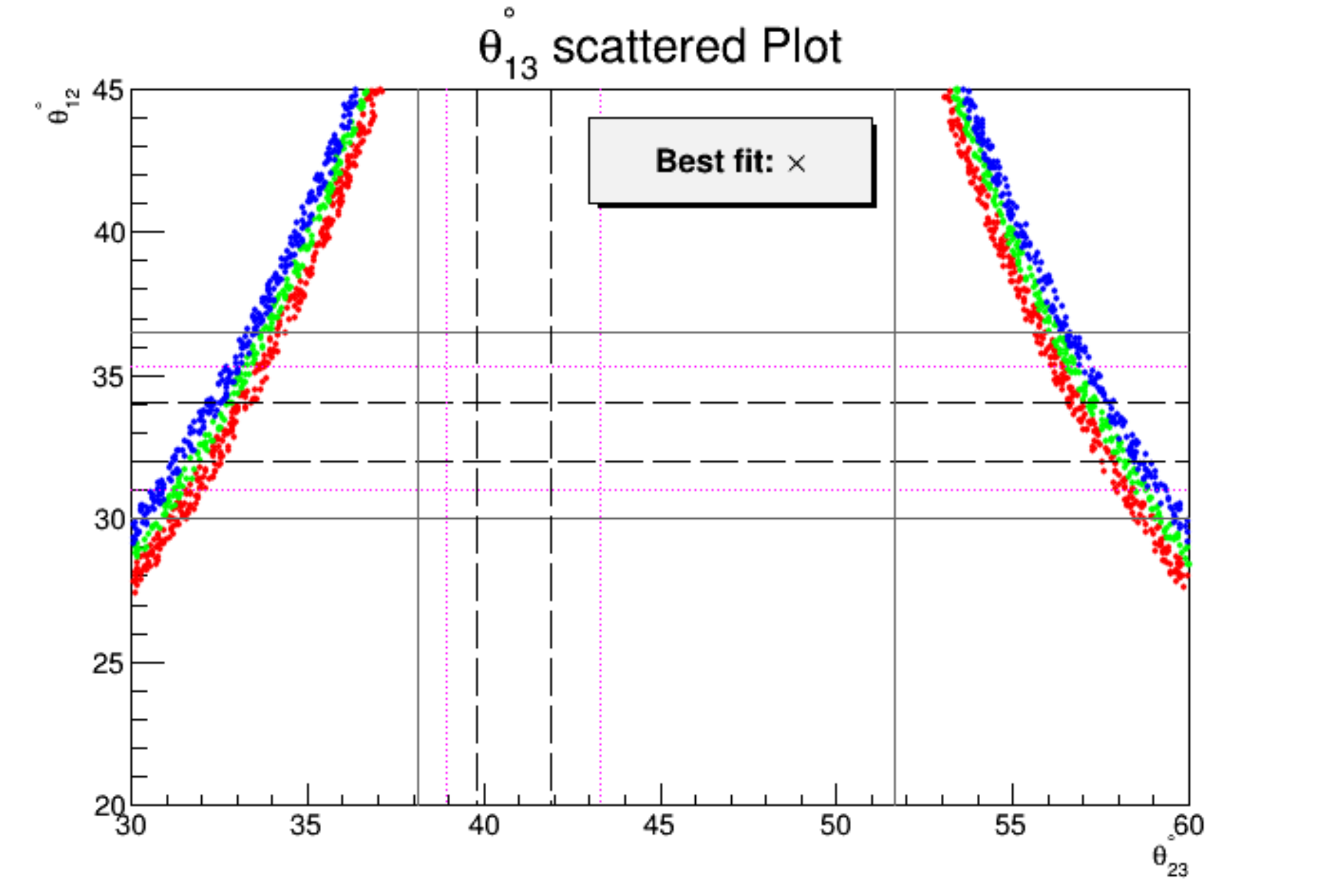}\\
\end{tabular}
\caption{\it{$U^{BMR}_{1223}$ scatter plot of $\chi^2$ (left fig.) over $\alpha-\beta$ (in radians) plane and $\theta_{13}$ (right fig.) 
over  $\theta_{23}-\theta_{12}$ (in degrees) plane. }}
\label{fig1223R.1}
\end{figure}

\begin{figure}[!t]\centering
\begin{tabular}{c c} 
\includegraphics[angle=0,width=80mm]{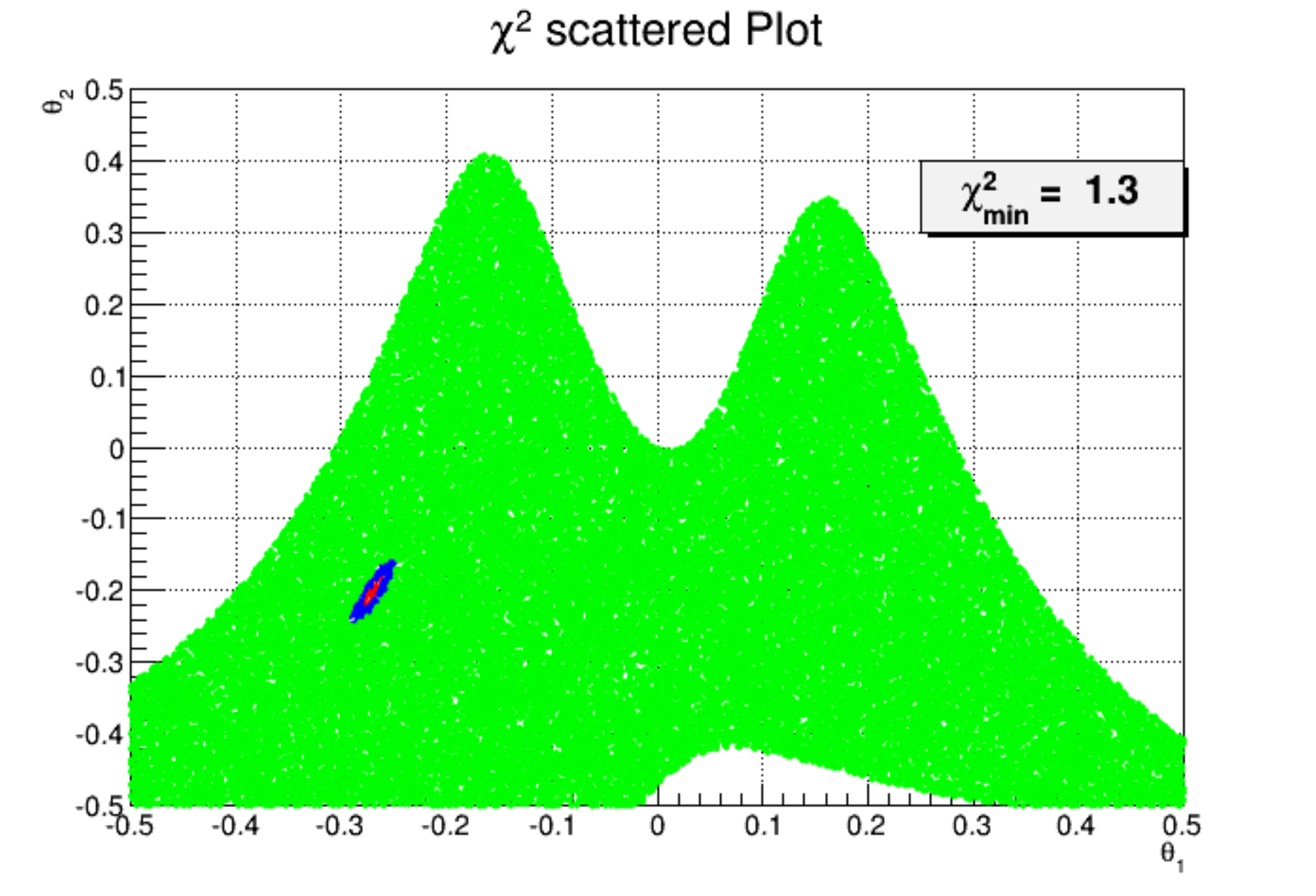} &
\includegraphics[angle=0,width=80mm]{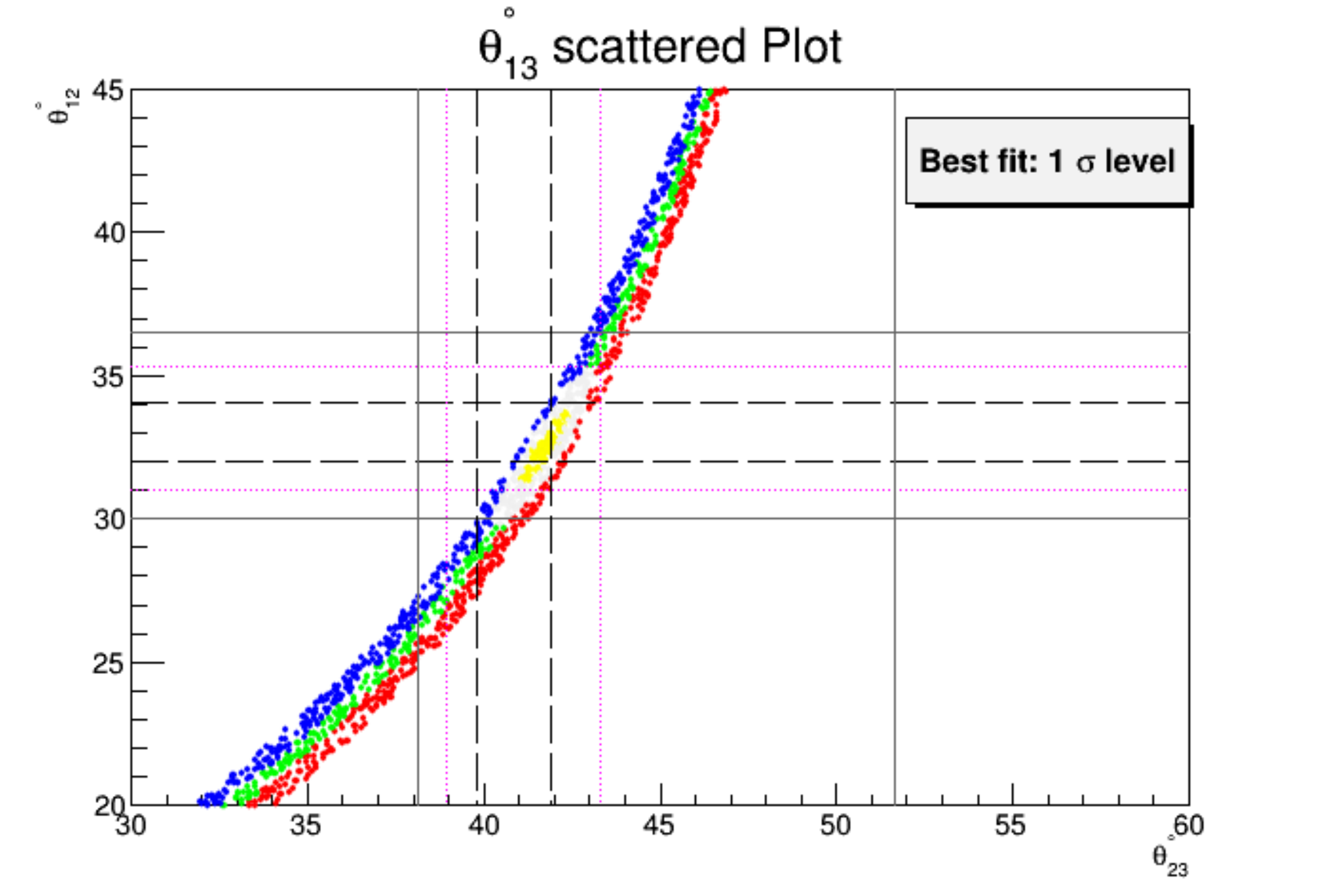}\\
\end{tabular}
\caption{\it{$U^{DCR}_{1223}$ scatter plot of $\chi^2$ (left fig.) over $\alpha-\beta$ (in radians) plane and $\theta_{13}$ (right fig.) 
over  $\theta_{23}-\theta_{12}$ (in degrees) plane. }}
\label{fig1223R.2}
\end{figure}

\begin{figure}[!t]\centering
\begin{tabular}{c c} 
\includegraphics[angle=0,width=80mm]{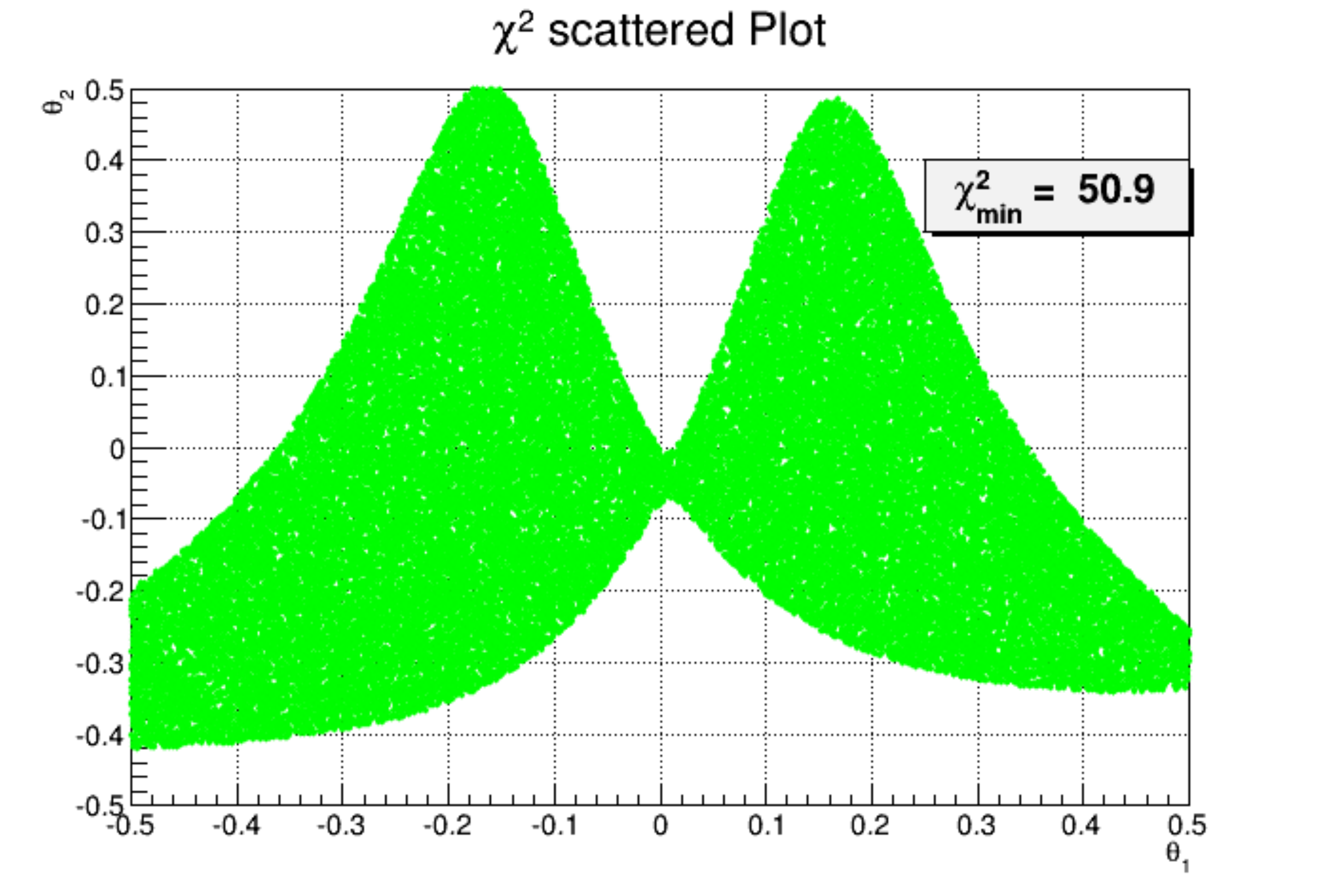} &
\includegraphics[angle=0,width=80mm]{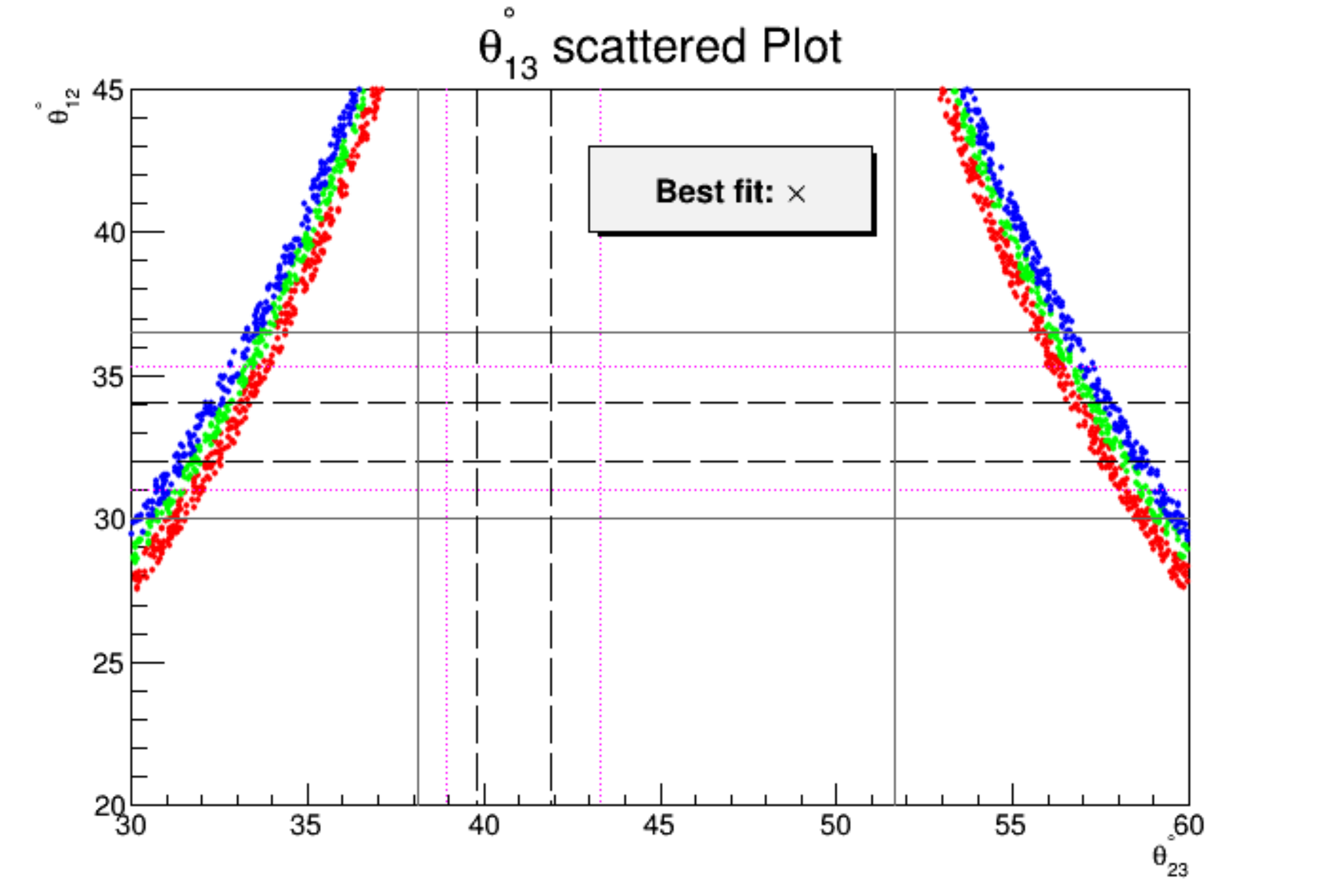}\\
\end{tabular}
\caption{\it{$U^{TBMR}_{1223}$ scatter plot of $\chi^2$ (left fig.) over $\alpha-\beta$ (in radians) plane and $\theta_{13}$ (right fig.) 
over $\theta_{23}-\theta_{12}$ (in degrees) plane. }}
\label{fig1223R.3}
\end{figure}

This case corresponds to rotations in 12 and 23 sector of  these special matrices.  
The neutrino mixing angles for small perturbation 
parameters $\alpha$ and $\beta$ are given by

\beqa
 \sin\theta_{13} &\approx&  |\beta U_{12} +\alpha\beta U_{11}|,\\
 \sin\theta_{23} &\approx& |\frac{U_{23}+ \beta U_{22}-\beta^2 U_{23} +\alpha\beta U_{21}}{\cos\theta_{13}}|,\\
 \sin\theta_{12} &\approx& |\frac{U_{12}+ \alpha U_{11}-(\alpha^2 + \beta^2)U_{12}}{\cos\theta_{13}}|.
\eeqa
Figs.~\ref{fig1223R.1}-\ref{fig1223R.3} corresponds to BM, DC and TBM case respectively with $\theta_1 = \beta$ and $\theta_2 = \alpha$. 
The main outlines of this scheme are given as:\\
{\bf{(i)}}The perturbation parameters enters at leading order into these mixing angles and hence show interesting correlations among themselves.\\
{\bf{(ii)}} This case is unfavorable for TBM and BM as $\chi^2 > 50$ in all parameter space. However for DC case it is possible to get
$\chi^2 < 2$ in a tiny parameter region.\\
{\bf{(iii)}} The minimum value of $\chi^2 \sim 50.8$, $1.3$ and $50.9$ for BM, DC and TBM case respectively.\\
{\bf{(iv)}} TBM and BM is completely excluded while DC case can fit mixing angles at 1$\sigma$ level.

\subsection{13-12 Rotation}

This case corresponds to rotations in 13 and 12 sector of  these special matrices. 
The neutrino mixing angles for small perturbation parameters $\alpha$ and $\gamma$ are given by

\beqa
 \sin\theta_{13} &\approx&  |\gamma U_{11} |,\\
  \sin\theta_{23} &\approx& |\frac{U_{23} + \gamma U_{21}-\gamma^2 U_{23}}{\cos\theta_{13}}|,\\
  \sin\theta_{12} &\approx& |\frac{U_{12} + \alpha U_{11}-\alpha^2 U_{12} }{\cos\theta_{13}}|.
\eeqa

\begin{figure}[!t]\centering
\begin{tabular}{c c} 
\includegraphics[angle=0,width=80mm]{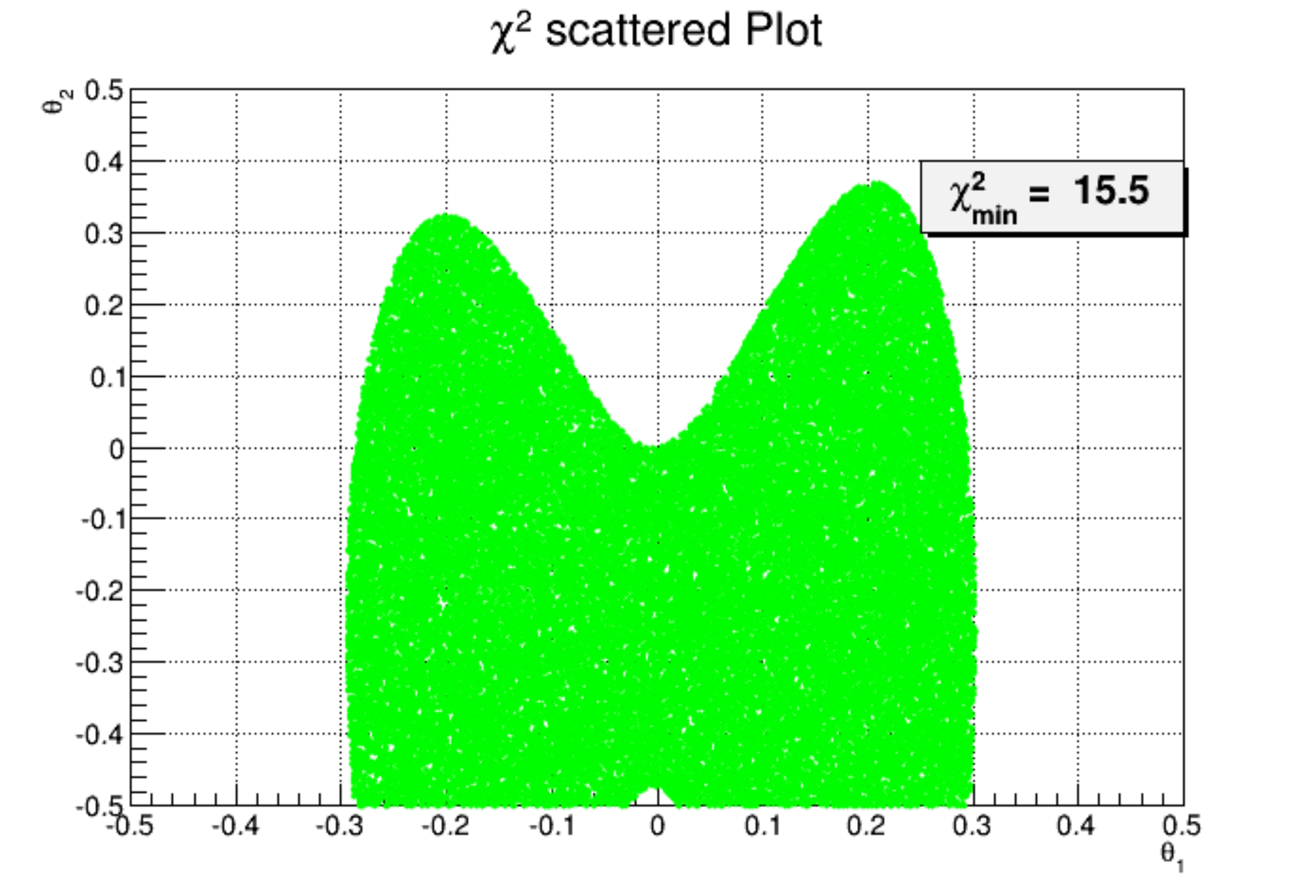} &
\includegraphics[angle=0,width=80mm]{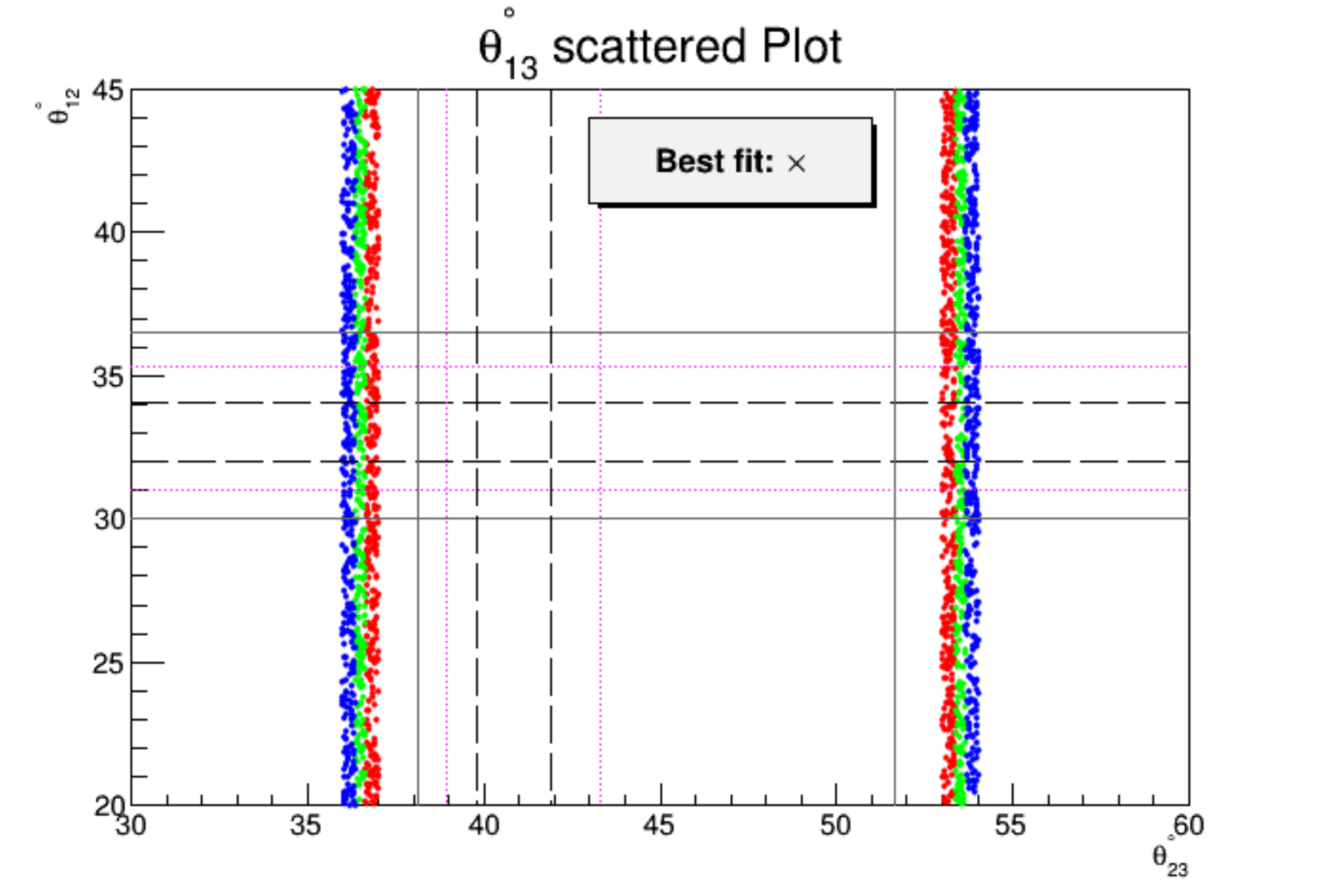}\\
\end{tabular}
\caption{\it{$U^{BMR}_{1312}$ scatter plot of $\chi^2$ (left fig.) over $\gamma-\alpha$ (in radians) plane and $\theta_{13}$ (right fig.) 
over  $\theta_{23}-\theta_{12}$ (in degrees) plane. }}
\label{fig1312R.1}
\end{figure}

\begin{figure}[!t]\centering
\begin{tabular}{c c} 
\includegraphics[angle=0,width=80mm]{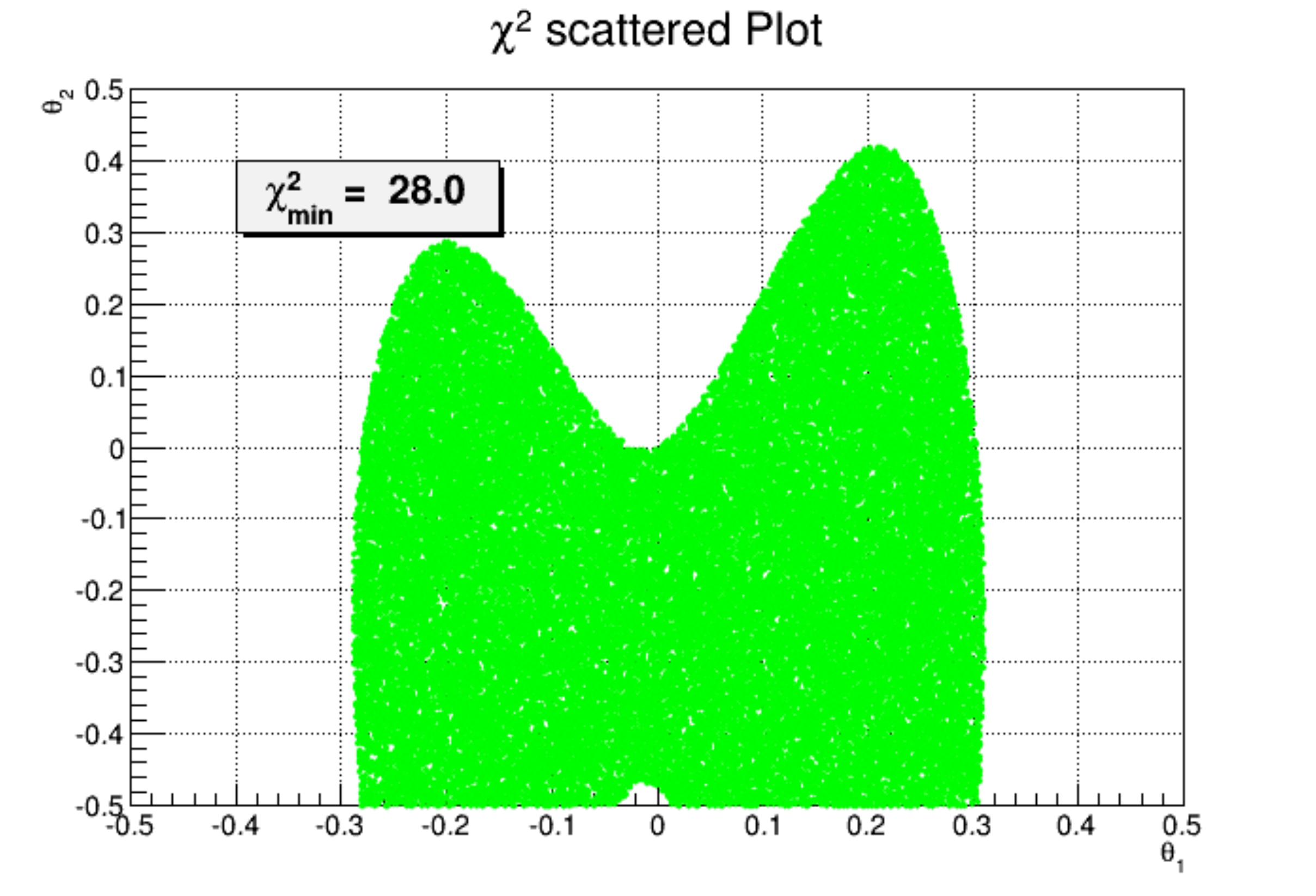} &
\includegraphics[angle=0,width=80mm]{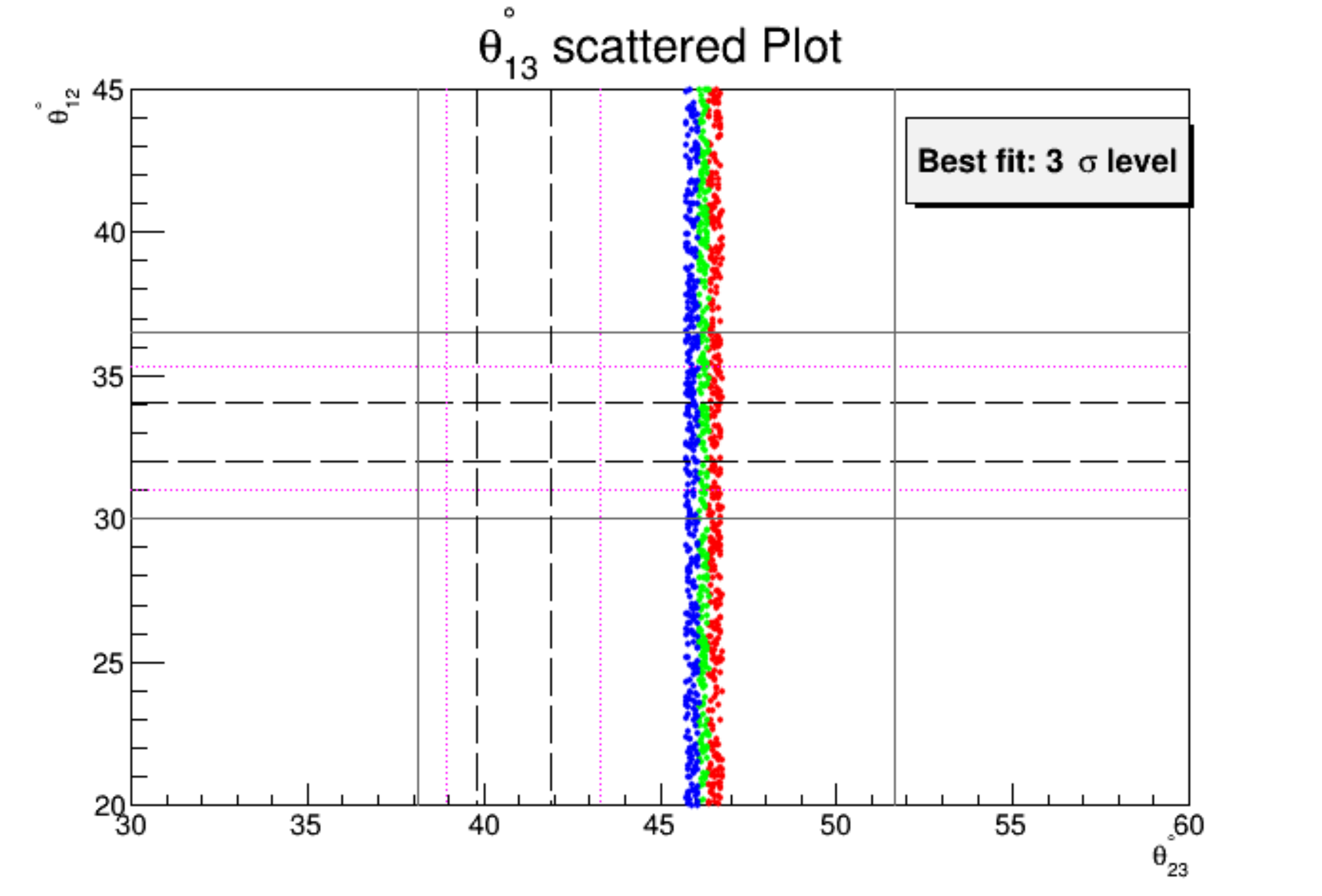}\\
\end{tabular}
\caption{\it{$U^{DCR}_{1312}$ scatter plot of $\chi^2$ (left fig.) over $\gamma-\alpha$ (in radians) plane and $\theta_{13}$ (right fig.) 
over  $\theta_{23}-\theta_{12}$ (in degrees) plane. }}
\label{fig1312R.2}
\end{figure}

\begin{figure}[!t]\centering
\begin{tabular}{c c} 
\includegraphics[angle=0,width=80mm]{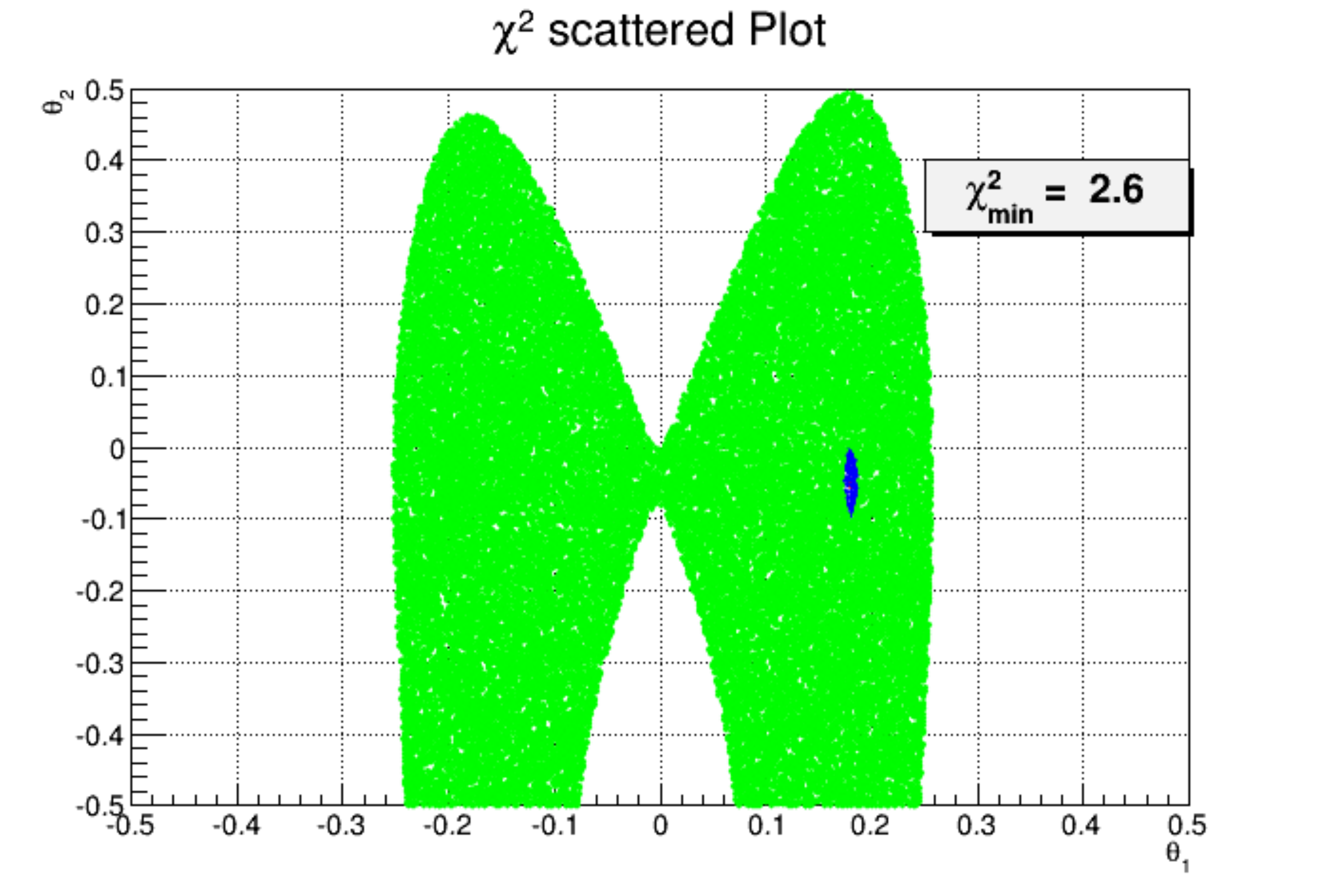} &
\includegraphics[angle=0,width=80mm]{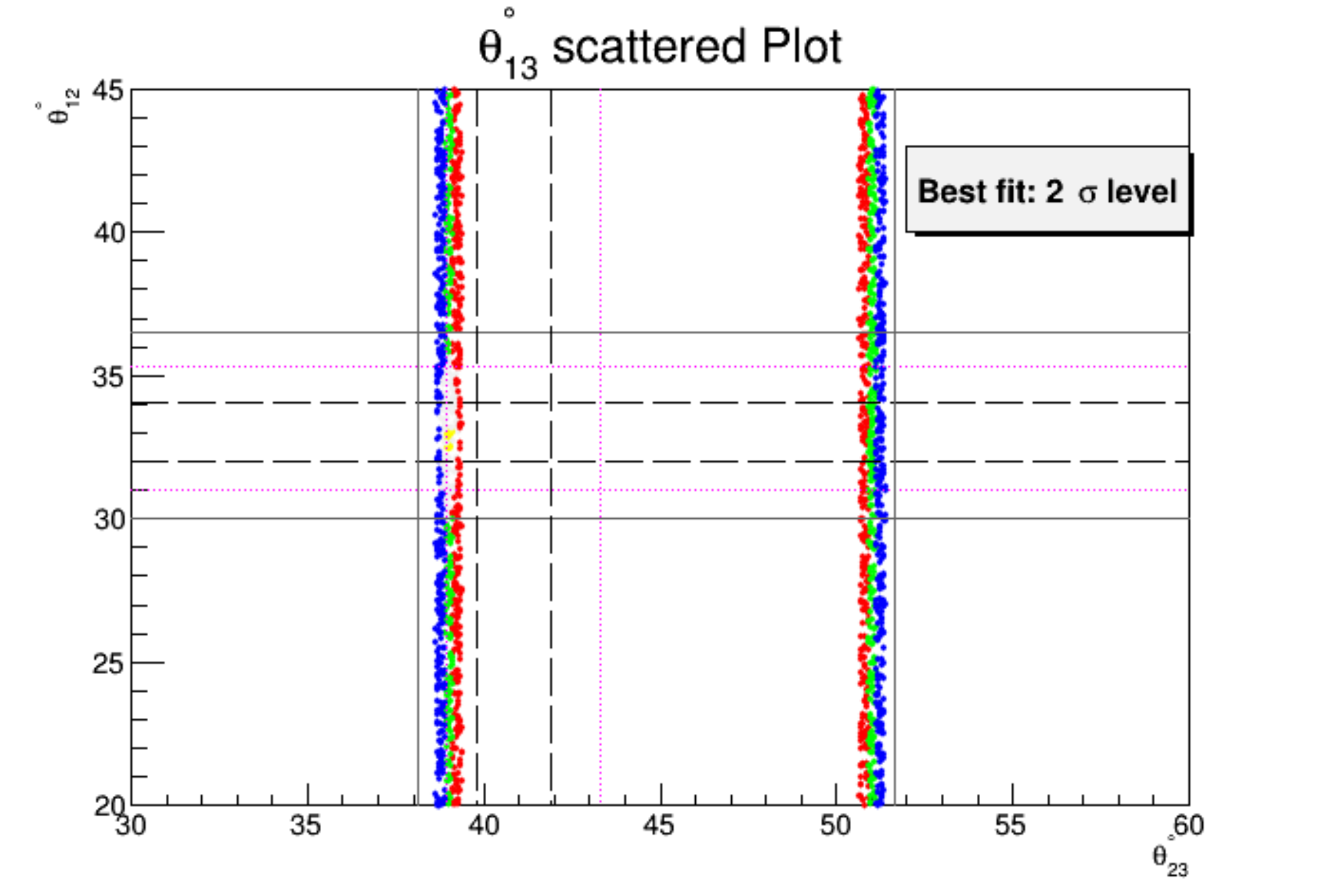}\\
\end{tabular}
\caption{\it{$U^{TBMR}_{1312}$ scatter plot of $\chi^2$ (left fig.) over $\gamma-\alpha$ (in radians) plane and $\theta_{13}$ (right fig.) 
over $\theta_{23}-\theta_{12}$ (in degrees) plane.  }}
\label{fig1312R.3}
\end{figure}

Figs.~\ref{fig1312R.1}-\ref{fig1312R.3} corresponds to BM, DC and TBM case respectively with $\theta_1 = \gamma$ and $\theta_2 =\alpha$.\\ 
{\bf{(i)}} The corrections to mixing angle $\theta_{13}$ and $\theta_{23}$ is only governed by perturbation parameter
$\gamma$.  Thus magnitude of parameter $\gamma$ is tightly constrainted from fitting of $\theta_{13}$. This in turn permits very narrow ranges
for $\theta_{23}$ corresponding to negative and positive values of $\gamma$ in parameter space.  However $\theta_{12}$ solely depends on $\alpha$ and thus
can have wide range of possible values in parameter space.\\\\
{\bf{(ii)}} The minimum value of $\chi^2 \sim 15.5$, $28.0$ and $2.6$ for BM, DC and TBM case respectively.\\
{\bf{(iii)}} DC and TBM are consistent at $3\sigma$ and $2\sigma$ level while BM is not consistent even at $3\sigma$ level.

\subsection{13-23 Rotation}

\begin{figure}[!t]\centering
\begin{tabular}{c c} 
\includegraphics[angle=0,width=80mm]{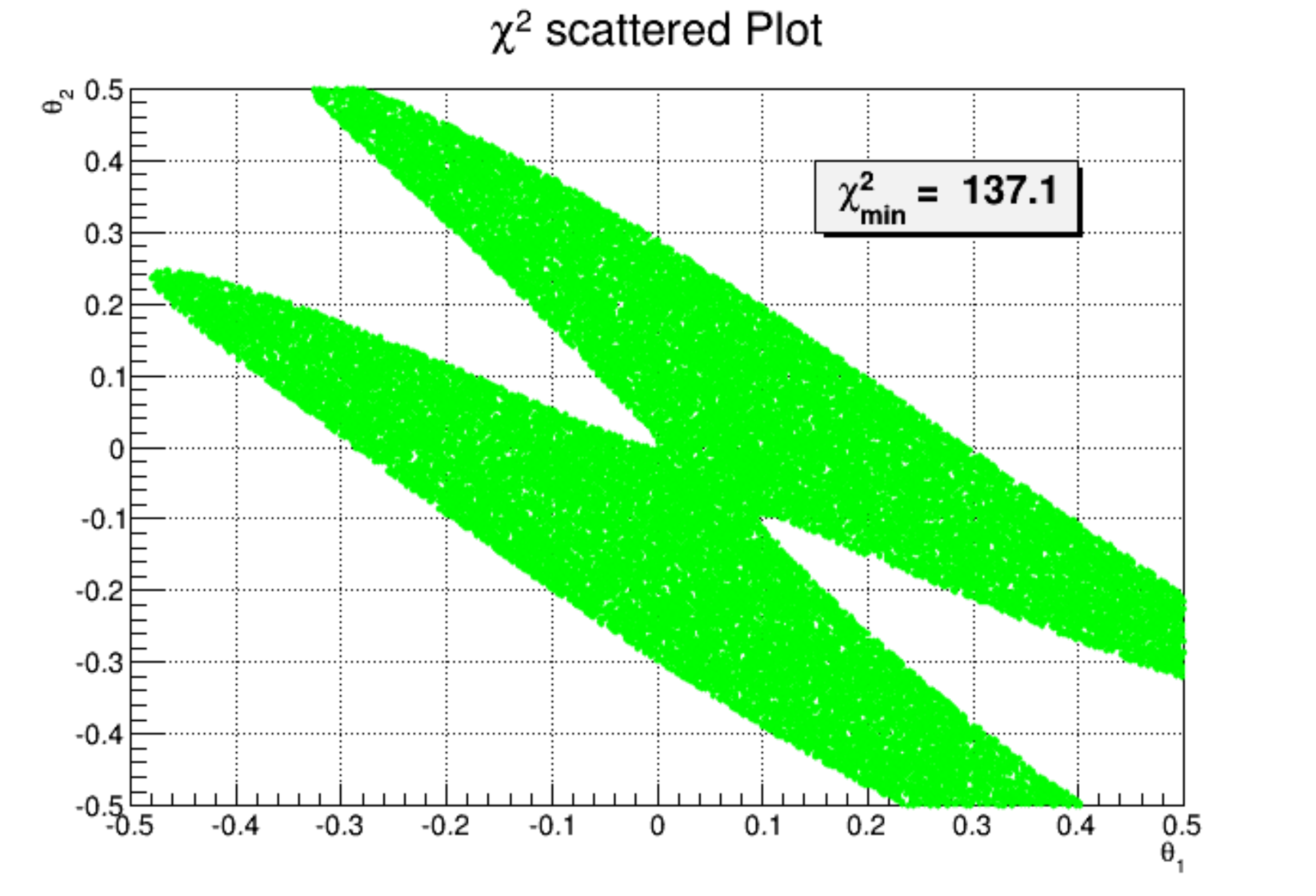} &
\includegraphics[angle=0,width=80mm]{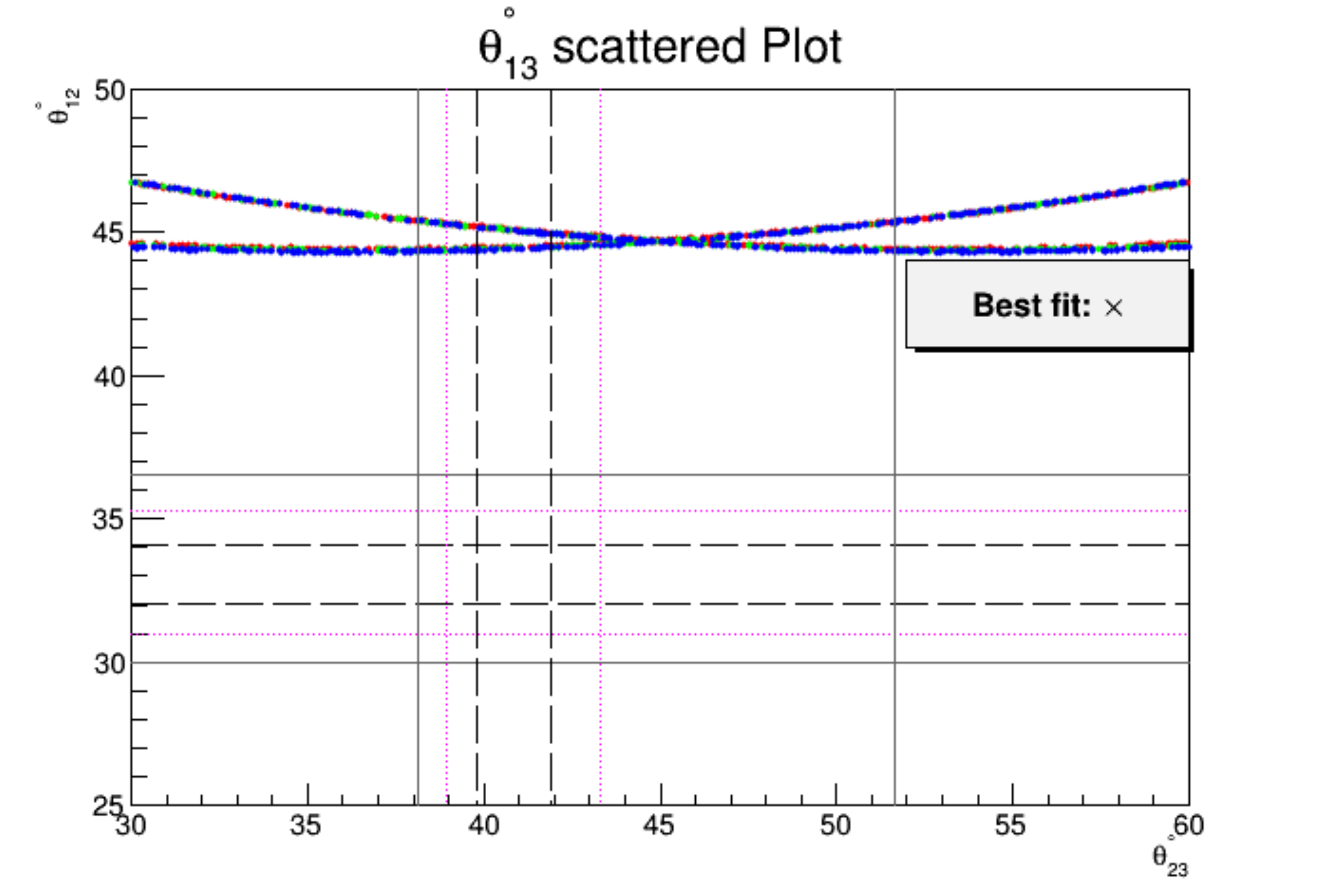}\\
\end{tabular}
\caption{\it{$U^{BMR}_{1323}$ scatter plot of $\chi^2$ (left fig.) over $\gamma-\beta$ (in radians) plane and $\theta_{13}$ (right fig.) 
over  $\theta_{23}-\theta_{12}$ (in degrees) plane. }}
\label{fig1323R.1}
\end{figure}

\begin{figure}[!t]\centering
\begin{tabular}{c c} 
\includegraphics[angle=0,width=80mm]{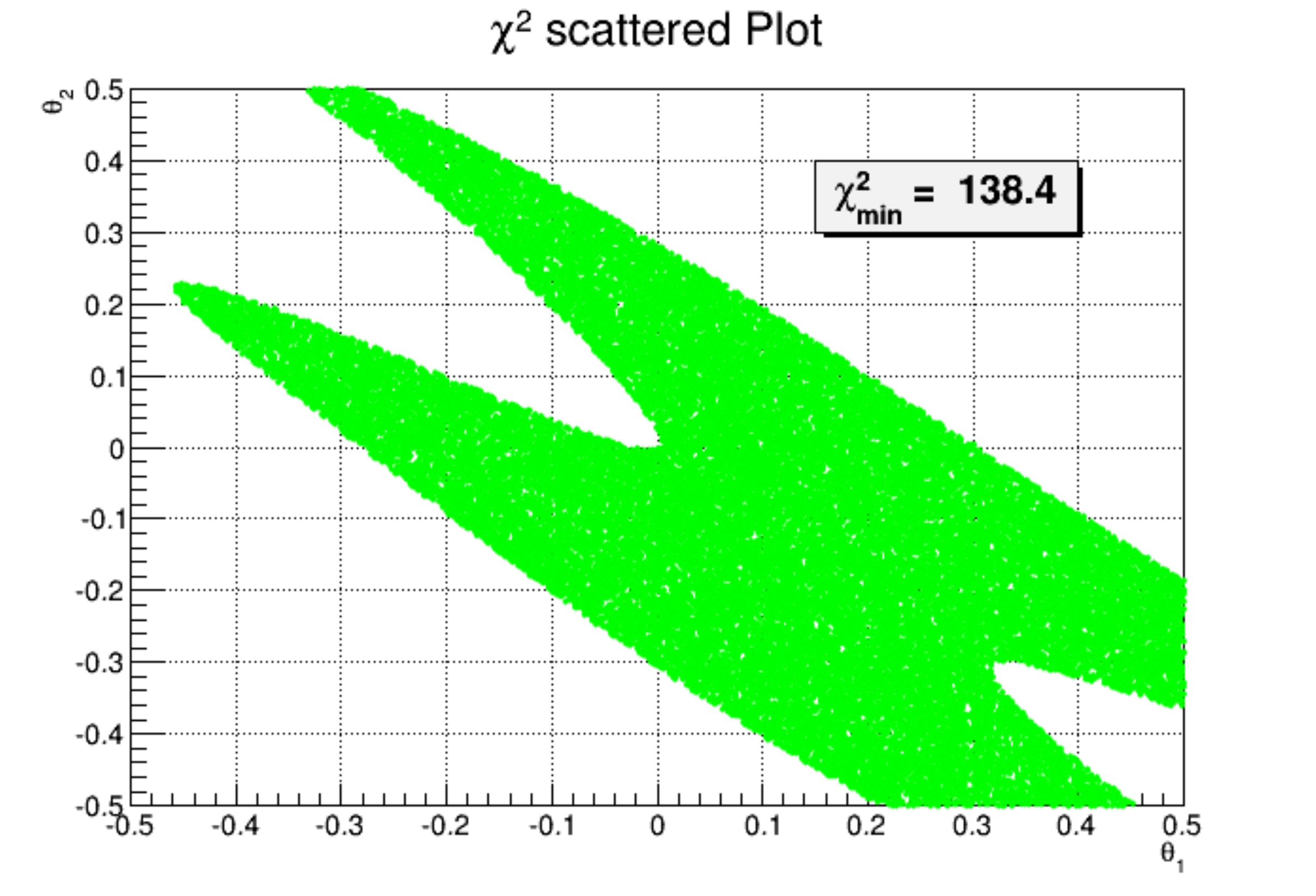} &
\includegraphics[angle=0,width=80mm]{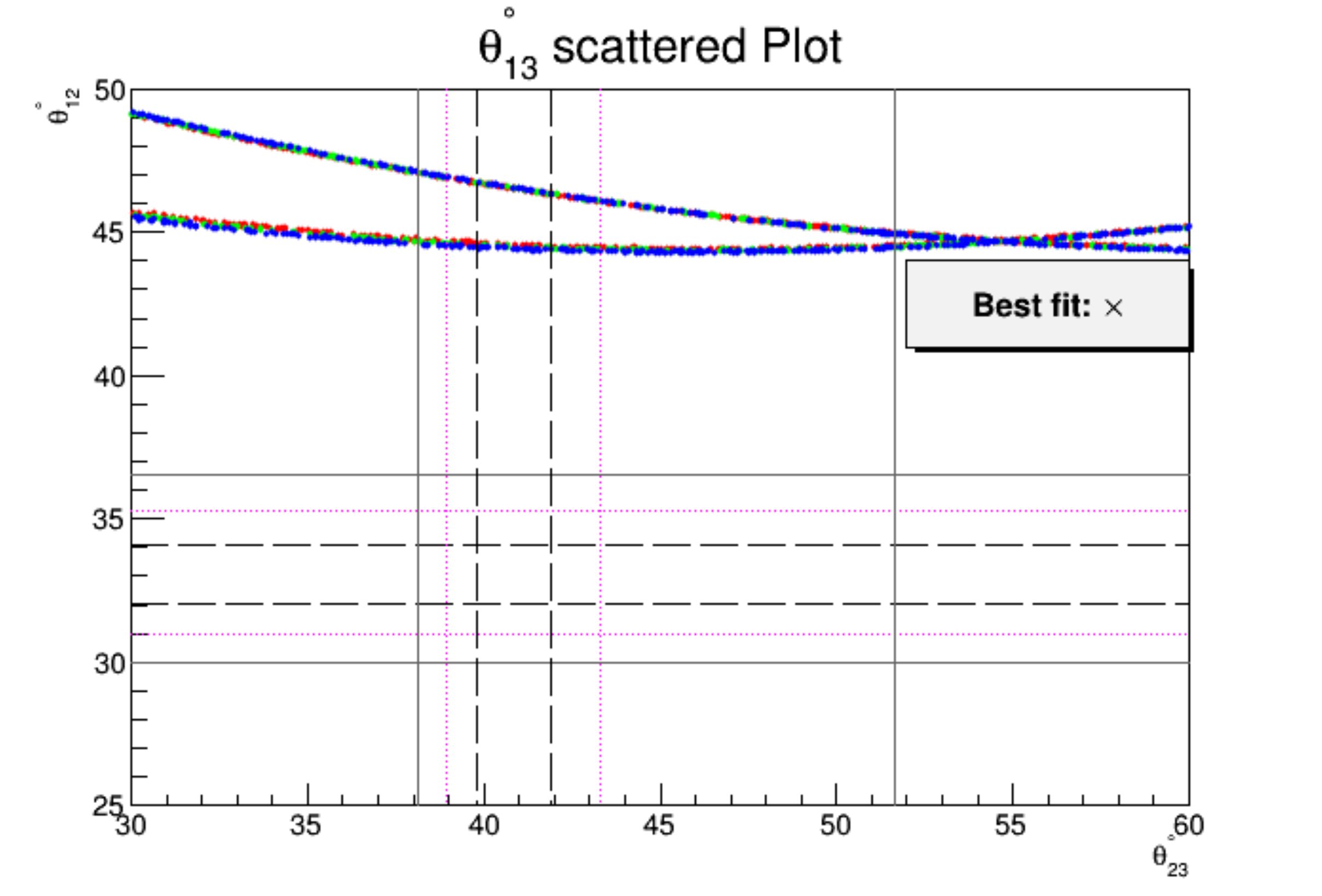}\\
\end{tabular}
\caption{\it{$U^{DCR}_{1323}$ scatter plot of $\chi^2$ (left fig.) over $\gamma-\beta$ (in radians) plane and $\theta_{13}$ (right fig.) 
over  $\theta_{23}-\theta_{12}$ (in degrees) plane. }}
\label{fig1323R.2}
\end{figure}

\begin{figure}[!t]\centering
\begin{tabular}{c c} 
\includegraphics[angle=0,width=80mm]{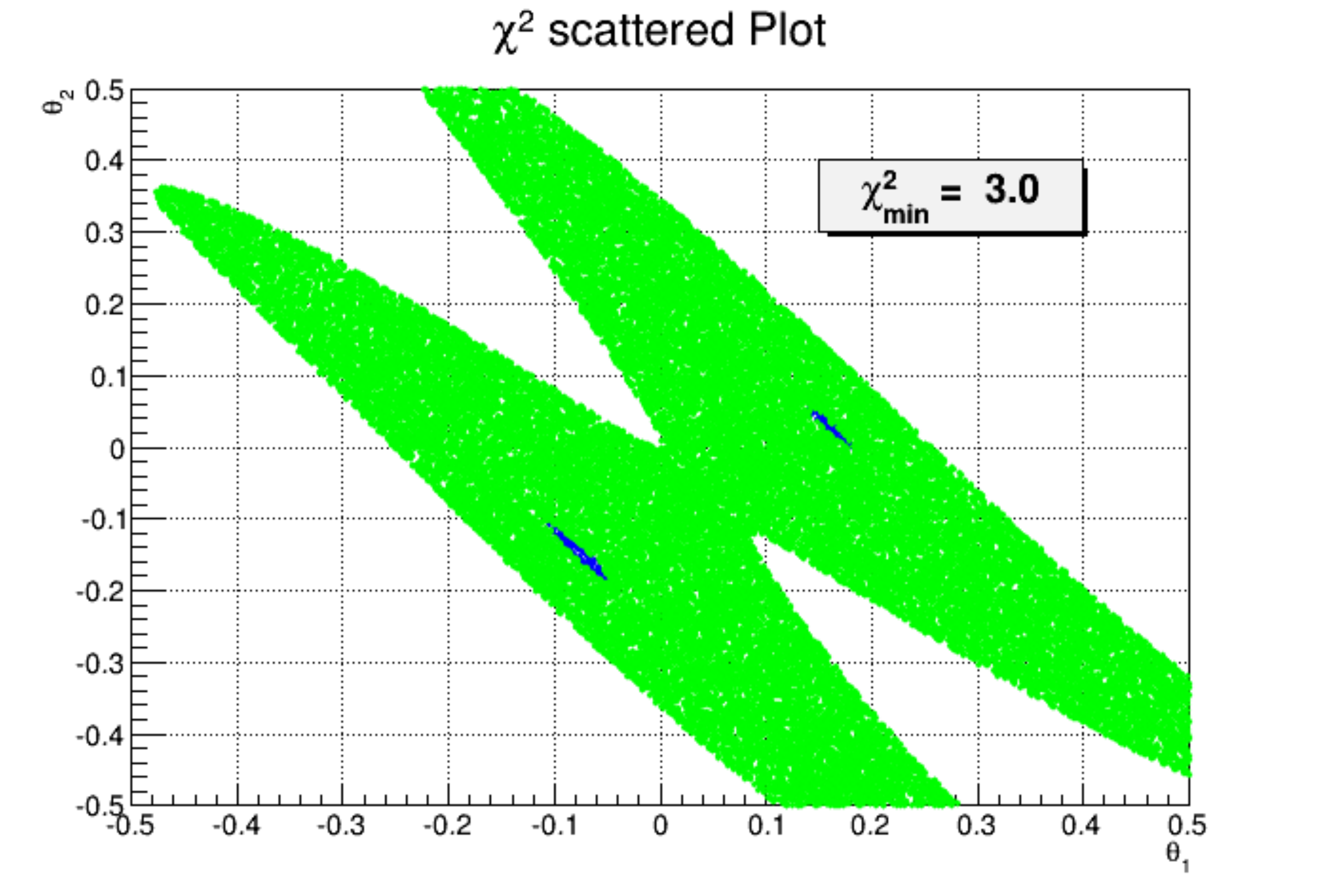} &
\includegraphics[angle=0,width=80mm]{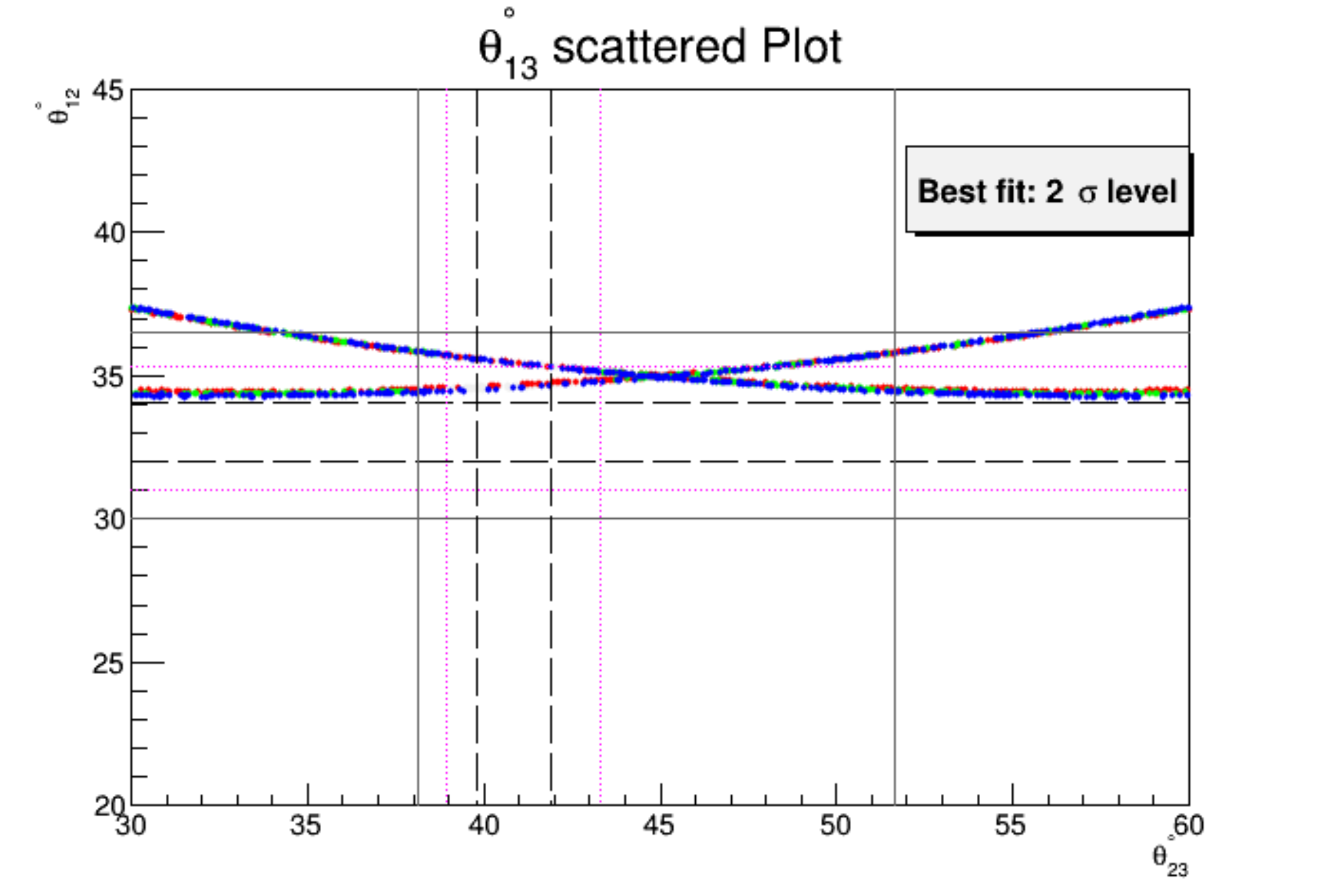}\\
\end{tabular}
\caption{\it{$U^{TBMR}_{1323}$ scatter plot of $\chi^2$ (left fig.) over $\gamma-\beta$ (in radians) plane and $\theta_{13}$ (right fig.) 
over $\theta_{23}-\theta_{12}$ (in degrees) plane. }}
\label{fig1323R.3}
\end{figure}

This case corresponds to rotations in 13 and 23 sector of  these special matrices. 
The neutrino mixing angles for small perturbation parameters $\gamma$ and $\beta$ are given by

\beqa
 \sin\theta_{13} &\approx&  |\beta U_{12} + \gamma U_{11}|,\\
 \sin\theta_{23} &\approx& |\frac{U_{23}+\beta U_{22}+\gamma U_{21}-(\beta^2 + \gamma^2)U_{23} }{\cos\theta_{13}}|,\\
 \sin\theta_{12} &\approx& |\frac{(1-\beta^2)U_{12} -\beta\gamma U_{11}}{\cos\theta_{13}}|.
 \eeqa
 Figs.~\ref{fig1323R.1}-\ref{fig1323R.3} corresponds to BM, DC and TBM case respectively with $\theta_1 = \gamma$ and $\theta_2 = \beta$.\\
{\bf{(i)}}For this rotation scheme, $\theta_{12}$ receives corrections only at O($\theta^2$) and thus its value remain close to its 
unperturbed value.\\
{\bf{(ii)}} The minimum value of $\chi^2 \sim 137.1$, $138.4$ and $3.0$ for BM, DC and TBM case respectively.\\
{\bf{(iii)}} All mixing angles can be fitted at 2$\sigma$ level in TBM while BM and DC case is excluded as their unperturbed $\theta_{12}$ value 
is quite away from  3$\sigma$ boundary.

\subsection{23-12 Rotation}

This case corresponds to rotations in 23 and 12 sector of  these special matrices.

\beqa
 \sin\theta_{13} &\approx&  |\beta U_{12}|,\\
 \sin\theta_{23} &\approx& |\frac{U_{23}+ \beta U_{22}-\beta^2 U_{23} }{\cos\theta_{13}}|,\\
 \sin\theta_{12} &\approx& |\frac{U_{12} + \alpha U_{11}-(\alpha^2 + \beta^2 )U_{12}}{\cos\theta_{13}}|.
\eeqa

\begin{figure}[!t]\centering
\begin{tabular}{c c} 
\includegraphics[angle=0,width=80mm]{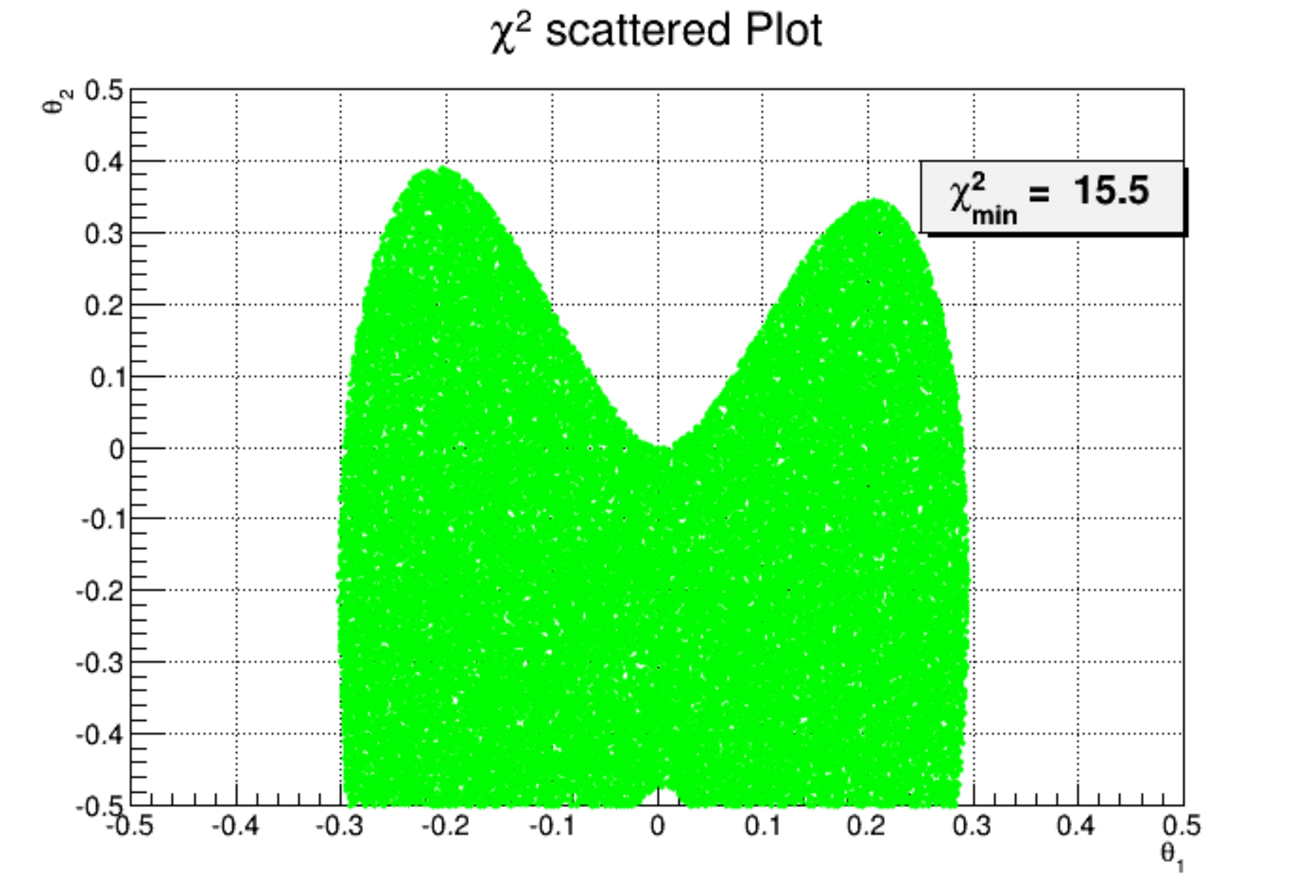} &
\includegraphics[angle=0,width=80mm]{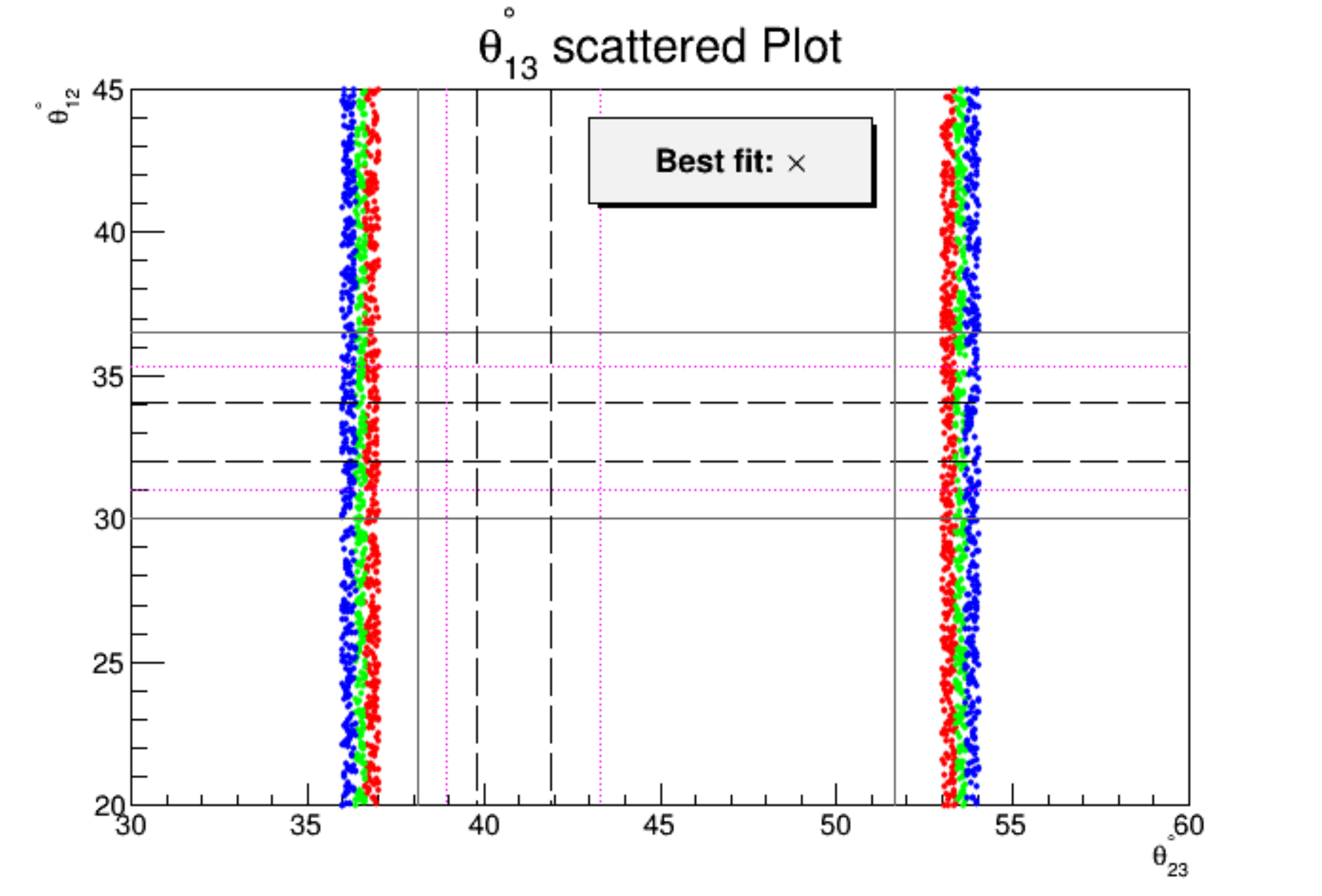}\\
\end{tabular}
\caption{\it{$U^{BMR}_{2312}$ scatter plot of $\chi^2$ (left fig.) over $\beta-\alpha$ (in radians) plane and $\theta_{13}$ (right fig.) 
over  $\theta_{23}-\theta_{12}$ (in degrees) plane. }}
\label{fig2312R.1}
\end{figure}

\begin{figure}[!t]\centering
\begin{tabular}{c c} 
\includegraphics[angle=0,width=80mm]{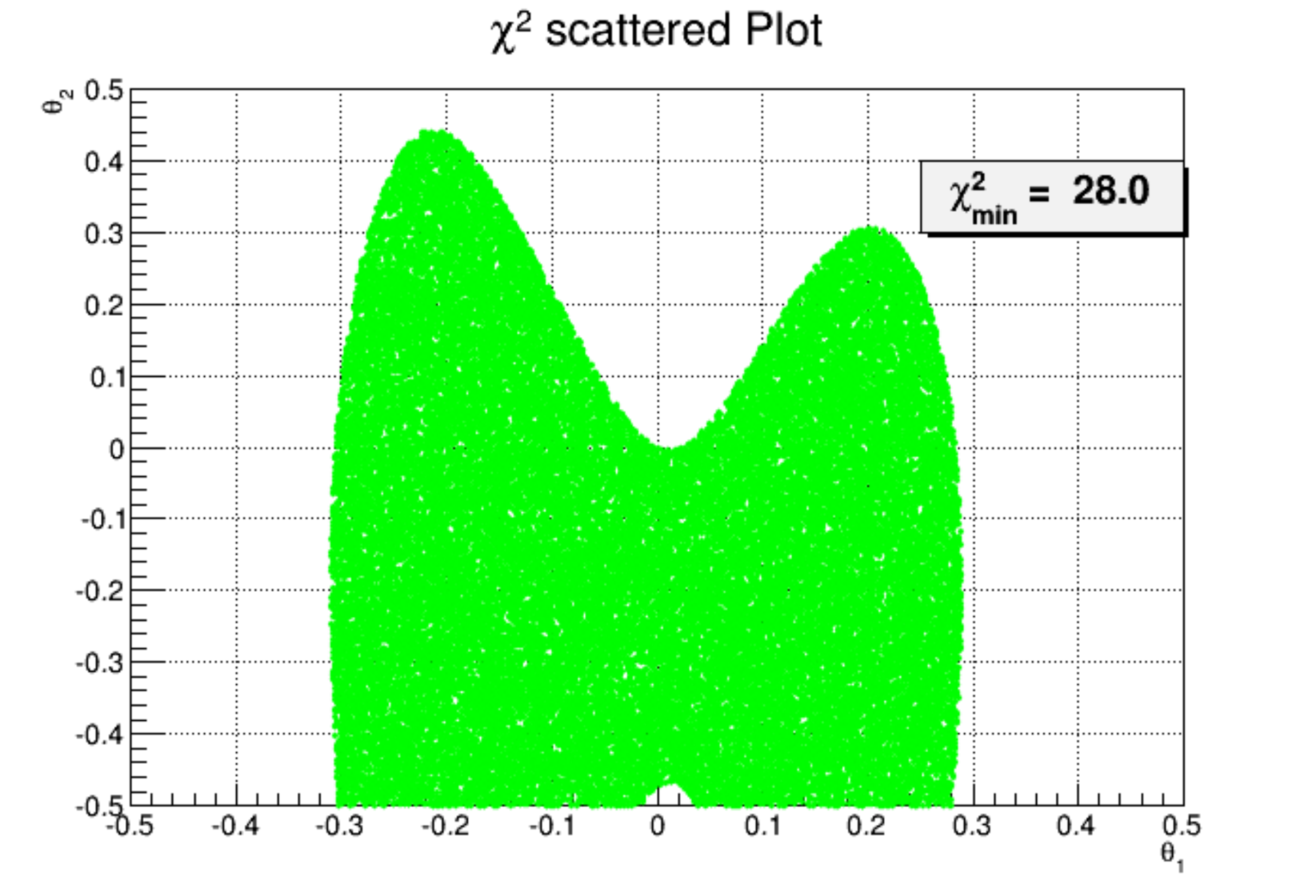} &
\includegraphics[angle=0,width=80mm]{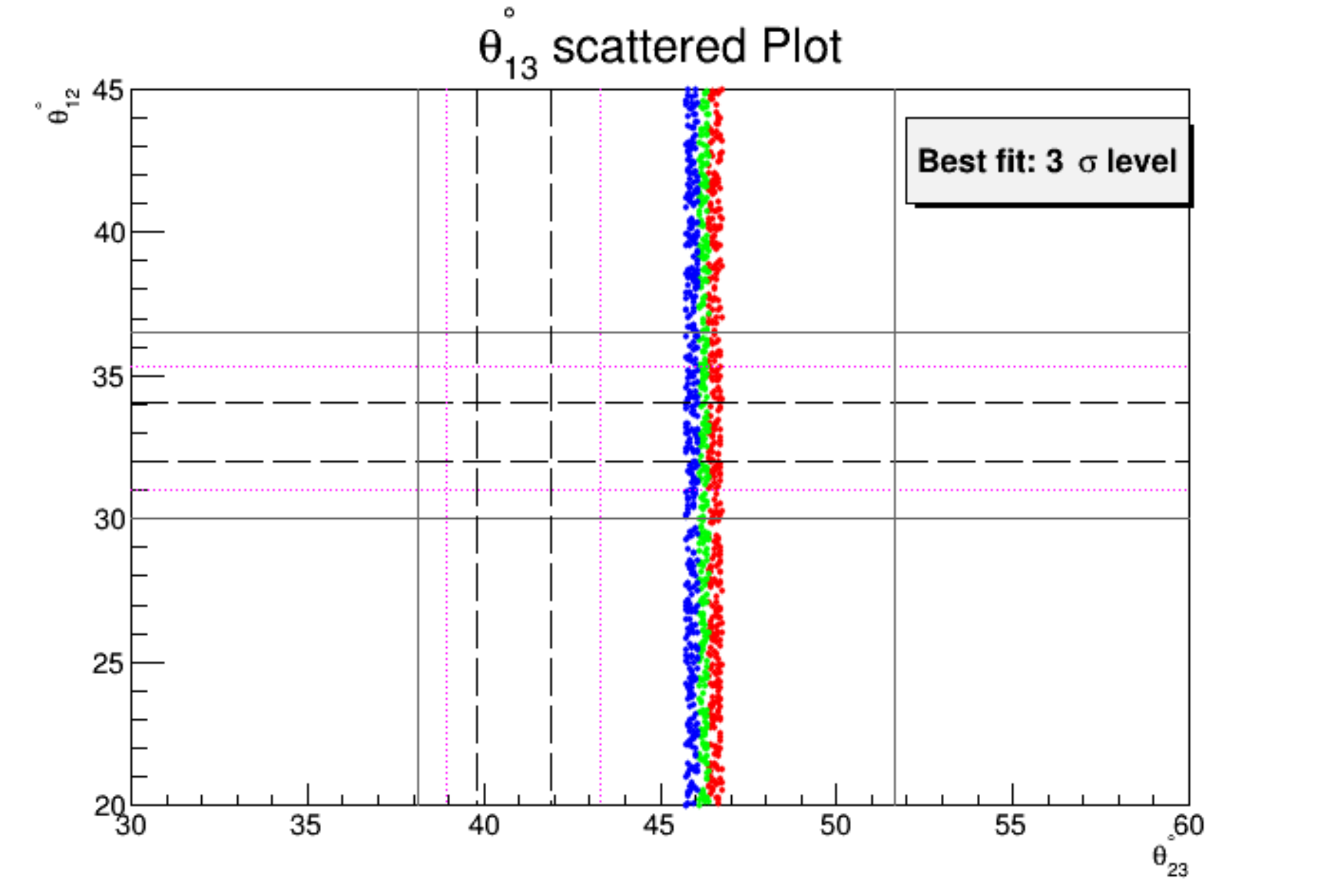}\\
\end{tabular}
\caption{\it{$U^{DCR}_{2312}$ scatter plot of $\chi^2$ (left fig.) over $\beta-\alpha$ (in radians) plane and $\theta_{13}$ (right fig.) 
over  $\theta_{23}-\theta_{12}$ (in degrees) plane. }}
\label{fig2312R.2}
\end{figure}

\begin{figure}[!t]\centering
\begin{tabular}{c c} 
\includegraphics[angle=0,width=80mm]{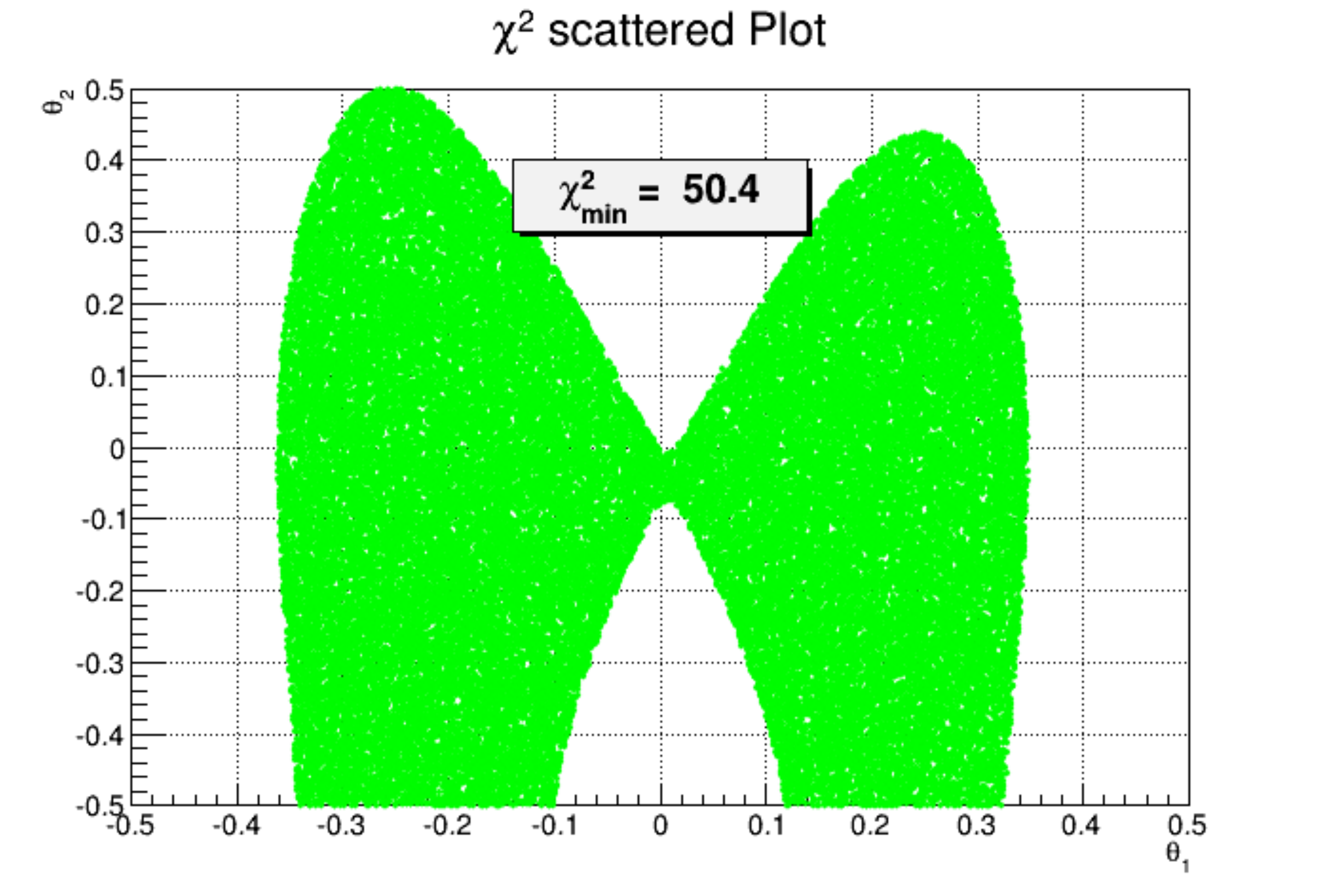} &
\includegraphics[angle=0,width=80mm]{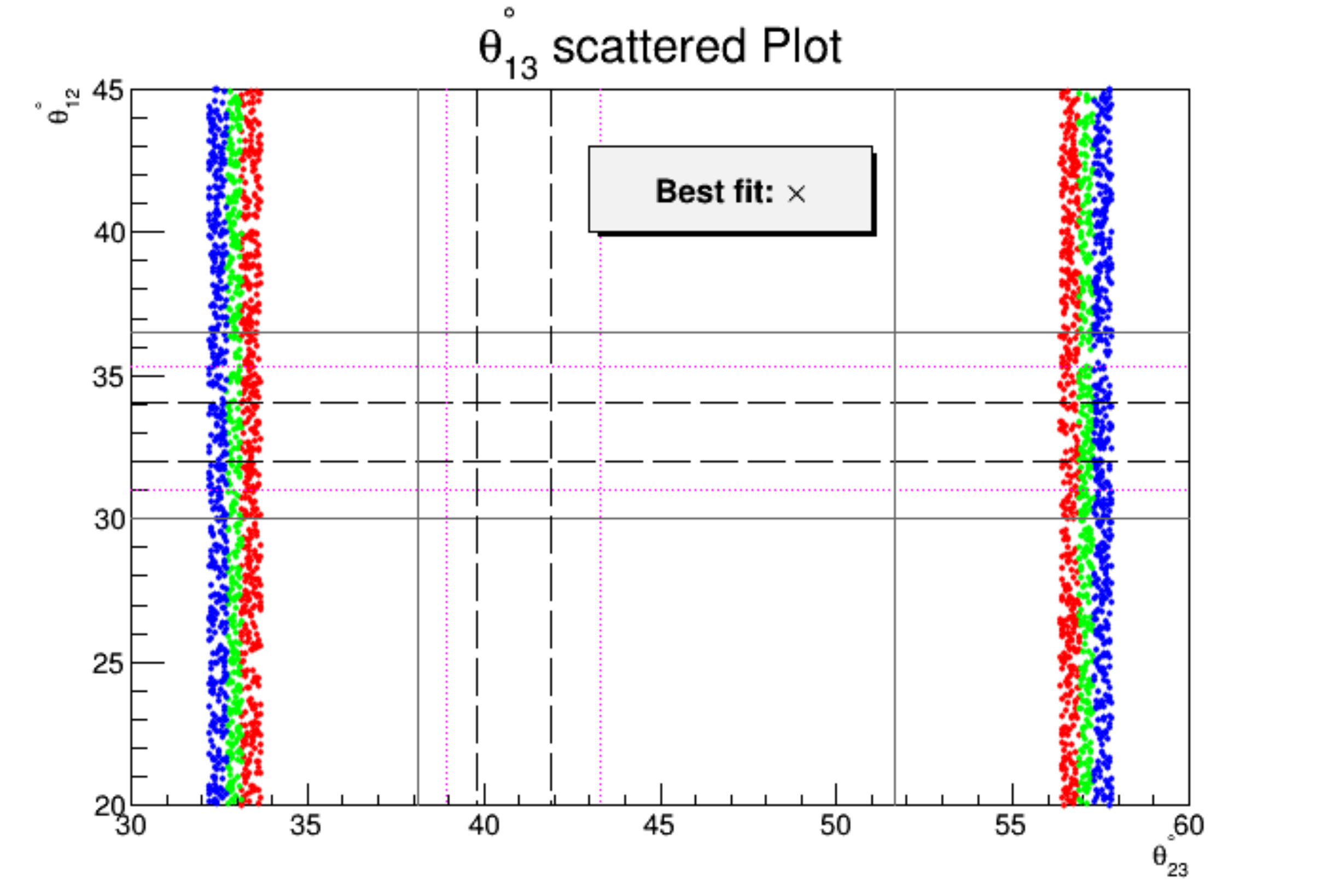}\\
\end{tabular}
\caption{\it{$U^{TBMR}_{2312}$ scatter plot of $\chi^2$ (left fig.) over $\beta-\alpha$ (in radians) plane and $\theta_{13}$ (right fig.) 
over $\theta_{23}-\theta_{12}$ (in degrees) plane. }}
\label{fig2312R.3}
\end{figure}

Figs.~\ref{fig2312R.1}-\ref{fig2312R.3} corresponds to BM, DC and TBM case respectively with $\theta_1 = \beta$ and $\theta_2 = \alpha$.\\
{\bf{(i)}}The corrections to mixing angle $\theta_{13}$ and $\theta_{23}$ is only governed by perturbation parameter
$\beta$.  Thus magnitude of parameter $\beta$ is tightly constrainted from fitting of $\theta_{13}$. This in turn allows only very narrow ranges
for $\theta_{23}$ corresponding to negative and positive values of $\beta$ in parameter space. However $\theta_{12}$ solely depends on $\alpha$ and thus
can have wide range of possible values in parameter space.\\
{\bf{(ii)}} The minimum value of $\chi^2 \sim 15.5$, $28.0$ and $50.4$ for BM, DC and TBM case respectively.\\
{\bf{(iii)}} BM and TBM case is completely excluded while DC is allowed at $3\sigma$ level.

\subsection{23-13 Rotation}

This case is much similar to 13-12 rotation with interchange of expressions for $\theta_{12}$ and
$\theta_{23}$ mixing angles. 
The neutrino mixing angles for small perturbation parameters $\beta$ and $\gamma$ are given by

\beqa
 \sin\theta_{13} &\approx&  |\beta U_{12} + \gamma U_{11}|,\\
 \sin\theta_{23} &\approx& |\frac{U_{23}+\beta U_{22}+\gamma U_{21}-(\beta^2 + \gamma^2)U_{23} }{\cos\theta_{13}}|,\\
 \sin\theta_{12} &\approx& |\frac{(\beta^2 -1)U_{12} }{\cos\theta_{13}}|.
 \eeqa

\begin{figure}[!t]\centering
\begin{tabular}{c c} 
\includegraphics[angle=0,width=80mm]{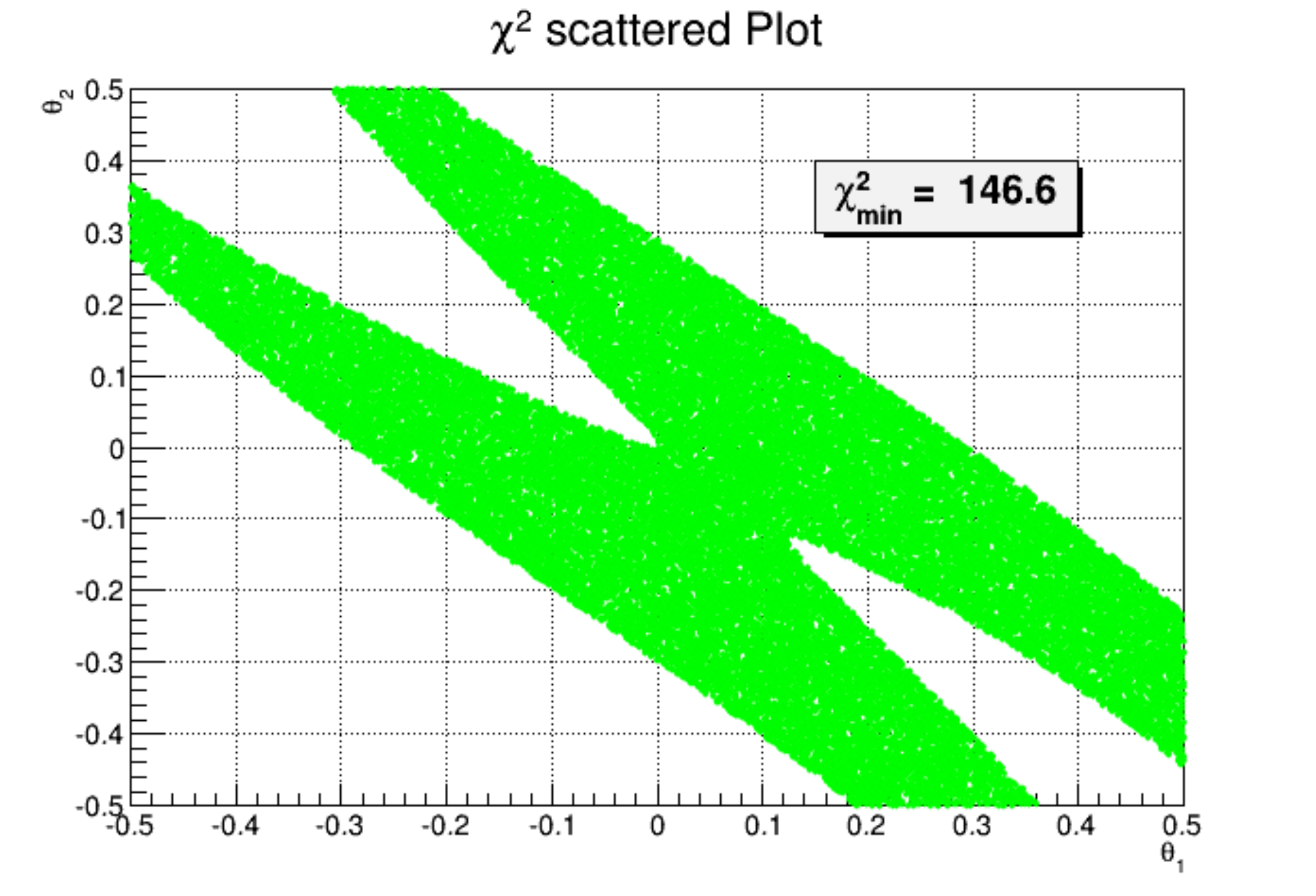} &
\includegraphics[angle=0,width=80mm]{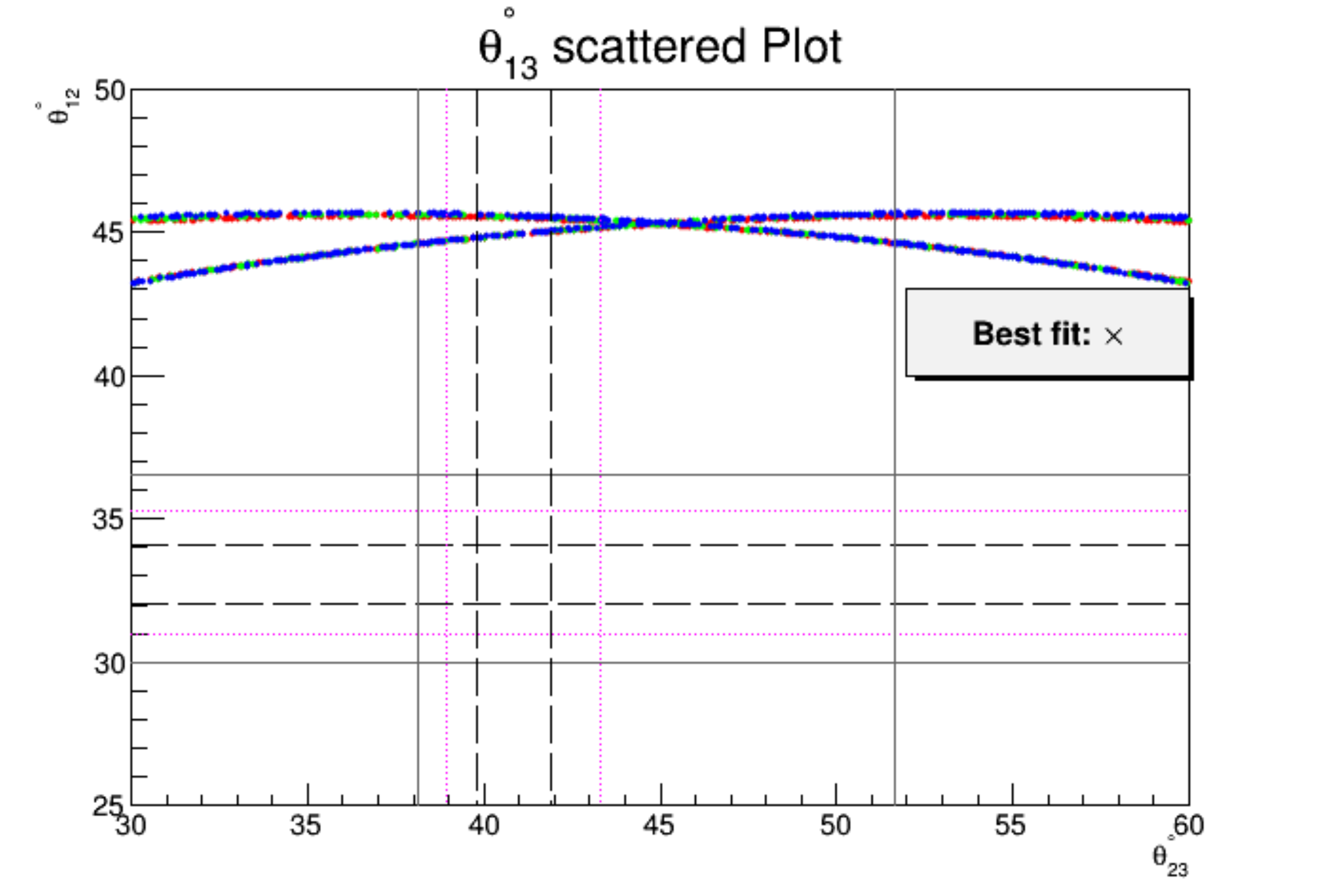}\\
\end{tabular}
\caption{\it{$U^{BMR}_{2313}$ scatter plot of $\chi^2$ (left fig.) over $\beta-\gamma$ (in radians) plane and $\theta_{13}$ (right fig.) 
over  $\theta_{23}-\theta_{12}$ (in degrees) plane. }}
\label{fig2313R.1}
\end{figure}

\begin{figure}[!t]\centering
\begin{tabular}{c c} 
\includegraphics[angle=0,width=80mm]{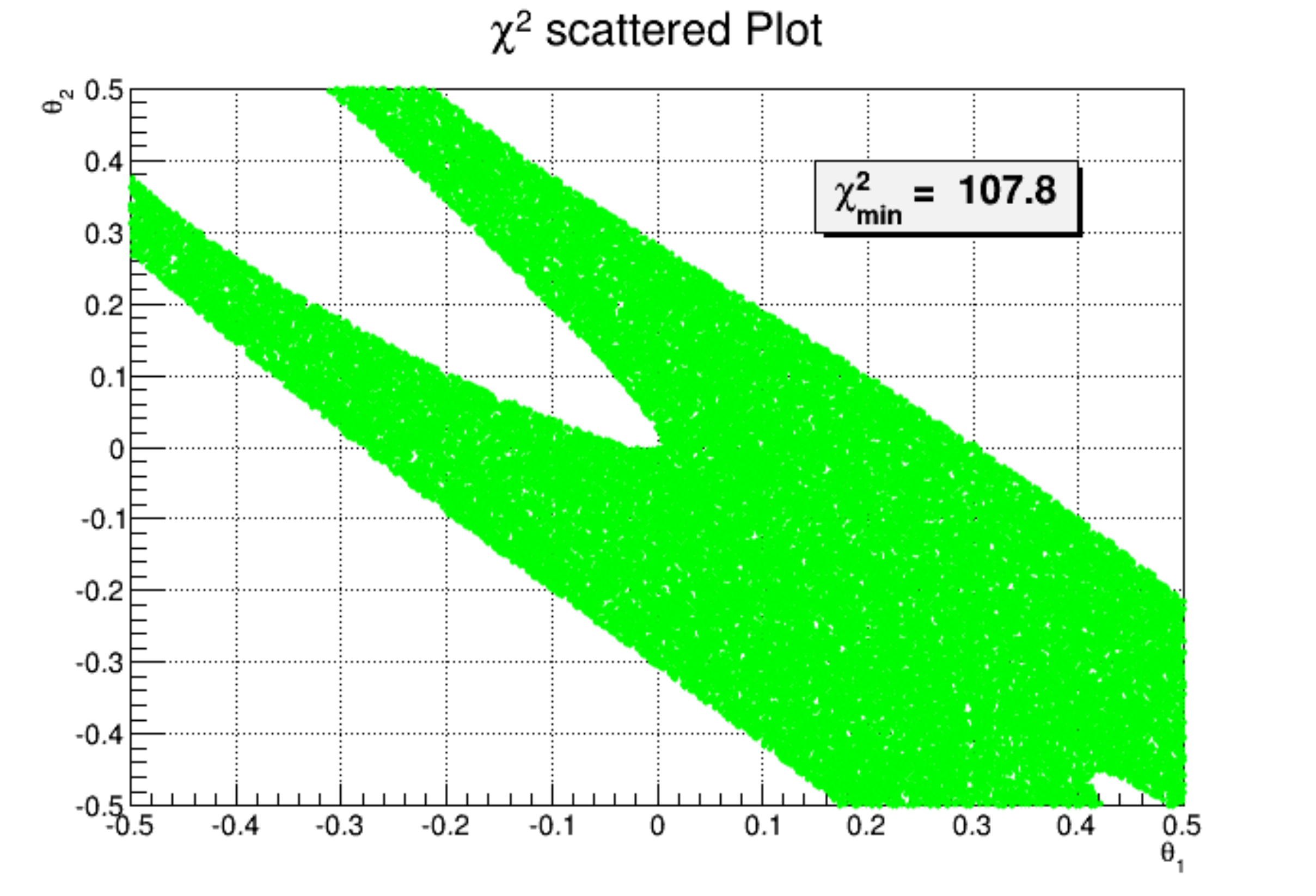} &
\includegraphics[angle=0,width=80mm]{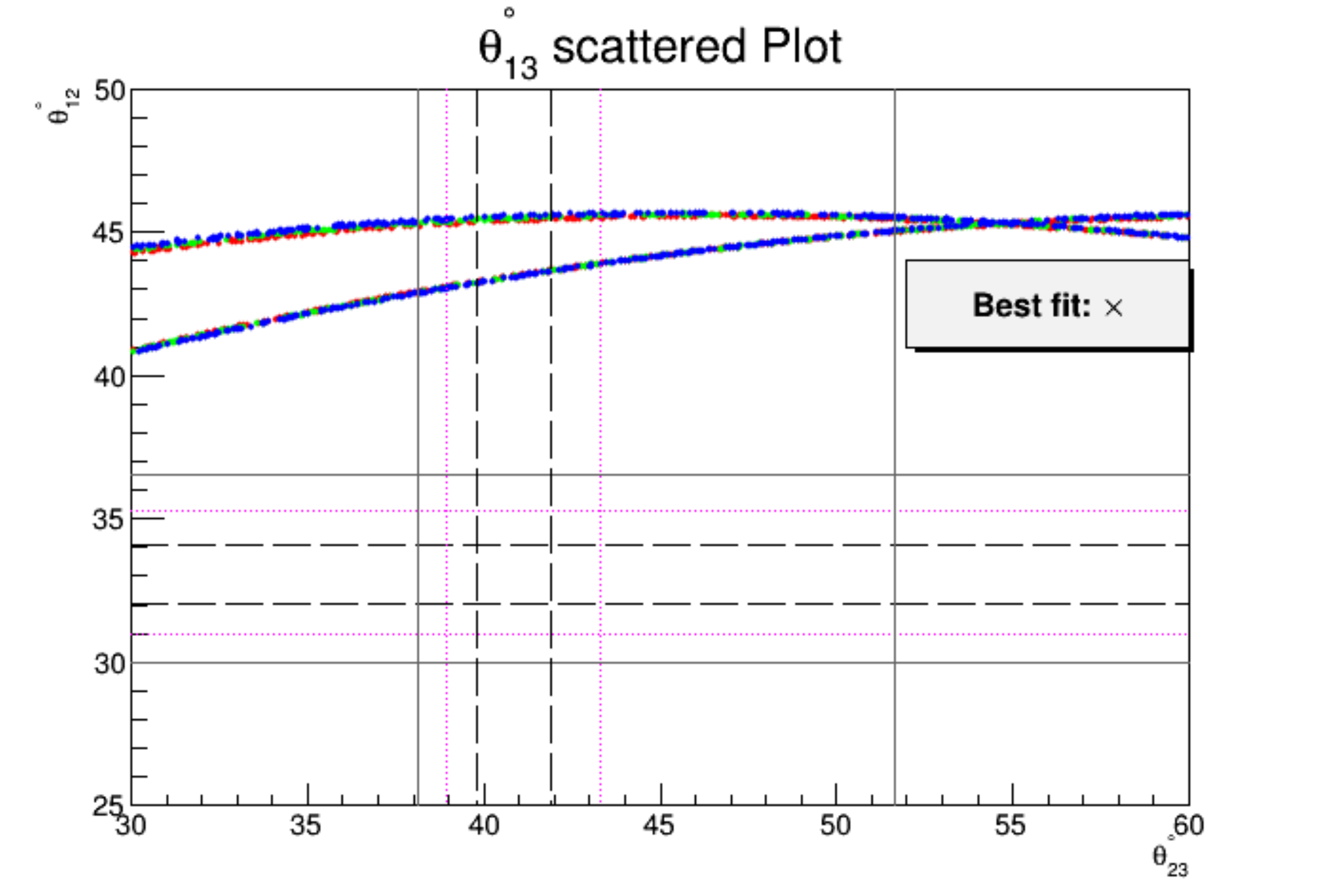}\\
\end{tabular}
\caption{\it{$U^{DCR}_{2313}$ scatter plot of $\chi^2$ (left fig.) over $\beta-\gamma$ (in radians) plane and $\theta_{13}$ (right fig.) 
over  $\theta_{23}-\theta_{12}$ (in degrees) plane. }}
\label{fig2313R.2}
\end{figure}

\begin{figure}[!t]\centering
\begin{tabular}{c c} 
\includegraphics[angle=0,width=80mm]{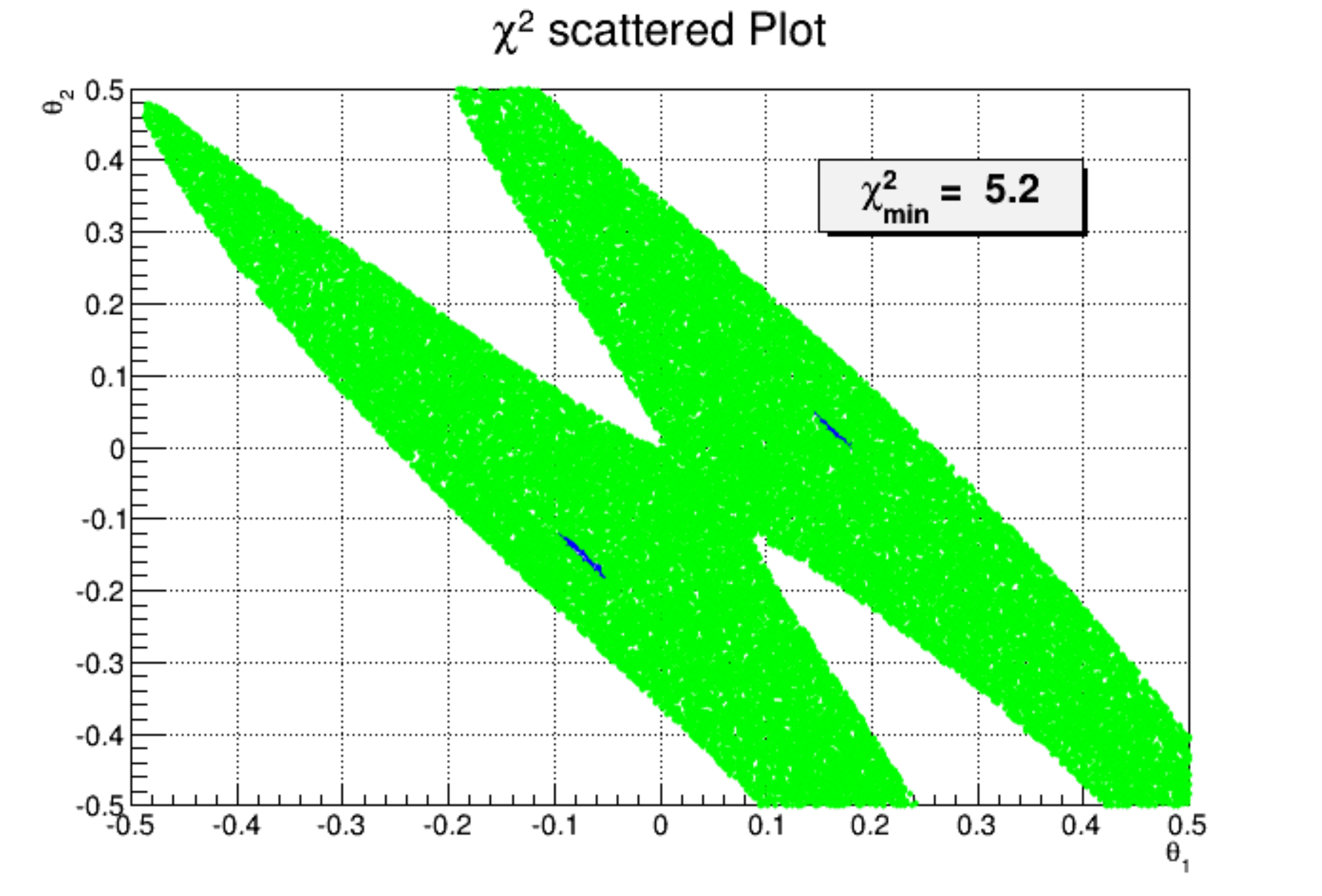} &
\includegraphics[angle=0,width=80mm]{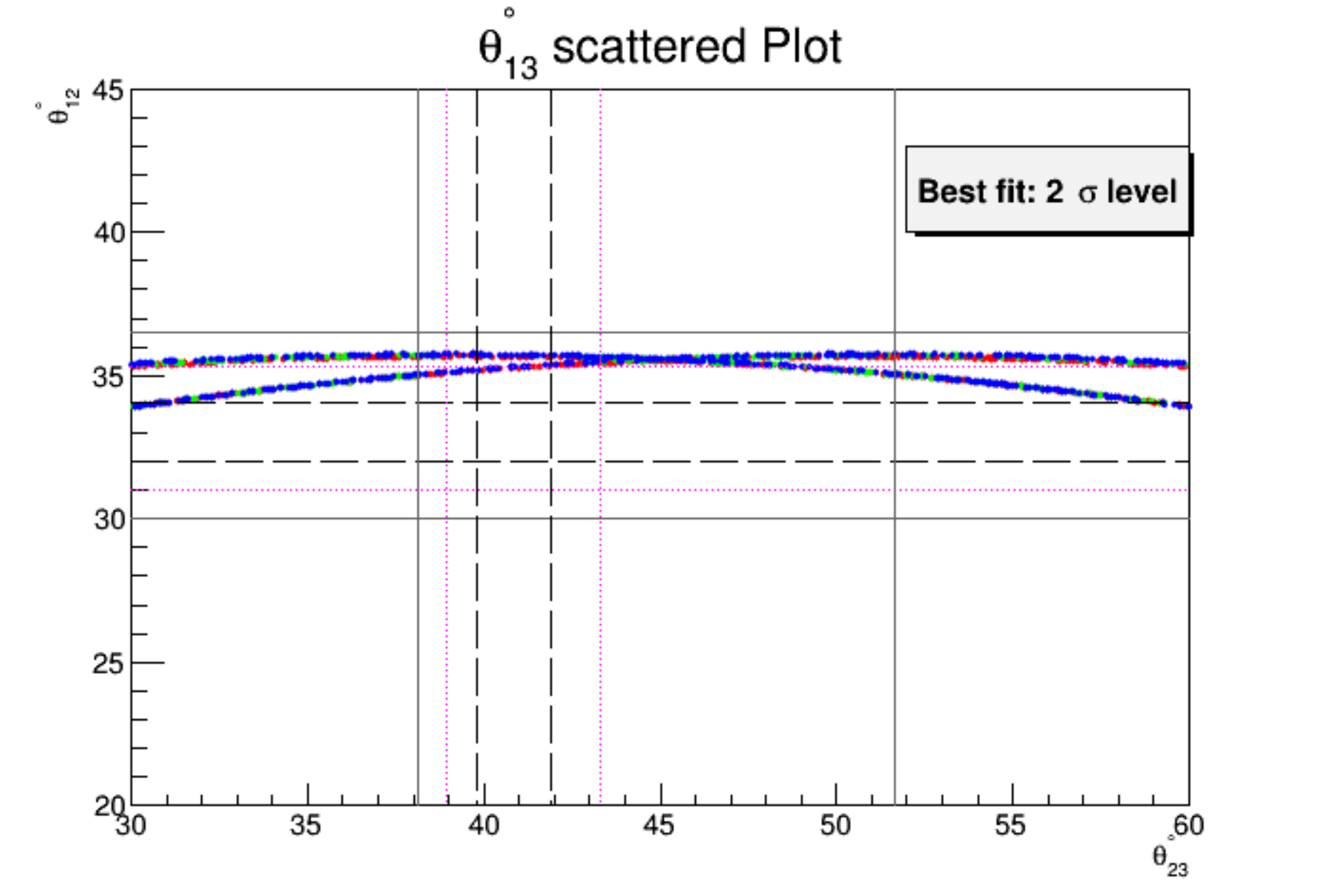}\\
\end{tabular}
\caption{\it{$U^{TBMR}_{2313}$ scatter plot of $\chi^2$ (left fig.) over $\beta-\gamma$ (in radians) plane and $\theta_{13}$ (right fig.) 
over $\theta_{23}-\theta_{12}$ (in degrees) plane.  }}
\label{fig2313R.3}
\end{figure}

Figs.~\ref{fig2313R.1}-\ref{fig2313R.3} corresponds to BM, DC and TBM case respectively with $\theta_1 = \gamma$ and $\theta_2 = \beta$.
The main characteristic features of this scheme are:\\
{\bf{(i)}} In this rotation scheme, $\theta_{12}$ angle receives corrections at O($\theta^2$) and thus its value remain close to its
unperturbed value. \\
{\bf{(ii)}} The minimum value of $\chi^2 \sim 146.6$, $107.8$ and $5.2$ for BM, DC and TBM case respectively.\\
{\bf{(iii)}} Thus BM and DC case is completely excluded as $\theta_{12}\sim 45^\circ$ which is well away from 3$\sigma$
line. However TBM is still consistent at 2$\sigma$ level.

\section{Summary and Conclusions}
Tribimaximal(TBM), bimaximal(BM) and Democratic(DC) mixing matrices offers to explain the neutrino mixing data with a common prediction of
vanishing reactor mixing angle. The atmospheric mixing angle($\theta_{23}$) is maximal in TBM and BM scenarios
while it takes the larger value of 54.7 $^{\circ}$ for DC mixing. The value of solar mixing angle ($\theta_{12}$) is maximal in BM and DC scenarios while its
 value is $35.3^{\circ}$ for TBM case. However experimental observation of non zero reactor mixing angle ($\theta_{13}\approx 8^\circ$) and departure 
of other two mixing angles  from maximality is asking for corrections in these mixing schemes. 

In this study, we investigated the 
perturbations around TBM, BM and DC mixing scenarios. These modifications are expressed in terms of 
three orthogonal rotation matrices R$_{12}$, R$_{13}$ and R$_{23}$ which acts on 12, 13 and 23 sector of unperturbed PMNS matrix respectively. 
We looked into various possible cases that are governed by one and two rotation matrices with corresponding modified PMNS matrices of the forms
\big($R_{ij}\cdot U,~R_{ij}\cdot R_{kl}\cdot U,~U\cdot R_{ij},~U \cdot R_{ij} \cdot R_{kl}$\big) where U is any one of these special matrices. As the 
form of PMNS matrix is given by $U_{PMNS} = U_l^{\dagger} U_\nu$ so these corrections may originate from charged lepton and neutrino
sector respectively. In this study, we restricted ourselves to CP conserving case by setting phases to be zero. The effects of CP violation on 
mixing angles will be reported somewhere else. For our analysis we constructed $\chi^2$ function which is a measure of deviation
from experimental best fit values of mixing angles. The numerical findings are presented 
in terms of $\chi^2$ vs perturbation parameters and as correlations among different neutrino mixing angles. The final results are summarized 
in Table~\ref{singRot2} and Table~\ref{DoubRot1} for single and double rotation cases respectively. For the sake of completion, we also included 
the results of $R_{ij}\cdot U\cdot R_{kl}$ perturbative
scheme~\cite{skgetal} and updated them for new mixing data in Table~\ref{DoubRot2}. 

The rotation $R_{12}\cdot U$, imparts negligible corrections to $\theta_{23}$ and thus its value remain close to its unperturbed prediction.  
Here BM case can achieve $\chi^2 \sim 24$ and thus allowed at $3\sigma$ level while other two cases are not consistent. In rotation 
$R_{13}\cdot U$,  $\theta_{23}$ receives very minor corrections through $\theta_{13}$ and thus stays close to its original prediction. The BM case is consistent at 
$3\sigma$ level while other two cases are not allowed. For $R_{23}\cdot U$ case,  $\theta_{13}$ is still zero 
and thus it was not investigated any further. The rotation $U\cdot R_{12}$, doesn't impart corrections to 13 mixing angle and thus we 
left its further discussion. For $U\cdot R_{13}$,
$\theta_{12}$ receives very minor corrections only through $\theta_{13}$ and thus its range remains quite close to its unperturbed value. The only perturbed TBM is allowed at $3\sigma$ level while 
other two mixing schemes are not consistent with mixing data. The $U\cdot R_{23}$ rotation scheme is not consistent with mixing data for all
three mixing schemes.

For rotation $R_{12}\cdot R_{13}\cdot U $, $\theta_{23}$ remain close to its unperturbed value. Thus TBM and BM cases can be
consistent at 3$\sigma$ level while DC is not allowed. The rotation $R_{12}\cdot R_{23}\cdot U $ for BM
and DC can fit all mixing angles within 2$\sigma$ level. However TBM case is completely
excluded for this rotation. The perturbation scheme $R_{13}\cdot R_{12}\cdot U $ scheme is much similar to $R_{12}\cdot R_{13}\cdot U $.
Here also  TBM and BM cases can be consistent at 3$\sigma$ level while DC is not consistent. For rotation $R_{13}\cdot R_{23}\cdot U $, 
TBM case is excluded  while BM and DC is consistent at 3$\sigma$ level. In 
$R_{23}\cdot R_{12}\cdot U $, TBM and DC is completely excluded while BM can only be viable at 3$\sigma$ level. Thus this case is not much preferable. 
The perturbation scheme  $R_{23}\cdot R_{13}\cdot U $ is much favorable for DC case as it successfully fits all mixing
angles at $1\sigma$ level with lowest value of $\chi^2$ among all mixing schemes. However TBM case is excluded for this rotation while BM can only be viable at 3$\sigma$ level.

 The perturbation scheme $ U \cdot R_{12}\cdot R_{13}$ is much preferable since it is possible
to fit all mixing angles within 1$\sigma$ range for TBM and BM case. However DC case can only be allowed at
3$\sigma$ level. The rotation scheme $ U \cdot R_{12}\cdot R_{23}$ is only preferable for DC while the other
two cases in scheme are excluded at 3$\sigma$ level. The mixing angles can be fitted at $1\sigma$ level in this rotation for
DC case. In rotation $ U \cdot R_{13}\cdot R_{12}$, it is possible to fit all mixing angles within 2$\sigma$ range
for perturbed TBM matrix and $3\sigma$ level for perturbed DC case. However BM is not viable as it failed to fit  mixing angles
even at $3\sigma$ level. The perturbed rotation 
 $ U \cdot R_{13}\cdot R_{23}$ imparts corrections to $\theta_{12}$ only at O($\theta^2$). Thus its value remain quite
 close to its original value and hence BM and DC cases are excluded while TBM can be consistent at 2$\sigma$ level. The
 rotation scheme $ U \cdot R_{23}\cdot R_{12}$ is only little preferable for DC case as it is possible to fit mixing angles
 at $3\sigma$ level while other two cases are not consistent. Like  $U \cdot R_{13}\cdot R_{23}$ case, rotation scheme
$U \cdot R_{23}\cdot R_{13}$ also imparts negligible corrections to $\theta_{12}$. Thus BM and DC case is excluded while perturbed
TBM case is still consistent at 2$\sigma$ level.

We also updated our previous analysis~\cite{skgetal} on \big($R_{ij}\cdot U\cdot R_{kl}$\big) PMNS 
matrices for new mixing data. To draw the comparison, we found  ($\chi^2_{min}$, Best fit level) for different mixing cases~\cite{skgetal} using
old mixing data~\cite{Gonzalez-Garcia:2014bfa} and compared  it with new obtained values. The corresponding results are given in Table~\ref{DoubRot2}. 
In all allowed cases, except \big($ R_{12}\cdot U_{BM}\cdot R_{13}$,~ $R_{23}\cdot U_{TBM}\cdot R_{13}$ and $R_{13}\cdot U_{TBM} \cdot R_{13}$\big), minimum $\chi^2$ value increases which imply fitting gets tougher for those cases. 
With further improvement of accuracy on mixing angles, many cases might be ruled out completely.

This completes our discussion on checking the consistency of various perturbative rotations with neutrino mixing data.
This study might turn out to be useful in restricting vast number of possible models which offers different corrections to these mixing schemes in neutrino model building physics. 
It thus can be a guideline for neutrino model building. However all such issues including the origin of these perturbations are left for future investigations. 

\section{Acknowledgements}
The author acknowledges the support provided by CERN ROOT group for data plotting. 

\appendix
\section{Results: Summary} \label{App:AppendixA}
In this appendix, we summarized all our results for easy reference on discussed rotation schemes. We also updated our previous analysis\cite{skgetal}
with new mixing data which was done for  \big($R_{ij}\cdot U\cdot R_{kl} $\big )  rotation scheme. 

In Table~\ref{singRot2}, we presented our result corresponding to \big($R_{ij}\cdot U,~U\cdot R_{ij} $\big ) PMNS
matrices. Table~\ref{DoubRot1} and Table~\ref{DoubRot2} contains the results corresponding to 
\big($R_{ij}\cdot R_{kl}\cdot U,~U\cdot R_{ij}\cdot R_{kl}$\big ) and \big($ R_{ij}\cdot U\cdot R_{kl} $\big) PMNS matrices respectively.

\begin{center}
\begin{tabular}{|l|l||l|l|l|l|l|l|}
\hline
 \multicolumn{4}{|l|}{$R_{ij}^l\cdot U$} &  \multicolumn{4}{l|}{$U \cdot R_{ij}^r$}\\
\cline{1-8} 
$R_{ij}^l$ & BM & DC & TBM & $R_{ij}^r$ & BM & DC & TBM  \\ 
 \hline\hline
$R_{12}^l$ & $(24.0,~3\sigma)$ & $(204.7,~\times)$ & $(45.0,~\times)$  & $R_{12}^r$ &$(960.7,~-)$ &$(1123.6,~-)$ & $(960.7,~-)$ \\
$R_{13}^l$ &$(34.3,~3\sigma)$ &$(202.6,~\times)$  &$(55.0,~\times)$ &$R_{13}^r$&$(183.3,~\times)$ &$(196.5,~\times)$ &$(9.6,~3\sigma)$  \\
$R_{23}^l$ &$(1094.7,~-)$ & $(1094.7,~-)$  &$(948.2,~-)$ &$R_{23}^r$ &$(151.1,~\times)$ &$(162.9,~\times)$ & $(52.2,~\times)$  \\
 \hline
\end{tabular}
\captionof{table}{\it{($\chi^2_{min}$, Best fit level) for perturbation schemes that are dictated by \big($R_{ij}\cdot U,~U\cdot R_{ij}$\big) PMNS
matrices. }}
\label{singRot2}
\end{center}

\begin{center}
\begin{tabular}{|l|l||l|l|l|l|l|l|}
\hline
  \multicolumn{4}{|l|}{$R_{ij}^l\cdot R_{kl}^l\cdot U$} &  \multicolumn{4}{l|}{$U \cdot R_{ij}^r\cdot R_{kl}^r$}\\
\cline{1-8} 
$R_{ij}^l\cdot R_{kl}^l$ & BM & DC & TBM & $R_{ij}^r\cdot R_{kl}^r$ & BM & DC & TBM \\ 
 \hline\hline
$R_{12}^l\cdot R_{13}^l$ &$(13.0,~3\sigma)$&$(173.1,~\times)$  &$(13.9,~3\sigma)$ & $R_{12}^r\cdot R_{13}^r$ & \fbox{${\bf{(1.1,~1\sigma)}}$} &$(65.8,~3\sigma)$ & \fbox{${\bf{(1.7,~1\sigma)}}$} \\
$R_{12}^l\cdot R_{23}^l$ &$(4.4,~2\sigma)$ &$(4.4,~2\sigma)$  &$(41.0,~\times)$ &$R_{12}^r\cdot R_{23}^r$ &$(50.8,~\times)$&\fbox{${\bf{(1.3,~1\sigma)}}$} &$(50.9,~\times)$ \\
$R_{13}^l\cdot R_{12}^l$ &$(8.8,~3\sigma)$ &$(140.0,~\times)$  &$(18.2,~3\sigma)$ &$R_{13}^r\cdot R_{12}^r$ &$(15.5,~\times)$ &$(28.0,~3\sigma)$ &$(2.6,~2\sigma)$ \\
$R_{13}^l\cdot R_{23}^l$ &$(21.0,~3\sigma)$ &$(21.0,~3\sigma)$  &$(19.7,~\times)$ &$R_{13}^r\cdot R_{23}^r$ &$(137.1,~\times)$ &$(138.4,~\times)$ &$(3.0,~2\sigma)$  \\
$R_{23}^l\cdot R_{12}^l$ &$(11.5,~3\sigma)$ &$(35.4,~\times)$  &$(32.3,~\times)$ &$R_{23}^r\cdot R_{12}^r$  &$(15.5,~\times)$ &$(28.0,~3\sigma)$ &$(50.4,~\times)$ \\
$R_{23}^l\cdot R_{13}^l$ &$(11.5,~3\sigma)$ &\fbox{${\bf{(0.02,~1\sigma)}}$}  &$(32.3,~\times)$ &$R_{23}^r\cdot R_{13}^r$ &$(146.6,~\times)$ &$(107.8,~\times)$ &$(5.2,~2\sigma)$\\
 \hline
\end{tabular}
\captionof{table}{\it{($\chi^2_{min}$, Best fit level) for rotation schemes that are governed by \big($R_{ij}\cdot R_{kl}\cdot U,~U\cdot R_{ij}\cdot R_{kl} $\big) PMNS 
matrices. }}
\label{DoubRot1}
\end{center}

\begin{center}
\begin{tabular}{|l|l||l|l|}
\hline
  \multicolumn{4}{|l|}{$(\chi^2_{min}[R_{ij}^l\cdot U\cdot R_{kl}^r])_{Old} \rightarrow (\chi^2_{min}[R_{ij}^l\cdot U\cdot R_{kl}^r])_{New}$}\\
\cline{1-4} 
$R_{ij}^l-R_{kl}^r$ & BM & DC & TBM \\ 
 \hline\hline
$R_{12}^l-R_{13}^r$ &\fbox{$(1.4,~1\sigma)\rightarrow {\bf{(0.8,~1\sigma)}}$}&$(0.8,~1\sigma)\rightarrow (6.3,~2\sigma)$ &$(0.8,~1\sigma)\rightarrow (6.3,~2\sigma)$\\
$R_{12}^l-R_{23}^r$ &$(0.1,~1\sigma)\rightarrow (23.2,~3\sigma)$&$(51.5,~\times)\rightarrow (147.9,~\times)$ &$(8.7,~2\sigma)\rightarrow (13.8,~3\sigma)$\\
$R_{13}^l-R_{12}^r$ &$(10.1,~3\sigma)\rightarrow (22.8,~3\sigma)$ &$(16.7,~\times)\rightarrow (202.6,~\times)$ &$(10.1,~3\sigma)\rightarrow (22.8,~3\sigma)$ \\
$R_{13}^l-R_{23}^r$ &\fbox{$(0.03,~1\sigma)\rightarrow {\bf{(0.3,~1\sigma)}}$} &$(11.7,~2\sigma)\rightarrow (26.0,~3\sigma)$ &$(11.7,~2\sigma)\rightarrow (26.0,~3\sigma)$ \\
$R_{23}^l-R_{12}^r$ &$(93.3,~-)\rightarrow (943.3,~-)$ &$(93.3,~-)\rightarrow (943.3,~-)$ &$(93.3,~-)\rightarrow (943.3,~-)$ \\
$R_{23}^l-R_{13}^r$ &$(278.5,~\times)\rightarrow (168.1,~\times)$ &$(278.5,~\times)\rightarrow (168.1,~\times)$ &$(9.7,~3\sigma)\rightarrow (7.0,~3\sigma)$\\
$R_{12}^l-R_{12}^r$ &$(5.9,~2\sigma)\rightarrow (12.6,~3\sigma)$ &$(9.1,~3\sigma)\rightarrow (169.4,~\times)$  &$(5.9,~2\sigma)\rightarrow (12.6,~3\sigma)$\\
$R_{13}^l-R_{13}^r$ &$(3.6,~2\sigma)\rightarrow (34.3,~3\sigma)$ &$(12.2,~3\sigma)\rightarrow (148.4,~\times)$ &\fbox{$(0.2,~1\sigma)\rightarrow {\bf{(0.2,~1\sigma)}}$}\\
$R_{23}^l-R_{23}^r$ &$(220.1,~\times)\rightarrow (135.4,~\times)$ &$(220.2,~\times)\rightarrow (135.4,~\times)$ &$(1.5,~2\sigma)\rightarrow (1.7,~2\sigma)$\\
 \hline
\end{tabular}
\captionof{table}{A comparison of \it{($\chi^2_{min}$, Best fit level) for perturbation scheme that are dictated by 
\big($R_{ij}\cdot U\cdot R_{kl} $\big) PMNS 
matrix with previous results\cite{skgetal}. Since previous mixing data~\cite{Gonzalez-Garcia:2014bfa} have two regions
of $\theta_{23}$ so here we mentioned best fit among those two possibilities. }}
\label{DoubRot2}
\end{center}



\end{document}